\documentclass[a4paper,12pt,twoside,openright]{report}

\usepackage[utf8x]{inputenc}
\usepackage[T1]{fontenc}

\usepackage{ae,aecompl} 

\usepackage{amsmath} 
\usepackage{amsfonts}
\usepackage{amssymb} 

\usepackage{array}
\usepackage{multirow} 

\usepackage[a4paper]{geometry} 
\usepackage{xspace} 
\usepackage{path} 
\usepackage{xcolor} 
\usepackage{colortbl} 
\usepackage{enumitem} 
\usepackage{fancyhdr} 
\usepackage{anysize} 
\usepackage{indentfirst}
\usepackage[small,hang]{caption} 
\usepackage{graphicx} 
\usepackage{float} 
\usepackage{pifont} 
\usepackage{textcomp} 
\usepackage{tocbibind} 
\usepackage{pdfpages} 
\usepackage{diagbox} 
\usepackage{ifthen} 
\usepackage{ifpdf} 
\usepackage{rotating}

\usepackage{eurosym} 

\ifpdf               
\usepackage{corporate/design_pdf}
\else
\usepackage{corporate/design_ps}
\fi

\ifpdf
\else
\fi
\usepackage{subfigure}

\ifpdf
	\usepackage[
		    pagebackref, 
                    ]{hyperref}
        \hypersetup{pdftitle={Numerical investigations of the fluid-structure interaction of a NACA0012 airfoil based on large-eddy simulations},
		    pdfsubject={FSI LES NACA0012},
		    pdfauthor={Larissa Bruna Streher},
                    linktocpage=true, 
		    bookmarks=true,
		    bookmarksopen=true,
		    bookmarksnumbered=true, 
		    pdfpagelabels=true,
		    pdfpagemode=None, 
		    plainpages=false, 
	            colorlinks=true, 
                    citecolor=blue,
		    urlcolor=Magenta,
		    filecolor=green,
		    linkcolor=red
                    }
\else
        \usepackage[bookmarks]{hyperref}
         \hypersetup{pdftitle={Numerical investigations of the fluid-structure interaction of a NACA0012 airfoil based on large-eddy simulations},
 		    pdfsubject={FSI LES NACA0012},
 		    pdfauthor={Larissa Bruna Streher},
                    linktocpage=true, 
                    breaklinks=true, 
 		    bookmarks=true,
 		    bookmarksopen=true,
 		    bookmarksnumbered=true, 
 		    pdfpagelabels=true,
 		    pdfpagemode=None, 
 	            colorlinks=true, 
                    citecolor=blue,
 		    urlcolor=Magenta,
 		    filecolor=green,
 		    linkcolor=red
                    }
\fi             

\definecolor{GreenYellow}   {cmyk}{0.15,0,0.69,0}
\definecolor{Yellow}        {cmyk}{0,0,1,0}
\definecolor{Goldenrod}     {cmyk}{0,0.10,0.84,0}
\definecolor{Dandelion}     {cmyk}{0,0.29,0.84,0}
\definecolor{Apricot}       {cmyk}{0,0.32,0.52,0}
\definecolor{Peach}         {cmyk}{0,0.50,0.70,0}
\definecolor{Melon}         {cmyk}{0,0.46,0.50,0}
\definecolor{YellowOrange}  {cmyk}{0,0.42,1,0}
\definecolor{Orange}        {cmyk}{0,0.61,0.87,0}
\definecolor{BurntOrange}   {cmyk}{0,0.51,1,0}
\definecolor{Bittersweet}   {cmyk}{0,0.75,1,0.24}
\definecolor{RedOrange}     {cmyk}{0,0.77,0.87,0}
\definecolor{Mahogany}      {cmyk}{0,0.85,0.87,0.35}
\definecolor{Maroon}        {cmyk}{0,0.87,0.68,0.32}
\definecolor{BrickRed}      {cmyk}{0,0.89,0.94,0.28}
\definecolor{Red}           {cmyk}{0,1,1,0}
\definecolor{OrangeRed}     {cmyk}{0,1,0.50,0}
\definecolor{RubineRed}     {cmyk}{0,1,0.13,0}
\definecolor{WildStrawberry}{cmyk}{0,0.96,0.39,0}
\definecolor{Salmon}        {cmyk}{0,0.53,0.38,0}
\definecolor{CarnationPink} {cmyk}{0,0.63,0,0}
\definecolor{Magenta}       {cmyk}{0,1,0,0}
\definecolor{VioletRed}     {cmyk}{0,0.81,0,0}
\definecolor{Rhodamine}     {cmyk}{0,0.82,0,0}
\definecolor{Mulberry}      {cmyk}{0.34,0.90,0,0.02}
\definecolor{RedViolet}     {cmyk}{0.07,0.90,0,0.34}
\definecolor{Fuchsia}       {cmyk}{0.47,0.91,0,0.08}
\definecolor{Lavender}      {cmyk}{0,0.48,0,0}
\definecolor{Thistle}       {cmyk}{0.12,0.59,0,0}
\definecolor{Orchid}        {cmyk}{0.32,0.64,0,0}
\definecolor{DarkOrchid}    {cmyk}{0.40,0.80,0.20,0}
\definecolor{Purple}        {cmyk}{0.45,0.86,0,0}
\definecolor{Plum}          {cmyk}{0.50,1,0,0}
\definecolor{Violet}        {cmyk}{0.79,0.88,0,0}
\definecolor{RoyalPurple}   {cmyk}{0.75,0.90,0,0}
\definecolor{BlueViolet}    {cmyk}{0.86,0.91,0,0.04}
\definecolor{Periwinkle}    {cmyk}{0.57,0.55,0,0}
\definecolor{CadetBlue}     {cmyk}{0.62,0.57,0.23,0}
\definecolor{CornflowerBlue}{cmyk}{0.65,0.13,0,0}
\definecolor{MidnightBlue}  {cmyk}{0.98,0.13,0,0.43}
\definecolor{NavyBlue}      {cmyk}{0.94,0.54,0,0}
\definecolor{RoyalBlue}     {cmyk}{1,0.50,0,0}
\definecolor{Blue}          {cmyk}{1,1,0,0}
\definecolor{Cerulean}      {cmyk}{0.94,0.11,0,0}
\definecolor{Cyan}          {cmyk}{1,0,0,0}
\definecolor{ProcessBlue}   {cmyk}{0.96,0,0,0}
\definecolor{SkyBlue}       {cmyk}{0.62,0,0.12,0}
\definecolor{Turquoise}     {cmyk}{0.85,0,0.20,0}
\definecolor{TealBlue}      {cmyk}{0.86,0,0.34,0.02}
\definecolor{Aquamarine}    {cmyk}{0.82,0,0.30,0}
\definecolor{BlueGreen}     {cmyk}{0.85,0,0.33,0}
\definecolor{Emerald}       {cmyk}{1,0,0.50,0}
\definecolor{JungleGreen}   {cmyk}{0.99,0,0.52,0}
\definecolor{SeaGreen}      {cmyk}{0.69,0,0.50,0}
\definecolor{Green}         {cmyk}{1,0,1,0}
\definecolor{ForestGreen}   {cmyk}{0.91,0,0.88,0.12}
\definecolor{PineGreen}     {cmyk}{0.92,0,0.59,0.25}
\definecolor{LimeGreen}     {cmyk}{0.50,0,1,0}
\definecolor{YellowGreen}   {cmyk}{0.44,0,0.74,0}
\definecolor{SpringGreen}   {cmyk}{0.26,0,0.76,0}
\definecolor{OliveGreen}    {cmyk}{0.64,0,0.95,0.40}
\definecolor{RawSienna}     {cmyk}{0,0.72,1,0.45}
\definecolor{Sepia}         {cmyk}{0,0.83,1,0.70}
\definecolor{Brown}         {cmyk}{0,0.81,1,0.60}
\definecolor{Tan}           {cmyk}{0.14,0.42,0.56,0}
\definecolor{Gray}          {cmyk}{0,0,0,0.50}
\definecolor{Black}         {cmyk}{0,0,0,1}
\definecolor{White}         {cmyk}{0,0,0,0}

\author{Larissa Bruna Streher}
\date{\Large{September, 2017}}
\title{Numerical investigations of the fluid-structure interaction of a NACA0012 airfoil based on large-eddy simulations}

\addresshead{Helmut-Schmidt-Universität}
\address{Universität der Bundeswehr Hamburg\\Postfach 70 08 22\\22008 Hamburg}

\faculty{Maschinenbau}
\department{Professur für Strömungsmechanik}
\professor{Univ.-Prof. Dr.-Ing. habil. Michael Breuer}

\def\drafttype{false}   

\def\FigPath{./Images}

\begin{document}

\newcommand{\squeezeup}{\vspace{-2.5mm}}
\renewcommand{\arraystretch}{1.2}

\ifpdf
   \DeclareGraphicsExtensions{.pdf,.png,.jpg,}
\fi

\DeclareGraphicsRule{.eps.gz}{eps}{.pdf}{`gunzip -c #1}

\pagestyle{empty}

\includepdf[pages=-]{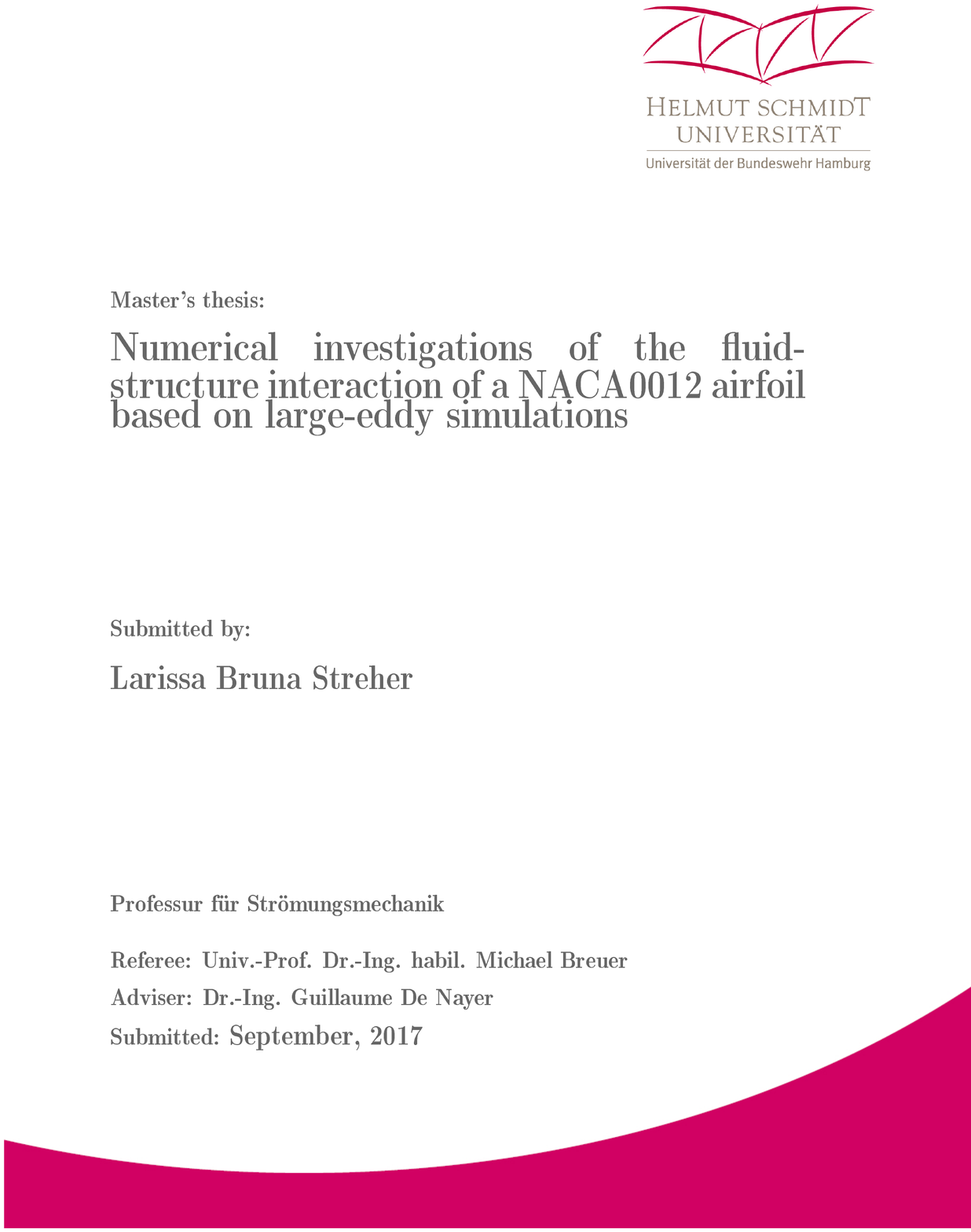}

\cleardoublepage
\chapter*{Declaration of originality}

\par I certify that the intellectual content of this thesis is the product of my own work and that all the assistance received in preparing this thesis and sources have been\break acknowledged.
\par 
\begin{tabular}{p{7.3cm} p{7.42cm}}
	&  \\
	\multicolumn{1}{l}{Place, Date} & \multicolumn{1}{r}{Signature}
\end{tabular}

\cleardoublepage
\addcontentsline{toc}{chapter}{Abstract}
\chapter*{Abstract\markboth{ABSTRACT}{}}
\subsection*{Numerical investigations of the fluid-structure interaction of a NACA0012 airfoil based on large-eddy simulations}

\par The purpose of this work is to expand the work of Streher \cite{Streher_2017} in order to investigate the aeroelastic instabilities generated by the flow around a moving NACA0012 airfoil. The profile has a chord length of $c=0.1\,\text{m}$ and is exposed to a flow at a Reynolds number of $Re=30{,}000$. The airfoil has only two degrees of freedom: Translation in relation to the vertical direction $x_2$ and a rotation around the span-wise axis $x_3$.
\par A partitioned approach based on two separate solvers and a fluid-structure interaction (FSI) coupling scheme is applied. The in-house CFD solver FASTEST-3D computes the fluid sub-problem according to the wall-resolved large-eddy simulation (LES) combined with the Smagorinsky model \cite{Smagorinsky_1963}. The structural sub-problem is solved by a rigid movement solver implemented by Viets \cite{Viets_2013}, which is based on the equations of motion for rigid bodies. The FSI coupling exchanges information between both solvers based on loose or strong coupling algorithms.
\par A thorough analysis of the problem is primarily performed in order to acquire a computational setup that provides the best compromise between accuracy and CPU-time requirements. Three system configurations are then tested and thoroughly investigated: One leads to an oscillatory behavior of the airfoil characterized by limited small amplitudes. The other is characterized by the presence of the torsional divergence aeroelastic instability. Finally, a configuration aimed at the flutter phenomenon is established. However, not enough data are currently available and therefore this dynamic aeroelastic instability can not be thoroughly investigated in the present work. 
\par In the near future, this study will serve as a base for the flutter experiments and computations performed at the Department of Fluid Mechanics located of the Helmut-Schmidt-University.
\cleardoublepage
\addcontentsline{toc}{chapter}{Abstract}
\chapter*{Kurzfassung\markboth{KURZFASSUNG}{}}
\subsection*{Numerische Untersuchungen zur Fluid-Struktur-Interaktion eines NACA0012-Tragflügels basierend auf Large-Eddy Simulationen}

\par Der Zweck dieser Arbeit ist es die aeroelastischen Instabilitäten, die durch die Umströmung eines beweglichen NACA0012 Tragflügels erzeugt werden, zu untersuchen und damit die Arbeit von Streher \cite{Streher_2017} zu erweitern. Das Profil hat eine Sehnenlänge von $c=0,1\,\text{m}$ und die Umströmung findet bei einer Reynolds-Zahl von $Re=30{.}000$ statt. Der Tragflügel hat nur zwei Freiheitsgrade: Eine Translation in Bezug auf die vertikale Richtung $x_2$ und eine Rotation um die Spannweiten-Achse $x_3$. 

\par Ein partitionierter Ansatz, der auf zwei getrennten Solvern und einem Kopplungs-schema für die Fluid-Struktur-Interaktion basiert, wird angewendet. Der hauseigene CFD-Löser FASTEST-3D berechnet das Strömungsproblem entsprechend einer wandaufgelösten Large-Eddy Simulation (LES) in Kombination mit dem Smagorinsky Modell \cite{Smagorinsky_1963}. Das Struktur-Problem wird durch einen starren Bewegungslöser gelöst, der von Viets \cite{Viets_2013} implementiert wurde. Dieser basiert auf den Bewegungsgleichungen für starre Körper. Die FSI-Kopplung tauscht Informationen zwischen beiden Solvern aus, die auf schwachen oder starken Kopplungsalgorithmen basieren.

\par Eine gründliche Analyse des Problems wird im ersten Kapitel durchgeführt, um ein rechnerisches Setup zu finden, das den besten Kompromiss zwischen Genauigkeit und CPU-Zeitanforderungen liefert. Drei Systemkonfigurationen werden dann getestet und sorgfältig untersucht: Die erste führt zu einem oszillatorischen Verhalten des Tragflügels, das durch begrenzte kleine Amplituden gekennzeichnet ist. Die andere ist durch das Vorhandensein einer aeroelastischen Instabilität gemäß einer Torsionsdivergenz gekenn-zeichnet. Schließlich wird eine auf das Flatterphänomen ausgerichtete Konfiguration gewählt. Allerdings sind derzeit nicht genügend Daten vorhanden und daher kann diese dynamische aeroelastische Instabilität in der vorliegenden Arbeit nicht gründlich untersucht werden. 

\par In naher Zukunft wird diese Studie als Basis für die Flatterexperimente und Berechnungen an der Professur für Strömungsmechanik der Helmut-Schmidt-Universität dienen.

\cleardoublepage
\pagestyle{empty}
\addcontentsline{toc}{chapter}{Acknowledgements}
\chapter*{Acknowledgements}
\par Foremost, I would like to thank Univ.-Prof.\ Dr.-Ing.\ habil.\ Michael Breuer for the introduction in the computational fluid dynamics area in 2013 and the support through the development of this project.
\par I also would like to express my sincere gratitude to my advisor Dr.-Ing.\ Guillaume De Nayer for the continuous support of my research project, for his patience, motivation, enthusiasm and immense knowledge. His guidance helped me in all the time of research and writing of this work.
\par I would also like to acknowledge the personal of the Institute of Fluid Mechanics of the Helmut-Schmidt University for the help during this thesis. 
\par Furthermore, I am tremendously grateful for the scholarship given by the Böttcher Stiftung. Without it, my entire master's degree would have not been possible.
\par Special thanks are given to Prof.\ Dr.\ Heinrich Kreye. Since 2012 he has helped and supported me thorough my studies. He played a crucial role in the processes of getting the Böttcher scholarship and coming back to Germany. Moreover, every time that I had a problem I could count on him and I knew that he would be really happy to help.
\par I would also like to thank Univ.-Prof.\ Dr.-Ing.\ Franz Joos for the job as student assistant since the beginning of my master's degree, which financially enabled my studies and gave me the opportunity of working with diverse softwares.  
\par Finally, I must express my very profound gratitude to my parents and to my boyfriend for providing me with unfailing support and continuous encouragement throughout my years of study ant through the process of researching and writing this work. This accomplishment would not have been possible without them.

\cleardoublepage
\pagestyle{empty}

\pagestyle{fancy}
\fancyhf{}
\fancyhead{}
\fancyfoot{}
\fancyhead[RO]{\rightmark}
\fancyhead[LO]{\leftmark}
\fancyfoot[C]{\thepage}


\pagenumbering{roman}

\setcounter{page}{1}

\setcounter{tocdepth}{3}      
\setcounter{secnumdepth}{3}   

\tableofcontents
\listoffigures
\listoftables
\chapter*{Nomenclature\markboth{NOMENCLATURE}{}}
\label{nomenclature}
\addcontentsline{toc}{chapter}{Nomenclature}

\noindent The abbreviations, notations and variables used in this report are explained below with the corresponding units. Notations and variables not included in this section are explained in the context of the report.

\section*{Abbreviations}
\begin{tabular}{ll}
ALE & Arbitrary Lagrangian-Eulerian \\
CC & Cell Center \\
CCS & Cartesian Coordinate System\\
CFD	& Computational Fluid Dynamics \\
CM & Center of Mass \\
CPU & Central Processing Unit \\
CSD & Computational Structural Dynamics \\
CV & Control Volume \\
DGCL & Discrete Geometric Conservation Law \\
DNS	& Direct Numerical Simulation \\
DOF & Degree of Freedom \\
LCO & Limit-Cycle Oscillation\\
LES	& Large-Eddy Simulation \\
FEM & Finite-Element Method \\
FSI	& Fluid-Structure Interaction \\
FVM & Finite-Volume Method \\
IDW & Inverse Distance Weighting \\
NACA & National Advisory Committee for Aeronautics \\
RANS & Reynolds-Averaged Navier Stokes \\
SCL & Space Conservation Law \\
TFI & Transfinite Interpolation \\
e.g. & exempli gratia \\
i.e. & id est \\
Eq. & Equation\\
Fig. & Figure \\
\end{tabular}
\section*{Notations}
\begin{tabular}{ll}
\(A_i\) & Index notation of vector A\\
\(B_{ij}\) & Index notation of second-order tensor B \\ 
\(\overline{C}\) & Large turbulence scale of C\\
\(\widetilde{D}\) & Time-averaged D\\
\({<}{E}{>}\) & Spatial and time-averaged E\\
\(F^*\) & Dimensionless value of F  \\\\\\\\\\\\\\\\\\\\\\\\\\\\\\\\\\\\\\\\\\\\\\\\\\\\\\\\\\\\\\\\\\\\\\\\\   
\end{tabular}

\section*{Variables}
\begin{tabular}{p{1.9cm}p{9.8cm}p{2.9cm}}
\(c\) & Airfoil chord length & m \\
\(C_D\) & Drag coefficient &  \\
\(c_{l,\,i}\) & Linear damping coefficient vector & $(\text{N}{\cdot}\text{s}{\cdot}\text{m}^{-1})$ \\
\(C_L\) & Lift coefficient &  \\
\(C_{rt,\,i}\) & Combined linear and torsional damping coefficient vector & \\
\(C_s\) & Smagorinsky constant &  \\
\(c_{t,\,i}\) & Torsional damping coefficient vector & $(\text{N}{\cdot}\text{m}{\cdot}\text{s}{\cdot}\text{rad}^{-1})$ \\
\(D\) & Damping ratio & \\
\(e_{m_{ex}}\) & Eccentricity of the mass & (m) \\
\(f\) & Frequency & (Hz)\\
\(f_n\) & Natural frequency & (Hz)\\
\(f_{skew}\) & Skew metric quality & \\
\(f_v\) & Vortex shedding frequency & (Hz) \\
\(F_{CSD,\,i}\) & Geometrically-scaled force & (N) \\
\(F_D\) & Drag force & (N) \\
\(F_{DA,\,i}\) & Damping force vector &  (N)  \\
\(F_{ext,\,i}\) & External force vector &  (N)  \\
\(F_{I,\,i}\) & Inertial force vector &  (N)  \\
\(F_L\) & Lift force &  (N)  \\
\(F_{p_{i}}\) & Pressure force vector &  (N)  \\
\(F_{rt,\,i}\) & Combined external forces and moment vector & \\
\(F_{S,\,i}\) & Spring force vector &  (N)  \\
\(F_{\tau_{ij}}\) & Shear force vector &  (N)  \\
\(I_{ij}\) & Identity tensor &  \\
\(J_{ij}\) & Mass moment of inertia tensor & $(\text{kg}{\cdot}\text{m}^2)$ \\
\(k_{l\,i}\) & Linear stiffness vector & $(\text{N}{\cdot}\text{m}^{-1})$ \\
\(k_{t\,i}\) & Torsional stiffness vector & $(\text{N}{\cdot}\text{m}{\cdot}\text{rad}^{-1})$ \\
\(K_{rt,\,i}\) & Combined linear and torsional stiffness vector & \\
\(l_k\) & Kolmogorov length  & (m) \\
\(L_3\) & Span-wise length of the mesh & (m) \\
\(L_{3,\,N}\) & Span-wise length of the NACA0012 model & (m) \\
\(L_{3,\,WT}\) & Span-wise length of wind tunnel & (m) \\
\(m^*\) & Mass ratio between body and displaced fluid & \\
\(m^{body}\) & Mass of the body & (kg) \\
\(m^{fluid}_disp\) & Mass of the displaced fluid & (kg) \\
\(m_{exc}\) & Eccentric mass & (kg) \\
\end{tabular}

\section*{Variables}
\begin{tabular}{p{1.9cm}p{9.8cm}p{2.9cm}}
\(m_{ij}\) & Mass tensor & (kg) \\
\(m^{N}\) & Mass of the NACA0012 model & (kg) \\
\(m^{tot}\) & Total mass: Airfoil plus supports & (kg) \\
\(M_{CSD,\,i}\) & Geometrically-scaled moment & ($\text{N}{\cdot}\text{m}$) \\
\(M_{ext,\,i}\) & External moment vector & $((\text{N}{\cdot}\text{m})$ \\
\(M_i\) & Moment around the $x_i$-axis & $(\text{N}{\cdot}\text{m})$ \\
\(M_{rt,\,ij}\) & Combined mass and mass moment of inertia tensor & \\
\(N\) & Number of nodes & \\
\(N_{L_3}\) & Number of nodes in the span-wise direction & \\
\(N_R\) & Number of nodes in the domain radius & \\
\(N_{SS}\) & Number of nodes in the suction side & \\
\(N_W\) & Number of nodes in the wake & \\
\(n\) & Iteration number &  \\
\(n_{FSI}\) & FSI sub-iteration number &  \\
\(n_i\) & Normal vector &  \\
\(p\) & Pressure & (Pa) \\
\(p^{\,corr}\) & Pressure correction & (Pa) \\
\(p^{\,pred}\) & Predicted pressure & (Pa) \\
\(p'\) & Pressure fluctuation & (Pa) \\
\(p_{t_0}\) & Initial pressure & (Pa) \\
\(p_{\infty}\) & Free-stream pressure & (Pa) \\
\(q_i\) & Quaternion component ($1\leq i\leq 4$) &  \\
\(q_\Phi\) & Source term of the transport variable $\Phi$ & \\
\(r_{CCS_{l},\,i}\) & Position of the body-fixed undeformed local Cartesian coordinate system in relation to the global Cartesian coordinate system  & (m)\\
\(r_{new,\,i}\) & Position of the deformed element node in relation to the global Cartesian coordinate system & (m) \\
\(r_{rot,\,i}\) & Rotated position vector in relation to the body-fixed deformed Cartesian coordinate system & (m) \\
\(r_{start,\,i}\) & Position of the undeformed element node in relation to the global Cartesian coordinate system & (m) \\
\(R\) & Domain radius & (m) \\
\(Re\) & Reynolds number &  \\
\(S\) & Area & $(\text{m}^2)$ \\
\(S_{ij}\) & Strain rate tensor & $(\text{s}^{-1})$ \\
\(S_{ref}\) & Reference area & $(\text{m}^2)$ \\
\end{tabular}

\section*{Variables}

\begin{tabular}{p{1.9cm}p{9.8cm}p{2.9cm}}
\(St\) & Strouhal number in relation to the airfoil chord &  \\	
\(t\) & Time & (s) \\	
\(t^*\) & Dimensionless time &  \\
\(t_n\) & Time step number  &  \\
\(T\) & Temperature  & (K) \\
\(T_v\) & Vortex shedding period  & (s) \\
\(U_{conv}\) & Mean convection velocity & $(m{\cdot}s^{-1})$ \\
\(u_i\) & Velocity vector in Cartesian coordinates & $(\text{m}{\cdot}\text{s}^{-1})$ \\
\(u_i^{\,pred,\,j}\) & Predicted velocity vector in Cartesian coordinates & $(\text{m}{\cdot}\text{s}^{-1})$ \\
\(u^*_i\) & Dimensionless velocity vector in Cartesian coordinates & \\	
\(u_{g,\,j}\) & Grid velocity vector in Cartesian coordinates & $(\text{m}{\cdot}\text{s}^{-1})$ \\
\(u_{in,\,i}\) & Inlet velocity vector in Cartesian coordinates & $(\text{m}{\cdot}\text{s}^{-1})$ \\
\(u_{sym,\,i}\) & Velocity vector in Cartesian coordinates at the symmetry boundary & $(\text{m}{\cdot}\text{s}^{-1})$ \\
\(u_{t_0,\,i}\) & Initial velocity vector in Cartesian coordinates & $(\text{m}{\cdot}\text{s}^{-1})$ \\
\(u_{\tau}\) & Shear stress velocity & $(\text{m}{\cdot}\text{s}^{-1})$ \\
\(v_i\) & Direction vector & \\
\(V\) & Volume & $(\text{m}^3)$ \\
\(V_N\) & Volume of the NACA0012 model & $(\text{m}^3)$ \\
\(V_{cell}\) & Cell volume & $(\text{m}^3)$ \\
\(W\) & Wake length & (m) \\
\(x_i\) & Cartesian coordinates & (m)\\
\(x_i^*\) & Dimensionless Cartesian coordinates &\\
\(X_0\) & Characteristic amplitude & (m) \\
\(X_i\) & Displacement vector & (m) \\
\(\dot{X_i}\) & Velocity vector & $(\text{m}{\cdot}\text{s}^{-1})$ \\
\(\ddot{X_i}\) & Acceleration vector & $(\text{m}{\cdot}\text{s}^{-2})$ \\
\(X_{N,\,i}\) & Position vector of the NACA0012 airfoil & (m) \\
\(X_{rt,\,i}\) & Combined translational and rotational displacement vector & \\
\(\dot{X}_{rt,\,i}\) & Combined translational and rotational velocity vector &  \\
\(\ddot{X}_{rt,\,i}\) & Combined translational and rotational acceleration vector & \\
\(y^+\) & Dimensionless wall distance & \\
\\ \\ \\ \\ \\ \\ \\ \\ \\ \\ \\ \\ \\ \\ \\ \\ \\ \\ \\ \\ \\ \\ \\ \\ \\
\end{tabular}

\section*{Greek variables}

\begin{tabular}{p{1.9cm}p{10.7cm}p{2cm}}
\(\alpha\) & Angle of attack & $(^\circ)$ \\
\(\alpha_{f}^s\) & Under-relaxation parameter for the displacement and velocity & \\
\(\alpha_{fxd}\) & IDW parameter for fixed boundaries & \\
\(\alpha_{m}^s\) & Under-relaxation parameter for the acceleration & \\
\(\alpha_{mv}\) & IDW parameter for moving boundaries & \\
\(\Gamma_{\Phi}\) & Diffusive flux of the transport variable $\Phi$  & \\
\(\delta_f\) & Under-relaxation parameter for the force & \\
\(\delta V_k\) & Swept volume of $k$ cell surface & $(\text{m}^3)$ \\
\(\Delta\) & Filter length & $(\text{m})$ \\
\(\Delta_i\) & Body displacement vector & $(\text{m})$ \\
\(\Delta t\) & Time step size &$(\text{s})$\\
\(\Delta y\) & Distance of the midpoint of the first cell & $(\text{m})$ \\
\(\Delta y^{first\,cell}\) & First cell wall distance & $(\text{m})$ \\
\(\epsilon_{FSI,\,disp}\) & Displacement residuum of the FSI sub-iteration & \\
\(\mu_f\) & Fluid dynamic viscosity & $(\text{Pa}{\cdot}\text{s})$ \\
\(\mu_T\) & Eddy viscosity & $(\text{Pa}{\cdot}\text{s})$ \\
\(\nu_f\)& Fluid kinematic viscosity &$(\text{m}^{2}{\cdot}\text{s}^{-1})$\\
\(\rho_f\) & Fluid density	& $(\text{kg}{\cdot}\text{m}^{-3})$\\
\(\sigma\) & Standard deviation &  \\
\(\tau_{ij}^{mol}\) & Molecular momentum transport  & $(\text{Pa})$ \\
\(\tau_{ij}^{SGS}\) & Subgrid-scale stress tensor & $(\text{Pa})$ \\
\(\tau_{ij}^{turb^*}\) & Dimensionless Reynolds stress tensor & \\
\(\tau_w\) & Wall shear stress & $(\text{Pa})$\\
\(\phi\) & Phase angle & $(\text{rad})$ \\
\(\Phi\) &  Transport variable&  \\
\(\varphi_i\) & Angular displacement vector& $(\text{rad})$ \\
\(\dot{\varphi_i}\) & Angular velocity vector & $(\text{rad}{\cdot}\text{s}^{-1})$ \\
\(\ddot{\varphi_i}\) & Angular acceleration vector & $(\text{rad}{\cdot}\text{s}^{-2})$ \\
\(\omega\) & Angular frequency & $(\text{rad}{\cdot}\text{s}^{-1})$ \\
\(\omega_{f}\) & Vorticity vector  & $(\text{s}^{-1})$ \\
\(\omega_{i,\,n}^s\) & Weighting function &  \\
\(\omega_n\) & Angular natural frequency & $(\text{rad}{\cdot}\text{s}^{-1})$ \\

\end{tabular}

\pagestyle{fancy}


\chapter*{Introduction\markboth{INTRODUCTION}{}}
\label{introduction}
\setcounter{page}{1}
\pagenumbering{arabic}
\addcontentsline{toc}{chapter}{Introduction}
\par The development of airplanes, alike most of the currently available technologies, started with preliminary observations of phenomena in nature. The first tentatives to build flying machines attempted to imitate the flapping-wing flight of the birds. For instance, already in 1485 Leonardo da Vinci designed a device in which the aviator lies down on a plank and works two large, membranous wings using hand levers, foot pedals and a system of pulleys. Despite of the early developments, the conception of currently available aircrafts which utilize fixed wings, propulsion systems and movable control surfaces, came up only in 1799 with the pioneer work of Sir George Cayley. He defined the lift and drag forces and presented the first scientific design for a fixed-wing aircraft: A kite mounted on a stick with a movable tail. This pioneer design enabled the concurrent development of aircrafts and new disciplines, such as aeroelasticty. 
\par Aeroelasticity is a young science that describes the influence of aerodynamic forces on elastic bodies. It combines features of fluid mechanics and solid mechanics and has become more important primarily in aeronautics due to the ever-increasing aircraft sizes and speeds \cite{Fung_2002}. An increase in speed represents a major challenge for the aeroelastic design. This leads to a rapid increase of the aerodynamics forces while the stiffness of the structure remains constant. This mis-balance can generate aeroelastic instabilities, which are characterized by an oscillatory behavior. A major problem during the early development of aircrafts was the wing divergence, which is a steady-state instability distinguished by an oscillation with zero frequency that causes a rapid wing twist and failure. For instance, this is probably the cause of the wing failure of Professor S. P. Langley's monoplane in 1903 \cite{Fung_2002}. For modern aircrafts, however, wing divergence does not represent a major challenge, since the critical speeds of flight at which divergence sets in are usually higher than those of other aeroelastic instabilities, such as flutter.
\par
\begin{figure}[H]
	\centering
	\subfigure[NASA Langley's Research Center: Lockheed Electra model following catastrophic flutter \cite{Chambers_2003}.]{\includegraphics[width=0.49\textwidth]{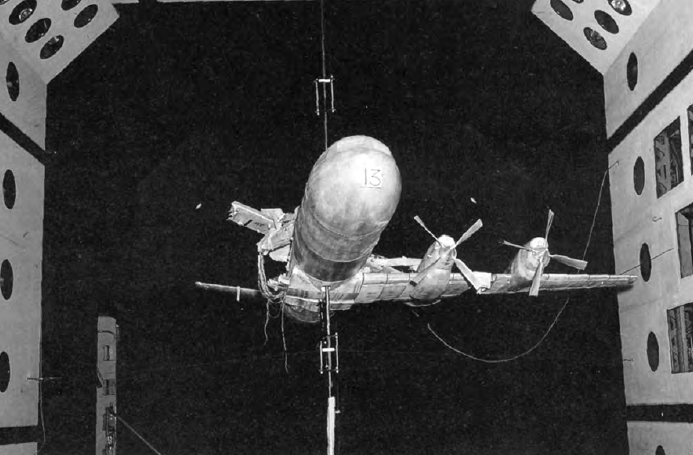} \label{fig:flutter_NASA}}\hfill
	\subfigure[Laboratory of Fluid Mechanics - Helmut Schmidt University: Experimental setup of the NACA0012 airfoil.  Courtesy of J. N. Wood]{\includegraphics[width=0.49\textwidth]{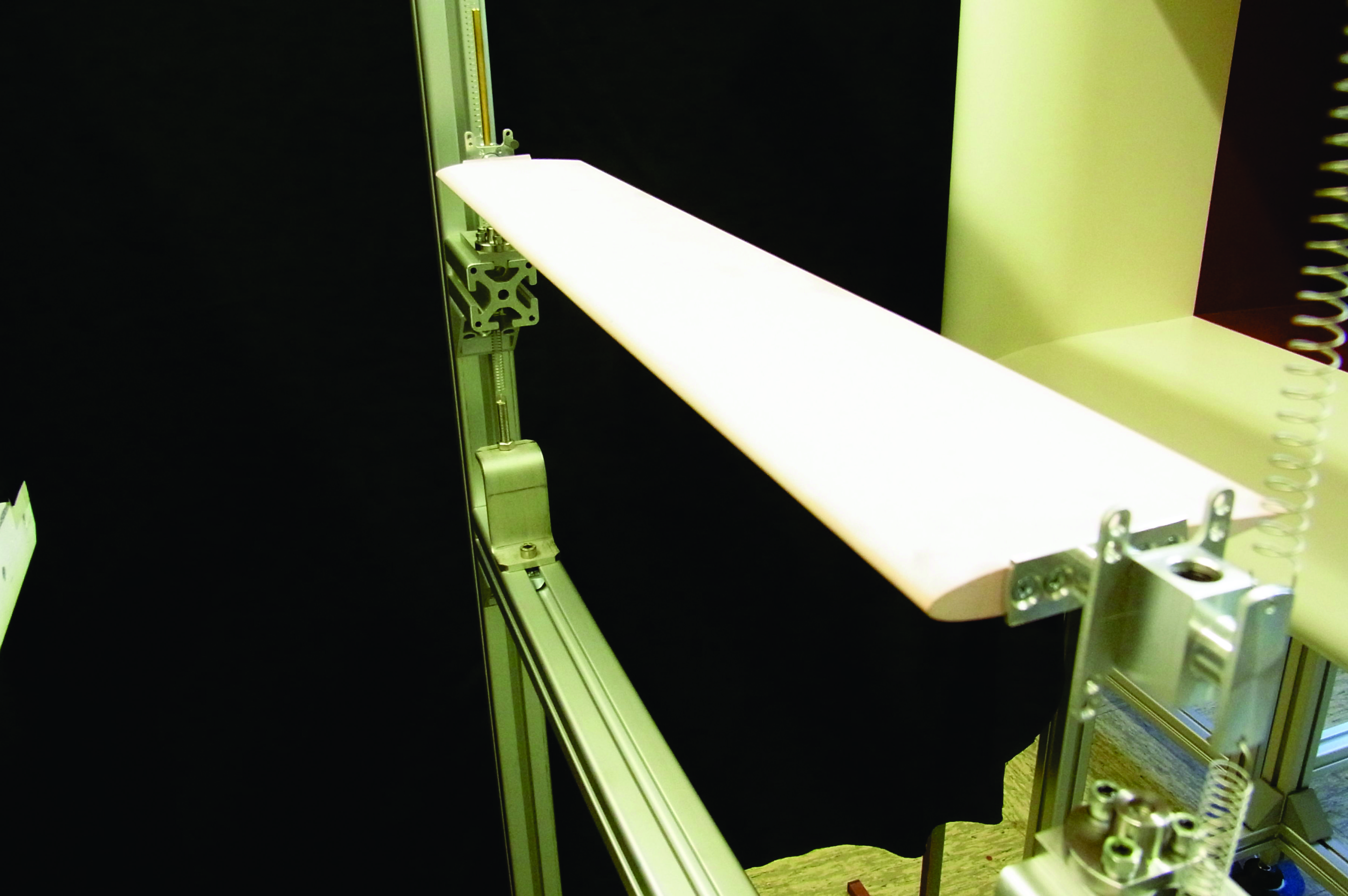}\label{fig:flutter_setup}}\hfill
	\caption{Aircraft and airfoil models used to experimentally study aeroelasticity.}
	\label{fig:drag_lift_history}
\end{figure}

 Flutter is a dynamic aeroelastic instability characterized by self-sustained exponentially growing oscillations. An intensive study of this phenomenon started during the arms race in the 1930s, since numerous accidents were caused by wing and tail flutter. In the past, this effect was also studied by flight testing. However, in 1938 a four-engined Junkers plane \mbox{Ju 90 VI} crashed during a flutter test, killing all scientists on-board \cite{Fung_2002}. Therefore, the theorical research on aeroelastic instabilities in the form of computations and experiments are since then emphasized. For instance, the NASA Langley Research Center has conducted the first transonic wind tunnel test focusing on a particular full-scale design already in 1960 \cite{Chambers_2003}. This aimed at the investigation of the aeroelastic properties of the Lockheed Electra, since this aircraft suffered a number of accidents, which evidence suggested that the wings of the airplane had failed and separated from the aircraft in flight (see Chambers \cite{Chambers_2003}). The tests in the wind tunnel confirmed that the cause of the accident was actually flutter, which build up and teared the airplane model apart in a matter of seconds, as illustrated in Fig.\ 
\ref{fig:flutter_NASA}.
\par Due to the increasing importance and the challenging nature of the theoretical studies of aeroelasticity, the Institute of Fluid Mechanics of the Helmut-Schmidt-University has broaden its main focuses, i$.$e$.$, fluid-structure interaction and multiphase flows, also to the studies of aeroelastic instabilities. A new experimental setup, as illustrated in Fig.\ \ref{fig:flutter_setup}, has been implemented in the Laboratory of Fluid Mechanics in order to enable the analysis of the aeroelastic properties of a symmetric NACA0012 airfoil, which is commonly used in airplanes for stability purposes, such as in vertical and horizontal stabilizers. The rigid airfoil model is composed of Sika Block M700 and is allowed to oscillate in two degrees of freedom, i$.$e$.$, translation in the $x_2$ direction and rotation around the $x_3$ axis.    
\par The computational part on the studies of the aeroelastics of the rigid\break NACA0012 airfoil with a chord length of \mbox{$c=0.1\,\text{m}$} are divided and executed in two parts: Firstly, wall-resolved large-eddy simulations (LES) on a fixed rigid NACA0012 profile at various angles of attack were performed for a Reynolds number of $Re=100{,}000$. The fluid domain is then thoroughly analyzed in relation to its flow characteristics and the aerodynamic properties. The results of this first investigation are presented in the work of \mbox{Streher \cite{Streher_2017}}.
\par Secondly, the previous studies are expanded to a moving NACA0012 in the current work, regarding that the applied Reynolds number is reduced to $Re=30{,}000$ in order to acquire reasonable CPU-time requirements. The fluid-structure interaction (FSI) of the rigid NACA0012 with two degrees of freedom (translation in relation to the vertical axis $x_2$ and rotation around the span-wise axis $x_3$) is computed according to a partitioned approach. The in-house sofware FASTED-3D, which applies a wall-resolved LES combined with the Smagorinsky model \cite{Smagorinsky_1963}, is utilized to compute the solution of the fluid domain. The solution of the structural sub-problem is acquired by the rigid movement solver implemented by Viets \cite{Viets_2013}, which is based on the equations of motion. The FSI coupling is performed by a partitioned approach relying on loose or strong coupling algorithms.

\par The current work aims at the investigation of the aeroelastic instabillities generated by the flow around a moving NACA0012 airfoil and is divided in the following manner: Firstly, in Chapter \ref{chap:numerical_methodology}, the applied numerical methodology with regard to the computational fluid dynamics (CFD), computational structural dynamics (CSD) and the FSI coupling is mentioned. Secondly, in Chapter \ref{chap:FSI_flow_NACA0012}, the test case is presented regarding the geometry and the computational setup for CFD, CSD and FSI coupling. Then, in \mbox{Chapter \ref{chap:preliminary_studies}}, preliminary studies are carried out to select the best mesh and the most effective mesh adaption method for the CFD subproblem. Pure CSD cases are also executed in order to validate the structural solver. Additionally, different FSI coupling algorithms are investigated and compared in order to find the best compromise between accuracy and CPU-time requirements. Afterwards, in Chapter \ref{chap:results}, the results obtained for the fluid-structure interaction of the NACA0012 with two degrees of freedom in turbulent flows are presented and discussed. Three principal configurations are evaluated: one aimed at limit-cycle oscillations and others aimed at torsional divergence and flutter. Finally, the achieved conclusions are summarized and further directions of studies are suggested.
\chapter{Numerical methodology}
\label{chap:numerical_methodology}

\par The numerical methods implemented to solve the fluid-structure interaction between the air flow and the rigid NACA0012 airfoil are presented in Sections \ref{sec:CFD} through \ref{sec:coupling_fluid_structure}. Foremost, the governing equations and methods to solve the CFD subproblem are explored. Then, the equations of motion for rigid bodies are mentioned as well as the applied numerical methods to approximate the solution of the CSD subproblem. Finally, the partitioned coupling scheme is presented.

\section{Computational fluid dynamics (CFD)}\markboth{CHAPTER 1.$\quad$NUMERICAL METHODOLOGY}{1.1$\quad$CFD}
\label{sec:CFD}

\par The CFD subproblem is solved with the in-house well-validated software \mbox{FASTEST-3D} (see Durst and Schäfer \cite{Durst_1996} and Breuer et al.\ \cite{Breuer_2012}). Since fluid-structure interaction (FSI) problems of rigid bodies are responsible for displacements of the structure, the CFD governing equations must consider this phenomenon. Therefore, conservation laws based on the arbitrary Lagrangian-Eulerian (ALE) approach are applied (see Section \ref{subsec:ALE}), regarding that turbulence is modeled according to the wall-resolved large-eddy simulation approach combined with the Smagorinsky model (see Section \ref{subsec:turbulence_modelling}). The governing equations are spatially and temporally discretized according to a finite-volume method (FVM) and a predictor-corrector scheme, respectively (see Section \ref{subsec:discretization}). The latter adopts a three sub-steps low-storage Runge-Kutta scheme as predictor and a pressure Poisson equation as corrector. Initial and boundary conditions (see Section \ref{subsec:initial_and_boundary_conditions}) are applied and the solution of the fluid domain is approximated. The computed fluid loads (see Section \ref{subsec:fluid_forces}) interact with the rigid body, resulting in a structure displacement and consequently a time-dependent computational domain. Hence, a mesh adaption process is carried out according to either a transfinite interpolation method or a hybrid inverse distance weighting - transfinite interpolation method (see Section \ref{subsec:mesh_adaptation}). 

\subsection{Arbitrary Lagrangian-Eulerian (ALE)}
\label{subsec:ALE}
\par The fluid-structure interaction between the flow and a rigid NACA0012 profile is characterized by an incompressible fluid submitted to a constant temperature and thus constant fluid properties. Moreover, the action of the fluid forces on the body is responsible for the displacement of the airfoil, resulting in a time-dependent computational domain. Hence, the arbitrary Lagrangian-Eulerian (ALE) formulation \cite{Hirt_1974} is applied for the conservations of mass and momentum.
\par The general form of the incompressible conservation laws as a function of the transport variable $\Phi$ is illustrated in Eq.\ (\ref{eq:Navier_Stokes}) for a non-varying domain. When $\Phi=1$, $\Gamma_{\Phi}=0$ and $q_{\Phi}=0$ the conservation of mass is represented, while when $\Phi=u_i$, $\Gamma_{\Phi}=\mu_f$ and $q_{\Phi}\neq0$ the conservation of momentum is depicted.
\begin{eqnarray}
\label{eq:Navier_Stokes}
\underbrace{\int_{V}\rho_f\frac{\partial \Phi}{\partial t}dV}_{\text{Local\,variation}}+\underbrace{\int_{V}\rho_f\frac{\partial}{\partial x_j}\left(u_j\,\Phi\right)dV}_{\text{Convective\,flux}}-\underbrace{\int_{V}\frac{\partial}{\partial x_j}\left(\Gamma_{\Phi}\frac{\partial \Phi}{\partial x_j}\right)dV}_{\text{Diffusive\,flux}}=\underbrace{\int_{V}q_{\Phi}\,dV}_{\text{Source\,term}}
\end{eqnarray} 
\par The ALE formulation is then applied to Eq.\ (\ref{eq:Navier_Stokes}), which basically re-formulates the conservation laws in order to acquire governing equations that are valid for temporally varying domains \cite{Breuer_2012}. This process is based on two steps: Firstly, the integral form of the conservation equations are adapted for a time-dependent fluid domain through the substitution of the fixed volumes by the time-dependent variable $V(t$). Secondly, the Gauss's divergence theorem and the Leibniz's rule are utilized, as described in the work of Donea et al.\ \cite{Donea_2004}. The former converts volume integrals into surface integrals and is utilized in the convective, diffusive and pressure terms. The latter transform the integral of temporal derivatives within time-dependent intervals into three other terms and is applied to the local variation term. 
\par The application of the previously described mathematical procedures results in\break \mbox{Eqs.\ (\ref{eq:ALE_mass}) and (\ref{eq:ALE_momentum})} for an incompressible fluid submitted to a constant temperature (see Breuer et al.\ \cite{Breuer_2012}). The former represents the conservation of mass and the latter stands for the conservation of momentum, regarding that the gravity force is negligible. 
\begin{eqnarray}
\frac{d}{dt}\int_{V(t)}\;dV+\int_{S(t)}(u_j-u_{g,j})\cdot n_j\;dS&=&0 \label{eq:ALE_mass} \\
\nonumber \frac{d}{dt}\int_{V(t)}\rho_f\;u_i\;dV+\int_{S(t)}\rho_f\,u_i (u_j-u_{g,j})\cdot n_j\;dS&=&-\int_{S(t)} \tau_{ij}^{mol} \cdot n_j\;dS\\ && - \int_{S(t)} p\cdot n_i\;dS \label{eq:ALE_momentum}
\end{eqnarray}
\par $\rho_f$, $u_i$, $p$, $\tau_{ij}^{mol}$ and $n_i$ represent respectively the fluid density, the velocity vector in Cartesian coordinates, the pressure, the molecular momentum transport tensor and the unit normal vector directed outward. The grid velocity $u_{g,j}$ emerges due to the deformable grid and describes the motion of the control volume surfaces. This velocity is computed according to the space conservation law (SCL) (see Demird{\v{z}}i{\'c} and Peri{\'c} \cite{Demirdvzic_1988}).

\subsection{Space conservation law (SCL)}
\label{subsec:SCL}

\par The space conservation law describes the change of position or shape of a control volume, assuring that no space is lost and therefore no artificial mass or momentum sources are generated \mbox{\cite{Breuer_2012,Demirdvzic_1988}}. This is represented by Eq.\ (\ref{eq:SCL}) and must be fulfilled simultaneously with the conversation of mass and momentum.
\begin{equation}
\frac{d}{dt}\int_{V(t)}\;dV=\int_{S(t)}u_{g,j}\cdot n_j\;dS. \label{eq:SCL}
\end{equation} 
\par When the SCL (\ref{eq:SCL}) is inserted in to the continuity equation for movable grids\break \mbox{(Eq.\ (\ref{eq:ALE_mass}))}, the conservation of mass for an incompressible fluid and a fixed grid is obtained, as illustrated in \mbox{Eq.\ (\ref{eq:SCL_mass_conservation})}:  
\begin{equation}
\int_{S(t)}u_j\cdot n_j\;dS=0 \label{eq:SCL_mass_conservation}
\end{equation}
\par This mathematical procedure illustrates that no additional grid flux must be calculated for the continuity equation \cite{Breuer_2012}, assuring that no artificial mass sources are produced (see Ferziger and Peri{\'c} \cite{Ferziger_1999}). 
\par In the case of the conservation of momentum, however, additional grid fluxes must be calculated. Therefore, a discrete geometric conservation law (DGCL) according to the work of Farhat et al.\ \cite{Farhat_2001} and of Breuer et al.\ \cite{Breuer_2012} is applied. This is consistent with the applied spatial and temporal discretization methods (see Sections \ref{subsec:spatial_discretization} and \ref{subsec:temporal_discretization}) and utilizes the volumes swept by each cell surface $\delta V_k$ in order to determine the additional grid flux according to the conservation of momentum, as shown in Eq.\ (\ref{eq:DGCL_convection_term}). The indices $e$, $w$, $n$, $s$, $t$ and $b$ represent respectively the east, west, north, south, top and bottom cell surfaces of the hexahedral control volumes (CV). 
\begin{equation}
\label{eq:DGCL_convection_term}
\int_{S(t)}(\rho_f\,u_i\,u_{g,j})\cdot n_j\;dS\;\approx\sum_{k=\{e, w,n,s,t,b\}}\left(\rho_f\,u_{i,k}\frac{\delta V_k^{n+1}}{\Delta t}\right)
\end{equation} 
\par The swept volumes $\delta V_k^{n+1}$ at the new time step are defined according to the work of Kordulla and Vinokur \cite{Kordulla_1983}, which decomposes the hexahedral control volumes into six tetrahedra holding the same diagonal. This method provides an accurate prediction of the swept volumes and assures that the sum of the volumes swept by each of the six surfaces are equal to the volume difference between the new $n+1$ and the old $n$ time steps, as illustrated in Eq.\ (\ref{eq:swept_volume}):
\begin{equation} \label{eq:swept_volume}
V^{n+1}-V^n=\sum_{k=e,w,n,s,t,b}\delta V_k^{n+1}.
\end{equation} 

\subsection{Turbulence modeling}
\label{subsec:turbulence_modelling}
The Reynolds number is a relation between the inertial and viscous forces acting on the fluid according to Eq.\ (\ref{eq:Reynolds_number}). $c$ and $u_{in}$ represent respectively the chord length of the NACA0012 profile and the inflow velocity, i$.$e$.$, the free-stream velocity.
\begin{equation}
Re=\frac{u_{in}\;\rho_f\;c}{\mu_f}
\label{eq:Reynolds_number}
\end{equation}
\par At large Reynolds numbers, the inertial forces overcome the viscous ones, resulting in a chaotic turbulent flow. Fluid turbulence is often visualized as a cascade of kinetic energy, which is characterized by the production of large scales of motion, the decay of the large scales into small scales through an inertial mechanism and the kinetic energy dissipation of the smallest scales (Kolmogorov length) in the form of thermal energy. 
\par In order to simulate turbulent flows, the energy cascade can be either directly solved by the Direct Numerical Simulation (DNS), partially modeled as done in the Large-Eddy Simulation (LES) or fully modeled according to Reynolds-Averaged Navier-Stokes Equations (RANS). In many applications, however, the full energy cascade cannot be directly computed since the required computational effort would exceed the available computing resources by many orders of magnitude. Therefore, LES is the most promising technique since it provides the best compromise between required computational effort and accuracy. Hence, this approach is utilized in the present thesis to perform the FSI simulations of the NACA0012 airfoil.   
\par Large-eddy simulation is based on the division of the turbulence spectrum into small and large scales. The former constitutes the low-energy contributions, is short-living, dissipative, universal (independent from geometry and boundary conditions) and nearly homogeneous and isotropic. Therefore this scale, as well as its influence on the large scales, is modeled. The latter constitutes the high-energy components, which are strongly problem-dependent. Thus, it is predicted directly by the spatially filtered Navier Stokes equations \cite{Breuer_2002}. 
\par Due to the direct usage of the conservation laws for the large eddies, the simulation time step and the grid resolution must be sufficiently small/fine in order to resolve the smallest eddies of the large scale. Even though this increases the accuracy, it also raises the numerical effort compared to RANS.
\par The governing equations for LES are acquired by the spatial filtering of the ALE conservation laws for an incompressible fluid. In the case of FASTEST-3D, as described by Durst and Schäfer \cite{Durst_1996}, this filtering process is performed by the grid itself, which has the advantage of coupling filtering and numerical method. However, it is susceptible to aliasing errors, which arise from the non-linearity of the equations.
\par The filtering process plus some mathematical procedures result in the LES equations according to the ALE formulation, which are subject to the commutation error caused by the approximation of the filtered partial derivatives.
\begin{eqnarray}
\frac{d}{dt}\int_{V(t)}\;dV+\int_{S(t)}(\overline{u}_j-\overline{u}_{g,j})\cdot n_j\;dS&=&0 \label{eq:LES_ALE_mass} \\
\nonumber \frac{d}{dt}\int_{V(t)}\rho_f\;\overline{u}_i\;dV+\int_{S(t)}\rho_f\,\overline{u}_i (\overline{u}_j-\overline{u}_{g,j})\cdot n_j\;dS&=&-\int_{S(t)} (\overline{\tau}_{ij}^{mol}+\tau_{ij}^{SGS}) \cdot n_j\;dS\\ && - \int_{S(t)} \overline{p}\cdot n_i\;dS \label{eq:LES_ALE_momentum}
\end{eqnarray}
\par Equations (\ref{eq:LES_ALE_mass}) and (\ref{eq:LES_ALE_momentum}) are respectively the conservation of mass and momentum for the large turbulent scales (characterized by an overbar) of an incompressible fluid subjected to constant temperature and a movable grid. ${\overline{u}_i}$, $\overline{p}$ and ${\overline{\tau}}_{ij}^{mol}$ represent, correspondingly, the large-scale velocity, pressure and molecular-dependent transport. The latter is a function of the dynamic viscosity $\mu_f$ and the large-scale strain rate tensor ${\overline {S}}_{ij}$, according to Eqs.\ (\ref{eq:molecule_dependent_transport_LES}) and (\ref{eq:strain_rate_tensor_LES}): 
\begin{eqnarray}
\label{eq:molecule_dependent_transport_LES}
{\overline{\tau}}_{ij}^{mol}&=&-2\mu_f {{\overline {S}}_{ij}}, \\
\label{eq:strain_rate_tensor_LES}
{\overline{S}}_{ij}&=&\frac{1}{2}{\left({\frac{\partial {\overline {u}}_i}{\partial x_j}+\frac{\partial {\overline {u}}_j}{\partial x_i}}\right)}.
\end{eqnarray}
\par The conservation of momentum (\ref{eq:LES_ALE_momentum}) comprises a new term, the subgrid-scale stress tensor $\tau_{ij}^{SGS}$. It is produced by the filtering process of the non-linear convective term, i.e$.$, ${\overline {u_i(u_j-u_{g,j})}}$, and symbolizes the following effects: The interaction of the large eddies that results in the production of small scales, the interaction between large and small scales and finally the interaction between small scales that culminates in the formation of large eddies in a process called backscatter. This term cannot be directly calculated, constituting the closure problem of turbulence.
\par In order to solve the governing equations, i.e$.$, to model the subgrid-scale stress tensor, various models have been proposed, as summarized by Breuer \cite{Breuer_2002}. In the present work, the Smagorinsky model \cite{Smagorinsky_1963} combined with a damping function near solid walls is used, since this was proven to be accurate enough for simulations of the flow around a fixed rigid NACA0012 airfoil at different angles of attack (see Streher \cite{Streher_2017}). This subgrid-scale model is based on the algebraic Boussinesq's approximation (see Eq.\ (\ref{eq:Boussinesq_approximation})) and states an analogy between the molecular-dependent transport for laminar flows and the subgrid-scale tensor, as per \mbox{Eq.\ (\ref{eq:Smagorinsky_equation})}:
\begin{eqnarray}
\label{eq:Boussinesq_approximation}
\tau _{ij}^{mol}&=&-2\,\mu_f\,{S}_{ij}, \\
\label{eq:Smagorinsky_equation}
\tau _{ij}^{SGS}&=&-2\,\mu_T\,{\overline S}_{ij}.
\end{eqnarray} 
\par The proportionality factor is the eddy viscosity $\mu_T$. It depends on the turbulence structure and may vary in space and time. Therefore, it is not a fluid property. $\mu_T$ can be estimated by the Smagorinsky model, which is based on the assumption that the modeled scales are isotropic, i.e$.$, the turbulence is in local equilibrium.
\par Due to a dimension analysis, Smagorinsky \cite{Smagorinsky_1963} stated that the eddy viscosity is proportional to a characteristic length $l_c$, velocity $v_c$ and density $\rho_c$ (this is assumed constant, i.e$.$, $\rho_c =\rho_{ref}$), according to Eq.\ (\ref{eq:eddy_viscosity_1}):
\begin{equation}
\label{eq:eddy_viscosity_1}
\mu_T\sim\rho_c\,l_c\,v_c.
\end{equation}
\par The characteristic velocity $v_c$ is approximated as a function of the characteristic length $l_c$ and the strain rate tensor of the large eddies ${\overline S}_{ij}$. The former ($l_c$) is estimated with the help of the filter width $\Delta$ and the Smagorinsky constant $C_s$. Furthermore, the Van-Driest damping function is applied near solid walls, which accurately evaluates the subgrid-scale stress tensor near fixed walls. Thus, $\mu_T$ is proportional to the Smagorinsky constant $C_s$, the filter length $\Delta$ and the dimensionless wall distance $y^+$, as demonstrated in \mbox{Eq$.$ (\ref{eq:eddy_viscosity_2})}:
\begin{equation}
\label{eq:eddy_viscosity_2}
\mu_T=C_s^2\Delta^2\sqrt{2{\overline S}_{ij}{\overline S}_{ij}}{\left[1-\exp{\left(\frac{-y^+}{25}\right)}^3\right]}.
\end{equation}
\par The Smagorinsky constant $C_s$ is empirically determined and the standard value of \mbox{$C_s=0.1$} is used for the NACA0012 airfoil simulations (see Streher \cite{Streher_2017}).
\par Since the filtering process and numerical method are implicitly coupled, the filter length $\Delta$ is defined as a function of each cell volume $V_{cell}$, according to Eq.\ (\ref{eq:cutoff_length}):
\begin{equation}
\label{eq:cutoff_length}
\Delta=\left(V_{cell}\right)^{\frac{1}{3}}.
\end{equation}
\par The dimensionless wall distance $y^+$ is defined as a function of the shear velocity $u_\tau$, the distance of the first cell middle point to the airfoil geometry $\Delta y$ and the kinematic viscosity of the fluid $\nu_f$, as stated in Eq.\ (\ref{eq:dimensionless_wall_distance}):
\begin{equation}
y^+=\frac{u_\tau\,\Delta y}{\nu_f} \label{eq:dimensionless_wall_distance}. \\
\end{equation}
\par The shear velocity $u_\tau$ is a function of the fluid density $\rho_f$ and the wall shear stress $\tau_w$, according to Eq.\ (\ref{eq:shear_velocity}). The latter, i.e$.$, $\tau_w$, is proportional to the dynamic viscosity $\mu_f$ and the velocity gradient $\partial u/\partial y$ near the wall $w$, as demonstrated in Eq.\ (\ref{eq:wall_shear_stress}):
\begin{eqnarray}
u_\tau&=&\sqrt{\frac{\tau_w}{\rho_f}} \label{eq:shear_velocity}, \\
\tau_w=\mu_f \left.\frac{\partial u}{\partial y}\right |_w &\approx& \mu_f \frac{\Delta u_i}{\Delta y}=\mu_f \frac{u_{first\,cell,\,avg,i}-0}{\Delta y}. \label{eq:wall_shear_stress}
\end{eqnarray}
\par In the present work, the dimensionless wall distance is approximated as a function of the time-averaged velocities of the cells located on the airfoil surface $u_{first\,cell,\,avg,\,i}$, the distance of the first cell middle point to the airfoil geometry $\Delta y$ and the kinematic viscosity $\nu_f$, as stated by Eq.\ (\ref{eq:dimensionless_wall_distance_used}):
\begin{equation}
\label{eq:dimensionless_wall_distance_used}
y^+=\sqrt{{\sqrt{\sum_{i=1}^{3}u_{first\,cell,\,avg,i}^2}}\cdot\frac{\Delta y}{\nu_f}}.
\end{equation}

\subsection{Discretization}
\label{subsec:discretization}
\par Since the conservation equations are non-linear and coupled, the solutions for complex problems are numerically approximated using spatial and temporal discretization methods. Moreover, the overbars utilized to specify the filtered properties within the governing equations are omitted.

\subsubsection{Spatial discretization}
\label{subsec:spatial_discretization}
\par A finite-volume method (FVM), which divides the computational domain into several control volumes through the generation of a mesh, is utilized to spatially discretize the governing equations in the integral form (filtered conservation of mass, momentum and space conservation law). In the present work block-structured meshes with hexahedral control volumes are utilized. The flow variables, such as the pressure $p$ and velocities $u_i$ are calculated and stored in the cell middle point.
\par A second-order accurate midpoint rule is applied to all control volumes in order to approximate the volume and surface integrals. While the former does not require interpolation methods, since all flow values are stored at the cell center and therefore already known, the latter requires an interpolation process in order to calculate the values on the cell faces. This is done according to a linear interpolation from the neighboring cells and is also second-order accurate.

\subsubsection{Temporal discretization}
\label{subsec:temporal_discretization}
\par Since a wall-resolved large-eddy simulation is used to describe the turbulent flow field, it is necessary to work with very small time steps. Therefore, an explicit temporal discretization is preferred. Although this approach demands a small time step $\Delta t$ in order to be stable, it is easier implemented on high-performance computers (vectorization and parallelization) and requires a smaller numerical effort per time step. It uses only known variable values from the old time step, i$.$e$.$, $n$, in order to calculate the current one, i.e$.$, $n+1$.
\par An explicit low-storage three sub-steps Runge-Kutta scheme, described by\break \mbox{Breuer et al$.$ \cite{Breuer_2012}}, is used in FASTEST-3D to predict the velocities $u_i^{\,pred}$ at the time step $n+1$, according to \mbox{Eq\ (\ref{eq:predictor_step})}. $u_i^{n}$ stands for the velocity at the last time step and $f\left(n,u_i^{n},p^{\,pred}\right)$ represents a function of the time step number, velocity and pressure, as well as of the geometry of the control volume. 
\begin{eqnarray}
\label{eq:predictor_step}
u_i^{\,pred,\,1}&=&u_i^{n}+\,\frac{\Delta t}{3}\;f\left(n,u_i^{n},p^{\,pred}\right)\nonumber\\
u_i^{\,pred,\,2}&=&u_i^{n}+\,\frac{\Delta t}{2}\;f\left(n,u_i^{\,pred,\,1},p^{\,pred}\right)\nonumber\\
u_i^{\,pred}&=&u_i^{n}+\;\Delta t\; f\left(n,u_i^{\,pred,\,2},p^{\,pred}\right)
\end{eqnarray}
\par The predicted velocities are calculated applying the values of a predicted pressure $p^{\,pred}$ (either a presumed value or an approximation based on the value at the last time step $p^{n}$) in the conservation of momentum. However, they may not fulfill the continuity equation. Hence, a corrector step must be executed in order to guarantee its fulfillment. 
\par The velocity field only satisfies both conservation laws when a correct pressure field is available. Thus, a Poisson equation (see Eq.\ (\ref{eq:Poisson})), deduced from the divergence of the momentum conservation, is used to calculate a pressure correction $p^{\,{corr}}$.
\begin{equation}
\label{eq:Poisson}
\frac{\partial}{\partial x_i} \left[\frac{\quad\partial p^{\,{corr}}}{\partial x_i}\right]=\frac{\rho}{\Delta t} \frac{\quad\partial u_i^{\,pred}}{\partial x_i}
\end{equation}
\par The pressure correction $p^{\,{corr}}$ is then used to estimate a new pressure $p^{n+1}$ and  velocity $u_i^{n+1}$, according to Eqs.\ (\ref{eq:neue_Druck}) and (\ref{eq:neue_Geschwindigkeit}), respectively. The latter, however, does not necessarily precisely fulfill the continuity equation at $n+1$. Therefore, the correction step must be repeated until a convergence criterion is achieved (see Section \ref{subsec:CFD_convergence_criterion}).
\begin{equation}
\label{eq:neue_Druck}
p^{n+1}=p^{\,pred}+p^{\,{corr}}
\end{equation}
\begin{equation}
\label{eq:neue_Geschwindigkeit}
u_i^{n+1}=u_i^{\,pred}-\frac{\Delta t}{\rho}\frac{\quad\partial p^{\,{corr}}}{\partial x_i}
\end{equation}

\subsection{Initial and boundary conditions}
\label{subsec:initial_and_boundary_conditions}
Since the discretized ALE equations constitute an initial boundary value problem, these conditions have to be specified in order to solve the CFD problem. Initial conditions for the whole computational domain and boundary conditions for all boundaries are required, since the conservation laws are parabolic in time and elliptic in space. 
\par The utilized initial and boundary conditions are thoroughly described in Sections \ref{sec:CFD_initial_conditions} and \ref{sec:CFD_boundary_conditions}, respectively. 

\subsection{Convergence criterion}
\label{subsec:CFD_convergence_criterion}
\par In the present work, the convergence criterion of the CFD solver is based on the fulfillment of the conservation of mass when the velocity $u_i^{n+1}$ calculated as a function of the predicted velocity $u_i^{pref}$ and the corrected pressure $p^{corr}$, as described in Eq.\ (\ref{eq:neue_Geschwindigkeit}), is applied. The utilized criterion is illustrated in Eq.\ (\ref{eq:CFD_convergence}):
\begin{equation}
\label{eq:CFD_convergence}
\int_{S(t)}u_j\cdot n_j\,dS\leq\mathcal{O}(10^{-10}).
\end{equation}  

\subsection{Fluid forces and moments}
\label{subsec:fluid_forces}

\par In order to calculate the forces and moments acting on the airfoil, FASTED-3D divides the surface located at the FSI interface $S$, into several areas $s$. Figure \ref{fig:forces_moments} illustrates this division, as well as the fluid forces and level arms:
\begin{figure}[H]
	\centering
	\centering
	\includegraphics[scale=1,draft=\drafttype]{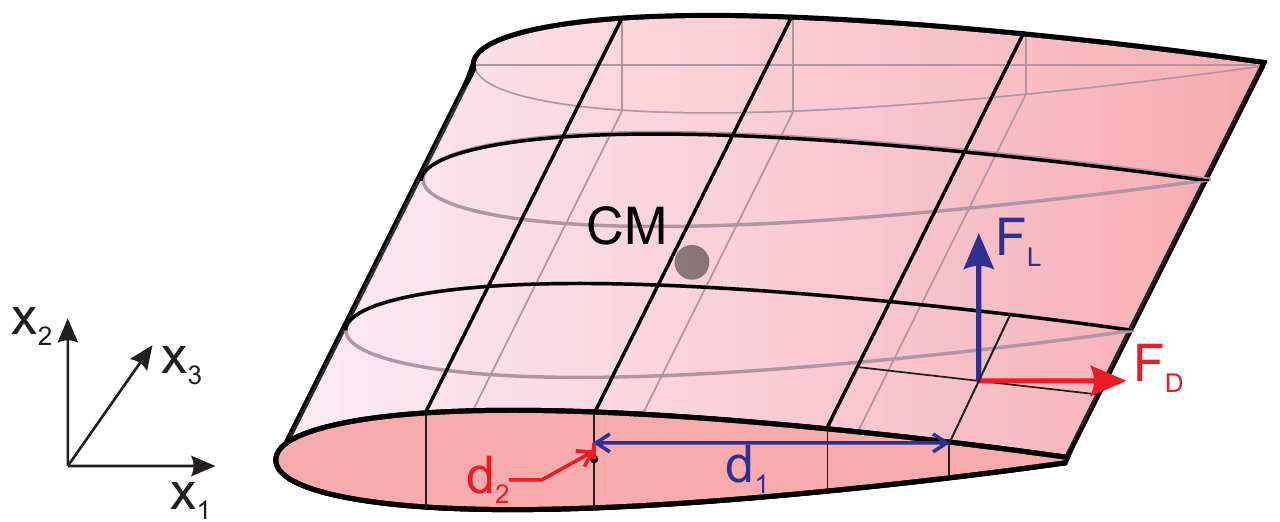}
	\caption{\label{fig:forces_moments}FSI surface discretization, fuid forces and level arms.}
\end{figure}
\par The total pressure and shear forces generated by the flow around the NACA0012 are then calculated according to Eqs.\ (\ref{eq:pressure_force}) to (\ref{eq:pressure_force_2}) (see Viets \cite{Viets_2013}). $F_{p_{i}}$ and $F_\tau{_{ij}}$ represent respectively the pressure and shear forces that act on the surface of the airfoil.
\begin{eqnarray}
\label{eq:pressure_force}
F_{p_{i}}&=&-\int_{S}p\;n_i\;dS,
\\ \label{eq:shear_force}
F_{\tau_{ij}}&=&-\int_{S}\tau_{ij}^{mol}\cdot n_j\;dS, 
\\ \label{eq:stress_tensor}
\tau_{ij}^{mol}&=&-\;\mu_f\frac{\partial u_i}{\partial x_j}.
\label{eq:pressure_force_2}
\end{eqnarray}
\par  The pressure and shear forces in the $x_1$ (chordwise) and $x_2$ directions generate respectively the drag $F_D$ and the lift $F_L$ forces:
\begin{eqnarray}
\label{eq:drag}
F_D=F_1=F_{\tau_{11}}+F_{p_1},\\
\label{eq:lift}
F_L=F_2=F_{\tau_{22}}+F_{p_2}.
\end{eqnarray}
\par The fluid forces are also responsible for the generation of a  moment in relation to the body-fixed coordinate system, which is located at the center of mass of the NACA0012 airfoil. Since the simulated airfoil has only two degrees of freedom (translation in \mbox{$x_2$-direction} and rotation around the $x_3$-axis), only the forces in the $x_1$ and $x_2$ directions, i$.$e$.$, $F_D$ and $F_L$, and the moment around the $x_3$ axis in relation to the center of mass $CM$, i$.$e$.$, $M_3^{CM}$, are exploited. The latter is approximated according to Eq.\ (\ref{eq:moment_z}), regarding that $d_1$ and $d_2$ are the lever arms (see Fig.\ref{fig:forces_moments}).
\begin{equation}
\label{eq:moment_z}
M_3^{CM}\approx\sum_{s}\left(F_L(s)\;d_1(s)-F_D(s)\;d_2(s)\right).\\
\end{equation}

\subsection{Mesh adaptation}
\label{subsec:mesh_adaptation} 
\par In the case of fluid-structure interaction, the computational domain varies in time due to the movement of the boundaries, requiring a grid displacement in order to fit the new configuration of the computational domain. Various mesh adaptation methods for structured grids are available and a review of diverse algorithms can be found in \mbox{Sen et al.\ \cite{Sen_2017}}.
\par A compromise between mesh quality and CPU-time requirements must be considered, while determining the utilized method, regarding that the mesh deformation algorithm should at least preserve the essential quality of the original mesh. Therefore, for cases utilizing the LES approach, the applied method must guarantee specially the conservation of the first cell height, as well as the orthogonality of the cells located at the vicinity of solid boundaries, where the flow gradients are large \cite{Sen_2017}. 
\par In the case of the NACA0012 simulations, the available time to perform the simulations and analyze the results is limited. Therefore, the transfinite interpolation method (TFI), which is an algebraic method that can be fully parallelized (see Breuer et al.\ \cite{Breuer_2012}), is preferred for the simulations characterized by small displacements. Nevertheless, a hybrid inverse distance weighting - transfinite interpolation method developed by \mbox{Sen et al.\ \cite{Sen_2017}} is also applied when large displacements and rotations are observed or foreseen. Although the latter demands more computational time, the deformed grid has a higher quality, allowing a more accurate solution of the FSI problem.

\subsubsection{Transfinite interpolation method (TFI)}
\label{subsubsec:transfinite_interpolation}
\par The transfinite interpolation method is only applicable for block-structured grids and is based on two steps: Firstly, the displacements calculated by the CSD solver are utilized in order to define the new position of the boundaries of each geometrical block (surface mesh nodes) and secondly, the distribution of the grid points located inside the blocks (volume mesh points) are computed according to shear mappings \mbox{(see Glück \cite{Glueck_2002} and Sen et al.\ \cite{Sen_2017}).} 
\par Although this technique allows to generate high-quality meshes in a single block requiring low computational time, which is proportional to the number of volume grid points, it is not suitable for cases characterized by large displacements due to two facts: Firstly, this method does not conserve the height of the cells located on the solid boundaries, which may lead to numerical difficulties if great translational displacements are observed. Secondly, it does not maintain the orthogonality of the cells in relation to the solid boundaries, which leads to computational errors and even to the formation of cells with negative volumes, in case of large body rotations.
\par Sen et al.\ \cite{Sen_2017} tested the TFI method for an airfoil submitted to large rotations and deformations caused by the fluid-structure interaction and proved that the utilization of this technique caused the arise of degenerated cells with cross-overs at the two extremities of the airfoil, destroying the mesh orthogonality also at boundary points and creating an artificial clustering of grid points on the surface of the airfoil. Therefore, the current work utilizes this algorithm only in the FSI cases characterized by small displacements and rotations.
\par
\begin{figure}[H]
 	\centering
 	\centering
 	\includegraphics[scale=0.63,draft=\drafttype]{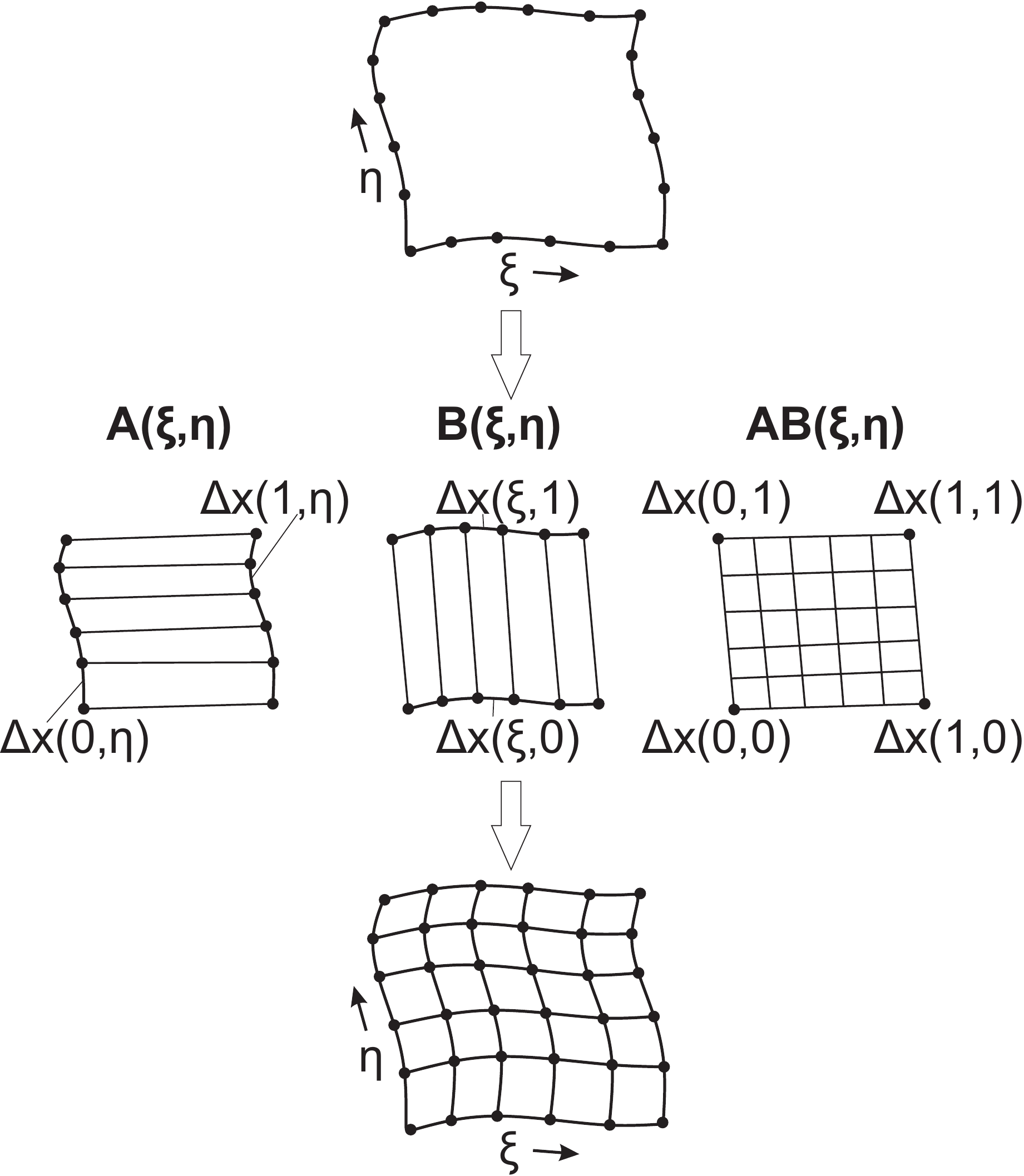}
 	\caption{\label{fig:TFI}Transfinite interpolation method (TFI) for a two-dimensional mesh.}
 \end{figure}

 The applied TFI method is illustrated in \mbox{Fig.\ \ref{fig:TFI}} for a two-dimensional grid, regarding that the location of the volume points are calculated only based on the position of the surface nodes, which are established according to the CSD displacements. 
\par Equations (\ref{eq:displacement}) through (\ref{eq:direct_sum_4}) demonstrate the applied TFI algorithm for a three-dimensional grid, which calculates the deformation at an arbitrary volume point $\Delta x_{i,j,k}$ and considers $1\leq i\leq I$, $1\leq j\leq J$ and \mbox{$1\leq k \leq K$ \cite{Sen_2017}}. $A$, $B$ and $C$ are the univariate Lagrange projectors, i$.$e$.$, shear mappings in the $\xi$, $\eta$ and $\zeta$ directions, respectively. The arc-length based computational coordinates $(\xi,\,\eta,\,\zeta)$ are computed only once (according to the undeformed grid) and are utilized in order to maintain the edge parameters of the initially undeformed grid, e$.$g$.$, the stretching factor $q$ (see Sen et al.\ \cite{Sen_2017}).
 \begin{equation}
 \label{eq:displacement}
 \Delta x_{i,j,k}=A\oplus B \oplus C = A + B + C - AB - BC - CA + ABC,
 \end{equation}
 \begin{eqnarray}
 \label{eq:shear_mapping_1} A(\xi_{i,j,k})&=&(1-\xi_{i,j,k})\;\Delta x_{1,j,k}+(\xi_{i,j,k})\;\Delta x_{I,j,k}, \\
 \label{eq:shear_mapping_2} B(\eta_{i,j,k})&=&(1-\eta_{i,j,k})\;\Delta x_{i,1,k}+(\eta_{i,j,k})\;\Delta x_{i,J,k}, \\
 \label{eq:shear_mapping_3}
 C(\zeta_{i,j,k})&=&(1-\zeta_{i,j,k})\;\Delta x_{i,j,1}+(\zeta_{i,j,k})\;\Delta x_{i,j,K},
 \end{eqnarray}
 \begin{eqnarray}
 \label{eq:direct_sum_1}
 AB(\xi_{i,j,k},\eta_{i,j,k})&=&(1-\xi_{i,j,k})(1-\eta_{i,j,k})\;\Delta x_{1,1,k}+\xi_{i,j,k}\,(1-\eta_{i,j,k})\;\Delta x_{I,1,k}\nonumber \\
 &&+(1-\xi_{i,j,k})\,\eta_{i,j,k}\;\Delta x_{i,J,k}+\xi_{i,j,k}\,\eta_{i,j,k}\;\Delta x_{I,J,k}, \\
 \label{eq:direct_sum_2}
 BC(\eta_{i,j,k},\zeta_{i,j,k})&=&(1-\eta_{i,j,k})(1-\zeta_{i,j,k})\;\Delta x_{i,1,1}+\eta_{i,j,k}\,(1-\zeta_{i,j,k})\;\Delta x_{i,J,1}\nonumber \\
 &&+(1-\eta_{i,j,k})\,\zeta_{i,j,k}\;\Delta x_{i,1,K}+\eta_{i,j,k}\,\zeta_{i,j,k}\;\Delta x_{i,J,K}, \\
 \label{eq:direct_sum_3}
 CA(\xi_{i,j,k},\zeta_{i,j,k})&=&(1-\zeta_{i,j,k})(1-\xi_{i,j,k})\;\Delta x_{1,j,1}+\zeta_{i,j,k}\,(1-\xi_{i,j,k})\;\Delta x_{1,j,K}\nonumber \\
 &&+(1-\zeta_{i,j,k})\,\xi_{i,j,k}\;\Delta x_{I,j,k}+\zeta_{i,j,k}\,\xi_{i,j,k}\;\Delta x_{I,j,K},
 \end{eqnarray}
 \begin{eqnarray}
 \label{eq:direct_sum_4}
 ABC(\xi_{i,j,k},\eta_{i,j,k},\zeta_{i,j,k})&=&(1-\xi_{i,j,k})(1-\eta_{i,j,k})(1-\zeta_{i,j,k})\;\Delta x_{1,1,1}+\xi_{i,j,k}\,(1-\eta_{i,j,k})\nonumber \\&&(1-\zeta_{i,j,k})\;\Delta x_{I,1,1}+(1-\xi_{i,j,k})\,\eta_{i,j,k}\,(1-\zeta_{i,j,k})\;\Delta x_{1,J,1}+\nonumber \\
 &&\xi_{i,j,k}\,\eta_{i,j,k}\,(1-\zeta_{i,j,k})\;\Delta x_{I,J,1}+(1-\xi_{i,j,k})(1-\eta_{i,j,k})\,\zeta_{i,j,k}\;\nonumber \\&&\Delta x_{1,1,K}+\xi_{i,j,k}(1-\eta_{i,j,k})\,\zeta_{i,j,k}\;\Delta x_{I,1,K}+(1-\xi_{i,j,k})\,\eta_{i,j,k}\,\nonumber \\&&\zeta_{i,j,k}\;\Delta x_{1,J,K}+\xi_{i,j,k}\,\eta_{i,j,k}\,\zeta_{i,j,k}\;\Delta x_{I,J,K}.
 \end{eqnarray}

\subsubsection{Hybrid inverse distance weighting - Transfinite interpolation method (IDW-TFI)}
\label{subsubsec:idw_tfi} 
\par A good compromise between CPU-time requirements and mesh quality for simulations characterized by large displacements is achieved by the hybrid IDW-TFI method, developed by \mbox{Sen et al.\ \cite{Sen_2017}.} This performs the grid deformation in two steps for block-structured grids: Firstly, the movement of the block-boundaries (surface nodes) are executed according to an inverse distance weighting interpolation method, which preserves the orthogonality of the control volumes and the height of the cells located on the airfoil surface. Secondly, the TFI technique is carried out for the displacement of the inner points, i$.$e$.$, volume nodes of each block. Although this procedure demands more computational time than the TFI algorithm, it  enables the generation of high-quality deformed grids even when large deformations are present. 
\par The utilized TFI method is explained in details in Section \ref{subsubsec:transfinite_interpolation}. The applied IDW algorithm is based on the work of Witteveen \cite{Witteveen_2010} and the improvements suggested by Luke et al.\ \cite{Luke_2012} and Maruyama et al.\ \cite{Maruyama_2014}. It calculates the displacements of the volume points $\Delta_{i,\,n}^v$ through an interpolation according to a weighting function $w_{i,\,n}^s$ of the surface displacements $\Delta_{i,\,n}^s$ \cite{Witteveen_2010} as shown in \mbox{Eq.\ (\ref{eq:IDW})}. The indices $i$ and $n$ represent the directions of the Cartesian coordinate system and an arbitrary grid node, respectively. 
\begin{equation}
\label{eq:IDW}
\Delta_{i,\,n}^v=\frac{\sum_{n=1}^{N_s}w_{i,\,n}^s\;\Delta_{i,\,n}^s}{\sum_{n=1}^{N_s}w_{i\,n}^s}
\end{equation}
\par In the present work, the applied weighting function is based on the work of\break \mbox{Luke et al.\ \cite{Luke_2012}}, which is illustrated in Eq.\ (\ref{eq:weighting_function}):
\begin{equation}
\label{eq:weighting_function}
w_{i,\,n}^s=A_n^s\,\left[\left(\frac{L_{def}}{||r_{i,\,n}-r_{i,\,n}^s||}\right)^3\left(\frac{\alpha\,L_{def}}{||r_{i,\,n}-r_{i,\,n}^s||}\right)^5\right].
\end{equation}
\par $r_{i,\,n}$ and $r_{i,\,n}^s$ are the position vectors of an arbitrary node and a grid point located on the block surface, respectively. $A_n^s$ is the area weight of the boundary node $n$ and $||\;.\;||$ denotes the Euclidean norm. The parameter $\alpha$ provides the relative weight of the nearby nodes in comparison to the distant \mbox{ones \cite{Sen_2017}.} This is thoroughly studied in Section \ref{appendix_mesh_adaption}, regarding that different parameters are applied for the fixed and the moving boundaries, i$.$e$.$, respectively $\alpha_{fxd}$ and $\alpha_{mv}$. $L_{def}$ describes the maximum distance of any mesh node to the mesh centroid and is defined according to Eq.\ (\ref{eq:L_def}):
\begin{equation}
\label{eq:L_def}
L_{def}=\max_{n=1}^{N_s}\,||\,r_{i,\,n}^s-r_{i\,n}^{centroid}\,||.
\end{equation}
\par The displacement of the nodes located on the body surface $\Delta_{i,\,n}^s$ are calculated according to the dismantled translational and rotational motions (see Sections \ref{subsubsec:translation} to \ref{subsec:body_disp}). The former is described with a vector, while the latter is described with quaternions (four-dimensional vectors) in order to avoid problems with the singularities of the rotation matrix \cite{Maruyama_2014}. Finally, the weighting function is designed based on a combination of parameters (see Sen et al.\ \cite{Sen_2017}), which are only estimated at the beginning of the interpolation process and therefore are based on the undeformed grid.      

\markboth{CHAPTER 1.$\quad$NUMERICAL METHODOLOGY}{1.1$\quad$CSD}
\section{Computational structural dynamics (CSD)}\markboth{CHAPTER 1.$\quad$NUMERICAL METHODOLOGY}{1.1$\quad$CSD}
\label{sec:CSD}

\par The current work investigates the fluid-structure interaction between the flow and a rigid NACA0012 airfoil. Therefore, the rigid body dynamics together with a time discretization method are exploited in order to approximate the displacement of the airfoil caused by the fluid forces and moments.
\par The utilized rigid movement solver was implemented by Viets \cite{Viets_2013} and is only validated for uncoupled systems characterized by constant external loads. Therefore, a validation regarding time-dependent external forces and coupled translational and rotational motions is also carried out in order to assure its accuracy while solving the NACA0012 FSI problem (see Section \ref{appendix_validation_CSD}). 

\subsection{Rigid body dynamics}
\label{subsec:rigid_body_dynamics}
\par A rigid body is characterized by the constant distance between two arbitrary points, even if forces and moments are acting on it. Therefore, the response of the body occurs only in the form of displacements, i$.$e$.$, the body does not suffer any deformations.
\par The rigid body dynamics is governed by Newton's and Euler's seconds laws for the translational and rotational motions, respectively. The former is illustrated in Eq.\ (\ref{eq:Newton_2_law}) for a translational motion regarding the acceleration of the body center of mass while the latter is demonstrated in Eq.\ (\ref{eq:Euler_2_law}) for a torque-free rotation, i$.$e$.$, a rotation around the center of mass (see Hibbeler \cite{Hibbeler_2001}). $F_i$, $m$, $\ddot{X_i}$, $M_i$, $J_{ij}$ and $\ddot{\varphi_i}$ represent the force vector, the mass, the translational acceleration vector, the moment vector, the mass moment of inertia tensor and the angular acceleration vector, respectively.   
\begin{eqnarray}
\label{eq:Newton_2_law}
F_i=m\;\ddot{X_i}\\
\label{eq:Euler_2_law}
M_i=J_{ij}\;\ddot{\varphi_i}
\end{eqnarray}
\par The motion of a rigid body can be modeled as a mass-spring-damper system, as illustrated in Fig.\ \ref{fig:mass_spring_damper_system} for a system with one translational degree of freedom in the $x_1$-direction (see Münsch \cite{Muensch_2015}):
\begin{figure}[H]
	\centering
	\centering
	\includegraphics[scale=0.7,draft=\drafttype]{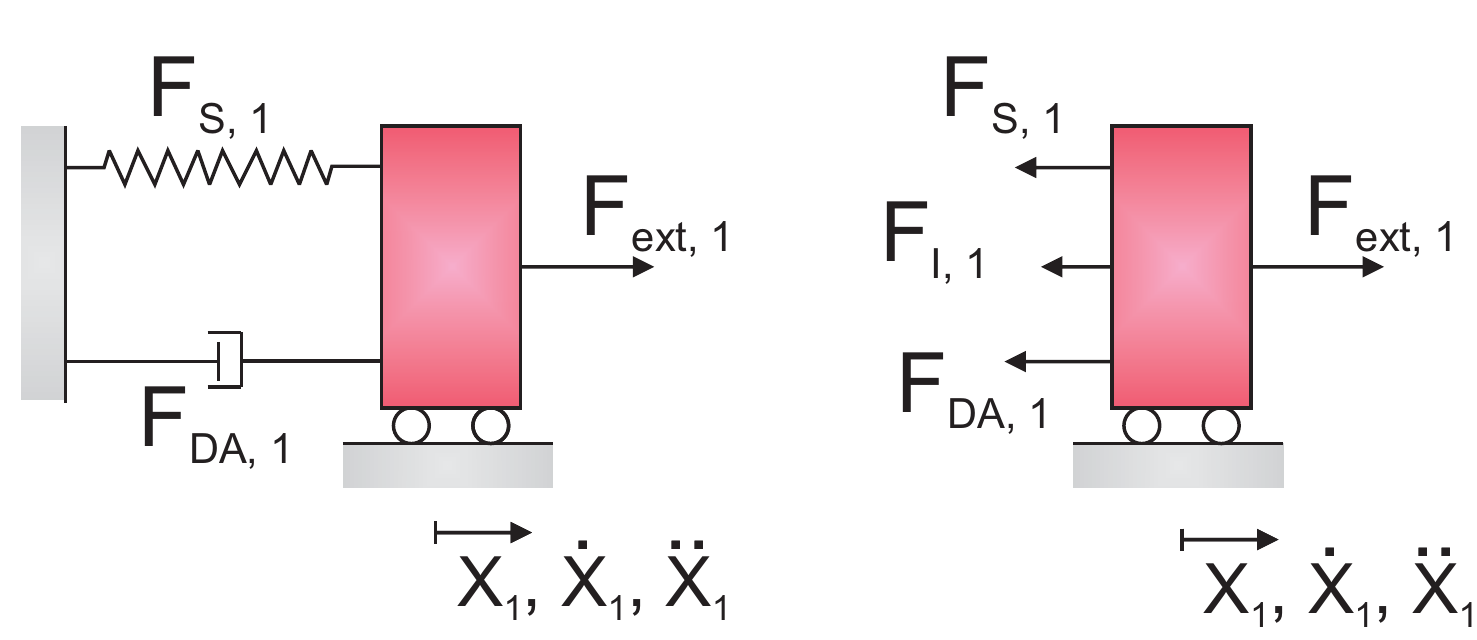}
	\caption{\label{fig:mass_spring_damper_system}Mass-spring-damper system and resultant forces acting on the mass.}
\end{figure}
\par $F_{S,\,1}$, $F_{DA,\,1}$, $F_{I,1}$ and $F_{ext,\,1}$ represent respectively the spring force as per Hooke's law, the damping force, the inertial force and the external force acting on the body in the $x_1$-direction. In the case of fluid-structure interaction, the external forces are the result of the pressure and the wall shear stress on the surface of the body, i$.$e$.$, the drag and lift forces (see Section \ref{subsec:fluid_forces}). Therefore, these forces are time-dependent.
\par While the spring force $F_{S,\,i}$ is proportional to the linear spring constant $k_{l,\,i}$ and the displacement $X_i$, the damping force $F_{DA,\,i}$ depends on the linear damping coefficient $c_{l,\,i}$ and the velocity $\dot{X_i}$. Furthermore, the inertial force is proportional to the mass $m_{ij}$ and the acceleration $\ddot{X_i}$. Hence, a mass-spring-damper system with only translational degrees of freedom in the three spatial directions ($i=1,\,2,\,3$) can be described by Eq.\ (\ref{eq:rigidmvt_translation}) \cite{Rao_2011}:
\begin{equation}
\label{eq:rigidmvt_translation}
\underbrace{m_{ij}\,\ddot{X_i}+c_{l,\,i}\,\dot{X_i}+k_{l,\,i}\,X_i}_{\text{internal\;forces}}=\underbrace{F_{ext,\,i}}_{\text{CFD\;forces}}.
\end{equation}
\par Similarly to the description of the translational motion, the rotational motion about the coordinate axes can also be modeled as a mass-spring-damper system, resulting in \mbox{Eq.\ (\ref{eq:rigidmvt_rotation})} \cite{Rao_2011}:
\begin{equation}
\label{eq:rigidmvt_rotation}
\underbrace{J_{ij}\,\ddot{\varphi_i}+c_{t,\,i}\,\dot{\varphi_i}+k_{t,\,i}\,\varphi}_{\text{internal\;moment}}=\underbrace{M_{ext,\,i}}_{\text{CFD\;moment}}.
\end{equation}
\par $J_{ij}$, $c_{t,\,i}$, $k_{t,\,i}$ and $M_{ext,\,i}$ are the mass moment of inertia, the torsional damping coefficient, the torsional spring stiffness and the external moment generated by the fluid forces (see Section \ref{subsec:fluid_forces}), respectively. ${\varphi_i}$, $\dot{\varphi_i}$ and $\ddot{\varphi_i}$ are the angular displacement, velocity and acceleration, respectively.
\par The translational and rotational motions are combined in a system of equations (see Eq.\ (\ref{eq:fsirigidmvt})), which represent the governing equations of the utilized CSD solver, according to Viets \cite{Viets_2013}. $I_{ij}$ is the identity tensor.   
\begin{equation}
\label{eq:fsirigidmvt}
\underbrace{\underbrace{\begin{bmatrix}
m  & 0 & 0 & 0 & 0 & 0\\
0 & m & 0 & 0 & 0 & 0\\
0 & 0 & m & 0 & 0 & 0\\
0 & 0 & 0 & J_{11} & J_{12} & J_{13}\\ 
0 & 0 & 0 & J_{21} & J_{22} & J_{23}\\
0 & 0 & 0 & J_{31} & J_{32} & J_{33}
\end{bmatrix}}_{M_{rt,\,ij}}
\underbrace{\begin{bmatrix}
\ddot{X_1}\\
\ddot{X_2}\\
\ddot{X_3}\\
\ddot{\varphi_1}\\
\ddot{\varphi_2}\\
\ddot{\varphi_3}
\end{bmatrix}}_{\ddot{X}_{rt,\,i}}
+I_{ij}
\underbrace{\begin{bmatrix}
c_{l,\,1} \\
c_{l,\,2}\\
c_{l,\,3}\\
c_{t,\,1}\\
c_{t,\,2}\\
c_{t,\,3}
\end{bmatrix}^T}_{C_{rt,\,i}}
\underbrace{\begin{bmatrix}
\dot{X_1}\\
\dot{X_2}\\
\dot{X_3}\\
\dot{\varphi_1}\\
\dot{\varphi_2}\\
\dot{\varphi_3}
\end{bmatrix}}_{\dot{X}_{rt,\,i}}
+I_{ij}
\underbrace{\begin{bmatrix}
k_{l,\,1} \\
k_{l,\,2}\\
k_{l,\,3}\\
k_{t,\,1}\\
k_{t,\,2}\\
k_{t,\,3} 
\end{bmatrix}^T}_{K_{rt,\,i}}
\underbrace{\begin{bmatrix}
{X_1}\\
{X_2}\\
{X_3}\\
{\varphi_1}\\
{\varphi_2}\\
{\varphi_3}
\end{bmatrix}}_{X_{rt,\,i}}}_{CSD}
=
\underbrace{\underbrace{\begin{bmatrix}
F_{ext,\,1}\\
F_{ext,\,2}\\
F_{ext,\,3}\\
M_{ext,\,1}\\
M_{ext,\,2}\\
M_{ext,\,3}\\
\end{bmatrix}}_{F_{rt,\,i}}}_{CFD}\\
\end{equation}
\par The mass moment of inertia $J_{ij}$ is time-independent for body-fixed coordinate systems. Moreover, if this fixed axis system is aligned with mutually orthogonal principal axes \mbox{(see Gere \cite{Gere_2004})}, which have the property of aligning the angular moment $M_{ext,\,i}$ with the angular velocity $\dot{\varphi_i}$, the mass moment of inertia tensor is diagonal and composed of only the principal moments of inertia, i$.$e$.$, $J_{11}$, $J_{22}$ and $J_{33}$ (see Hibbeler \cite{Hibbeler_2001}). Therefore, the system of equations (\ref{eq:fsirigidmvt}) becomes uncoupled.
\par In the case of bodies with constant density, the axes characterized by a plane symmetry are principal ones, which are distinguished by the vanishment of the products of inertia \mbox{(see Gere \cite{Gere_2004})}. Moreover, any axis perpendicular to a principal axis is also a principal one. Therefore, since the body fixed Cartesian coordinate system of the current work is located at the NACA0012 center of mass (see Section \ref{sec:CSD_setup}) and this airfoil model is characterized by a constant density (see Appendix \ref{appendix_experimental_setup})
and two symmetry planes, i$.$e$.$, $x_1x_3$ and $x_1x_2$, the mass moment of inertia tensor is characterized by the suppression of the products of inertia $J_{12}$, $J_{13}$ and $J_{23}$ (see Section \ref{subsec:moment_inertia}).

\subsection{Temporal discretization}
\label{subsec:temporal_discretization_CSD}
\par The solution of the system presented in Eq.\ (\ref{eq:fsirigidmvt}) is acquired by implicit numerical time integration methods for second-order differential equations, such as the standard Newmark \cite{Newmark_1959} and the generalized-$\alpha$ \cite{Wood_1980} methods. Implicit approaches utilize not only known variables (at the last time step $n$), but also unknown variables (at the current time step $n+1$) in order to approximate the solution. Therefore, these techniques are generally stable and can use larger time step sizes, when comparing to explicit time integration methods (see V{\'a}zquez \cite{Vazquez_2007}).

\subsubsection{Standard Newmark method}
\label{sec:newmark_method}
\par The standard Newmark method is an implicit time integration method utilized to compute the dynamic response of a body. Firstly, it applies the structural responses computed at the last time step $n$, i$.$e$.$, displacement $X_{rt,\,i}^n$, velocity $\dot{X}_{rt,\,i}^n$ and acceleration $\ddot{X}_{rt,\,i}^n$, together with the external (fluid) forces and moments at the current time step ($F_{rt,\,i}^{n+1}$) in order to approximate the new displacement vector $X_{rt,\,i}^{n+1}$ \cite{Newmark_1959} (see Eq.\ (\ref{eq:newmark_displacement})). Thereto, the inverse matrix $\left[M_{rt,\,ij}\;a_0+C_{rt,\,i}\,a_1+K_{rt,\,i}\right]^{-1}$ is calculated according to Cramer's rule, which states that the inverse matrix is directly proportional to the adjoint matrix and inversely proportional to the matrix determinant.
\begin{eqnarray}
\nonumber {X_{rt,\,i}^{n+1}}&=&\left[M_{rt,\,ij}\;a_0+C_{rt,\,i}\,a_1+K_{rt,\,i}\right]^{-1}\left(F_{rt,\,i}^{n+1}\right.\\ \nonumber
&+&X_{rt,\,i}^n\,\left[M_{rt,\,ij}\;a_0+C_{rt,\,i}\,a_1\right]\\\nonumber&+&\dot{X}_{rt,\,i}^n\,\left[M_{rt,\,ij}\;a_2+C_{rt,\,i}\,a_4\right]\\
&+&\ddot{X}_{rt,\,i}^n\,\left[M_{rt,\,ij}\;a_3+C_{rt,\,i}\,a_5\right]\left)\right.
\label{eq:newmark_displacement}
\end{eqnarray}
\par The velocity and acceleration vectors at the new time steps $n+1$ are then computed through Eqs.\ (\ref{eq:newmark_velocity}) and (\ref{eq:newmark_acceleration}), respectively \cite{Viets_2013}:
\begin{eqnarray}
\label{eq:newmark_velocity}
\dot{X}_{rt,\,i}^{n+1}&=&a_1\,\left[X_{rt,\,i}^{n+1}-X_{rt,\,i}^n\right]-a_4\,\dot{X}_{rt,\,i}^n-a_5\,\ddot{X}_{rt,\,i}^n, \\
\label{eq:newmark_acceleration}
\ddot{X}_{rt,\,i}^{n+1}&=&a_0\,\left[X_{rt,\,i}^{n+1}-X_{rt,\,i}^n\right]-a_2\,\dot{X}_{rt,\,i}^n-a_3\,\ddot{X}_{rt,\,i}^n.
\end{eqnarray}
\par The coefficients $a_0$ through $a_5$ are calculated according to Eq.\ (\ref{eq:newmark_coefficients}):
\begin{eqnarray}
\label{eq:newmark_coefficients} \nonumber
a_0&=&\frac{1}{\beta\,\Delta t^2},\quad a_1=\frac{\gamma}{\beta\,\Delta t},\quad
a_2=\frac{1}{\beta\,\Delta t},\quad a_3=\frac{1}{2\beta}-1,\\ a_4&=&\frac{\gamma}{\beta}-1,\quad a_5=\frac{\gamma\,\Delta t}{2\beta}-\Delta t.
\end{eqnarray} 
\par The selection of the Newmark parameters $\beta$ and $\gamma$ influences the stability, accuracy and dissipation of the method. The current work utilizes the standard (trapezoidal rule) Newmark method, which is characterized by $\beta=0.25$ and $\gamma=0.5$. This is second-order accurate and unconditionally stable since the stability condition $2\beta\geq \gamma$ is fulfilled \mbox{(see Sieber \cite{Sieber_2002})}. However, this method is characterized by the presence of high frequencies generated by the time discretization scheme (see V\'azquez \cite{Vazquez_2007}). These artificial frequencies can be dissipated either with the utilization of a Newmark parameter $\gamma>0.5$, which leads to accuracy reduction (see Münsch \cite{Muensch_2015}), or with the utilization of other methods based on minor changes of the Newmark technique, such as the generalized-$\alpha$ \mbox{method \cite{Vazquez_2007}.}

\subsubsection{Generalized-$\alpha$ method}
\label{sec:generalized_alpha}
\par The generalized-$\alpha$ method is an enhancement of the Newmark approach proposed by Chung and Hulbert \cite{Chung_1993}, which dissipates the numerically generated high-frequencies while minimizing the unwanted low frequency dissipation and maintaining the order of accuracy. 
\par Firstly, the variables $X_{rt,\,i}$, $\dot{X}_{rt,\,i}$, $\ddot{X}_{rt,\,i}$ and $F_{rt,\,i}$ are replaced by an under-relaxation according to Eqs.\ (\ref{eq:gen_alpha_under_relaxation_disp}) to (\ref{eq:gen_alpha_under_relaxation_accel}). In the present work, the under-relaxation factor for the forces $\delta_{f}$ is only applied if the simulation diverges for $\delta_f=1$. 
\begin{eqnarray}
\label{eq:gen_alpha_under_relaxation_disp}
X_{rt,\,i}^u&=&(1-\alpha_f^s)\,\,\,X_{rt,\,i}^n+\alpha_f^s \,\,X_{rt,\,i}^{n+1}\\
\label{eq:gen_alpha_under_relaxation_vel}
\dot{X}_{rt,\,i}^u&=&(1-\alpha_f^s)\,\,\,\dot{X}_{rt,\,i}^n+\alpha_f^s \,\,\dot{X}_{rt,\,i}^{n+1}\\
\label{eq:gen_alpha_under_relaxation_accel}
\ddot{X}_{rt,\,i}^u&=&(1-\alpha_m^s)\,\,\ddot{X}_{rt,\,i}^n+\alpha_m^s \,\ddot{X}_{rt,\,i}^{n+1}\\
\label{eq:gen_alpha_under_relaxation_forces}
F_{rt,\,i}^u&=&(1-\delta_{f})\;\;F_{rt,\,i}^n\,\,+\delta_{f}\,\,\,F_{rt,\,i}^{n+1}
\end{eqnarray} 
\par The generalized-$\alpha$ parameters, i$.$e$.$, $\alpha_f^s$ and $\alpha_m^s$, are calculated according to Eq. (\ref{eq:parameters_generalized_alpha_low_frequency}) for optimal low frequency dissipation \cite{Vazquez_2007} and according to Eq.\ (\ref{parameter_generalized_alpha_high_frequency}) for a second-order accurate method with dissipation of the high frequencies. When $\rho_\infty^s=1$, a minimal damping of the low frequencies is present. However, these frequencies are still damped and therefore are not directly comparable to the non dissipated low frequencies of the standard Newmark method. This trapezoidal rule Newmark method is achieved only if no under-relaxation of the displacement, velocity and acceleration vectors are considered, i$.$e$.$, \mbox{$\alpha_f^s=\alpha_m^s=1$}.
\begin{equation}
\label{eq:parameters_generalized_alpha_low_frequency}
\alpha_f^s=\frac{1}{1+\rho_\infty^s},\quad \alpha_m^s=\frac{2-\rho_\infty^s}{1+\rho_\infty^s}, \quad \rho_\infty^s\in[0,1]. 
\end{equation}
\begin{equation}
\label{parameter_generalized_alpha_high_frequency}
\beta=\frac{1}{4}\left(1+\alpha_m^s-\alpha_f^s\right)^2,\quad\gamma=\frac{1}{2}+\alpha_m^s-\alpha_f^s.
\end{equation}
\par The under-relaxed structural responses and the fluid forces calculated according to \mbox{Eqs.\ (\ref{eq:gen_alpha_under_relaxation_disp}) to (\ref{eq:gen_alpha_under_relaxation_forces})} are then applied to the under-relaxed equation of motion, i$.$e$.$,\break \mbox{Eq.\ (\ref{eq:eq_motion_underrelaxed})}, resulting in the generalized-$\alpha$ main equation \cite{Viets_2013}. This approximates the unknown displacement at the current time step $X_{rt,\,i}^{n+1}$ as a function of the known displacement, velocity and acceleration vectors at the last time step $n$, as well as of the under-relaxed fluid forces and moments (see \mbox{Eq. (\ref{eq:displacement_generalized_alpha}))}.
\begin{eqnarray}
\label{eq:eq_motion_underrelaxed}
F_{rt,\,i}^u&=&M_{rt,\,ij}\,\ddot{X}_{rt,\,i}^u+C_{rt,\,i}\,\dot{X}_{rt,\,i}^u+K_{rt,\,i}\,X_{rt,\,i}^u \\\nonumber
X_{rt,\,i}^{n+1}&=&\left[M_{rt,\,ij}\;b_0+C_{rt,\,i}\,b_1+\alpha_f^s\,K_{rt,\,i}\right]^{-1}\left.\right(F_{rt,\,i}^{u}\\\nonumber
&+&X_{rt,\,i}^n\,[M_{rt,\,ij}\;b_0+C_{rt,\,i}\,b_1+K_{rt,\,i}\,(\alpha_f^s-1)]\\\nonumber
&+&\dot{X}_{rt,\,i}^n\,\left[M_{rt,\,ij}\;b_2+C_{rt,\,i}\,b_4\right]\\
&+&\ddot{X}_{rt,\,i}^n\,[M_{rt,\,ij}\;b_3+C_{rt,\,i}\,b_5]\left)\right.
\label{eq:displacement_generalized_alpha} 
\end{eqnarray}
\par The coefficients $b_0$ to $b_5$ are calculated according to Eq.\ (\ref{eq:gen_alpha_coefficients}):
\begin{eqnarray}
\label{eq:gen_alpha_coefficients} \nonumber
b_0&=&\frac{\alpha_m^s}{\beta\,\Delta t^2},\quad b_1=\frac{\gamma\,\alpha_f^s}{\beta\,\Delta t},\quad
b_2=\frac{\alpha_m^s}{\beta\,\Delta t},\quad b_3=\frac{\alpha_m^s}{2\beta}-1,\\ b_4&=&\frac{\gamma\, \alpha_f^s}{\beta}-1,\quad b_5=\Delta t \,\alpha_f^s\left(\frac{\gamma}{2\beta}-1\right).
\end{eqnarray} 
\par Aiming at the solution of Eq.\ (\ref{eq:displacement_generalized_alpha}), the system is divided into a translational and a rotational part (see Sections \ref{subsubsec:translation} and \ref{subsubsec:rotation}). Since the considered airfoil is rigid, the governing equations (see Eq.\ (\ref{eq:fsirigidmvt})) are linear and therefore can be directly solved. Moreover, when the generalized-$\alpha$ parameters are set to \mbox{$\alpha_f^s=\alpha_m^s=1$}, the standard (trapezoidal rule) Newmark method is reproduced. 
\par The current work applies the generalized-$\alpha$ method with the parameters set to\break \mbox{$\alpha_f^s=\alpha_m^s=1$}, which actually represent the standard Newmark method, in order to solve the fluid-structure interaction of a NACA0012 airfoil.
\subsection{Translation}
\label{subsubsec:translation}
\par The three first rows of Eq.\ (\ref{eq:fsirigidmvt}) represent the translational motion of a rigid body, as shown in Eq.\ (\ref{eq:fsirigidmvt_translation}): 
\begin{equation}
\label{eq:fsirigidmvt_translation}
	\underbrace{\begin{bmatrix}
		m  & 0 & 0\\
		0 & m & 0\\
		0 & 0 & m\\
		\end{bmatrix}}_{M_{t,\,ij}}
	\underbrace{\begin{bmatrix}
		\ddot{X_1}\\
		\ddot{X_2}\\
		\ddot{X_3}\\
		\end{bmatrix}}_{\ddot{X}_{t,\,i}}
	+I_{ij}
	\underbrace{\begin{bmatrix}
		c_{l,\,1} \\
		c_{l,\,2}\\
		c_{l,\,3}\\
		\end{bmatrix}^T}_{C_{t,\,i}}
	\underbrace{\begin{bmatrix}
		\dot{X_1}\\
		\dot{X_2}\\
		\dot{X_3}\\
		\end{bmatrix}}_{\dot{X}_{t,\,i}}
	+I_{ij}
	\underbrace{\begin{bmatrix}
		k_{l,\,1} \\
		k_{l,\,2}\\
		k_{l,\,3}\\
		\end{bmatrix}^T}_{K_{t,\,i}}
	\underbrace{\begin{bmatrix}
		{X_1}\\
		{X_2}\\
		{X_3}\\
		\end{bmatrix}}_{X_{t,\,i}}
=
	\underbrace{\begin{bmatrix}
		F_{ext,\,1}\\
		F_{ext,\,2}\\
		F_{ext,\,3}\\
		\end{bmatrix}.}_{F_{t,\,i}}\\
\end{equation}
\par The solution of this uncoupled system at the current time step, i$.$e$.$, $X_{t,\,i}^{n+1}$, is achieved through the utilization of the \mbox{generalized-$\alpha$} method (see Eq.\ (\ref{eq:displacement_generalized_alpha})) for \mbox{$\alpha_f^s=\alpha_m^s=1$}, which actually corresponds to the standard Newmark approach. This requires only the inversion of the $\left[M_{t,\,ij}\;b_0+C_{t,\,i}\,b_1+\alpha_f^s\,K_{t,\,i}(\alpha_f^s-1)\right]$ term according to Cramer's rule (see Viets \cite{Viets_2013}). The calculated translational displacement vector $X_{t,\,i}^{n+1}$ can then be directly added to the initial position vector $r_{start,\,i}$ of each element node located on the body surface in order to assist the calculation of the current position vector $r_{new,\,i}$ \mbox{(see Section \ref{subsec:body_disp})}.
\subsection{Rotation}
\label{subsubsec:rotation}
\par The rotational motion of a rigid body is represented by the last three rows of \mbox{Eq.\ (\ref{eq:fsirigidmvt})}, i$.$e$.$, by Eq.\ (\ref{eq:fsirigidmvt_rotation}):
\begin{equation}
\label{eq:fsirigidmvt_rotation}
	\underbrace{\begin{bmatrix}
		J_{11} & J_{12} & J_{13}\\ 
		J_{21} & J_{22} & J_{23}\\
		J_{31} & J_{32} & J_{33}
		\end{bmatrix}}_{M_{r,\,ij}}
	\underbrace{\begin{bmatrix}
		\ddot{\varphi_1}\\
		\ddot{\varphi_2}\\
		\ddot{\varphi_3}
		\end{bmatrix}}_{\ddot{X}_{r,\,i}}
	+I_{ij}
	\underbrace{\begin{bmatrix}
		c_{t,\,1}\\
		c_{t,\,2}\\
		c_{t,\,3}
		\end{bmatrix}^T}_{C_{r,\,i}}
	\underbrace{\begin{bmatrix}
		\dot{\varphi_1}\\
		\dot{\varphi_2}\\
		\dot{\varphi_3}
		\end{bmatrix}}_{\dot{X}_{r,\,i}}
	+I_{ij}
	\underbrace{\begin{bmatrix}
		k_{t,\,1}\\
		k_{t,\,2}\\
		k_{t,\,3} 
		\end{bmatrix}^T}_{K_{r,\,i}}
	\underbrace{\begin{bmatrix}
		{\varphi_1}\\
		{\varphi_2}\\
		{\varphi_3}
		\end{bmatrix}}_{X_{r,\,i}}
=
	\underbrace{\begin{bmatrix}
		M_{ext,\,1}\\
		M_{ext,\,2}\\
		M_{ext,\,3}\\
		\end{bmatrix}}_{F_{r,\,i}}\\
\end{equation}
\par Different from the translational motion, the rotation of a rigid body is frequently characterized by a coupled system due to the mass moment of inertia tensor $M_{r,\,ij}$. In this work, however, this system is uncoupled, since the body-fixed Cartesian coordinate system is aligned with mutually orthogonal principal axes (see Section \ref{subsec:moment_inertia}).
\par The angles of rotation around the three Cartesian coordinate axes at the current time step $X_{r,\,i}^{n+1}$, which are calculated according to the standard Newmark method, indicate a rotation of the body-fixed local Cartesian coordinate system $r_{CCS_{l},\,i}$ and therefore cannot be directly added to the initial position vector $r_{start,\,i}$. In order to allow this addition, a rotated position vector $r_{rot,\,i}$ is required. This is calculated based on quaternions, which are utilized instead of Euler angles in order to save memory as well to avoid problems with the existing singularities of the Euler angles. 
\par Quaternions $Q$ are four-dimensional vectors with one real and three imaginary components, i$.$e$.$, \mbox{$Q=q_1+i\,q_2+j\,q_3+k\,q_4$}, which represent rotations without the utilization of Euler angles \mbox{(see De Nayer \cite{De_Nayer_2008}}). The utilized quaternions and the rotated position vector $r_{rot,\,i}$ are calculated according to Eqs.\ (\ref{eq:norm_angular_disp}) to (\ref{eq:rotated_position_vector}). Firstly, the norm of the rotational displacement vector $|X_{r,\,i}|$ is calculated. Secondly, the direction vector $v_i$ is determined, which is necessary for the calculation of the quaternion components $q_i$. Finally, the rotated position vector $r_{rot,\,i}$ for the element nodes located on the surface of the body can be computed based on the initial position $r_{start,\,i}$ and the position of the undeformed local Cartesian coordinate system $r_{CCS_l,\,i}$, according to \mbox{Viets \cite{Viets_2013}}.
\begin{eqnarray}
\label{eq:norm_angular_disp}
|X_{r,\,i}|&=&\sqrt{X_{r,\,1}^2+X_{r,\,2}^2+X_{r,\,3}^2}\\\nonumber \\
v_i&=&\frac{1}{|X_{r,\,i}|}X_{r,\,i} \\\nonumber \\\nonumber 
q_1&=&\cos\left(\frac{|X_{r,\,i}|}{2}\right)\\ \nonumber
q_2&=&v_1\,\sin\left(\frac{|X_{r,\,i}|}{2}\right)\\\nonumber
q_3&=&v_2\,\sin\left(\frac{|X_{r,\,i}|}{2}\right)\\
q_4&=&v_3\,\sin\left(\frac{|X_{r,\,i}|}{2}\right)
\end{eqnarray}   
\begin{eqnarray}
\nonumber
r_{rot,\,1}&=&(r_{start,\,1}-r_{CCS_l,\,1})(1-2q_3^2-2q_4^2)+(r_{start,\,2}-r_{CCS_l,\,2})(2q_2q_3-2q_1q_4)\\\nonumber
&+&(r_{start,\,3}-r_{CCS_l,\,3})(2q_2q_4+2q_1q_3)\\ \nonumber
r_{rot,\,2}&=&(r_{start,\,1}-r_{CCS_l,\,1})(2q_2q_3+2q_1q_4)+(r_{start,\,2}-r_{CCS_l,\,2})(1-2q_2^2-2q_4^2)\\\nonumber
&+&(r_{start,\,3}-r_{CCS_l,\,3})(2q_3q_4-2q_1q_2)\\ \nonumber
r_{rot,\,3}&=&(r_{start,\,1}-r_{CCS_l,\,1})(2q_2q_4-2q_1q_3)+(r_{start,\,2}-r_{CCS_l,\,2})(2q_3q_4+2q_1q_2)\\
&+&(r_{start,\,3}-r_{CCS_l,\,3})(1-2q_2^2+2q_3^2)\label{eq:rotated_position_vector}
\end{eqnarray}

\subsection{Body displacement}
\label{subsec:body_disp}
\par
\begin{figure}[H]
	\centering
	\includegraphics[scale=0.58,draft=\drafttype]{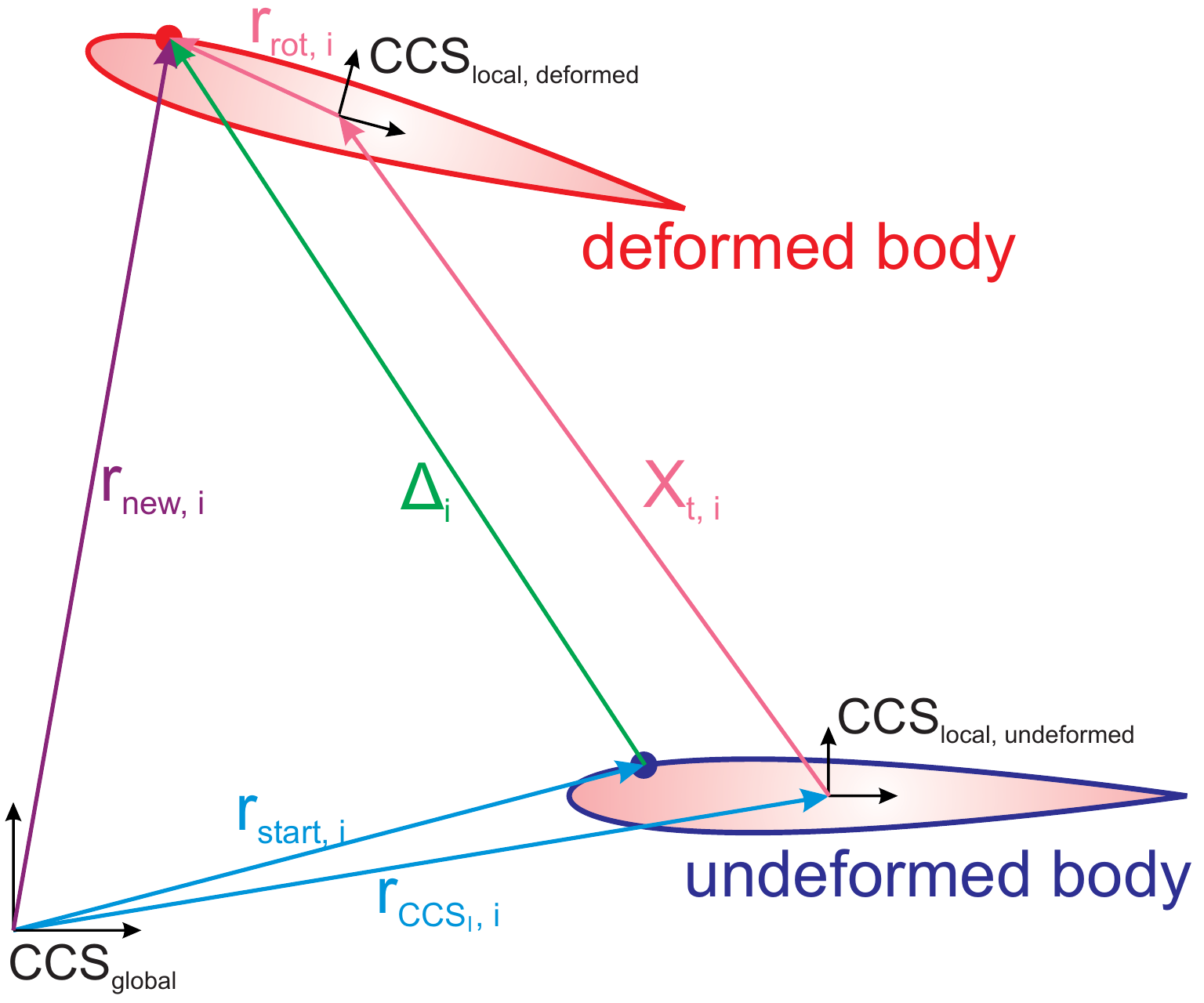}
	\caption{\label{fig:body_disp}Vectors of the coupled rotation-translation system.}
\end{figure}

 The body displacement $\Delta_i$ (see Fig.\ \ref{fig:body_disp}), i$.$e$.$, the difference vector between the deformed and undeformed (initial) states (see Eq.\ (\ref{eq:body_disp_simple})), is generated by the coupled rotational and translational motions. This is calculated based on a vectorial sum of the translational displacement $X_{t,\,i}$, the rotated position vector $r_{rot,\,i}$, the location of the undeformed body-fixed local Cartesian coordinate system (CCS) $r_{CCS_l,\,i}$ and the location of the considered element node in relation to the global coordinate system $r_{start,\,i}$, i$.$e$.$, the position vector in the initial (undeformed) state, as stated in Eq. (\ref{eq:total_body_displacement}).
\par This variable represents the structural response of the body to the external forces and moments and therefore is calculated only for the nodes located on the structure surface, i$.$e$.$, on the interface between CFD and CSD subproblems. As the displacement of all nodes located on this interface are calculated, these are transfered to the CFD solver by the FSI coupling algorithm, finalizing the routine executed by the rigid movement solver for each time step.
\begin{eqnarray}
\label{eq:body_disp_simple}
\Delta_i&=&r_{new,\,i}-r_{start,\,i} \\
\label{eq:total_body_displacement}
\Delta_i&=&r_{CCS_l,\,i}+X_{t,\,i}+r_{rot,\,i}-r_{start,\,i}
\end{eqnarray}
\squeezeup\squeezeup\squeezeup

\markboth{CHAPTER 1.$\quad$NUM. MET.}{1.1$\quad$COUP. BETWEEN FLUID AND STRUCTURE}
\section{Coupling between fluid and structure}\markboth{CHAPTER 1.$\quad$NUM. MET.}{1.1$\quad$COUP. BETWEEN FLUID AND STRUCTURE}
\label{sec:coupling_fluid_structure}
\par The fluid flow around a rigid body generates pressure and shear stress forces, which are responsible for the displacement of the structure. This displacement, in turn, leads to a change of the fluid domain. Hence, both flow and structure act on each other, generating a coupled fluid-structure problem, i$.$e$.$, a fluid-structure interaction problem.
\par Fluid-structure interaction problems can be solved according to a monolithic or a partitioned approach. The former is based on the simultaneous solution of the fluid and structural subproblems and is characterized by the stability and specificity (the applicable numerical methods are restricted). Moreover, it requires massive programming efforts and utilizes the same spatial discretization method (for instance FVM, FEM, etc$.$) for both the fluid and the structural domains \cite{Muensch_2015}. Therefore, the current work utilizes the partitioned approach, which combines well-validated solvers for fluid and structure and requires programming efforts only for subroutines in order to exchange information between both softwares. Furthermore, this technique is characterized by its generality and the presence of convergence problems, which restricts the choice of the utilized time-step \cite{Breuer_2012, Sieber_2002}.\break Nevertheless, the convergence properties can be improved by exchanging data at the interface more than once per time-step, as performed with predictor-corrector schemes (see Breuer et al. \cite{Breuer_2012}) or with strongly coupled algorithms.
\par The implicit coupling algorithm, i$.$e$.$, strong coupling, assures the energy conservation and the dynamic equilibrium between flow and structure at every time step due to the usage of more than one FSI sub-iteration ($n_{FSI}>1$) \cite{Muensch_2015}. However, for aeroelastic problems, which usually do not experience convergence problems due to the negligible added mass effect \cite{Causin_2005} (see Section \ref{subsec:added_mass_effect}), this approach leads only to a higher required computational time. Therefore, the best compromise between accuracy and required computational effort is achieved by loose (explicit) coupled algorithms for such cases. For the comparison of the loose and strong coupled algorithms of the FSI between flow and NACA0012 airfoil see Section \ref{subsec:investigation_coupling_algorithm}
.
\par
\begin{figure}[H]
	\centering
	\centering
	\includegraphics[scale=0.75,draft=\drafttype]{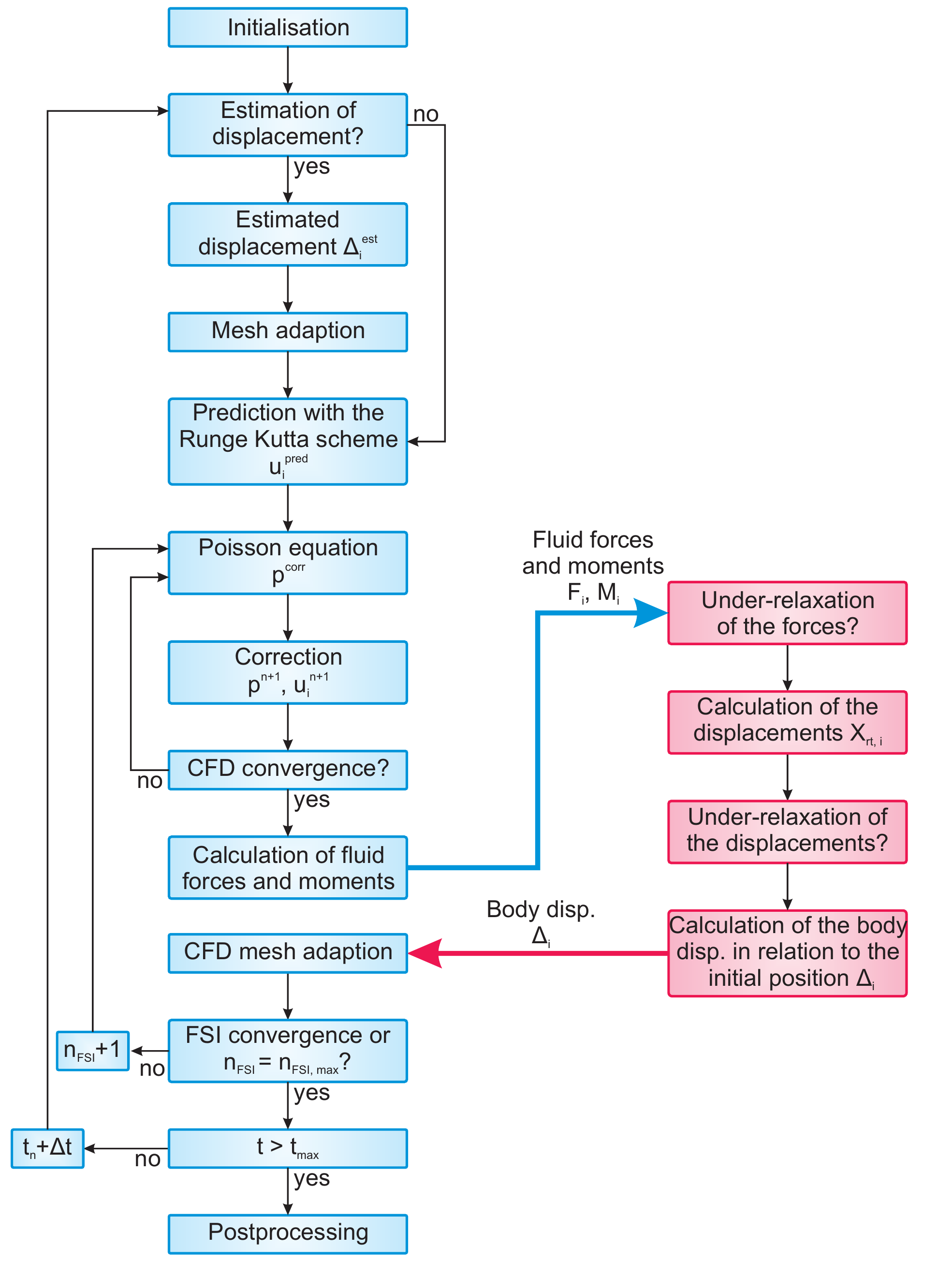}
	\caption{\label{fig:FSI_coupling}Flow chart of the utilized coupling between CFD and CSD.}
\end{figure}

 The applied partitioned approach is illustrated in Fig.\ \ref{fig:FSI_coupling}, regarding that the applied setup is thoroughly described in Chapter \ref{chap:FSI_flow_NACA0012}. It solves the fluid subproblem with the in-house software FASTEST-3D, while the structural subproblem is computed by the in-house rigid movement solver implemented by \mbox{Viets \cite{Viets_2013}}.
\par Firstly, the solver is initialized. An estimation of the displacement is performed\break (see Section \ref{subsec:estimation_disp}) and the mesh is adapted (as per \mbox{Section \ref{subsec:mesh_adaptation})}. Secondly, FASTEST-3D approximates the solution of the fluid domain based on a FVM (according to\break \mbox{Section \ref{subsec:spatial_discretization})} and a predictor-corrector scheme (see \mbox{Section \ref{subsec:temporal_discretization})}. When the CFD solution converges (as stated in Section \ref{subsec:CFD_convergence_criterion}), the forces and moments acting on the body are calculated (see Section \ref{subsec:fluid_forces}). This information is subsequently exchanged with the CSD solver, which does not under-relax the loads in the current work. The translational and rotational displacements $X_{rt,\,i}$ at the current time step are then computed according to a standard Newmark time integration scheme (see Sections \ref{subsec:temporal_discretization_CSD}, \ref{subsubsec:translation} and \ref{subsubsec:rotation}). Afterwards, the body displacement vector in relation to the initial position (undeformed state) $\Delta_i$ is calculated (as stated in \mbox{Section \ref{subsec:body_disp})} and transferred to FASTEST-3D, which performs the adaption of the mesh (according to \mbox{Section \ref{subsec:mesh_adaptation})}. If the flow and structure reach a dynamic equilibrium (see \mbox{Section \ref{subsec:FSI_convergence_criterion})} or if the predetermined maximal number of FSI sub-iterations $n_{FSI}$ is exceeded, the calculation of the next time step is started. Finally, when the maximal simulation time $t_{max}$ is reached, the results are post-processed by the software TECPLOT 360 and subsequently analyzed.

\subsection{Estimation of displacement}
\label{subsec:estimation_disp}
\par A linear extrapolation of the translational and rotational displacements $X_{rt,\,i}$ is utilized in order to reduce the number of FSI sub-iterations $n_{FSI}$ required to achieve a dynamic equilibrium. Consequently, it accelerates the convergence speed while decreasing the computational time (see Glück \cite{Glueck_2002}). The applied first-order implicit extrapolation method (Euler method) is calculated according to Eq.\ (\ref{eq:estimation_dips}): 
\begin{equation}
\label{eq:estimation_dips}
X_{rt,\,i}^{est,\,n+1}=2X_{rt,\,i}^n-X_{rt,\,i}^{n-1}.
\end{equation}
\par A second-order extrapolation is not considered, since De Nayer and Breuer \cite{DeNayer_2014} observed that it does not improve the convergence properties for small time step sizes. Therefore, a study of the benefits acquired due to the application of only a linear extrapolation is carried out and presented in Section \ref{subsec:investigation_estimation_displacement}.

\subsection{Under-relaxation of forces and displacements}
\label{subsec:underrelaxation_forces_disp}
\par The under-relaxation of the forces and displacements is applied in order to improve the convergence properties of the FSI solver. Specially when large time step sizes are utilized, an over-prediction of the fluid forces is possible \cite{Muensch_2015}. Therefore, the application of  under-relaxation parameters is required in order to guarantee the convergence. 
\par Since the present work is characterized by the utilization of small time step sizes ($\Delta t \leq \mathcal{O}(10^{-5})\,s$), this technique is only utilized if the simulation diverges during the computation of FSI problem. The under-relaxed displacements and loads are then calculated according to Eqs.\ (\ref{eq:gen_alpha_under_relaxation_disp}) and (\ref{eq:gen_alpha_under_relaxation_forces}).

\subsection{FSI convergence criterion}
\label{subsec:FSI_convergence_criterion}
\par The residuum of displacement is utilized as the convergence criterion for the FSI solver (see Breuer \cite{Breuer_2008} and Viets \cite{Viets_2013}). This is calculated according to Eq.\ (\ref{eq:residuum_disp}):
\begin{eqnarray}
\label{eq:residuum_disp}
\epsilon_{FSI,\,disp}&=&\frac{\sqrt{\sum_{i=1}^{6}(X_{rt,\,i}^{n,\,n_{FSI}}-X_{rt,\,i}^{n,\,n_{FSI}-1})^2}}{\sqrt{\sum_{i=1}^{6}(X_{rt,\,i}^{n,\,n_{FSI}}-X_{rt,\,i}^{n-1})^2}}.
\end{eqnarray}
\par The convergence criterion applied in the current work is described in Section \ref{sec:FSI_setup}.

\chapter{NACA0012 airfoil with two degrees of freedom in turbulent flows: Geometry and setup}
\label{chap:FSI_flow_NACA0012}

\par The NACA0012 airfoil is a symmetric profile usually utilized for stability purposes in airplanes, such as in horizontal and vertical stabilizers. Therefore, it is commonly investigated by numerical and experimental studies and thus a vast literature is available, enabling the establishment of the model ability to reproduce the physics, i$.$e$.$, the model validation.
\par The current work is a continuation of the NACA0012 airfoil investigations performed by Streher \cite{Streher_2017}, which aimed at the investigation of the flow characteristics for a fixed airfoil at different angles of attack and submitted to a Reynolds number of $Re=100{,}000$. The wall-resolved LES simulations utilizing the Smagorinsky model were validated through a comparison with numerical and experimental results available in the literature. The present work, then, expands this study in order to perform fluid-structure interaction simulations.

\section{Test section geometry}\markboth{CHAPTER 2.$\quad$FREE RIG. NACA0012 AIR. IN TURB. FLOWS}{2.1$\quad$TEST SEC. GEO.}
\label{sec:geometry}
\par The NACA0012 profile stands for the National Advisory Committee for Aeronautics four digit series airfoil profiles. The first two digits, i.e$.$, $00$, represent a symmetrical airfoil and the last two stand for a profile with a maximal thickness of twelve percent of the chord length $c$, located at thirty percent of the chord \cite{Jacobs_1935}. 
\par The modeled airfoil has a chord length of $c=0.1\,m$ and a sharp trailing-edge. This is described by NASA \cite{NASA_2016} through Eq.\ (\ref{eq:NACA0012}). The origin of the global coordinate system is located at the leading edge. $X_{N,\,2}$ and $X_{N,\,1}$ represent the vertical and horizontal coordinates of the airfoil, respectively. 
\begin{eqnarray}
\nonumber
\label{eq:NACA0012}
X_{N,\,2}&=&\pm 0.59468918c\left[0.29822277\sqrt{\frac{X_{N,\,1}}{c}}-0.12712523\frac{X_{N,\,1}}{c}-0.35790791\left(\frac{X_{N,\,1}}{c}\right)^2\right.\\
&&+0.29198497\left(\frac{X_{N,\,1}}{c}\right)^3\left.-0.10517461\left(\frac{X_{N,\,1}}{c}\right)^4\right]
\end{eqnarray}
\par The two-dimensional geometry of the computational fluid domain is established with the help of artificial boundaries, which are carefully chosen according to the work of Almutari \cite{Almutari_2010}. This presents the results of various wall-resolved large-eddy simulations for a fixed rigid NACA0012 profile at diverse configurations of Reynolds numbers and angles of attack, which are validated by a-posteriori studies against the results of DNS simulations achieved by \mbox{Jones et al. \cite{Jones_2008}}, as well as by the experimental results of Rinoiei and Takemura \cite{Takemura_2004}. Moreover, this test section geometry is already tested with the in-house software FASTEST-3D for the case of wall-resolved LES of a fixed rigid NACA0012 at a Reynolds number of $Re=100{,}000$ (see Streher \cite{Streher_2017}), delivering satisfactory results.
\par
\begin{figure}[H]
	\centering
	\centering
	\includegraphics[scale=0.85,draft=\drafttype]{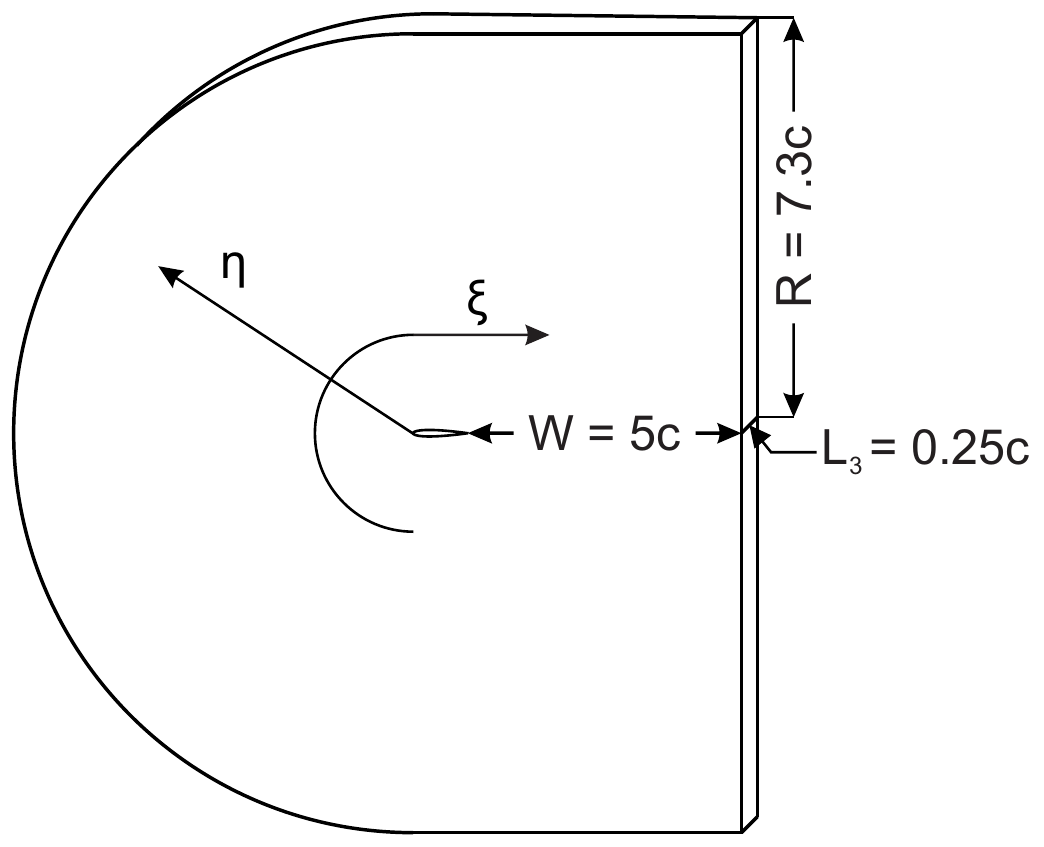}
	\caption{\label{fig:fluid_domain_3d}Fluid domain of the NACA0012 airfoil profile.}
\end{figure}

 Figure \ref{fig:fluid_domain_3d} illustrates the simulated fluid domain and the curvilinear coordinates $\eta$ and $\xi$ utilized in order to simplify the description of the meshes. The test section is characterized by the wake length $W=5\,c$ and the domain radius $R=7.3\,c$. Due to a contradiction between the span-wise lengths suggested by Almutari \cite{Almutari_2010} and by \mbox{Schmidt \cite{Schmidt_2016}}, i$.$e$.$, respectively $L_3=0.5\,c$ and \mbox{$L_3=0.25\,c$}, the influence of this parameter is investigated and discussed in Section \ref{sec:study_span_wise_length}. Since the best compromise between accuracy and computational time is achieved for the narrowest span-wise domain, i$.$e$.$, $L_3=0.25\,c$, this is utilized for the further investigations. 

\section{Computational meshes}\markboth{CHAPTER 2.$\quad$FREE RIG. NACA0012 IN TURB. FLOWS}{2.2$\quad$COMPT. MESHES}
\label{sec:meshes}
\par
\begin{figure}[H]
	\centering
	\subfigure[Overall view.]{\includegraphics[width=0.495\textwidth]{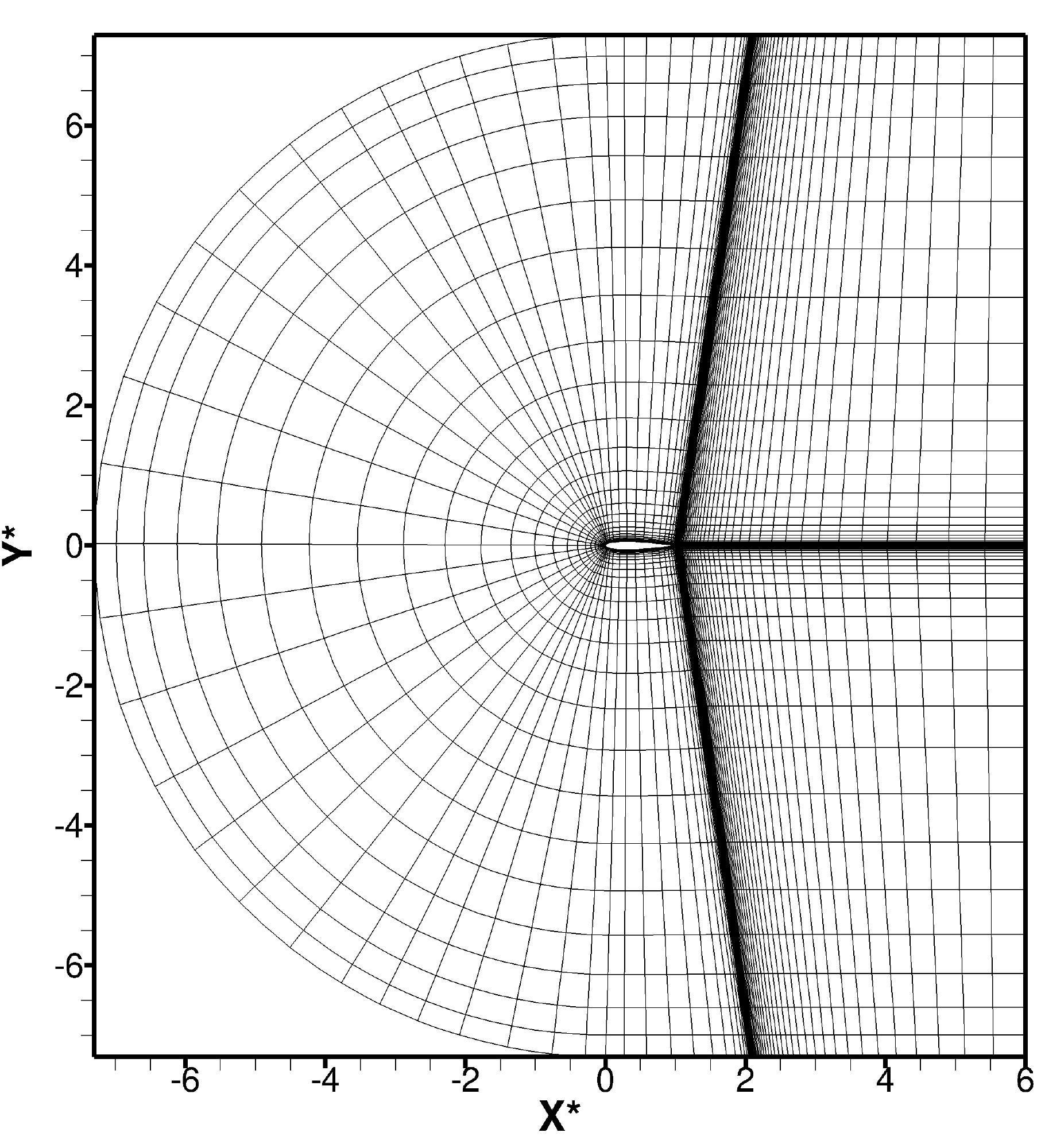}\label{fig:mesh}}\hfill
	\subfigure[Mesh near the airfoil.]{\includegraphics[width=0.495\textwidth]{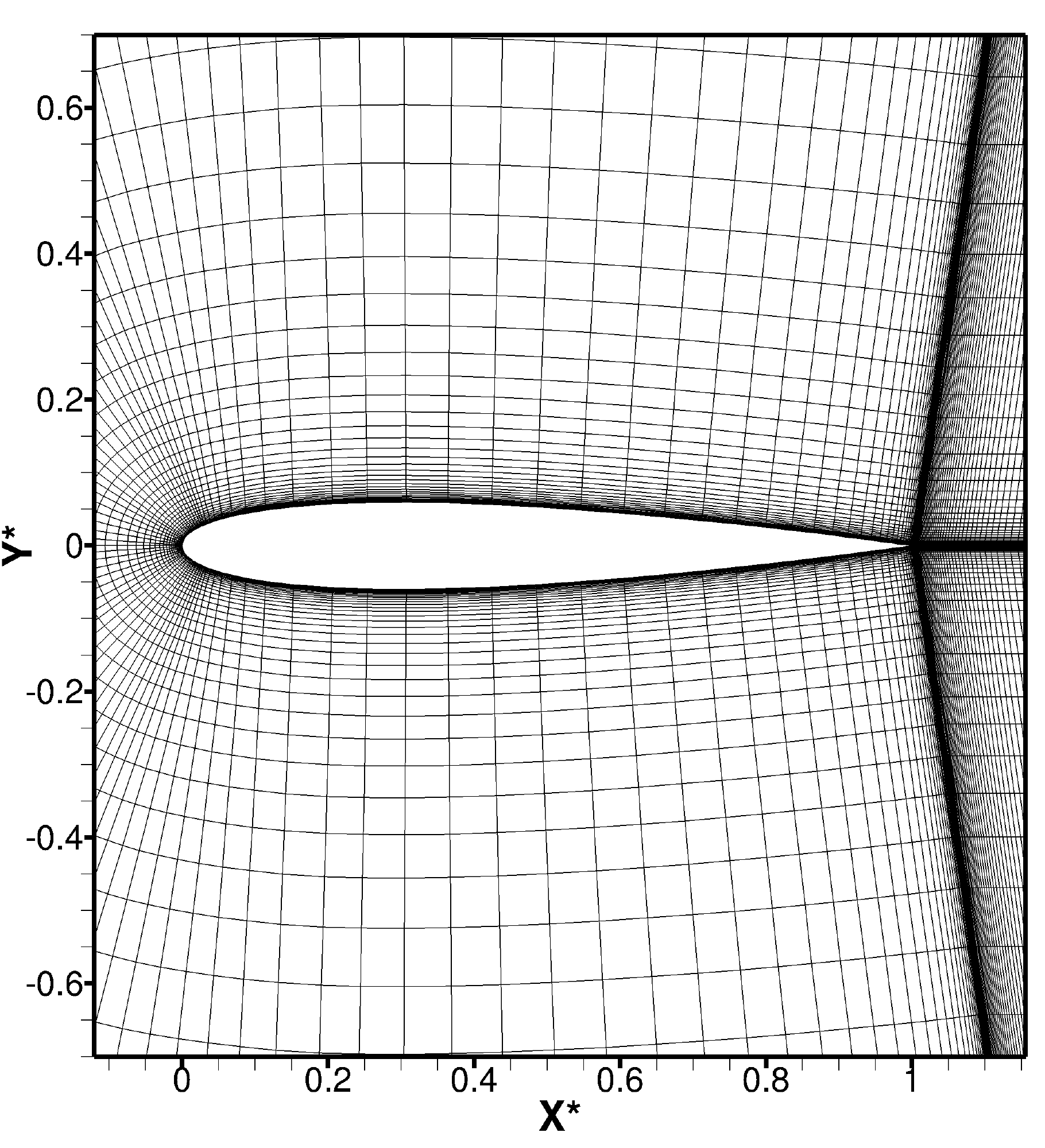} \label{fig:mesh_zoom}}\hfill
	\caption{Mesh $m{-}L_3^{min}{-}y^{+}_{max}$.}
	\label{fig:coarse_grid}
\end{figure}

 The computational fluid domain is discretized into control volumes by the software ANSYS ICEM CFD. Three different block-structured hexahedral C-grids with varying numbers of nodes $N$ and first cell heights $\Delta y^{\,first\;cell}$ are generated. \mbox{Table \ref{table:mesh_parametes}} indicates the different grid parameters. Figure \ref{fig:coarse_grid} illustrates the $m{-}L_3^{min}{-}y^{+}_{max}$ grid \mbox{(see Table \ref{table:mesh_parametes}),} regarding that only one out of two mesh lines is shown in the $\xi$ and the $\eta$ axes in order to achieve a good quality image.
\par $N_W$, $N_R$, $N_{L_3}$ and $N_{SS}$ represent the number of nodes utilized to discretize the edges of the wake $W$, domain radius $R$, span-wise length $L_3$ and suction side of the airfoil, respectively (see Fig.\ \ref{fig:fluid_domain_3d}). Since the NACA0012 airfoil is symmetric, an equal number of nodes is applied at the pressure and suction side edges.
\par The generated meshes are composed of twelve geometrical blocks characterized by smooth transitions at the interfaces. The mesh quality is evaluated according to the determinant of the Jacobi-Matrix and the internal angles of every control volume, regarding that target values of respectively $1$ and $90^\circ$ are aimed at in order to accomplish a mesh composed of only orthogonal cells with a perfect shape. Since all cells of the generated grids have determinants of the Jacobi-Matrix greater than $0.85$ and internal angles greater than $81^\circ$, these are considered high-quality meshes.
\begin{table}[H]
	\centering
	\begin{tabular}{p{2.5cm} p{2.3cm} p{3.5cm} p{1.7cm} p{0.5cm} p{0.5cm} p{0.5cm} p{0.5cm}}
		\hline
		\multicolumn{1}{c}{\multirow{2}{*}{\centering{\bf{Mesh}}}} & \centering{\bf{Span-wise length $L_3$}} & \centering{\bf{First cell height $\Delta y^{\,first\;cell}$ (m)}} & \centering{\bf{Control volumes}} & \multicolumn{1}{c}{\multirow{2}{*}{\centering{\bf{$N_W$}}}} & \multicolumn{1}{c}{\multirow{2}{*}{\centering{\bf{$N_R$}}}} & \multicolumn{1}{c}{\multirow{2}{*}{\centering{\bf{$N_{L_3}$}}}} & \multicolumn{1}{c}{\multirow{2}{*}{\centering{\bf{$N_{SS}$}}}} \tabularnewline 
		\hline
		\centering{$m{-}L_3^{min}{-}y^{+}_{min}$} & \centering{0.25\,c}& \centering{$1.8\cdot10^{-5}$} & \centering{1{,}065{,}600} & \centering{91} & \centering{61} & \centering{61} & \centering{63} \tabularnewline
		\centering{$m{-}L_3^{min}{-}y^{+}_{max}$} & \centering{0.25\,c}& \centering{$5.0\cdot10^{-5}$} & \centering{1{,}065{,}600} & \centering{91} & \centering{61} & \centering{61} & \centering{63} \tabularnewline
		\centering{$m{-}L_3^{max}{-}y^{+}_{min}$} & \centering{0.50\,c}& \centering{$1.8\cdot10^{-5}$} & \centering{2{,}131{,}200} & \centering{91} & \centering{61} & \centering{121} & \centering{63} \tabularnewline
		\hline
	\end{tabular}
	\caption{\label{table:mesh_parametes}Parameters applied for the grid generation.}
\end{table}
\par Mesh independence studies are carried out for all three grids in \mbox{Sections \ref{sec:study_span_wise_length} and \ref{sec:investigation_first_cell_height}}. These are based on the analysis of the aerodynamic coefficients, as well as  of the dimensionless wall distance. They assure that even the $m{-}L_3^{min}{-}y^{+}_{max}$ mesh, which requires the smallest computational effort, does not affect the results. Therefore, the $m{-}L_3^{min}{-}y^{+}_{max}$ grid is further applied for the FSI test cases.

\section{CFD setup}\markboth{CHAPTER 2.$\quad$FSI BETWEEN FLOW AND NACA0012 AIR.}{2.3$\quad$CFD SETUP}
\label{sec:CFD_setup}
\par The numerical methodologies described in Section \ref{sec:CFD} are applied in conjunction with the following CFD setup in order to solve the computational problem. The fluid and simulation parameters are summarized in Tables \ref{table:flow_parameters} and \ref{table:simulation_parameters}. The temperature is presumed to be constant and the fluid is incompressible. The inflow velocities in the $x_1$, $x_2$ and $x_3$ directions are $u_{in,\,1}=4.47\,\text{m}{\cdot}\text{s}^{-1}$ and $u_{in,\,2}=u_{in,\,3}=0\,\text{m}{\cdot}\text{s}^{-1}$, respectively. The Reynolds number is calculated using the free-stream conditions and the airfoil chord length, i$.$e$.$, $c=0.1\,\text{m}$.
\par Although Streher \cite{Streher_2017} applied a Reynolds number of $Re=100{,}000$ for the wall-resolved simulations of a fixed rigid NACA0012 airfoil, this work approximates the computational domain of the same airfoil at a Reynolds number of $Re=30{,}000$ in order to acquire less computational intensive simulations. This reduction of the Reynolds number allows the computation of the fluid-structure interaction with a coarser grid, i$.$e$.$, $m{-}L_3^{min}{-}y^{+}_{max}$, which is composed of about $14\%$ of the nodes of the coarser grid applied by \mbox{Streher \cite{Streher_2017}}, i$.$e$.$, $m{-}0{-}y^+_{max}$. Moreover, when comparing the $m{-}0{-}y^+_{max}$ grid from Streher \cite{Streher_2017} with the $m{-}L_3^{min}{-}y^{+}_{max}$ mesh from the current work, the height of the cells located on the airfoil increased from \mbox{$\Delta y^{first\,cell}=1.8{\cdot}10^{-5}\,\text{m}$} to \mbox{$\Delta y^{first\,cell}=5{\cdot}10^{-5}\,\text{m}$} and the time step size increased from \mbox{$\Delta  t=1{\cdot}10^{-6}\,\text{s}$} to \mbox{$\Delta t=1{\cdot}10^{-5}\,\text{s}$}. Therefore, the required computational time by reduced in $95.5\%$.
\begin{table}[!htbp]
	\centering
	\begin{tabular}{l l}
		\hline
		Temperature & $T=300\,\text{K}$ \tabularnewline 
		Inflow velocity & $u_{in,\,in}=4.47\;\text{m}{\cdot}\text{s}^{-1}$ \tabularnewline
		Fluid density & $\rho_f=1.225\;\text{kg}{\cdot}\text{m}^{-3}$ \tabularnewline
		Dynamic fluid viscosity & $\displaystyle{\mu_f=18.27\cdot10^{-6}\;\text{Pa}{\cdot}\text{s}}$ \tabularnewline
		Reynolds number & $\displaystyle{Re=\frac{u_{in,\,1}\;\rho_f\;c}{\mu_f}=30{,}000}$ \tabularnewline
		\hline	
	\end{tabular}
	\caption{\label{table:flow_parameters}Fluid parameters.}
\end{table}
\begin{table}[!htbp]
	\centering
	\begin{tabular}{l l}
		\hline
		Spatial discretization & FVM \tabularnewline
		Temporal discretization & Predictor-corrector scheme \tabularnewline
		Turbulence approach & LES \tabularnewline
		Sub-grid scale model & Smagorinsky \tabularnewline
		Damping function & Van-Driest \tabularnewline
		Smagorinsky constant & $C_s=0.1$ \tabularnewline 
		Wall function & None \tabularnewline
		\multirow{2}{*}{Mesh adaption} & Small displacements: TFI \tabularnewline
		& Large displacements: Hybrid IDW-TFI\tabularnewline
		\hline	
	\end{tabular}
	\caption{\label{table:simulation_parameters}Simulation parameters.}
\end{table}
\par For the thorough description of the CSD and FSI setups see Sections  \ref{sec:CSD_setup} and \ref{sec:FSI_setup}, respectively.

\subsection{Initial conditions}
\label{sec:CFD_initial_conditions}
\par The conservation laws are parabolic in time. Therefore, initial conditions for the whole fluid domain are required. In the current work, these conditions are applied for the velocity vector and pressure in relation to a Reynolds number of $Re=30{,}000$, according to \mbox{Eqs.\ (\ref{eq:initial_velocity_i})} - (\ref{eq:initial_pressure}). The initial velocity vector is characterized only by a velocity in the main flow direction (chord-wise direction), which is equal to the inlet velocity vector \mbox{(see \mbox{Section \ref{subsubsec:inflow}})}.
\begin{eqnarray}
u_{t_0,\,1}&=&4.47\;\text{m}{\cdot}\text{s}^{-1} \label{eq:initial_velocity_i}\\
u_{t_0,\,2}&=&0\;\text{m}{\cdot}\text{s}^{-1} \label{eq:initial_velocity_j}\\
u_{t_0,\,3}&=&0\;\text{m}{\cdot}\text{s}^{-1} \label{eq:initial_velocity_k}\\
p_{t_0}&=&0\;Pa \label{eq:initial_pressure}
\end{eqnarray}

\subsection{Boundary conditions}
\label{sec:CFD_boundary_conditions}
\par The incompressible governing equations are elliptical in space. Therefore, the definition of boundary conditions for all boundaries is mandatory.

\subsubsection{Inlet}
\label{subsubsec:inflow}
\par The inlet velocity, i$.$e$.$, the velocity vector of the cells located at the inlet boundary, is given in Dirichlet form according to \mbox{Eqs.\ (\ref{eq:inlet_velocity_1})} - (\ref{eq:inlet_velocity_3}) for a Reynolds number of \mbox{$Re=30{,}000$}:
\begin{eqnarray}
u_{in,\,1}&=&4.47\;\text{m}{\cdot}\text{s}^{-1} \label{eq:inlet_velocity_1}, \\
u_{in,\,2}&=&0\;\text{m}{\cdot}\text{s}^{-1} \label{eq:inlet_velocity_2}, \\
u_{in,\,3}&=&0\;\text{m}{\cdot}\text{s}^{-1} \label{eq:inlet_velocity_3}. \\
\end{eqnarray}
\par A pressure boundary condition (Eq. (\ref{eq:inlet_pressure})) in the Neumann form is utilized for the pressure correction $p^{\,corr}$, according to the work of Ferziger and Peri{\'c} \cite{Ferziger_1999}:
\begin{equation}
\left. \frac{\quad \partial p^{\,corr}}{\partial x_1}\right |_{in}=0.
\label{eq:inlet_pressure}
\end{equation}

\subsubsection{Symmetry boundary}
\label{subsubsec:symmetry_boundary_condition}
\par A symmetry boundary can be used if both geometry and flow behavior have mirror symmetry, which is never the case for turbulent flows. However, since the edges at the top and the bottom of the fluid domain are located sufficiently far from the airfoil wake, these can be approximated as symmetric. These edges are characterized by zero velocities normal to the boundary, according to Eqs.\ (\ref{eq:symmetry_1}) to (\ref{eq:symmetry_3}): 
\begin{eqnarray}
u_{sym,\,1}&\neq&0\;\text{m}{\cdot}\text{s}^{-1}, \label{eq:symmetry_1} \\ 
u_{sym,\,2}&=&0\;\text{m}{\cdot}\text{s}^{-1}, \label{eq:symmetry_2} \\ 
u_{sym,\,3}&\neq&0\;\text{m}{\cdot}\text{s}^{-1}. \label{eq:symmetry_3}
\end{eqnarray}
\par Furthermore, only a velocity gradient of $u_2$ in relation to the $x_2$ direction is available (see Ferziger and Peri{\'c} \cite{Ferziger_1999}). Therefore, only the normal stress component $\tau_{22}$ is present at the symmetry edges, according to Eqs.\ (\ref{eq:symmetry_flux_1}) - (\ref{eq:symmetry_flux_3}):
\begin{eqnarray}
\tau_{sym,\,12}&=&-\;\;\mu\left.\frac{\partial u_1}{\partial x_2}\right|_{sym}=0, \label{eq:symmetry_flux_1} \\ 
\tau_{sym,\,22}&=&-2\mu\left.\frac{\partial u_2}{\partial x_2}\right|_{sym}\neq0, \label{eq:symmetry_flux_2} \\ 
\tau_{sym,\,32}&=&-\;\;\mu\left.\frac{\partial u_3}{\partial x_2}\right|_{sym}=0. \label{eq:symmetry_flux_3}
\end{eqnarray}

\subsubsection{No-slip wall}
\label{subsubsec:wall}
\par The no-slip wall condition states that a fluid adheres to the wall and consequently moves with its velocity. For the case of fluid-structure interaction, a boundary condition in the form of Dirichlet is applied. This indicates that the flow velocity at the moving wall $u_{w,\,i}$ is equal to the body velocity $u_{body,\,i}$, as stated in Eq.\ (\ref{eq:wall_velocity}):
\begin{equation}
\label{eq:wall_velocity}
u_{w,\,i}=u_{body,\,i}.
\end{equation}
\par The structure velocity is approximated according to a first-order accurate difference approximation scheme. This is based on the time step size $\Delta t$, as well as on the structural displacements $X_i$ at the last time step $n$ and at the current time step $n+1$ \mbox{(see Münsch \cite{Muensch_2015})}:
\begin{equation}
\label{eq:body_velocity}
u_{body,\,i}^{n+1}=\frac{X_i^{n+1}-X_i^n}{\Delta t}.
\end{equation}

\subsubsection{Outlet}
\label{subsubsec:outlet}
\par The outlet boundary must be located where the eddies can pass through the outlet in an undisturbed manner and without reflection, so that this edge has no or only a minor influence on the solution of the domain, as described by Breuer \cite{Breuer_2013}. 
\par The simulation of the flow around the NACA0012 profile requires the usage of a convective outflow boundary condition for the outlet boundary:
\begin{equation}
\label{eq:outlet_boundary}
\frac{\partial u_i}{\partial t}+U_{conv}\left.\frac{\partial u_i}{\partial x_1}\right|_{out}=0.
\end{equation}
\par The convective velocity $U_{conv}$ is parallel to the stream-wise direction and therefore normal to the outlet boundary. In the present case $U_{conv}$ is set to the free-stream velocity, i.e$.$, $U_{conv}=u_{in,\,1}=4.47\,\text{m}{\cdot}\text{s}^{-1}$.

\subsubsection{Periodic boundary}
\label{subsubsec:periodic_boundary_condition}
\par The periodic boundary condition allows the reduction of the fluid domain and consequently the decrease of the computational effort. It is based on the simultaneous exchange of the boundary values at two corresponding domain boundaries, as stated by Breuer \cite{Breuer_2002}.
\par The usage of this type of boundary is, however, only possible in homogeneous directions, which are characterized by no variation of the statistically averaged flow. The period length must be carefully chosen in order to guarantee that the largest eddies are captured. This can be assured by the calculation of the turbulent two-point correlations, which must tend to zero in the half domain size (see Breuer \cite{Breuer_2002}).
\par In the case of the FSI investigations of the NACA0012 airfoil, the span-wise direction is homogeneous and characterized by \mbox{$\widetilde{u}_{i}(x_1,x_2,x_3,t)=\widetilde{u}_i(x_1,x_2,x_3+\Delta x_3,t)$} and \mbox{$\widetilde{p}(x_1,x_2,x_3,t)=\widetilde{p}(x_1,x_2,x_3+\Delta x_3,t)$.} A periodic boundary condition is, then, applied in this direction. Two period lengths of $L_3=0.5\,c$ and $L_3=0.25\,c$ are tested and thoroughly discussed in Section \ref{sec:study_span_wise_length}. Since the best compromise between accuracy and computational time is achieved with $L_3=0.25\,c$, this is applied for the computation of the FSI test cases.

\section{CSD setup}
\label{sec:CSD_setup}
\par Although the CFD solver computes the fluid domain based on a span-wise length of $L_3=0.25c$, the CSD setup is configured according to the rigid NACA0012 model utilized by the experimental setup ($L_{3,\,N}=0.6c$), which is thoroughly discussed in Appendix \ref{appendix_experimental_setup}. This process is carried out in order to enable a direct comparison with the experimental results and requires the scaling of the forces and moments computed by the CFD solver (see Section \ref{subsec:scaling_fluid_forces}).
\par The structural properties, as well as the initial and boundary conditions are set in the $csdvalues.dat$ file, which is compiled by the rigid movement solver during the initialization sub-routine. This file is described in Table \ref{table:csdvalues} for the FSI cases with one translational and one rotational degree of freedom, i$.$e$.$, translational degree of freedom in the $x_2$-direction and rotational degree of freedom around the $x_3$-axis ($\varphi_3$), respectively.
\par Boolean algebra is utilized in order to define the system degrees of freedom (DOF): the first three inputs are regarding the three translational degrees of freedom in the $x_1$, $x_2$ and $x_3$ directions, while the last three stand for the rotations about the $x_1$, $x_2$ and $x_3$ axes ($\varphi_1$, $\varphi_2$ and $\varphi_3$, respectively). If a degree of freedom is not considered \mbox{(set to false $F$)}, all variables of this column, with an exception of the mass moment of inertia, are set to zero by the initialization subroutine. 
\par The initial conditions are given in the form of initial displacements, velocities and accelerations, while the boundary conditions are given by the external (fluid) forces and moments computed by the CFD solver. Since the body is initially in dynamic equilibrium, all this conditions are set to zero at $t=0\,\text{s}$.
\par The body properties are given by the mass and the mass moment of inertia tensor, according to lines 5 and 9 of the initialization file. The spring stiffness and the damping ratio (see Eq.\ (\ref{eq:damping_ratio})) are given in the sixth and seventh lines, respectively. The location of the body-fixed undeformed Cartesian coordinate system in relation to origin of the Cartesian coordinate system established by the CFD mesh (global CCS), is given by the last line of the $csdvalues.dat$ file. In order to guarantee an uncoupled system, the body-fixed undeformed Cartesian coordinate system (CCS) is located at the airfoil center of mass (see Section \ref{subsec:moment_inertia}).
\begin{table}[H]
	\centering
	\begin{tabular}{p{2.8cm} | p{1.7cm} p{1.7cm} p{1.7cm} p{1.7cm} p{1.7cm} p{1.7cm}}
		\cline{1-7}
		\centering{\bf{$\,$}}  & \centering{\bf{$x_1$}} & \centering{\bf{$x_2$}} & \centering{\bf{$x_3$}} & \centering{\bf{$\varphi_1$}} & \centering{\bf{$\varphi_2$}} & \centering{\bf{$\varphi_3$}} \tabularnewline \hline
		\centering{\bf{DOF}} & \centering{F} &\centering{T} & \centering{F} & \centering{F} & \centering{F} & \centering{T} \tabularnewline
		\centering{\bf{Displacement}} & \centering{0.00E+00} & \centering{0.00E+00} & \centering{0.00E+00} & \centering{0.00E+00} & \centering{0.00E+00} & \centering{0.00E+00} \tabularnewline
		\centering{\bf{Velocity}} & \centering{0.00E+00} & \centering{0.00E+00} & \centering{0.00E+00} & \centering{0.00E+00} & \centering{0.00E+00} & \centering{0.00E+00} \tabularnewline
		\centering{\bf{Acceleration}} & \centering{0.00E+00} & \centering{0.00E+00} & \centering{0.00E+00} & \centering{0.00E+00} & \centering{0.00E+00} & \centering{0.00E+00} \tabularnewline
		\centering{\bf{Moment of inertia}} & \multirow{2}{*}{\centering{1.23E-02}} & \multirow{2}{*}{\centering{0.00E+00}} & \multirow{2}{*}{\centering{0.00E+00}} & \multirow{2}{*}{\centering{1.25E-02}} & \multirow{2}{*}{\centering{0.00E+00}} & \multirow{2}{*}{\centering{1.97E-04}} \tabularnewline
		\centering{\bf{Spring stiffness}} & \multirow{2}{*}{\centering{0.00E+00}} & \multirow{2}{*}{\centering{$\quad\; k_{l,\,eq}$}} & \multirow{2}{*}{\centering{0.00E+00}} & \multirow{2}{*}{\centering{0.00E+00}} & \multirow{2}{*}{\centering{0.00E+00}} & \multirow{2}{*}{\centering{$\quad \;k_{t,\,eq}$}} \tabularnewline
		\centering{\bf{Damping ratio}} & \multirow{2}{*}{\centering{0.00E+00}} & \multirow{2}{*}{\centering{0.00E+00}} & \multirow{2}{*}{\centering{0.00E+00}} & \multirow{2}{*}{\centering{0.00E+00}} & \multirow{2}{*}{\centering{0.00E+00}} & \multirow{2}{*}{\centering{0.00E+00}} \tabularnewline
		\centering{\bf{Loads}} & \centering{0.00E+00} & \centering{0.00E+00} & \centering{0.00E+00} & \centering{0.00E+00} & \centering{0.00E+00} & \centering{0.00E+00} \tabularnewline
		\centering{\bf{Mass}} & \centering{4.32E-01} & \centering{$\,$} & \centering{$\,$} & \centering{$\,$} & \centering{$\,$} & \centering{$\,$} \tabularnewline
		\centering{\bf{CCS}} & \centering{4.17E-02} & \centering{0.00E+00} & \centering{3.00E-01} & \centering{$\,$} & \centering{$\,$} & \centering{$\,$} \tabularnewline		
		\hline	
	\end{tabular}
	\caption{\label{table:csdvalues}Initialization file $csdvalues.dat$: Initial and boundary conditions as well as system properties of the NACA0012 airfoil.}
\end{table}
\subsection{System stiffness}
\label{subsec:system_stiffness}
\par
\begin{figure}[H]
	\centering
	\includegraphics[width=0.4\textwidth]{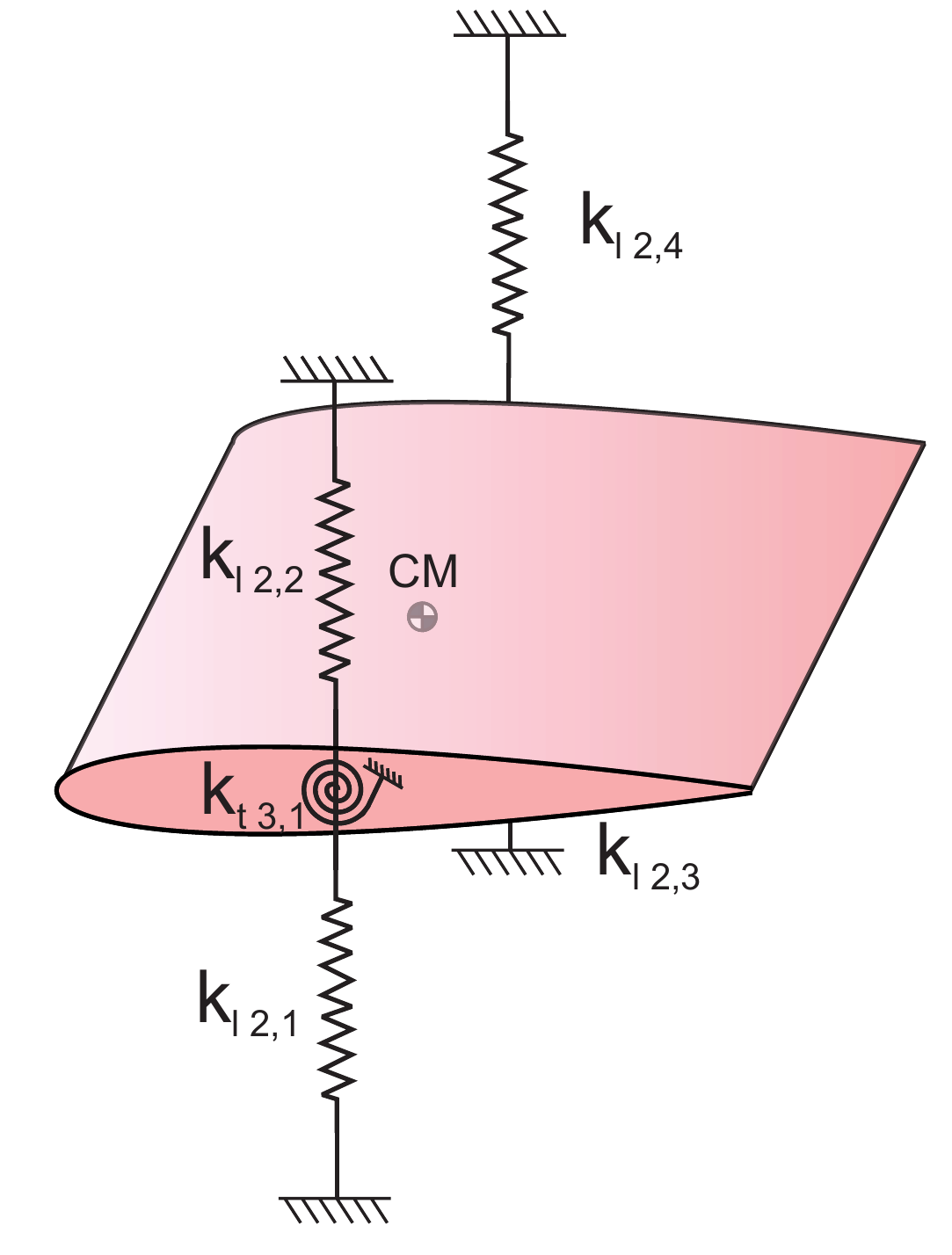}
	\caption{Configuration of the mass-spring system of the experimented NACA0012 airfoil.}
	\label{fig:springs_NACA0012}
\end{figure}

 The experimental NACA0012-spring system is characterized by the presence of four linear and two torsional springs, as illustrated in Fig.\ \ref{fig:springs_NACA0012}. The torsional stiffness $k_{t\,3,\,2}$ do not appear in the image since it is located at the back of the airfoil.  
\par Due to the presence of more than one spring per degree of freedom, the total stiffness of the up and down and pitch movements is calculated by the equivalent stiffness.\break Equations (\ref{eq:equivalent_linear_stiffness}) and (\ref{eq:equivalent_torsional_stiffness}) illustrate respectively the equivalent linear and torsional stiffnesses of the $x_2$ and $\varphi_3$ degrees of freedom, regarding parallel mounted springs. 
\begin{eqnarray}
\label{eq:equivalent_linear_stiffness}
k_{l\,2,\,eq}&=&k_{l\,2,1}+k_{l\,2,2}+k_{l\,2,3}+k_{l\,2,4}\\
\label{eq:equivalent_torsional_stiffness}
k_{t\,3,\,eq}&=&k_{t\,3,1}+k_{t\,3,2}
\end{eqnarray}

\subsection{Total mass}
\label{subsec:total_mass}
\par The mass of the airfoil is calculated according to Eq.\ (\ref{eq:mass_airfoil}). The chord and span-wise lengths as well as the density of the experimental airfoil model are respectively $c=0.1\,\text{m}$,  $L_{3,\,N}=0.6\,\text{m}$ and $\rho_{N}=700\,\text{kg}{\cdot}\text{m}^{-3}$ (see Appendix \ref{appendix_experimental_setup}). $X_{N,\,2}$ describes the airfoil profile as a function of the chord length and the  $x_1$ coordinate (see Eq.\ (\ref{eq:NACA0012})). 
\begin{eqnarray}
m_{N}&=&2\;\rho_{N}\;L_{3,\,N}\int_{0}^{c}X_{N,\,2}\,dx_1
\label{eq:mass_airfoil}\\
m_{N}&=&0.3392\;\text{kg}\label{eq:mass_airfoil_value}
\end{eqnarray}
\par The experimental setup is characterized by the presence of two aluminum supports fixed on the airfoil and two guiding rods in order to guide the movement in the $x_2$-direction, as illustrated in Fig.\ \ref{fig:guiding_rod}:
\begin{figure}[H]
	\centering
	\includegraphics[width=0.4\textwidth]{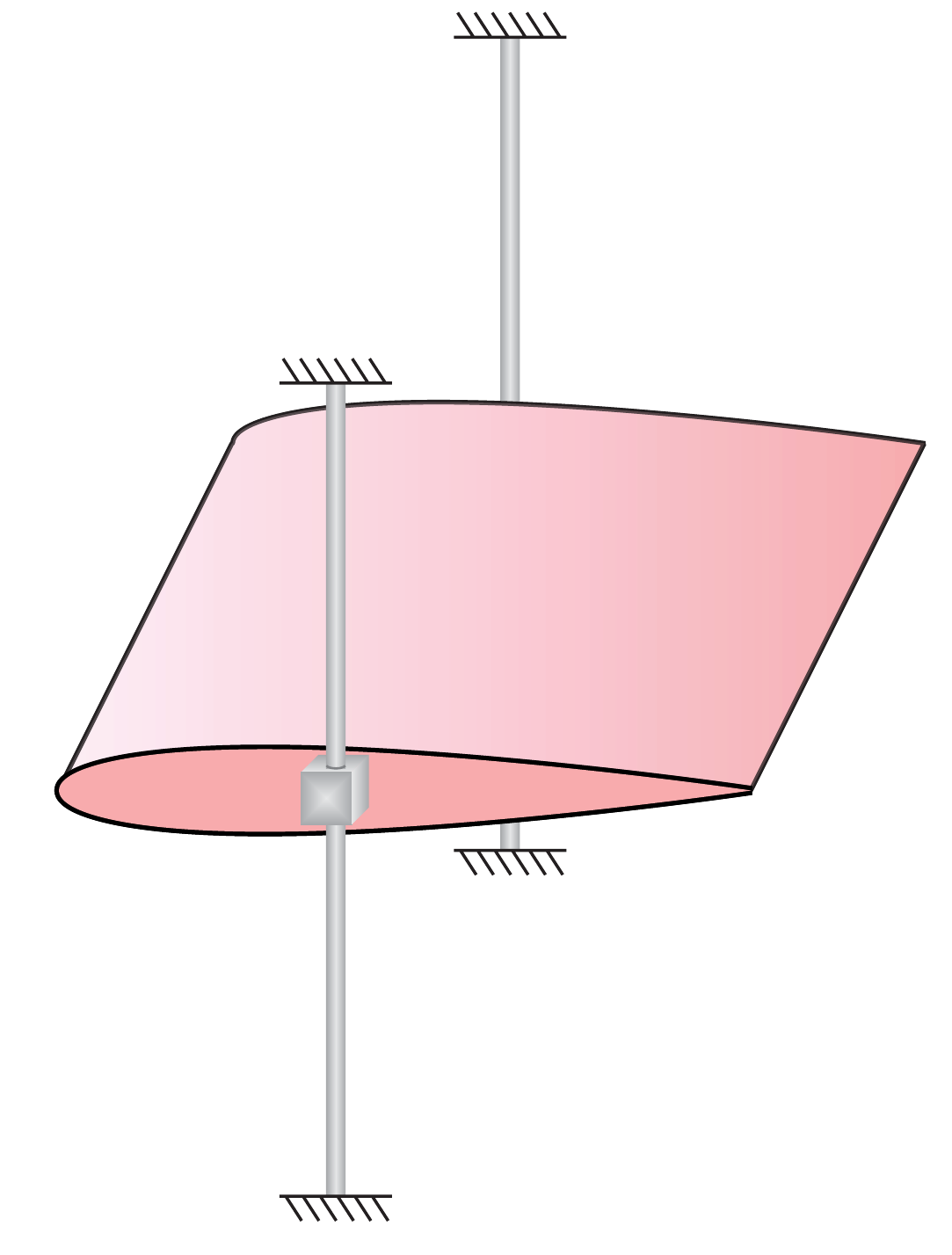}
	\caption{Guiding rods and supports present in the experimental setup.}
	\label{fig:guiding_rod}
\end{figure} 
\par Since the aluminum supports are fixed on the airfoil, these influence the dynamic equilibrium of the body. Therefore, their masses ($m_\text{{supports}}=0.09299\,\text{kg}$) must be considered by the CSD solver. Hence, the total mass $m_{tot}$ utilized in the $csdvalues.dat$ file is:
\begin{eqnarray}
\label{eq:total_mass}
m_{tot}&=&m_{N}+m_\text{{supports}}, \\
\label{eq:total_mass_value}
m_{tot}&=&0.43219\,\text{kg}.
\end{eqnarray}
\subsection{Center of mass}
\label{subsec:center_mass}
\par The undeformed body-fixed Cartesian coordinate system is located at the center of mass of the structure. Since the NACA0012 profile is symmetrical and the airfoil model has a span-wise length of $L_{3,\,N}=0.6\,\text{m}$, the $x_2$ and $x_3$ coordinates of the center of mass in relation to the origin of the global Cartesian coordinate system are respectively $x_{CM,\,2}=0\,\text{m}$ and $x_{CM,\,3}=0.3\,\text{m}$ (see Fig.\ \ref{fig:points_O_CM}). The $x_1$ coordinate of the center of mass is calculated according to \mbox{Eq.\ (\ref{eq:mass_center_x})}, regarding that the density of the airfoil is constant.
\begin{eqnarray}
\label{eq:mass_center_x}
x_{CM,\,1}=\frac{\int_{0}^{c}x_1\,|X_{N,\,2}|\,dx_1}{\int_{0}^{c}|X_{N,\,2}|\,dx_1}=0.0417\,\text{m}
\end{eqnarray}
\par Due to the fact that the supports fixed on the airfoil are symmetric in relation to the center of mass (see Fig.\ \ref{fig:guiding_rod}), their presence do not affect the location of this center.

\subsection{Mass moment of inertia}
\label{subsec:moment_inertia}
\par In order to guarantee an uncoupled system, the origin of the body-fixed Cartesian coordinate system of the NACA0012 profile must be coincident with a Cartesian coordinate system constituted of only orthogonal to each other principal axes. These have the property of aligning the angular moment $M_{ext,\,i}$ with the angular velocity $\dot{\varphi_i}$ and therefore are distinguished by the extinction of the products of inertia $J_{12}$, $J_{13}$ and $J_{23}$ \mbox{(see Hibbeler \cite{Hibbeler_2001}).} Since the airfoil is characterized by symmetries in the $x_1x_2$ and $x_1x_3$ planes, when the body-fixed coordinate system is located at the body center of mass, the $x_1$, $x_2$ and $x_3$ axes are considered principal ones (see Gere \cite{Gere_2004}). Hence, only the main diagonal of the mass moment of inertia tensor in relation to the center of mass of the NACA0012 airfoil remains, i$.$e$.$, the $J_{11}^{CM}$, $J_{22}^{CM}$ and $J_{33}^{CM}$ terms.
\par
\begin{figure}[H]
	\centering
	\centering
	\includegraphics[scale=1,draft=\drafttype]{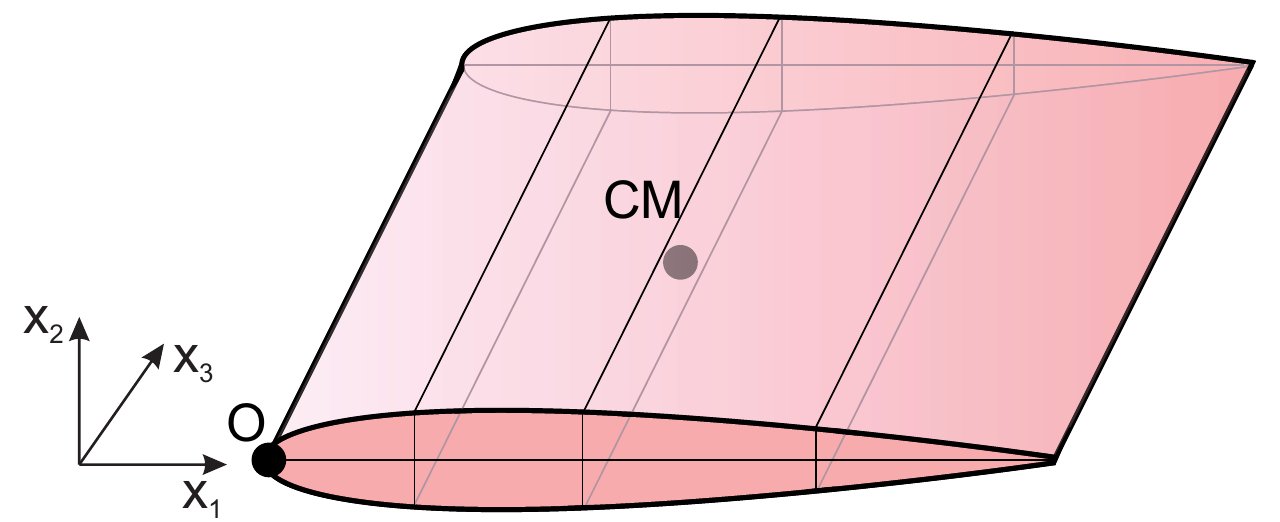}
	\caption{\label{fig:points_O_CM}O and CM: Origin of the global and local Cartesian coordinate systems, respectively.}
\end{figure}

 The mass moment of inertia is the quantity that indicates the object resistance to a rotational acceleration about a particular axis, i$.$e$.$, it indicates the distribution of the body mass in relation to a determined axis. The diagonal components of this tensor in relation to the global coordinate system ($J_{N,\,ii}^o$), which origin is located at the point O (see Fig.\ \ref{fig:points_O_CM}), are calculated according to Eqs.\ (\ref{eq:inertia_NACA_xx}) to (\ref{eq:inertia_NACA_zz}) for the NACA0012 airfoil. $X_{N,\,2}$ stands for the NACA0012 profile function (see Eq.\ \ref{eq:NACA0012}).
\begin{eqnarray}
\nonumber
J_{N,\,11}^o&=&\int_{V}\rho_N(x_2^2+x_3^2)\,dV=2\rho_N\int_{0}^{c}\int_{0}^{X_{N,\,2}}\int_{0}^{L_{3,\,N}}(x_2^2 + x_3^2)\;dx_3\,dx_2\,dx_1 \\
&=&\frac{2\,\rho_{N}}{3}\int_{0}^{c}\left(\left|X_{N,\,2}(x_1)\right|^3\,L_{3,\,N}+\left|X_{N,\,2}(x_1)\right|\,L_{3,\,N}^3\right)\,dx_1 \label{eq:inertia_NACA_xx}\\ \nonumber
\label{eq:inertia_NACA_yy}
J_{N,\,22}^o&=&\int_{V}\rho_N(x_1^2+x_3^2)\,dV=2\rho_N\int_{0}^{c}\int_{0}^{X_{N,\,2}}\int_{0}^{L_{3,\,N}}(x_1^2 + x_3^2)\;dx_3\,dx_2\,dx_1 \\
&=&2\,\rho_{N}\int_{0}^{c}\left(x_1^2\,\left|X_{N,\,2}(x_1)\right|\,L_{3,\,N}+\left|X_{N,\,2}(x_1)\right|\,\frac{L_{3,\,N}^3}{3}\right)\,dx_1\\ \nonumber
\label{eq:inertia_NACA_zz}
J_{N,\,33}^o&=&\int_{V}\rho_N(x_1^2+x_2^2)\,dV=2\rho_N\int_{0}^{c}\int_{0}^{X_{N,\,2}}\int_{0}^{L_{3,\,N}}(x_1^2 + x_2^2)\;dx_3\,dx_2\,dx_1 \\
&=&2\,\rho_{N}\,L_{3,\,N}\int_{0}^{c}\left(x_1^2\,|X_{N,\,2}(x_1)|+\frac{|X_{N,\,2}(x_1)|^3}{3}\right)\,dx_1
\end{eqnarray}
\par Since the undeformed body-fixed coordinate system (see Fig.\ \ref{fig:body_disp}) is located at the airfoil center of mass CM, the mass moment of inertia tensor is then calculated with reference to this point. Therefore, the parallel axis theorem, i$.$e$.$, Huygens-Steiner theorem, is applied \mbox{(see Hibbeler \cite{Hibbeler_2001})}. This utilizes the perpendicular distances between the considered axis in the global and local coordinate systems, which origins are located respectively at points O and CM, in order to calculate the mass moment of inertia in relation to the airfoil center of mass, as stated in Eqs.\ (\ref{eq:inertia_NACA_CM_xx}) to (\ref{eq:inertia_NACA_CM_zz}):
\begin{eqnarray}
\label{eq:inertia_NACA_CM_xx}
J_{N,\,11}^{CM}=J_{N,\,11}^o-m_{N}\,x_{CM,\,3}^2=1.0200{\cdot}10^{-2}\;\text{kg}{\cdot}\text{m}^2,\\
\label{eq:inertia_NACA_CM_yy}
J_{N,\,22}^{CM}=J_{N,\,22}^o-m_{N}\,x_{CM,\,3}^2=1.1000{\cdot}10^{-2}\;\text{kg}{\cdot}\text{m}^2,\\
\label{eq:inertia_NACA_CM_zz}
J_{N,\,33}^{CM}=J_{N,\,33}^o-m_{N}\,x_{CM,\,1}^2=1.8687{\cdot}10^{-4}\;\text{kg}{\cdot}\text{m}^2.
\end{eqnarray}
\par The products of inertia $J_{N,\,ij}^{CM}$ for $i\neq j$ at the center of mass are zero due to the fact that the coordinate axes are principal ones, as demonstrated in \mbox{Eqs.\ (\ref{eq:inertia_NACA_CM_xy}) to (\ref{eq:inertia_NACA_CM_yz})}:
\begin{eqnarray}
 \label{eq:inertia_NACA_CM_xy}\nonumber
 &&\quad\quad\;\, J_{N,\,12}^{CM}\;\;=\;\;\int_{A}x_1\,x_2\,dA=\int_{-x_{CM,\,1}}^{c-x_{CM,\,1}}\int_{-X_{N,\,2}}^{X_{N,\,2}}x_1\,x_2\;dx_2\,dx_1\\
 &&\quad\quad\quad\quad\quad\;\;=\;\;\frac{1}{2}\int_{-x_{CM,\,1}}^{c-x_{CM,\,1}}x_1\left(X_{N,\,2}^2-X_{N,\,2}^2\right)dx_1=0\,kg{\cdot}m^2,\\ 
\label{eq:inertia_NACA_CM_xz} \nonumber J_{N,\,13}^{CM}&=&\int_{A}x_1\,x_3\,dA=\rho_N\int_{V}x_1\,x_3\,dV=\rho_N\int_{-x_{CM,\,1}}^{c-x_{CM,\,1}}\int_{-X_{N,\,2}}^{X_{N,\,2}}\int_{-\frac{L_{3,\,N}}{2}}^{\frac{L_{3,\,N}}{2}}x_1\,x_3\;\,dx_3\,dx_2\,dx_1\\
&=&\frac{\rho_N}{2}\int_{-x_{CM,\,1}}^{c-x_{CM,\,1}}x_1\left(\frac{L_{3,\,N}^2}{4}-\frac{L_{3,\,N}^2}{4}\right)dx_1=0\;\text{kg}{\cdot}\text{m}^2,\\ \nonumber
\label{eq:inertia_NACA_CM_yz}
J_{N,\,23}^{CM}&=&\int_{A}x_2\,x_3\,dA=\rho_N\int_{V}x_2\,x_3\,dV=\rho_N\int_{-x_{CM,\,1}}^{c-x_{CM,\,1}}\int_{-X_{N,\,2}}^{X_{N,\,2}}\int_{-\frac{L_{3,\,N}}{2}}^{\frac{L_{3,\,N}}{2}}x_2\,x_3\;\,dx_3\,dx_2\,dx_1\\
&=&\frac{\rho_N}{2}\int_{-x_{CM,\,1}}^{c-x_{CM,\,1}}x_2\left(\frac{L_{3,\,N}^2}{4}-\frac{L_{3,\,N}^2}{4}\right)dx_1=0\;\text{kg}{\cdot}\text{m}^2.\\ \nonumber
\end{eqnarray}
\par The calculated airfoil mass moments of inertia are in great agreement with the ones calculated by a CAD software (see Fig.\ \ref{fig:airfoil_model_specifications}). Nevertheless, the presence of the supports fixed on the airfoil (see Fig.\ \ref{fig:guiding_rod}) also influence the rotational motion. Therefore, the mass moment of inertia tensor of the system composed of airfoil and supports must be considered. This has slightly higher moments for the diagonal components \mbox{(see Fig.\ \ref{fig:airfoil_model_properties})} while the products of inertia remain zero since the location of the principal axes are not altered, as presented in \mbox{Eq.\ (\ref{eq:inertia_total_CM})}:
\begin{equation}
\label{eq:inertia_total_CM}
J_{tot,\,ij}^{CM}=
\begin{bmatrix}
		1.22740{\cdot}10^{-2} & 0 & 0\\ 
		0 & 1.24656{\cdot}10^{-2} & 0\\
		0 & 0 & 1.97173{\cdot}10^{-4}
\end{bmatrix}
\;\text{kg}{\cdot}\text{m}^2.
\end{equation}

\subsection{Scaling of the fluid forces and moments}
\label{subsec:scaling_fluid_forces}
\par Since the simulated airfoil has a span-wise length of either $L_3=0.25\,c$ or $L_3=0.5\,c$ due to the utilization of a periodic boundary condition in this homogeneous direction and the system stiffness is not scaled in relation to the experimental setup, the fluid loads (forces and moments) must be geometrically scaled. This mathematical procedure assures the comparability of both numerical and experimental results and therefore enables a possible validation of the simulations. 
\begin{figure}[H]
	\centering
	\centering
	\includegraphics[scale=1.3,draft=\drafttype]{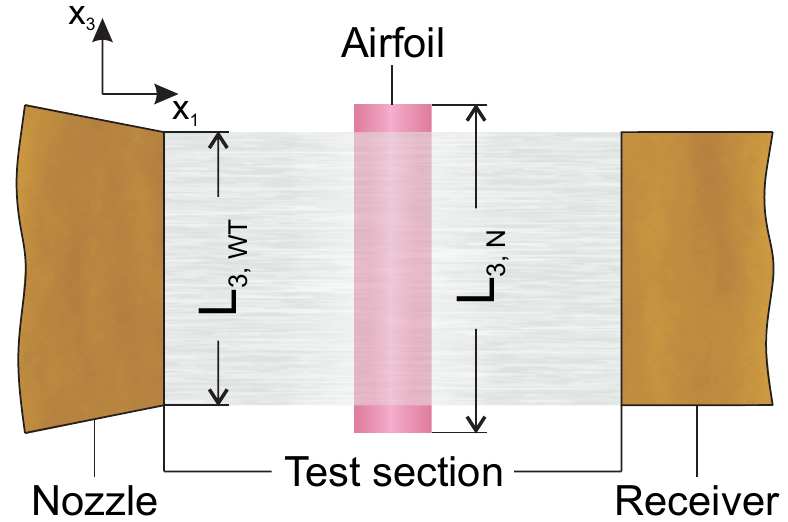}
	\caption{\label{fig:lengths_airfoil_wind_tunnel} Airfoil length submitted to the flow.}
\end{figure} 
\par Whereas the airfoil model has a span-wise length of $L_{3,\,N}=0.6\,\text{m}$, the test section of wind tunnel has a width of $L_{3,\,WT}=0.5\,\text{m}$ in order to guarantee the investigation of an infinite airfoil (without the presence of wing tip turbulence), as illustrated in Fig.\ \ref{fig:lengths_airfoil_wind_tunnel}. The flow moves, then, only around the airfoil within the wind tunnel length $L_{3,\,WT}$. Therefore, as the CSD solver receives the fluid loads from FASTEST-3D, it primarily scales these values in relation to the length $L_{3,\,WT}$, as stated in \mbox{Eqs.\ (\ref{eq:scaling_forces}) and (\ref{scaling_moment})}:
\begin{eqnarray}
\label{eq:scaling_forces}F_{CSD,\,i}&=&F_i\cdot\frac{L_{3,\,WT}}{L_3},\\
\label{scaling_moment}
M_{CSD,\,i}&=&M_i\cdot\frac{L_{3,\,WT}}{L_3}.
\end{eqnarray}  
\par $F_{CSD,\,i}$ and $M_{CSD,\,i}$ represent the scaled forces and moments utilized by the CSD solver, while $F_i$ and $M_i$ represent the forces and moments calculated by the CFD solver. The scaling factor is composed of the length of the airfoil that is submitted to the flow $L_{3,\,WT}$ and the span-wise length of the airfoil in the CFD mesh $L_3$. 
\section{FSI setup}
\label{sec:FSI_setup}
\par The coupling algorithm explained in Section \ref{sec:coupling_fluid_structure} applies the FSI setup described in Table \ref{table:FSI_setup}: 
\begin{table}[!htbp]
	\centering
	\begin{tabular}{p{6.9cm} p{0.1cm} p{3.7cm}}
		\hline
		Under-relaxation of the fluid loads & & No \tabularnewline
		Under-relaxation of the displacements & & No \tabularnewline
		Coupling method & & Loose ($n_{FSI}=1$)\tabularnewline
		Convergence criterion & & $\epsilon_{FSI,\,disp}<1{\cdot}10^{-6}$ \tabularnewline
		Estimation of displacements & & Linear extrapolation \tabularnewline
		\hline	
	\end{tabular}
	\caption{\label{table:FSI_setup}FSI setup.}
\end{table} 
\par This setup is selected according to the results of preliminary studies, which are thoroughly described in Section \ref{appendix_FSI_coupling_studies}. The forces and displacements are not under-relaxed since the added mass effect is negligible (see Section \ref{subsec:added_mass_effect}) and convergence problems are not present. The negligibility of this effect also allows the utilization of a loose coupling method, which is proven to be accurate as well as less computational intensive, as thoroughly described in Section \ref{subsec:investigation_coupling_algorithm}. Furthermore, a prediction of the displacements according to a first-order accurate linear extrapolation is utilized since this reduces CPU-time requirements (see Section \ref{subsec:investigation_estimation_displacement}). 

\chapter{Preliminary studies}
\label{chap:preliminary_studies}

\par The fluid setup, the CSD solver and the FSI coupling are thoroughly studied and validated aiming at a model that provides the best compromise between accuracy and computational effort. Firstly, the generated meshes are investigated according to the span-wise length, the mapping strategy for the geometrical block-structure to the parallel block-structure and the first cell height. Then, the available mesh adaption methods are compared. Secondly, the CSD solver is validated, regarding time-dependent external forces and coupled systems. Finally, the FSI coupling is analyzed according to the algorithm responsible for the estimation of displacements, the added mass effect and the coupling scheme itself.

\markboth{CHAPTER 3.$\quad$PREL. STUDIES}{3.1$\quad$CFD STUDIES}
\section{CFD studies}\label{appendix_mesh_adaption}\label{appendix_mesh_studies}

\par The ability of the mesh to provide an accurate solution of the fluid domain in relation to the computational time is thoroughly investigated considering a fixed, as well as a moving airfoil. The three generated meshes, i$.$e$.$, $m{-}L_3^{min}-y^{+}_{min}$, $m{-}L_3^{min}-y^{+}_{max}$ and $m{-}L_3^{max}-y^{+}_{min}$ (see Section \ref{sec:meshes}), are analyzed according to the span-wise length $L_3$, the mapping strategy and the first cell height $\Delta y^{first\,cell}$. The mesh that provides the best comprise between accuracy and required computational time is then utilized to simulate the coupled problem. Finally, different mesh adaption methods are investigated: the TFI, IDW and hybrid IDW-TFI schemes are tested.  

\subsection{Influence of the span-wise length of the computational domain}
\label{sec:study_span_wise_length}
\par A periodic boundary condition is applied in the $x_3$-direction due to its homogeneity. Since the choice of the span-wise length represents a  compromise between accuracy and numerical effort, the lengths $L_3=0.5\,c$ and $L_3=0.25\,c$ are investigated, as stated in \mbox{Table \ref{table:span_wise_length}}. The former is selected according to the work of Almutari \cite{Almutari_2010}, in which the two-point correlation for the span-wise direction is calculated and is approximately zero for the half domain size, while the latter is based on the work of Schmidt \cite{Schmidt_2016} and \mbox{Visbal et al.\ \cite{Visbal_2009}.}  
\begin{table}[H]
	\centering
	\begin{tabular}{p{3cm} p{2.2cm} p{2cm} p{2cm} p{2.5cm}}
		\hline
		\multicolumn{1}{c}{\multirow{2}{*}{\centering{\bf{Mesh}}}} & \centering{\bf{Span-wise length (m)}} & \centering{\bf{Mapping strategy}} &\centering{\bf{Time steps}} & \centering{\bf{Time step size (s)}}\tabularnewline \hline
		\centering{$m{-}L_3^{max}-y^{+}_{min}$} & \centering{$0.50\,c$} & \centering{3} & \centering{$170{,}000$} & \centering{$6\cdot10^{-6}$} \tabularnewline
		\centering{$m{-}L_3^{min}{-}y^+_{min}$} & \centering{$0.25\,c$} & \centering{3} &\centering{$170{,}000$} & \centering{$6\cdot10^{-6}$} \tabularnewline			
		\hline	
	\end{tabular}
	\caption{\label{table:span_wise_length}Comparison of the simulations performed for the meshes with different span-wise lengths $L_3$.}
\end{table}
\par The simulations are performed for a fixed airfoil at a Reynolds number of \mbox{$Re=30{,}000$} utilizing the mapping strategy 3 (see Section \ref{subsubsec:study_mapping_strategy}). The lift and drag coefficients are calculated according to Eqs.\ (\ref{eq:lift_coefficient}) and (\ref{eq:drag_coefficient}). These are time-averaged during a dimensionless time of $t^*=45.6$ and are summarized in Table \ref{table:drag_lift_coefficients}:
\begin{table}[H]
	\centering
	\begin{tabular}{c p{2.4cm} p{3cm} p{2.4cm} p{3cm}}
		\hline
		\multicolumn{1}{c}{\multirow{3}{*}{\textbf{Mesh}}} & \multicolumn{2}{c}{\textbf{Lift coefficient}} & \multicolumn{2}{c}{\textbf{Drag coefficient}} \tabularnewline
		\cline{2-5}
		& \multicolumn{1}{c}{\multirow{2}{*}{\textbf{Mean}}} & \centering{\textbf{Standard deviaton}} & \multicolumn{1}{c}{\multirow{2}{*}{\textbf{Mean}}} & \centering{\textbf{Standard deviaton}} \tabularnewline \hline
		\centering{$m{-}L_3^{max}{-}y^+_{min}$} & \centering{$\widetilde{C}_L=-0.002$} & \centering{$\sigma_{C_L}=0.021$} & \centering{$\widetilde{C}_D=0.025$} & \centering{$\sigma_{C_D}=0.001$} \tabularnewline
		\centering{$m{-}L_3^{min}{-}y^+_{min}$} & \centering{$\widetilde{C}_L= 0.000$} & \centering{$\sigma_{C_D}=0.022$} & \centering{$\widetilde{C}_D=0.025$} & \centering{$\sigma_{C_D}=0.001$} \tabularnewline
		\hline
	\end{tabular}
	\caption{\label{table:drag_lift_coefficients}Time-averaged lift and drag coefficients at $Re=30{,}000$. $m{-}L_3^{max}{-}y^+_{min}$ and $m{-}L_3^{min}{-}y^+_{min}$ meshes.}	
\end{table}
\par The mean and standard deviation values of the averaged drag coefficient are the\break same for both meshes, indicating that the mesh does not influence the\break computation of this aerodynamic property. Since the NACA0012 airfoil is symmetric\break and submitted to an angle of attack of $\alpha=0^\circ$, the expected mean value of the lift\break coefficient is $\widetilde{C}_L=0$. A minimal deviation of this value is, however, present when\break the computations are performed on the $m{-}L_3^{max}{-}y^+_{min}$ mesh. Moreover, minor discrepancies of the standard deviations of the lift coefficient computed by both meshes are also present.
\par
\begin{figure}[H]
	\centering
	\includegraphics[width=0.6\textwidth]{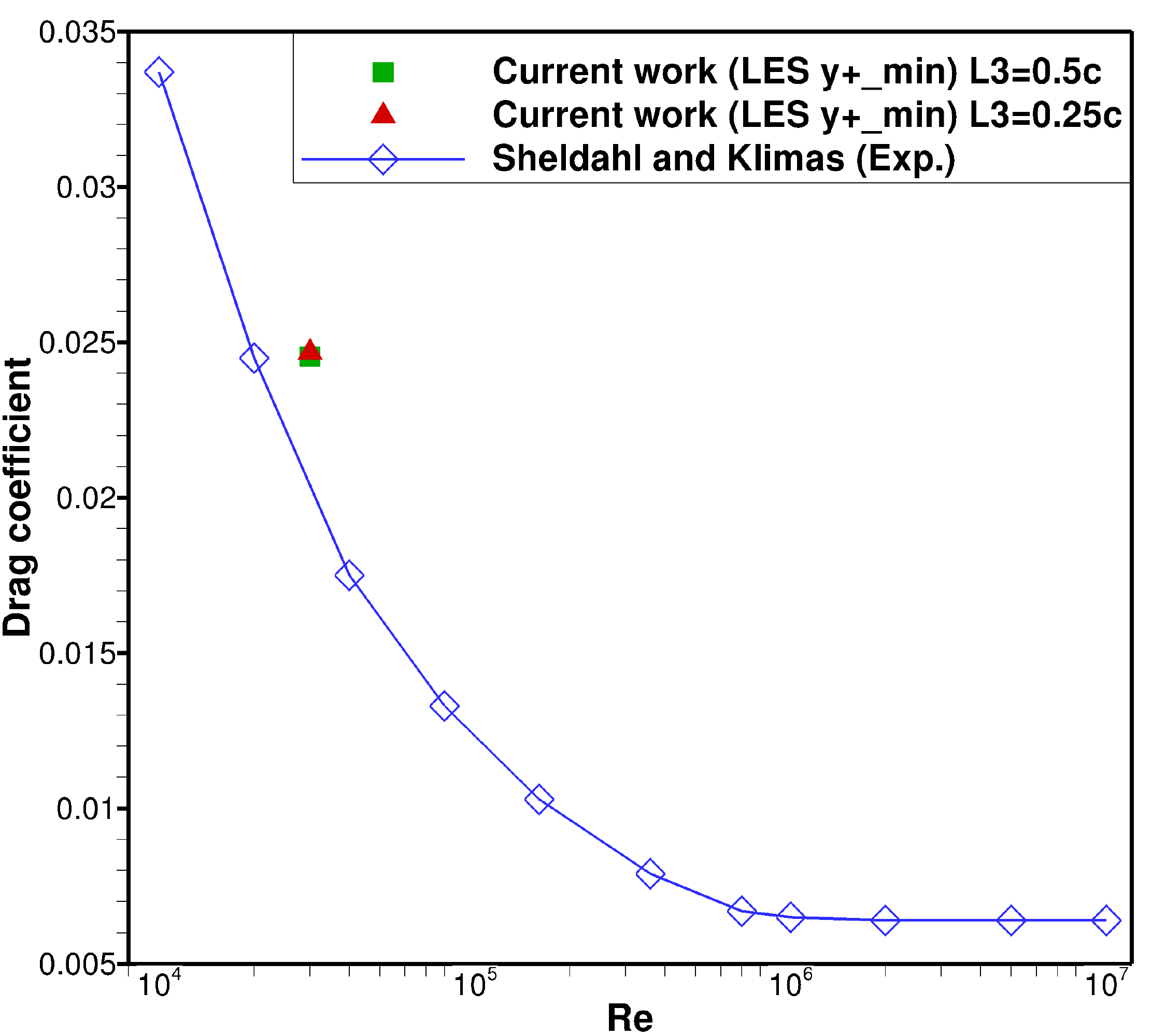} 
	\caption{Drag coefficients of the airfoil at an incidence of $\alpha=0^\circ$. Current work: NASA NACA0012 \cite{NASA_2016}. \mbox{Sheldahl and Klimas \cite{Sheldahl_1981}}: NACA0012 Eppler model.}
	\label{fig:comparison_lz}
\end{figure}

\begin{figure}[H]
	\centering
	\includegraphics[width=0.6\textwidth]{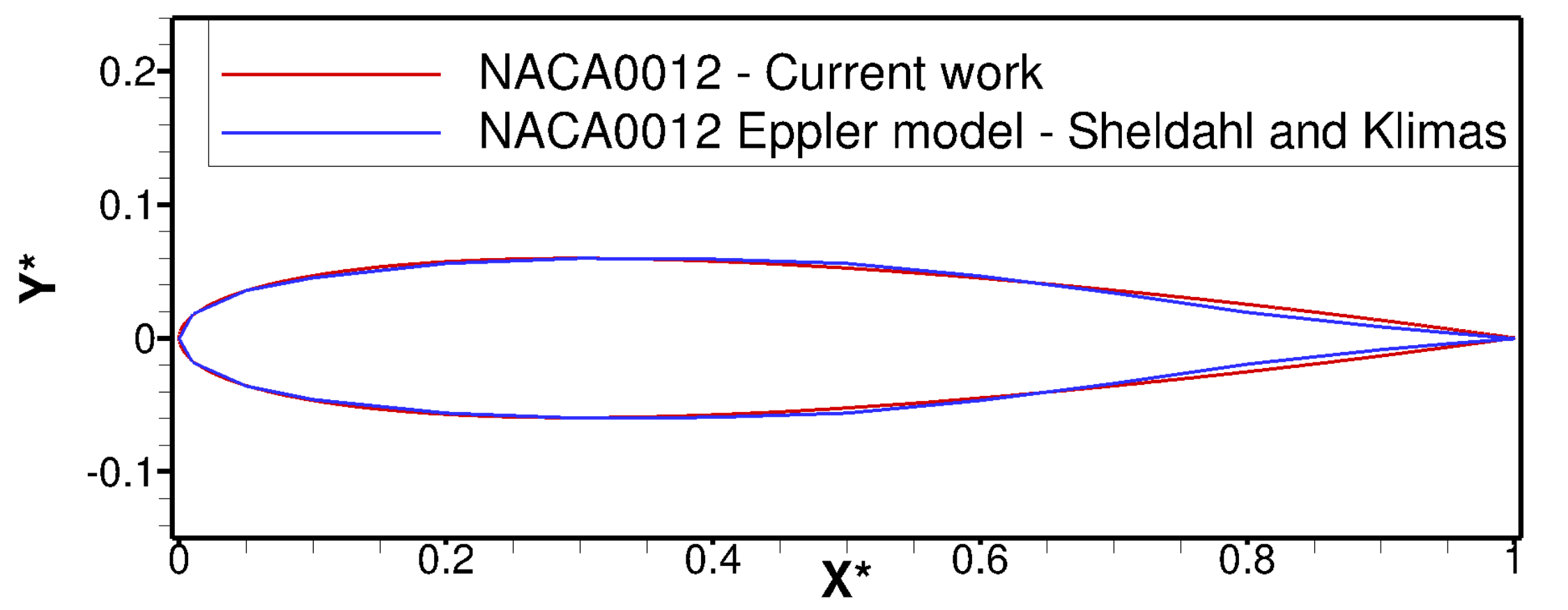} 
	\caption{NASA NACA0012 \cite{NASA_2016} profile used in the current work and NACA0012 Eppler model profile from Sheldahl and Klimas \cite{Sheldahl_1981}.}
	\label{fig:comparison_NACA0012_NASA_Eppler}
\end{figure}

 The calculated drag coefficients are also quantitatively compared to the experimental results of Sheldahl and Klimas \cite{Sheldahl_1981} for an incidence of $\alpha=0^\circ$, as illustrated in \mbox{Fig.\ \ref{fig:comparison_lz}}. These experiments were performed in a wind tunnel for a NACA0012 Eppler model profile at a Reynolds number range of $1\cdot10^4\leq Re\leq1\cdot10^7$ and an incidence range of \mbox{$0^\circ \leq \alpha \leq 180^\circ$}. Since the simulated airfoil profile (see NASA \cite{NASA_2016}) is slightly different from the NACA0012 Eppler profile (see Fig.\ \ref{fig:comparison_NACA0012_NASA_Eppler}), the results achieved by the simulations are overestimated compared to the experimental results.  
\par Due to the minor differences in the results achieved by both simulated meshes\break ($m{-}L_3^{max}{-}y^+_{min}$ and $m{-}L_3^{min}{-}y^+_{min}$), even the mesh with the smallest span-wise\break length is able to accurately simulate the fluid domain. Moreover, the $m{-}L_3^{min}{-}y^+_{min}$ mesh uses only $53.05\%$ of the required computational time of the $m{-}L_3^{max}{-}y^+_{min}$ grid. Therefore, the meshes with a span-wise length of $L_3=0.25\,c$ are applied to compute the FSI cases. 

\subsection{Mapping strategy}
\label{subsubsec:study_mapping_strategy}
\par The twelve geometrical blocks generated by ICEM CFD are divided and combined in a process called mapping. This aims at the generation of parallel blocks with almost the same number of control volumes in order to guarantee the fastest computation of the flow domain. Each parallel block is solved by a different processor with information exchange occurring only at its faces, so that the computation of all blocks can be performed at the same time (parallel computing).
\par Four mapping strategies for the $m{-}L_3^{min}{-}y^+_{min}$ mesh are investigated, regarding that cubic parallel blocks with 70$,$000 control volumes are aimed at. Moreover, the blocks are not divided in the $\eta$-direction in order to avoid numerical problems arising due to the stiffening of block boundaries. Simulations for the fluid-structure interaction of a NACA0012 airfoil with two degrees of freedom (pitch and up and down) are carried out for a Reynolds number of $Re=30{,}000$ and a linear stiffness of $k_{l,\,2}=144\,\text{N}{\cdot}\text{m}^{-1}$. Thousand time steps ($n=1000$) are computed in order to compare the efficiency of the different mappings. The properties of the applied mapping strategies are compared in Table \ref{table:mapping_strategy}: 
\begin{table}[H]
	\centering
	\begin{tabular}{p{1.69cm} p{2.2cm} p{3.5cm} p{1.9cm} p{3.9cm}}
		\hline
		\multicolumn{1}{c}{\multirow{3}{1.69cm}{\centering{\bf{Mapping strategy}}}} & \multicolumn{1}{c}{\multirow{3}{2.2cm}{\centering{\bf{Required processors}}}} & \centering{\bf{Averaged number of CVs per processor}} &\centering{\bf{Load-balancing efficiency}} & \centering{\bf{Averaged computational time per time step}} \tabularnewline \hline
		\centering{1} & \centering{18} & \centering{59$,$200} & \centering{54.81\%} & \centering{0.93 s} \tabularnewline
		\centering{2} & \centering{23} & \centering{46$,$330}&\centering{64.35\%} & \centering{0.90 s} \tabularnewline
		\centering{3} & \centering{13} & \centering{81$,$969} & \centering{75.90\%} & \centering{0.72 s} \tabularnewline
		\centering{4} & \centering{15} & \centering{71$,$040}& \centering{87.70\%} & \centering{0.72 s} \tabularnewline				
		\hline	
	\end{tabular}
	\caption{\label{table:mapping_strategy}Comparison of the mapping strategies.}
\end{table}
\par Load-balancing efficiency indicates the effectiveness of the work (computation of the whole fluid domain) division into separate tasks performed by each processor. Generally, the higher the load-balancing efficiency, the lower the required computational time, since the processor idle time is optimized. Although the mapping strategy 4 has the highest load-balancing efficiency, the required computational time is equal to the one achieved by mapping strategy 3. Therefore, the processor idle time, the information exchange between processors and the computation of the parallel blocks are optimized for the mapping \mbox{strategies 3 and 4}. Nonetheless, strategy 4 is utilized to  investigate the fluid-structure interaction between the flow and the NACA0012 airfoil due to the highest load-balancing efficiency.
\begin{figure}[H]
	\centering
	\subfigure[Top view.]{\includegraphics[width=0.44\textwidth]{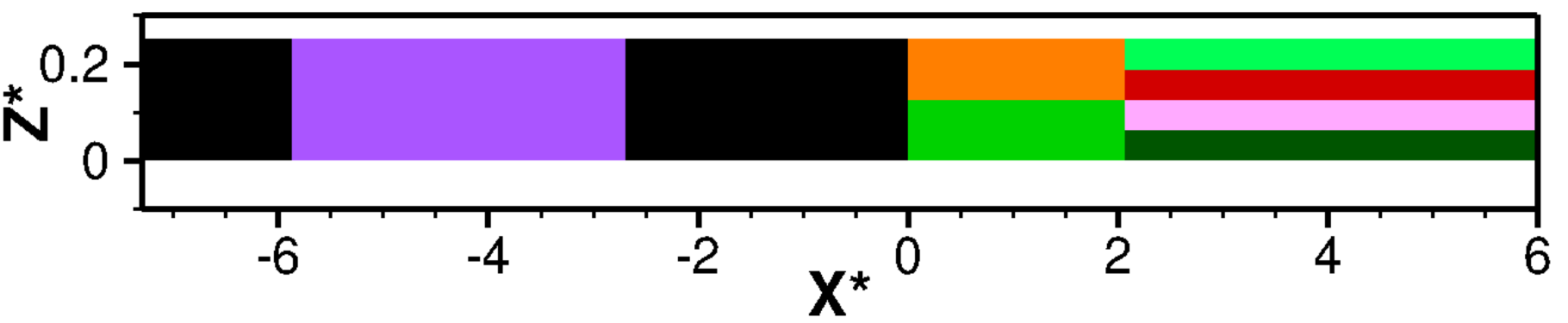}\label{fig:blocking_top_view}}\\
	\centering	
	\subfigure[Front view.]{\includegraphics[width=0.44\textwidth]{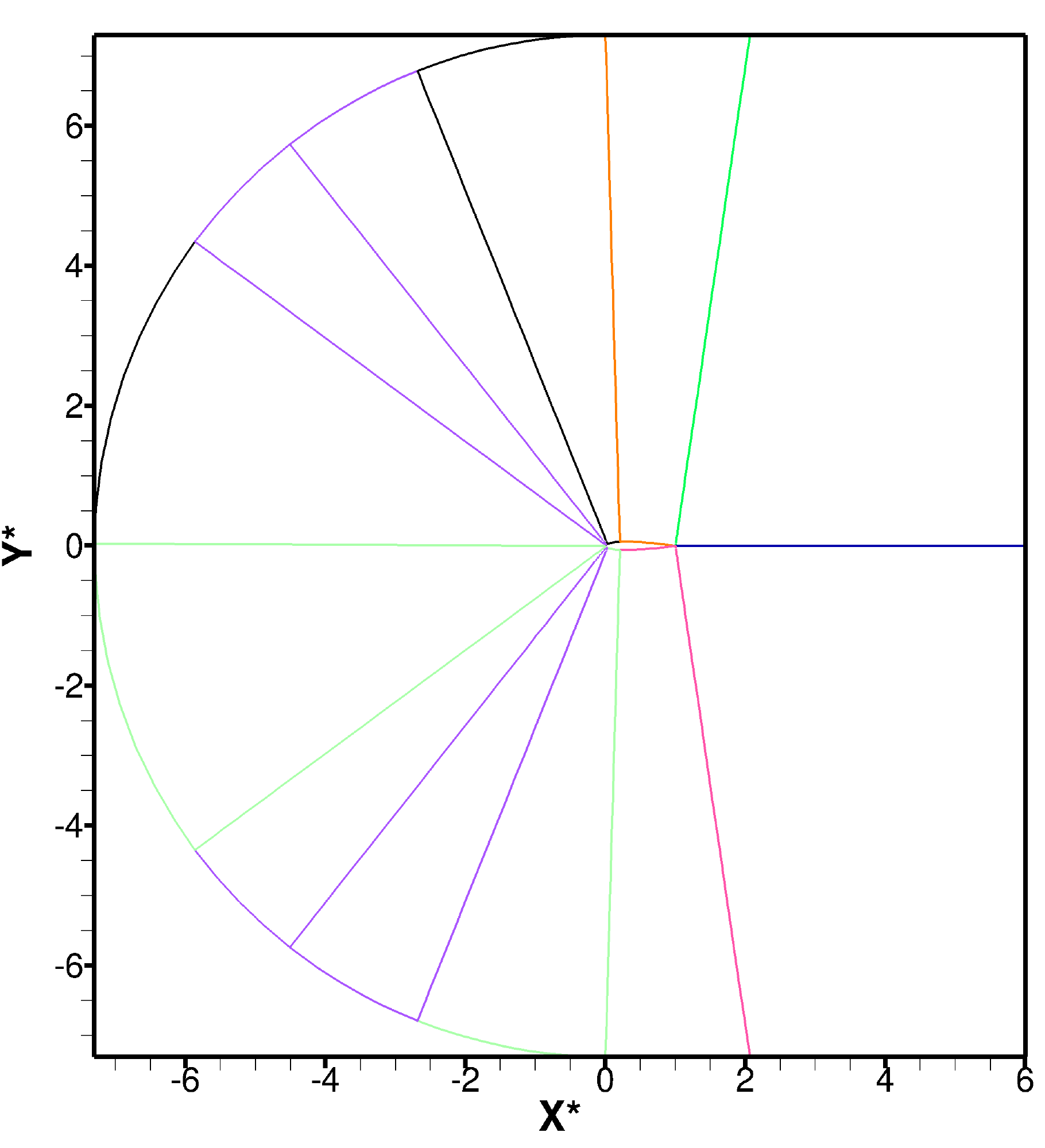} \label{fig:blocking_front_view}}
\end{figure}
\begin{figure}[H]
	\centering
	\subfigure[Bottom view.]{\includegraphics[width=0.44\textwidth]{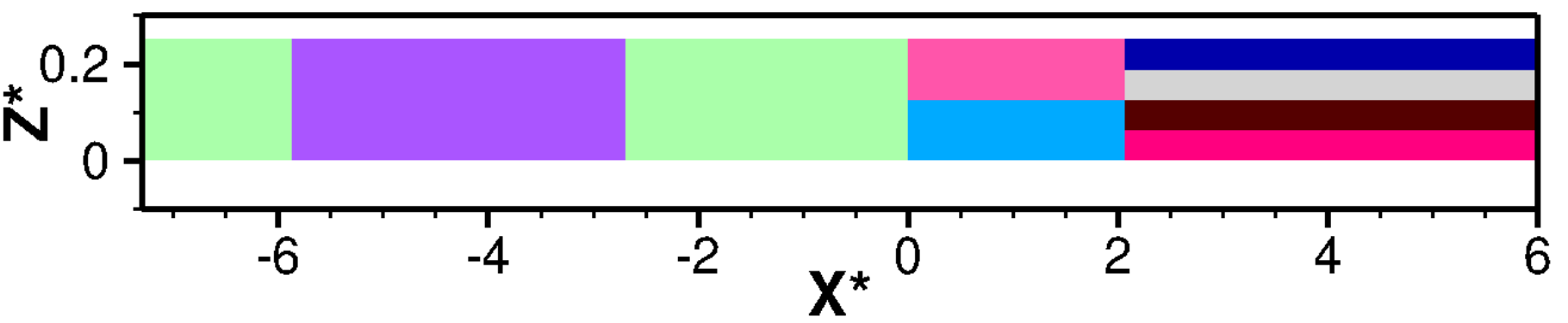}\label{fig:blocking_bottom_view}}
	\caption{Parallel blocks of the mapping strategy 4.}
	\label{fig:mapping_grid}
\end{figure} 
\par The distribution of the parallel blocks of mapping strategy 4 is illustrated in \mbox{Fig.\ \ref{fig:mapping_grid}}. Each color represents the blocks computed by a separate processor. That leads to a total number of 15 processors for the computation of the whole fluid domain.

\subsection{Selection of the first cell height}
\label{sec:investigation_first_cell_height}
\par The no-slip condition on the airfoil surface is responsible for the generation of a velocity gradient in the cells located at this interface. In order to guarantee the accuracy of the wall-resolved LES, the gradients present in a cell must be small. Therefore, the maximal allowed height of the cells in contact with no-slip walls is limited. 
\par Since the first cell height is directly related to the maximal applicable time step used to obtain a converged solution, a best compromise between accuracy and required computational time has to be found. Therefore, two meshes with the same number of control volumes, the same mapping strategy and different first cell heights are compared, as illustrated in Table \ref{table:first_cell_height}. The applied first cell heights are based on the work of Streher \cite{Streher_2017}. The simulations are performed for a fixed airfoil at a Reynolds number of $Re=30{,}000$ until a dimensionless time of $t^*=t\,u_{1,\,in}/c=200$ is achieved. Thereto, a total of $n=450{,}000$ and $n=750{,}000$ time steps are computed for the $m{-}L_3^{min}-y^{+}_{max}$ and $m{-}L_3^{min}{-}y^+_{min}$ meshes, respectively. The time-averaging process is started when the flow is completely developed, i$.$e$.$, after $t^*_{init}=67$. 
\begin{table}[H]
	\centering
	\begin{tabular}{p{3cm} p{2.09cm} p{2.3cm} p{2.8cm}}
		\hline
		\multicolumn{1}{c}{\multirow{2}{*}{\centering{\bf{Mesh}}}} & \centering{\bf{Span-wise width (m)}} &\centering{\bf{First cell height (m)}} & \centering{\bf{Time step size (s)}}\tabularnewline \hline
		\centering{$m{-}L_3^{min}-y^{+}_{max}$} & \centering{$0.25\,c$} & \centering{$5.0\cdot10^{-5}$} & \centering{$1\cdot10^{-5}$} \tabularnewline
		\centering{$m{-}L_3^{min}{-}y^+_{min}$} & \centering{$0.25\,c$} &\centering{$1.8\cdot10^{-5}$} & \centering{$6\cdot10^{-6}$} \tabularnewline			
		\hline	
	\end{tabular}
	\caption{\label{table:first_cell_height}Comparison of the meshes with different first cell heights.}
\end{table}
\par
\begin{figure}[H]
	\centering
	\subfigure[$m{-}L_3^{min}-y^{+}_{max}$.]{\includegraphics[width=0.49\textwidth]{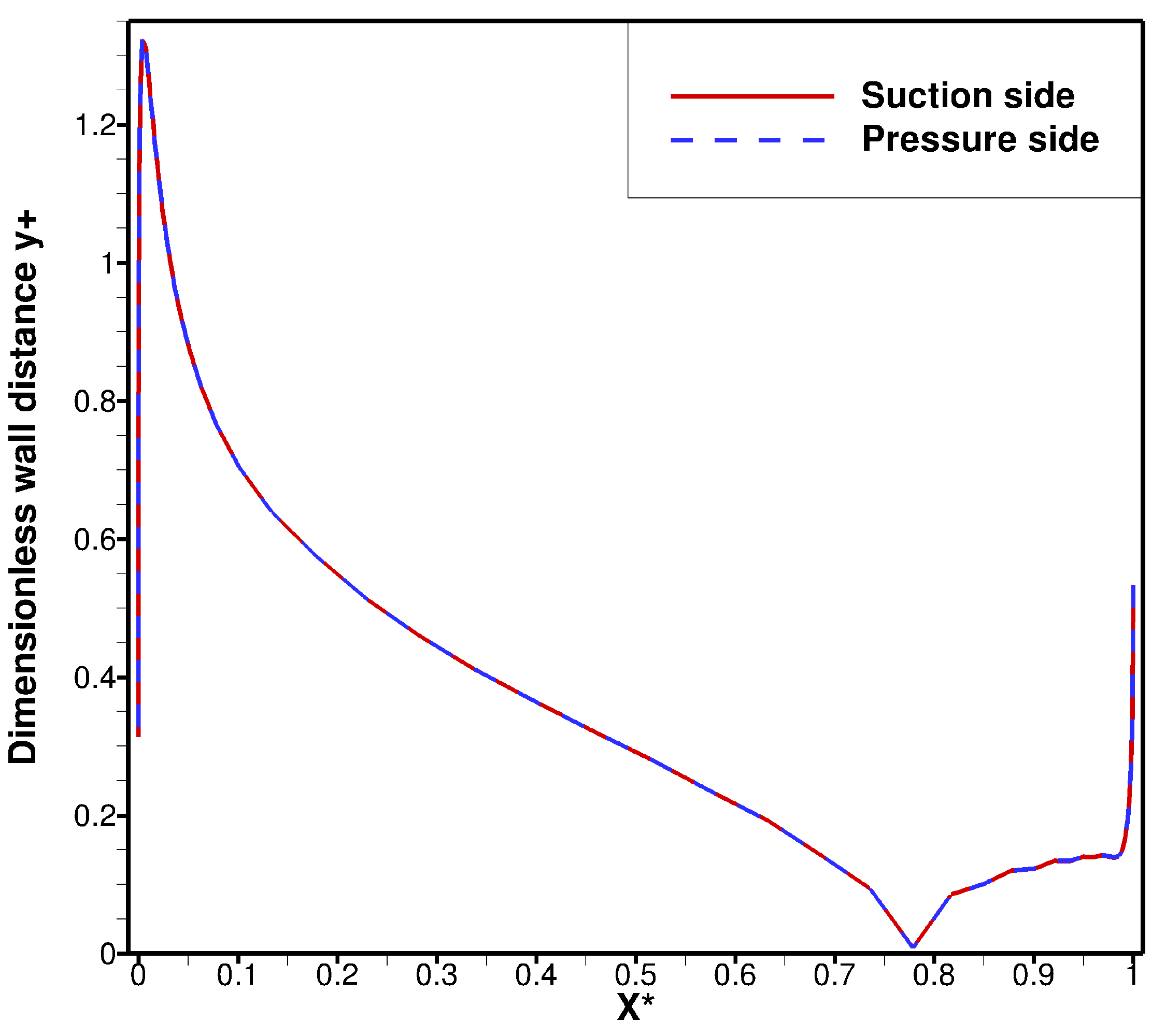} \label{fig:y_plus_5}}\hfill
	\subfigure[$m{-}L_3^{min}{-}y^+_{min}$.]{\includegraphics[width=0.49\textwidth]{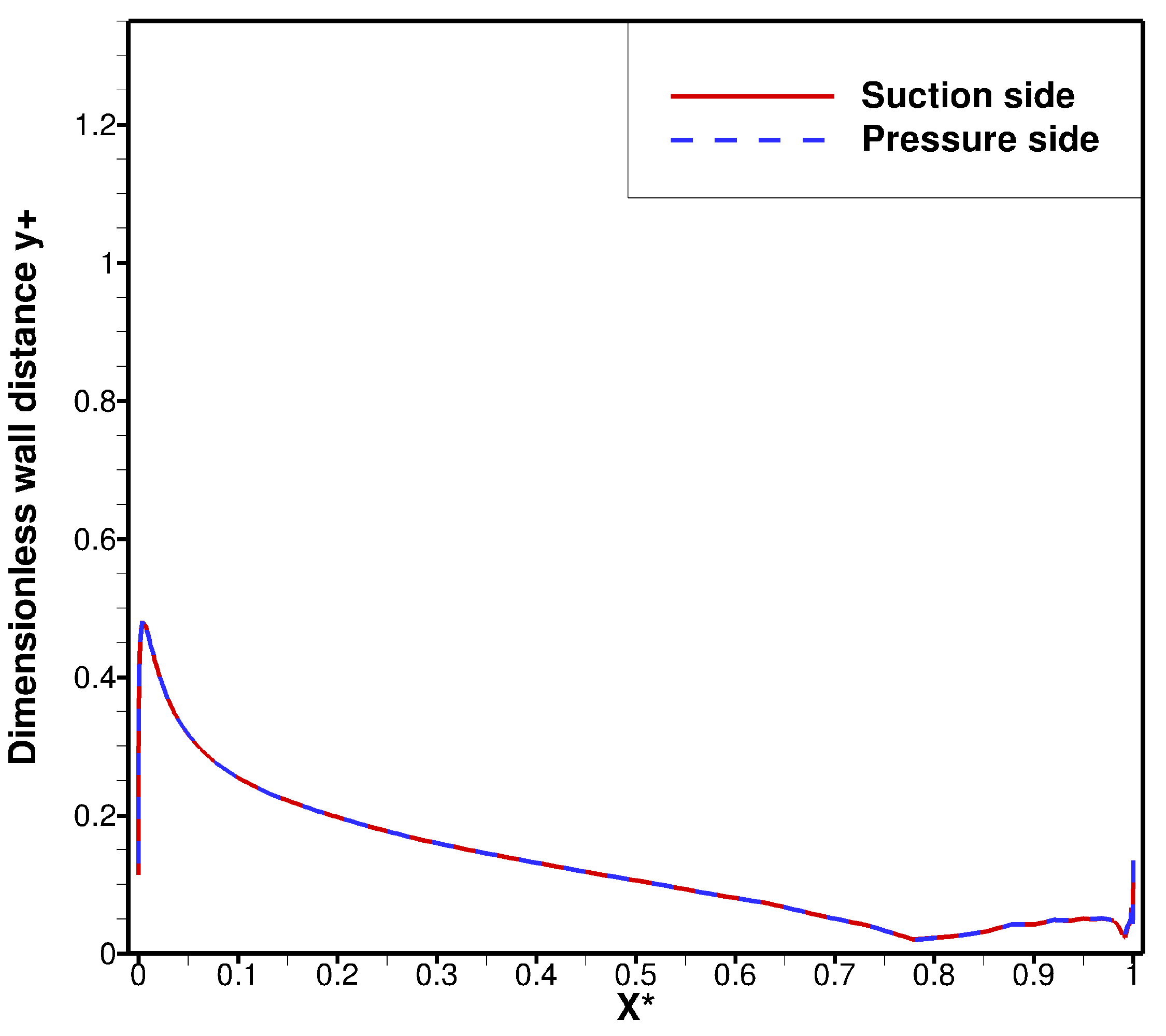}\label{fig:y_plus_1pt8}}\hfill
	\caption{Time-averaged dimensionless wall distances of the meshes \mbox{$m{-}L_3^{min}-y^{+}_{max}$} and  \mbox{$m{-}L_3^{min}-y^{+}_{max}$} with $L_3=0.25\,c$.}
	\label{fig:y_plus_lz_0pr25}
\end{figure}

 The simulation with the $m{-}L_3^{min}-y^{+}_{max}$ mesh is approximately two times faster than the one with the $m{-}L_3^{min}-y^{+}_{min}$ grid. The accuracy of the simulations is compared based on the time-averaged dimensionless wall distances $y^+$, as illustrated in \mbox{Fig.\ \ref{fig:y_plus_lz_0pr25}}. 
\par The maximal dimensionless wall distance is achieved for the \mbox{$m{-}L_3^{min}-y^{+}_{max}$} mesh. Since these values are within a range of $0\leq y^+\leq 1.4$, the viscous sublayer can still be resolved and only minor velocity differences are present on the cells. Hence, the \mbox{$m{-}L_3^{min}-y^{+}_{max}$} grid is selected for the future FSI simulations, as it represents the best compromise between accuracy and required computational time.

\subsection{Influence of the mesh adaption algorithm}
\label{appendix_mesh_adaption}

\par The $m{-}L_3^{min}-y^{+}_{max}$ mesh with the mapping strategy 4 is applied in order to thoroughly analyze the available mesh adaption algorithms. Since the largest displacements are achieved for the test case characterized by extremely small linear and torsional stiffnesses (see \mbox{Section \ref{sec:torsional_divergence})}, the investigations are based on the maximal airfoil response achieved by this computation.
\par Various simulations carried out with distinct mesh adaption algorithms, as well as various re-meshing parameters are performed in order to establish the method that provides the better quality of the mesh while requiring the smallest computational time. These test cases are simulated considering only a translational displacement of $X_2=0.3\,\text{m}$, which is slightly greater than the maximal translational displacement achieved by the coupled simulation with the TFI method (see Section \ref{subsubsec:flutter_displacements}), since the rotational displacement is not considered. The latter is not thoroughly studied, since the main cause of the numerical difficulties is actually the maintenance of the height of the cells located on the airfoil surface and not the preservation of the mesh orthogonality \mbox{(see Section \ref{subsubsec:flutter_mesh_quality})}.
\par The translational displacement of $X_2=0.3\,\text{m}$ is gradually applied as a boundary condition according to Eq.\ (\ref{eq:disp_func}), regarding that $A=0.3\,$m. This procedure must be undertaken, since the IDW and hybrid IDW-TFI algorithms adapt the mesh in relation to the last time step. A total of $n=1{,}000$ time steps are computed with a time step size of \mbox{$\Delta t=1{\cdot}10^{-3}\,\text{s}$}. The fluid domain is not solved and therefore the required computational time represents only the time needed to adapt the mesh in relation to the new boundaries.  
\begin{equation}
\label{eq:disp_func}
X_{2,\,mesh\,adapt.}=A\,t
\end{equation}
\par The results achieved for all test cases are summarized in Table \ref{table:comp_re_meshing_algorithms}, for the TFI, IDW and hybrid IDW-TFI methods examined. Although the former algorithm is not applicable for this test case (see \mbox{Section \ref{sec:torsional_divergence})}, it is also computed in order to enable a comparison of the required computational time and the mesh quality. The weighting function used by the IDW and IDW-TFI methods (see Section \ref{subsubsec:idw_tfi}), can be adapted according to the parameters $\alpha_{fxd}$ and $\alpha_{mv}$. Their increase leads to a stiffer mesh adaption (see \mbox{Sen et al.\ \cite{Sen_2017})}. The former parameter ($\alpha_{fxd}$) describes the adaption of the fixed boundaries, i$.$e$.$, inlet, outlet and symmetry while the latter ($\alpha_{mv}$) describes the adaption of the moving boundaries, that is, the FSI no-slip wall boundary.
\begin{table}[H]
	\centering
	\begin{tabular}{p{3cm} p{3cm} p{3.5cm} p{4cm}}
		\hline
		\multicolumn{2}{c}{\bf{Re-meshing}} & \multicolumn{1}{c}{\multirow{2}{3cm}{\centering{\bf{Computational time (h)}}}} & \multicolumn{1}{c}{\multirow{2}{4cm}{\centering{\bf{First cell height $\Delta y^{first\,cell}$ (m)}}}} \tabularnewline\cline{1-2}
		\centering{\bf{Method}} & \centering{\bf{Parameters}} & & \tabularnewline \hline
		\multicolumn{1}{c}{\multirow{1}{*}{\centering{TFI}}} & \multicolumn{1}{c}{\multirow{1}{*}{\centering{-}}} & \multicolumn{1}{c}{\multirow{1}{*}{\centering{0.020}}} & \centering{$\left[2.8{\cdot}10^{-5},\; 7.0{\cdot}10^{-5}\right]$} \tabularnewline
		\multicolumn{1}{c}{\multirow{2}{*}{\centering{IDW}}} & \centering{$\alpha_{fxd}=0.1$ $\alpha_{mv}=0.3$} & \multicolumn{1}{c}{\multirow{2}{*}{\centering{9.639}}} & \multicolumn{1}{c}{\multirow{2}{*}{\centering{$5.0{\cdot}10^{-5}$}}} \tabularnewline
		\multicolumn{1}{c}{\multirow{2}{*}{\centering{IDW-TFI}}} & \centering{$\alpha_{fxd}=0.05$ $\alpha_{mv}=0.3$} & \multicolumn{1}{c}{\multirow{2}{*}{\centering{3.035}}} & \multicolumn{1}{c}{\multirow{2}{*}{\centering{$5.0{\cdot}10^{-5}$}}} \tabularnewline
		\multicolumn{1}{c}{\multirow{2}{*}{\centering{IDW-TFI}}} & \centering{$\alpha_{fxd}=0.1$ $\alpha_{mv}=0.3$} & \multicolumn{1}{c}{\multirow{2}{*}{\centering{3.030}}} & \multicolumn{1}{c}{\multirow{2}{*}{\centering{$5.0{\cdot}10^{-5}$}}} \tabularnewline
		\multicolumn{1}{c}{\multirow{2}{*}{\centering{IDW-TFI}}} & \centering{$\alpha_{fxd}=0.13$ $\alpha_{mv}=0.3$} & \multicolumn{1}{c}{\multirow{2}{*}{\centering{3.031}}} & \multicolumn{1}{c}{\multirow{2}{*}{\centering{$5.0{\cdot}10^{-5}$}}} \tabularnewline
		\multicolumn{1}{c}{\multirow{2}{*}{\centering{IDW-TFI}}} & \centering{$\alpha_{fxd}=0.15$ $\alpha_{mv}=0.3$} & \multicolumn{1}{c}{\multirow{2}{*}{\centering{3.053}}} & \multicolumn{1}{c}{\multirow{2}{*}{\centering{$5.0{\cdot}10^{-5}$}}} \tabularnewline
		\multicolumn{1}{c}{\multirow{2}{*}{\centering{IDW-TFI}}} & \centering{$\alpha_{fxd}=0.2$ $\alpha_{mv}=0.3$} & \multicolumn{1}{c}{\multirow{2}{*}{\centering{3.034}}} & \multicolumn{1}{c}{\multirow{2}{*}{\centering{$5.0{\cdot}10^{-5}$}}} \tabularnewline
		\multicolumn{1}{c}{\multirow{2}{*}{\centering{IDW-TFI}}} & \centering{$\alpha_{fxd}=0.5$ $\alpha_{mv}=0.3$} & \multicolumn{1}{c}{\multirow{2}{*}{\centering{3.032}}} & \multicolumn{1}{c}{\multirow{2}{*}{\centering{$5.0{\cdot}10^{-5}$}}} \tabularnewline
		\multicolumn{1}{c}{\multirow{2}{*}{\centering{IDW-TFI}}} & \centering{$\alpha_{fxd}=0.1$ $\alpha_{mv}=0.1$} & \multicolumn{1}{c}{\multirow{2}{*}{\centering{3.033}}} & \multicolumn{1}{c}{\multirow{2}{*}{\centering{$5.0{\cdot}10^{-5}$}}} \tabularnewline
		\multicolumn{1}{c}{\multirow{2}{*}{\centering{IDW-TFI}}} & \centering{$\alpha_{fxd}=0.13$ $\alpha_{mv}=0.1$} & \multicolumn{1}{c}{\multirow{2}{*}{\centering{3.069}}} & \multicolumn{1}{c}{\multirow{2}{*}{\centering{$5.0{\cdot}10^{-5}$}}} \tabularnewline
		\multicolumn{1}{c}{\multirow{2}{*}{\centering{IDW-TFI}}} & \centering{$\alpha_{fxd}=0.13$ $\alpha_{mv}=0.5$} & \multicolumn{1}{c}{\multirow{2}{*}{\centering{3.031}}} & \multicolumn{1}{c}{\multirow{2}{*}{\centering{$5.0{\cdot}10^{-5}$}}} \tabularnewline
		\hline
	\end{tabular}
	\caption{\label{table:comp_re_meshing_algorithms}Comparison of the re-meshing algorithms: Translational displacement of \mbox{$X_2=0.3\,m$}.}
\end{table}
\par
\begin{figure}[H]
	\centering
	\subfigure[Adapted mesh - Overall view.]{\includegraphics[width=0.49\textwidth]{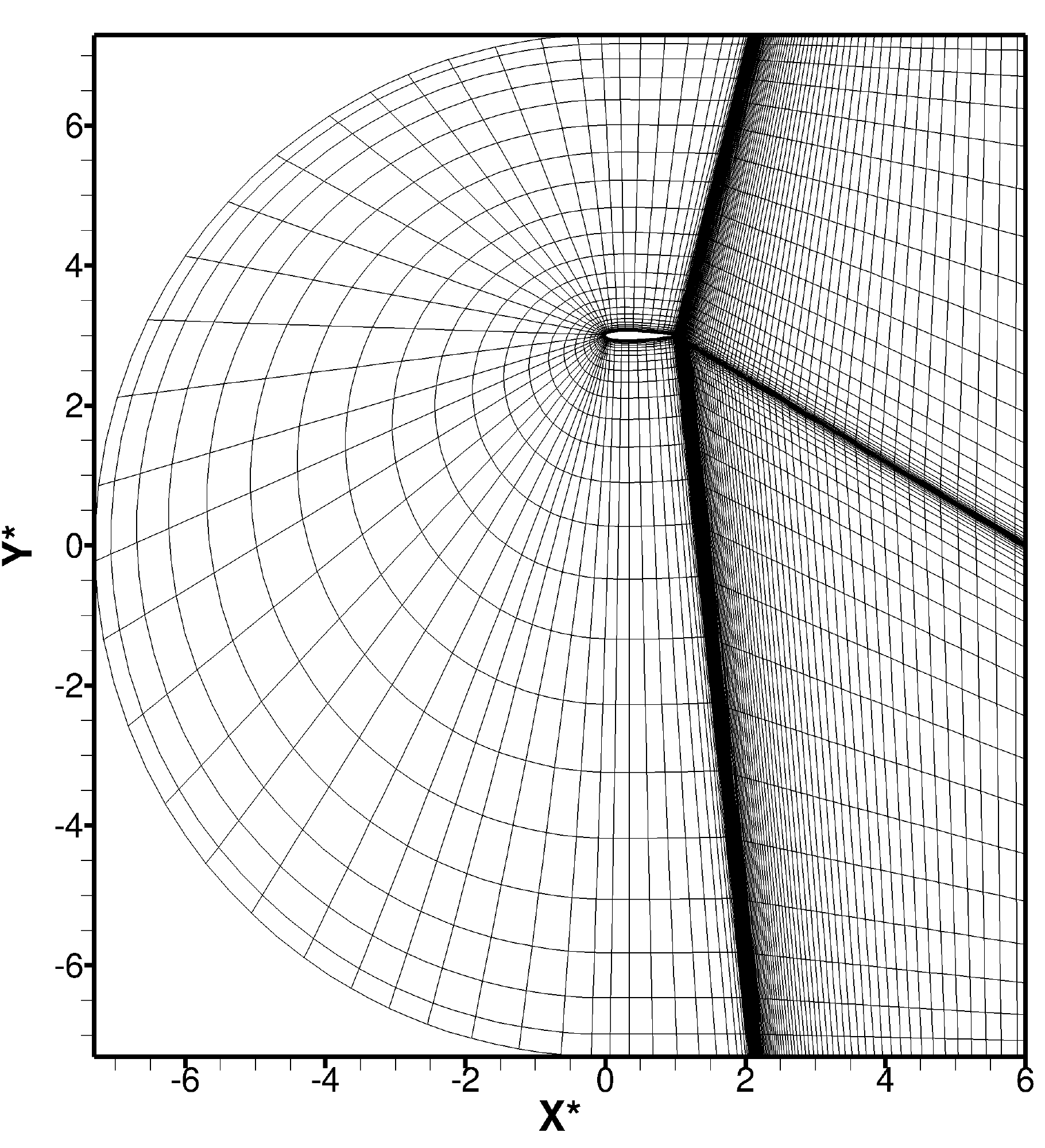} \label{fig:mesh_overall_tfi_re_meshing}}\hfill
	\subfigure[Adapted mesh - Focus on the trailing-edge.]{\includegraphics[width=0.49\textwidth]{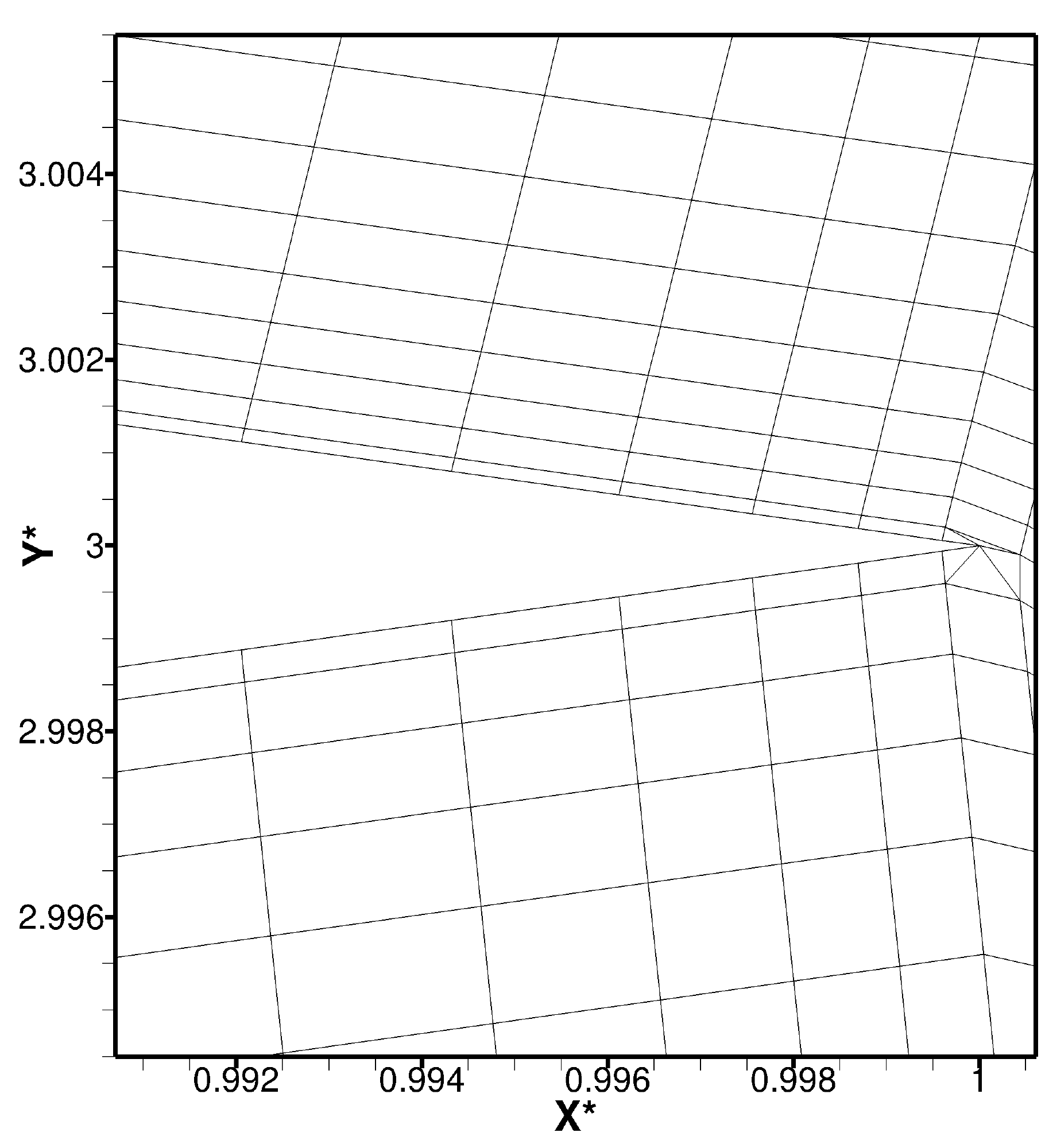} \label{fig:mesh_trailing_edge_tfi_re_meshing}}\hfill
	\caption{TFI algorithm: Mesh adaption considering a translational displacement of \mbox{$X_2=0.3\,\text{m}$}. Note that instead of the grid, the grid connecting the cell centers is shown.}
	\label{fig:mesh_tfi_re_meshing}
\end{figure}

 The smallest computational time is achieved utilizing the TFI algorithm. The adapted mesh is illustrated in Fig.\ \ref{fig:mesh_tfi_re_meshing} for the considered displacement, regarding that one out of two mesh lines is displayed for both the $\xi$-axis on the front domain and the $\eta$-axis on the whole fluid domain. The mesh orthogonality is not deeply affected due to the fact that no airfoil rotation is present. The first cell height is not maintained, varying between $2.8{\cdot}10^{-5}\,\text{m}\leq \Delta y^{first\,cell}\leq 7.0{\cdot}10^{-5}\,\text{m}$ for the control volumes located on the airfoil surface,\break where the initial value is $\Delta y^{first\,cell}= 5.0{\cdot}10^{-5}\,\text{m}$. The cell height is also not maintained at the fixed boundaries, i$.$e$.$, the control volumes located at the bottom of the fluid domain are larger than those located at the top. Moreover, degenerated cells occur at the\break trailing-edge. Therefore, the utilization of this adaptation method always leads to numerical difficulties if large displacements are observed, even if no rotation is present.
\par The IDW  and hybrid IDW-TFI methods are characterized by the maintenance of the first cell height, i$.$e$.$, $\Delta y^{first\,cell}=5{\cdot}10^{-5}\,\text{m}$. The required computational time of the pure IDW method is about 3.2 and 481.9 times higher than the hybrid IDW-TFI and TFI algorithms, respectively. Since the achieved mesh quality is exactly the same for the hybrid method, the IDW scheme is not considered for further studies.
\par
\begin{figure}[H]
	\centering
	\subfigure[Adapted mesh - Overall view.]{\includegraphics[width=0.49\textwidth]{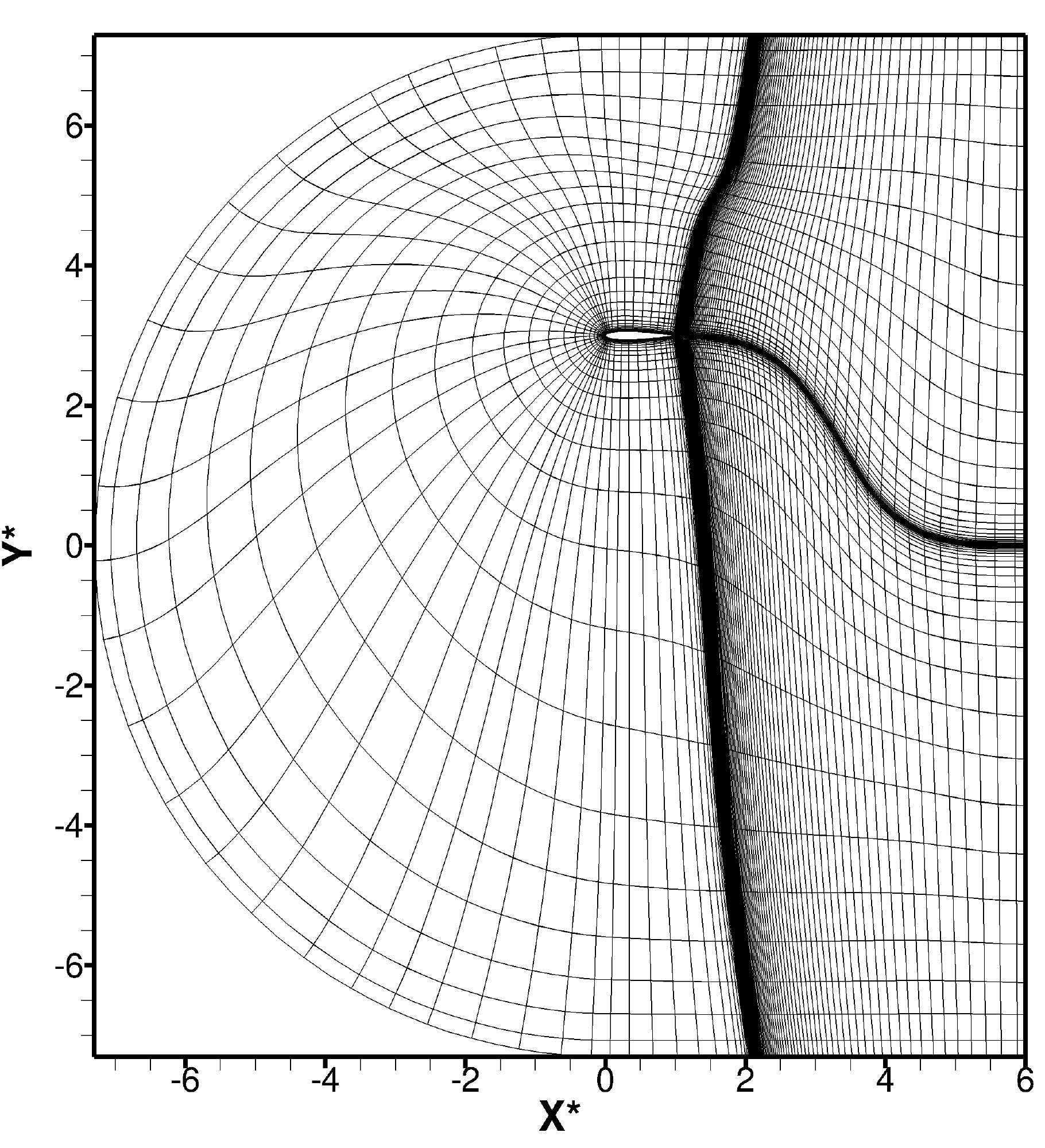} \label{fig:mesh_overall_idw_tfi_re-meshing}}\hfill
	\subfigure[Adapted mesh: Focus on the trailing-edge.]{\includegraphics[width=0.49\textwidth]{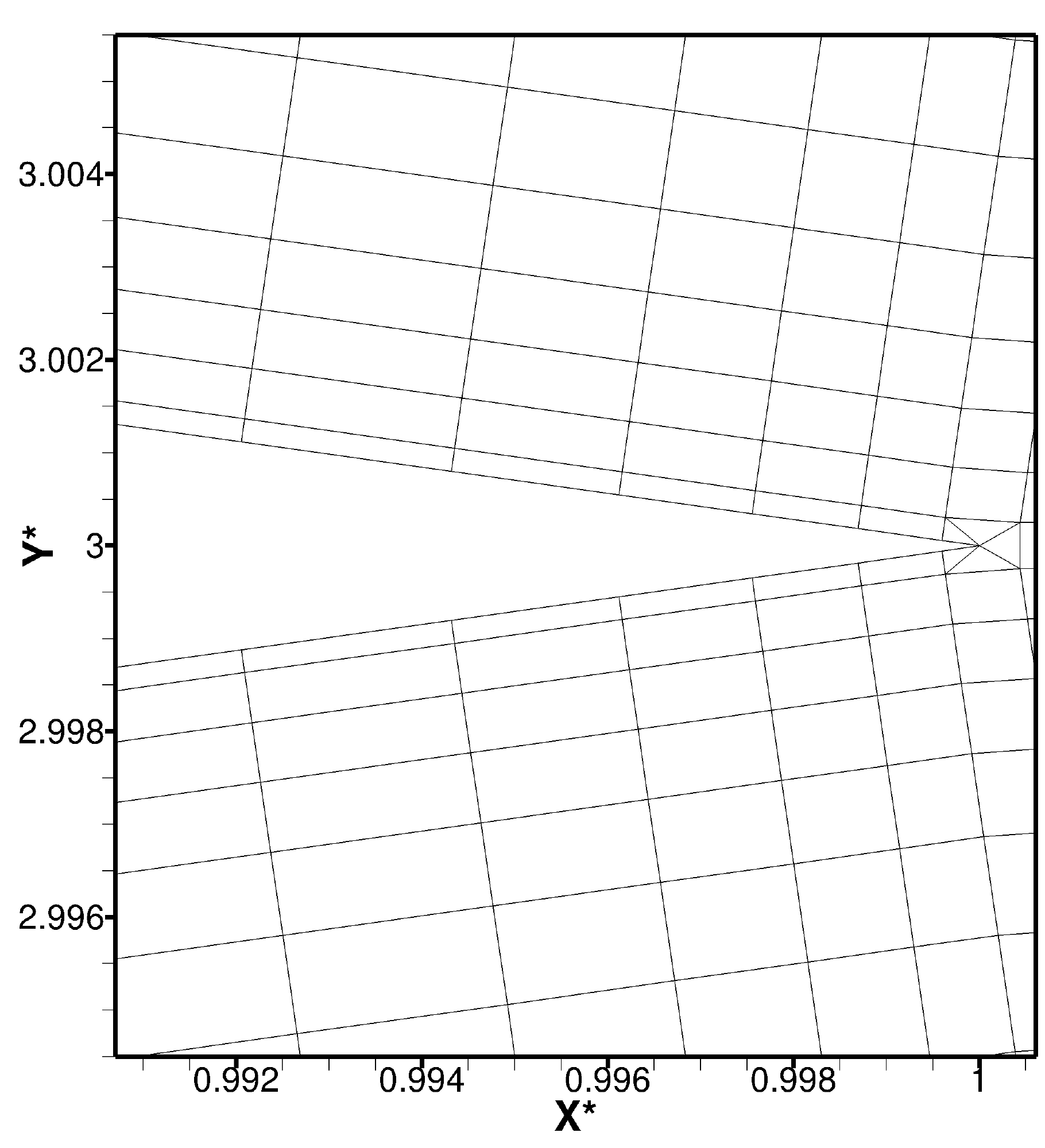}\label{fig:mesh_trailing_edge_idw_tfi_re_meshing}}\hfill
	\caption{Hybrid IDW-TFI algorithm with $\alpha_{fxd}=\alpha_{mv}=0.1$: Mesh adaption considering a translational displacement of $X_2=0.3\,\text{m}$. Grid connecting the cell centers is depicted.}
	\label{fig:mesh_idw_tfi_re_meshing}
\end{figure}

 Nine different combinations of the parameters that determine the weighting function of the fixed and the movable boundaries, i$.$e$.$, $\alpha_{fxd}$ and $\alpha_{mv}$, are investigated for the hybrid IDW-TFI algorithm. Since the best mesh quality is achieved for the parameters $\alpha_{fxd}=\alpha_{mv}=0.1$, the mesh adapted utilizing these values is illustrated in Fig.\ \ref{fig:mesh_idw_tfi_re_meshing}.
\par
\begin{figure}[H]
	\centering
	\centering
	\includegraphics[scale=0.8,draft=\drafttype]{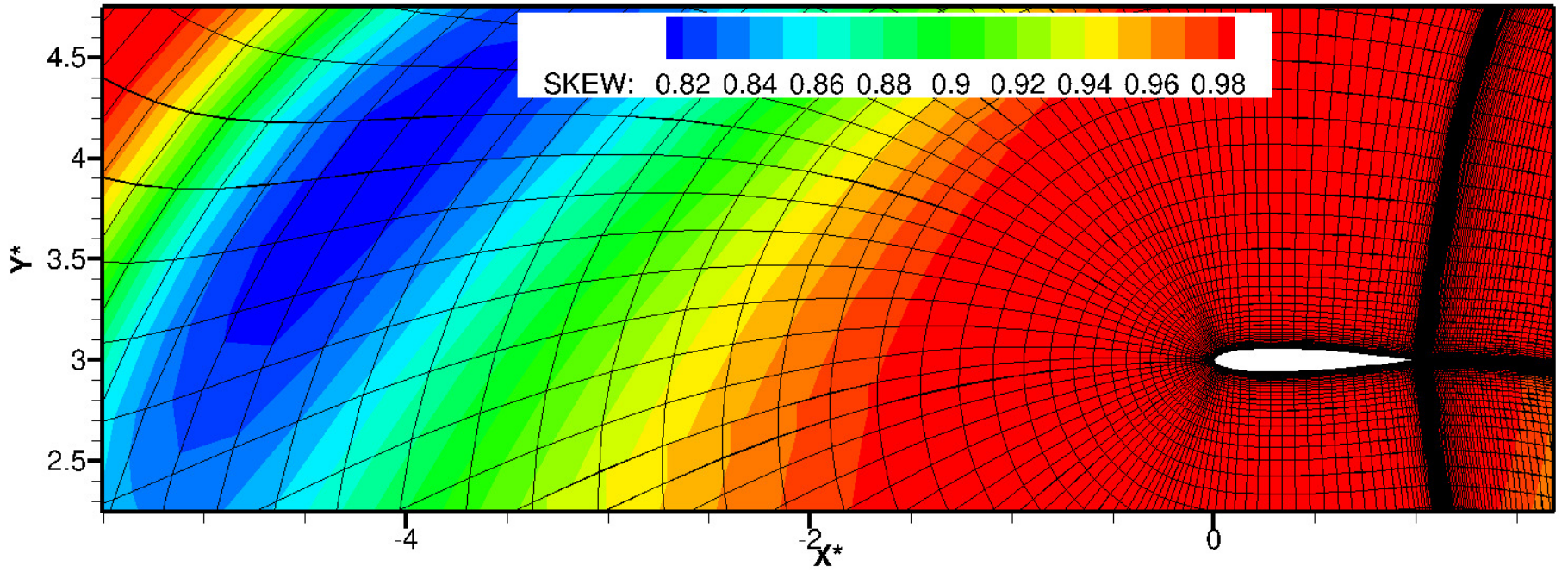}
	\caption{\label{fig:mesh_skew_idw_tfi_re_meshing}Hybrid IDW-TFI algorithm with $\alpha_{fxd}=\alpha_{mv}=0.1$: Mesh quality based on the skewness of the grid considering a translational displacement of $X_2=0.3\,\text{m}$. Grid connecting the cell centers is depicted.}
\end{figure}

 The problems generated by the TFI algorithm are overcome through the utilization of this method, i$.$e$.$, no degenerated cells are generated at the trailing-edge, as well as the first cell height is maintained for all cells located on the airfoil surface, enabling the computation with constant time step sizes. Moreover, the required computational time is reasonable considering the one needed by the pure IDW method. Although the achieved skew metric quality is low, i$.$e$.$, $f_{skew}\geq0.599$, Fig.\ \ref{fig:mesh_skew_idw_tfi_re_meshing} illustrates that the cells with lower quality are located far away from the airfoil, which does not lead to accuracy problems since the critical flow phenomena occur near the airfoil.

\subsection{Conclusions}
\label{sec:conclusions_mesh_investigations}
\par The thorough investigations of the three generated meshes and the four\break mapping strategies pointed out that the minimal computational time is achieved with the \mbox{$m{-}L_3^{min}-y^{+}_{max}$} mesh while utilizing the mapping strategy 4. Since this configuration also provides a reasonable accuracy, this is applied to approximate the solution of the coupled test cases.
\par Moreover, the TFI method is applied in case of small displacements, since this requires small computational times in order to generate good quality meshes. Furthermore, the hybrid IDW-TFI method is chosen to simulate cases characterized by large airfoil translational and rotational displacement. Although this requires a higher computational time than the TFI method, it does not lead to numerical difficulties due to the maintenance of the height of the cells located on the airfoil surface and therefore can also be utilized for large rotations without degenerated cells with negative volumes. In addition, the parameters $\alpha_{fxd}=\alpha_{mv}=0.1$ are applied since the generation of the best mesh quality considering the orthogonality of the grid is achieved.

\markboth{CHAPTER 3.$\quad$PRELIMINARY STUDIES}{3.1$\quad$CSD STUDIES}
\section{CSD studies}\label{appendix_validation_CSD}

\par In the case of fluid-structure interaction between the flow and the NACA0012 profile with two degrees of freedom (pitch and up and down), the external forces acting on the rigid airfoil, i$.$e$.$, the fluid forces, are time-dependent. Since Viets \cite{Viets_2013} validated the applied in-house rigid movement solver (see Section \ref{sec:CSD}) only for the case of a constant external force, this software is also verified regarding systems with a time-dependent external force and a coupled translational and rotational motion.

\subsection{System under rotating unbalance}
\label{sec:rotating_unbalance}
\par
\begin{figure}[H]
	\centering
	\includegraphics[scale=0.53,draft=\drafttype]{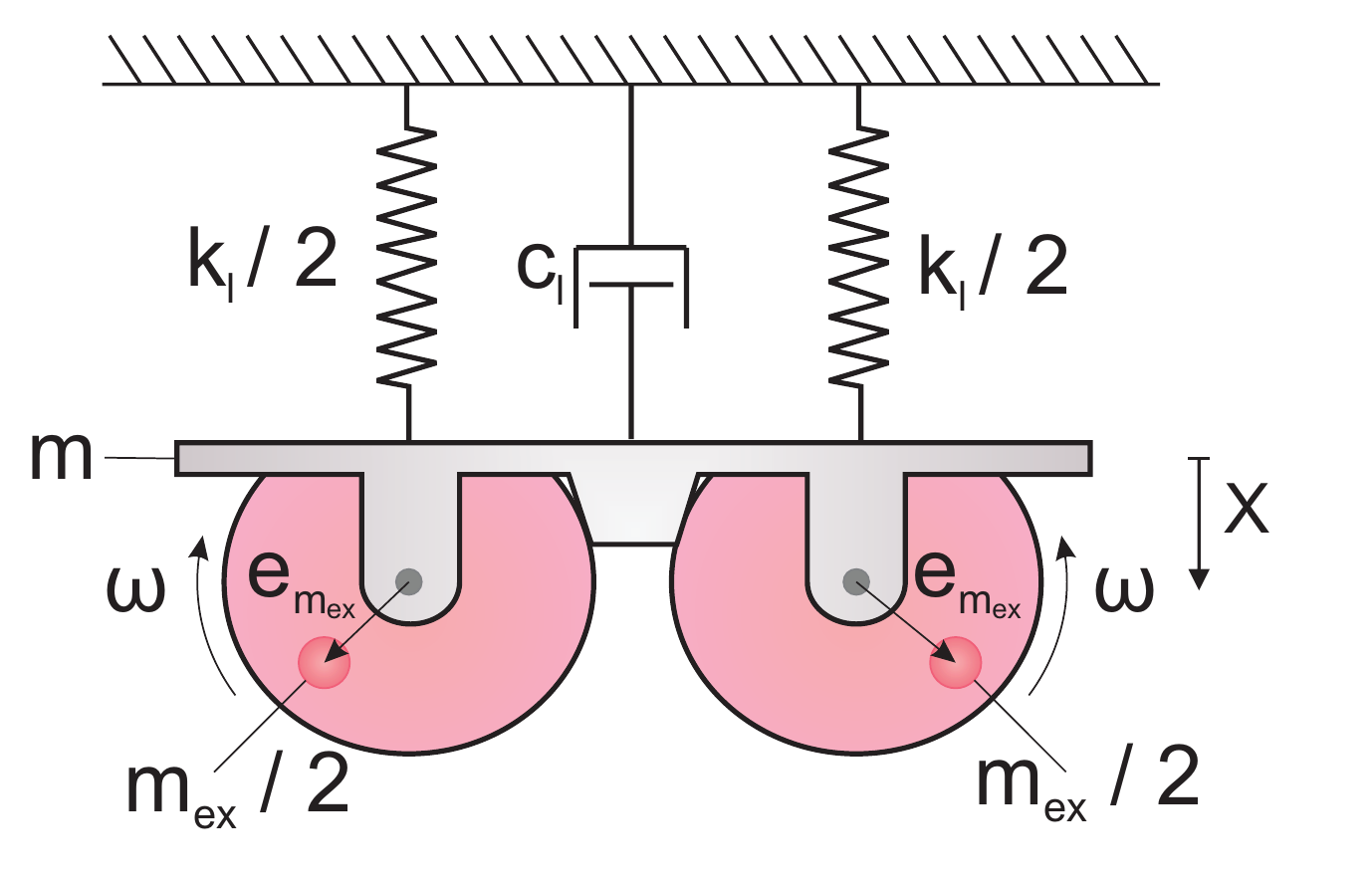}
	\caption{\label{fig:rotating_unbalance}System under rotating unbalance.}
\end{figure}

 Rotating unbalance is one of the main causes of vibration in rotating machinery, such as turbines and rotors. A simplified model, as illustrated in Fig.\ \ref{fig:rotating_unbalance}, is utilized in order to compare the analytical and numerical results. 
\par Two eccentric masses $m_{ex}/2$ rotate with the same constant angular frequency $\omega$ in opposite directions. These are located at a distance $e_{m_{ex}}$ from the center of mass. The generated centrifugal force ($m_{ex}\,e_{m_{ex}}\,\omega^2$) is decomposed into the horizontal and vertical components. While the former is canceled due to the rotation of the eccentric masses in opposite directions, the latter contributes to the vibration of the machine \cite{Rao_2011}. Therefore, displacements of the rigid body occurs only in the vertical direction.
\par Equations (\ref{eq:rotating_unbalance_undamped}) and (\ref{eq:rotating_unbalance_damped}) describe the motion of the body for the undamped and damped cases (see Rao \cite{Rao_2011}), respectively. $m$, $c_l$, $k_l$ and $t$ represent the total mass of the system, the linear damping coefficient, the linear spring stiffness and the time, respectively. $\ddot{X}$, $\dot{X}$ and $X$ are the translational acceleration, velocity and displacement of the body (the subscripted $t$ and directional index $i=2$, for instance in $X_{t,\,2}$, are omitted for the sake of brevity). The left side of the equation stands for the internal forces while the right one represents the time-dependent external force caused by the rotation of the eccentric masses.
\begin{eqnarray}
\label{eq:rotating_unbalance_undamped}
m\,\ddot{X}+k_l\,X=m_{ex}\,e_{m_{ex}}\,\omega^2\,\sin(\omega t)\\
\label{eq:rotating_unbalance_damped}
m\,\ddot{X}+c_l\,\dot{X}+k_l\,X=m_{ex}\,e_{m_{ex}}\,\omega^2\,\sin(\omega t)
\end{eqnarray} 
\par The analytical solutions of these differential equations are based on the sum of the homogeneous and the particular solutions (see Rao \cite{Rao_2011}). Equations (\ref{eq:rotating_unbalance_solution_undamped}) and (\ref{eq:rotating_unbalance_solution_underdamped}) represent respectively the solution of the undamped and the underdamped (D\,<\,1) systems:
\begin{eqnarray}
\label{eq:rotating_unbalance_solution_undamped}
X(t)&=&X_0\,\cos(\omega_n t + \phi_1)+X_0\,\sin(\omega t + \phi_2), \\\nonumber
\label{eq:rotating_unbalance_solution_underdamped}
X(t)&=&e^{-D\,\omega_n\,t}\,[X_0\,\cos(\omega_n t+\phi_1)+D\,X_0\,\sin(\omega_n t+\phi_1)]\\&+&X_0\,\sin(\omega t+\phi_2).
\end{eqnarray}
\par $X_0$ represents a characteristic amplitude and is calculated according to Eq.\ (\ref{eq:rotating_unbalance_amplitude}). The damping ratio $D$ is a relation between the utilized damping constant $c_l$ and the critical damping coefficient $c_c$, according to Eq.\ (\ref{eq:damping_ratio}). The angular natural frequency $w_n$ is a relation between the spring stiffness $k_l$ and the total mass of the system $m$ (see Eq.\ (\ref{eq:natural_frequency})). $\phi_1$ and $\phi_2$ represent the phase angles. These are calculated through a system of equations, which are generated by the application of the initial conditions in the governing equation \mbox{(either Eq.\ (\ref{eq:rotating_unbalance_undamped}) or Eq.\ (\ref{eq:rotating_unbalance_damped}))}. The current analysis utilizes a zero displacement and zero velocity as initial conditions, i$.$e$.$, $X(t=0)=0$ and $\dot{X}(t=0)=0$.
\begin{eqnarray}
\label{eq:rotating_unbalance_amplitude}
X_0&=&\frac{m_{ex}\,e_{m_{ex}}\,\omega^2}{\sqrt{(k_l-m\,\omega^2)^2+(c_l\,\omega)^2}}\\
\label{eq:damping_ratio}
D&=&\frac{c_l}{c_c}=\frac{c_l}{2\sqrt{k\,m}}\\
\label{eq:natural_frequency}
\omega_n&=&\sqrt{\frac{k_l}{m}}
\end{eqnarray}
\par When the angular frequency $\omega$ of the eccentric masses and the angular natural frequency $\omega_n$ have the same value, resonance occurs. The current analysis is performed for an angular frequency near the natural frequency for damping ratios of $D=0$ and $D=0.25$. The setup of these test cases is presented in Table \ref{table:rotating_unbalance_variables}, while the analytical solution is indicated in \mbox{Table \ref{table:rotating_unbalance_analytical}}:
\begin{table}[H]
	\centering
	\begin{tabular}{p{3cm} p{3.5cm} p{3cm}}
		\hline
		{\bf{Variable}} & \bf{Undamped case} & \bf{Damped case} \tabularnewline \hline
		$m\,(\text{kg})$ &\multicolumn{1}{c}{10} & \multicolumn{1}{c}{10}  \tabularnewline
		$c_l\,(\text{N}{\cdot}\text{s}{\cdot}\text{m}^{-1})$ & \multicolumn{1}{c}{0} & \multicolumn{1}{c}{25} \tabularnewline
		$D$& \multicolumn{1}{c}{0}& \multicolumn{1}{c}{0.25} \tabularnewline
		$k_l\,(\text{N}{\cdot}\text{m}^{-1})$& \multicolumn{1}{c}{250}& \multicolumn{1}{c}{250} \tabularnewline
		$m_{ex}\,(\text{kg})$& \multicolumn{1}{c}{1}& \multicolumn{1}{c}{1} \tabularnewline
		$e_{m_{ex}}\,(\text{m})$& \centering{0.45}& \centering{0.45} \tabularnewline	
	\end{tabular}
\end{table}
\begin{table}[H]
	\centering
	\begin{tabular}{p{3cm} p{3.5cm} p{3cm}}
		$\omega\,(\text{rad}{\cdot}\text{s}^{-1})$& \centering{4.5} & \centering{4.5}  \tabularnewline
		$\omega_n\,(\text{rad}{\cdot}\text{s}^{-1})$& \centering{5}& \centering{5} \tabularnewline
		\hline	
	\end{tabular}
	\caption{\label{table:rotating_unbalance_variables}Setup of the systems submitted to rotating unbalance.}
\end{table}
\begin{table}[H]
	\centering
	\begin{tabular}{l c c}
		\hline
		\multicolumn{1}{c}{\bf{Variable}} & \bf{Undamped case} & \bf{Damped case} \tabularnewline \hline
		$X_0\,(\text{m})$& 0.182& 0.075 \tabularnewline
		$\phi_1\,(\text{rad})$& $\pi$& -0.133 \tabularnewline
		$\phi_2\,(\text{rad})$& $\pi/2$ & -1.438 \tabularnewline
		\hline	
	\end{tabular}
	\caption{\label{table:rotating_unbalance_analytical}Analytical solutions - Rotating unbalance test case.}
\end{table}
\par Two numerical simulations are performed for each system configuration: One with the ode45 solver of Matlab Simulink and the other with the rigid movement solver implemented by Viets \cite{Viets_2013}. The former is based on the Dormand-Prince method \cite{Dormand_1980}, which is a fifth-order accurate explicit variable-step solver applicable for models with continuous states. Therefore, it is suitable for mass-spring-damper systems. The specified minimal and maximal step sizes are $\Delta t_{min}=10^{-4}\,\text{s}$ and $\Delta t_{max}=10^{-3}\,\text{s}$, regarding that the time step size is reduced or increased within this interval using a local error control to achieve the specified tolerance \cite{Matlab_2015}, i$.$e$.$, $10^{-3}$. The rigid movement solver (see Section \ref{sec:CSD}) applies the implicit standard Newmark time discretization method with a time step size of $\Delta t=10^{-3}\,\text{s}$, since this is the largest value that does not influence the solution. The system is allowed to oscillate \mbox{only in one degree of freedom at a time.}
\par
\begin{figure}[H]
	\centering
	\subfigure[Undamped system $D=0$.]{\includegraphics[width=0.49\textwidth]{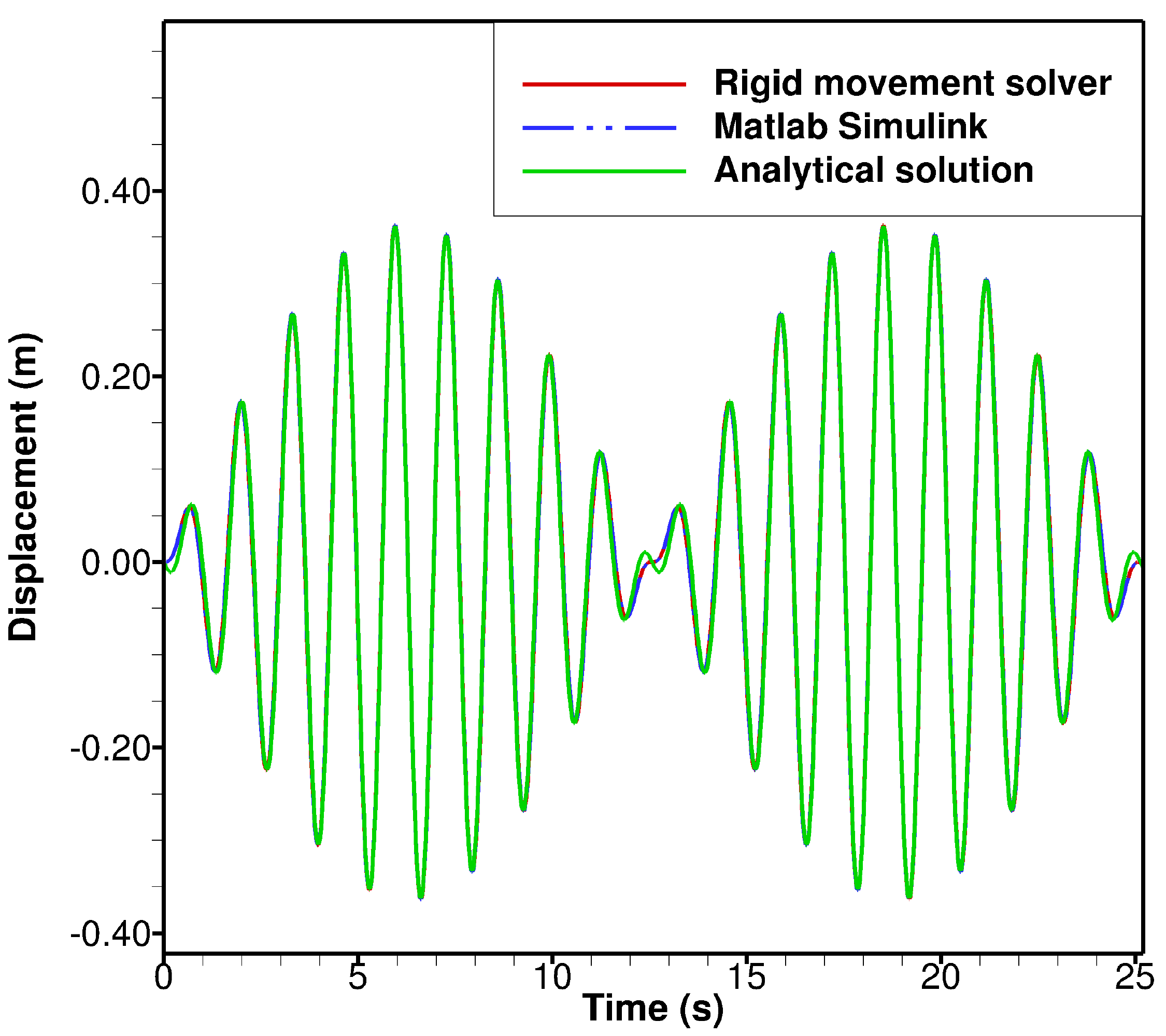}\label{fig:disp_rotating_unbalance_undamped}}\hfill
	\subfigure[Damped system $D=0.25$.]{\includegraphics[width=0.49\textwidth]{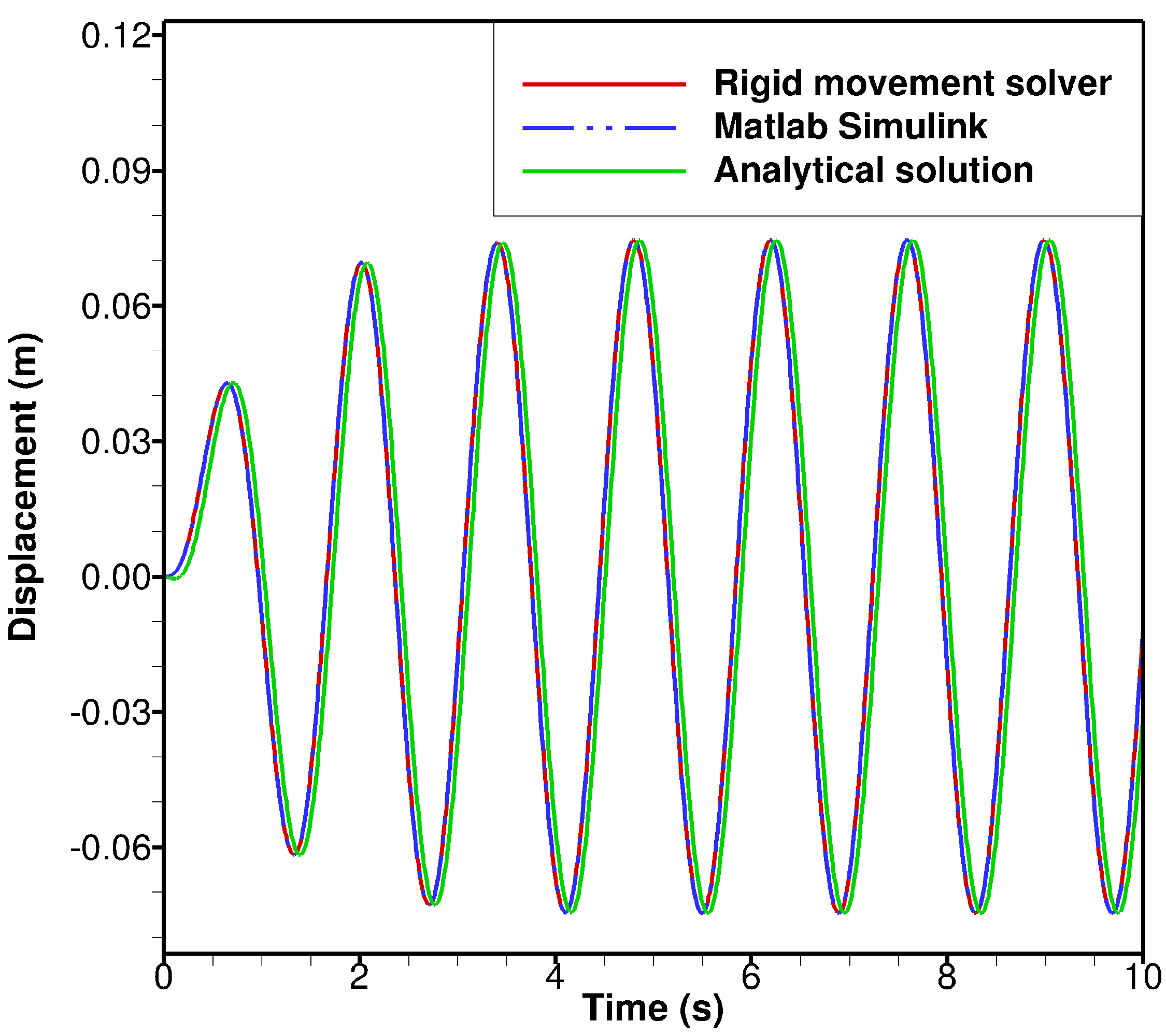} \label{fig:disp_rotating_unbalance_damped}}\hfill
	\caption{Displacements of the undamped and damped rotating unbalance cases: Numerical and analytical solutions.}
	\label{fig:disp_rotating_unbalance}
\end{figure}

 The comparison between analytical and numerical solutions are presented in\break \mbox{Fig.\ \ref{fig:disp_rotating_unbalance}} for both undamped and damped cases. All six degrees of freedom are tested and deliver the same solution.   
\par The numerical solutions achieved with Matlab Simulink and the rigid movement solvers deliver exactly the same results. Although the utilized initial conditions and the maximal amplitude are the same for both numerical and analytical solutions, a small discrepancy is present at the beginning of the oscillation for the undamped case and a minimal phase shift is present for the damped case. Since the angular frequency of the eccentric masses are near the resonance frequency, a considerable amplitude of oscillation is achieved for $D=0$ and the damping coefficient is responsible for the constant amplitude achieved for $D=0.25$.

\subsection{System under coupled translation and rotation}
\label{sec:coupled_translation_rotation}

\par
\begin{figure}[H]
	\centering
	\includegraphics[scale=0.8,draft=\drafttype]{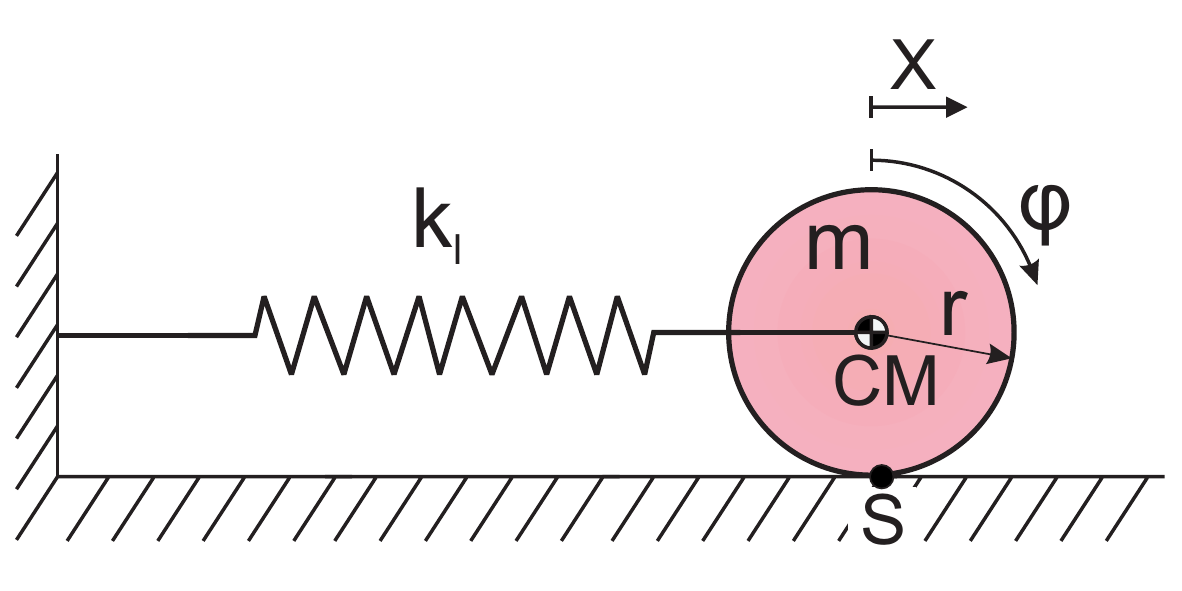}
	\caption{\label{fig:coupled_rotation_translation}System with coupled rotation and translation.}
\end{figure}

 The disc of the disc-spring system rolls over its instantaneous center of rotation $S$ without sliding (see Fig.\ \ref{fig:coupled_rotation_translation}). Its movement is characterized by a coupled translational and rotational motion, since the instantaneous center of rotation does not coincide with the center of mass $\mbox{CM}$.
\par The governing equations for the rotational and translational motions (see Eqs.\ (\ref{eq:disc_spring_system_rotation}) and (\ref{eq:disc_spring_system_translation})) are given by Euler's equation of motion for a rotation outside the center of mass (see Gere \cite{Gere_2004}) and through the relation between the rotational and translational motions (see Rao \cite{Rao_2011}). $\ddot{\varphi}$ and $\varphi$ represent the angular acceleration and displacement while $k_l$, $m$, $r$ and $X$ represent the linear spring stiffness, the disc mass, the disc radius and the translational displacement, respectively. 
\begin{eqnarray}
\label{eq:disc_spring_system_rotation}
\ddot{\varphi}+\frac{2\,k_l}{3\,m}\varphi&=&0\\
\label{eq:disc_spring_system_translation}
X&=&r\,\varphi
\end{eqnarray}
\par The setup of the present test case is described in Table \ref{table:coupled_rotation_translation} for both the rigid movement solver and the ode45 solver of Matlab Simulink. The initial conditions are set to \mbox{$\varphi (t=0)=1$} and $\dot{\varphi} (t=0)=0$.
\begin{table}[H]
	\centering
	\begin{tabular}{p{3.5cm} p{4.5cm} p{4cm}}
		\hline
		\multicolumn{1}{c}{\multirow{2}{*}{\centering{\bf{Variable}}}} & \centering{\bf{Rigid movement solver}} & \centering{\bf{Matlab Simulink ode45 solver}} \tabularnewline \hline
		$m\,(\text{kg})$& \centering{1}& \centering{1} \tabularnewline
		$r\,(\text{m})$& \centering{0.5} & \centering{0.5} \tabularnewline
		$k_l\,(\text{N}{\cdot}\text{m}^{-1})$& \centering{6} & \centering{6} \tabularnewline
		$\Delta t\,(\text{s})$ & \centering{$10^{-3}$} & \centering{$10^{-4}\leq \Delta t\leq 10^{-3}$} \tabularnewline
		Time discretization method& \multicolumn{1}{c}{\multirow{2}{*}{\centering{Standard Newmark}}} & \multicolumn{1}{c}{\multirow{2}{*}{\centering{Dormand-Prince}}} \tabularnewline
		\hline	
	\end{tabular}
	\caption{\label{table:coupled_rotation_translation}Setup of the numerical solutions. Case: Coupled rotation and translation.}
\end{table}
\par
\begin{figure}[H]
	\centering
	\subfigure[Rotational displacement.]{\includegraphics[width=0.49\textwidth]{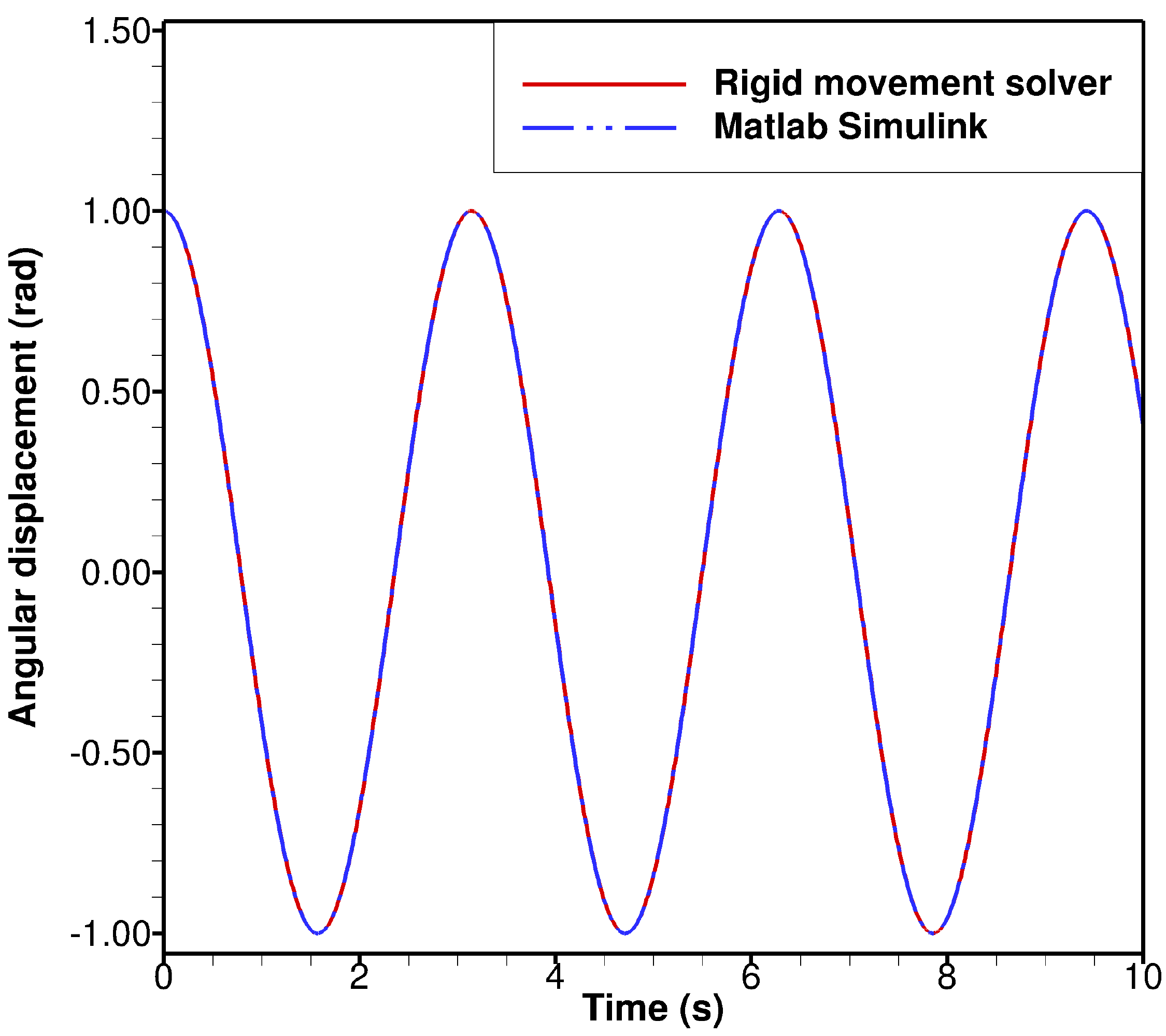} \label{fig:coupled_system_rotation}}\hfill
	\subfigure[Translation displacement.]{\includegraphics[width=0.49\textwidth]{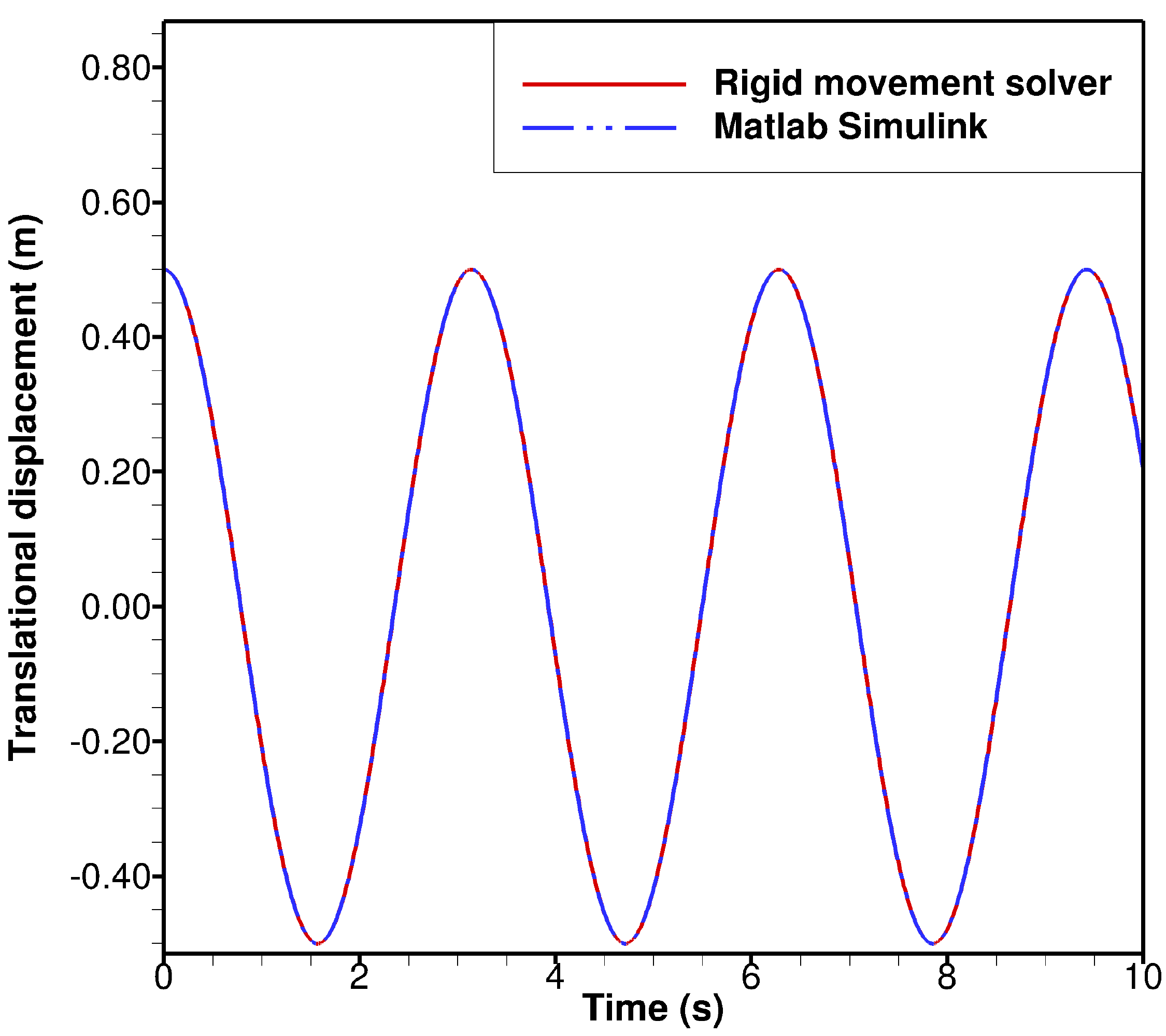}\label{fig:coupled_system_translation}}\hfill
	\caption{Translational and rotational displacements of the coupled system.}
	\label{fig:coupled_system_oscillation}
\end{figure}

 For the simulations performed with the rigid movement solver, two degrees of freedom (one translational and one rotational) are tested at a time and all nine possible combinations deliver the same results. The rotational and translational oscillations are illustrated in Fig.\ \ref{fig:coupled_system_oscillation}. The results achieved with the ode45 solver of Matlab Simulink and with the in-house rigid movement solver are exactly the same. 

\subsection{Conclusions}
\label{sec:conclusions_CSD_validation}
\par The executed test cases assure the accuracy of the rigid movement solver implemented by Viets \cite{Viets_2013} through comparisons with reference analytical and numerical solutions. The former is based on Rao \cite{Rao_2011} and applicable only for the rotating unbalance test case, while the latter is achieved through simulations performed in the reference solver \emph{ode45} from Matlab Simulink. Due to the great coincidence between all comparable solutions, the rigid movement code is thus also validated for problems characterized by time-dependent external forces and coupled motions. Therefore, it should be able to accurately compute the structural response of the NACA0012 airfoil, when it is submitted to fluid-structure interaction.

\markboth{CHAPTER 3.$\quad$PRELIMINARY STUDIES}{3.1$\quad$FSI COUPLING STUDIES}
\section{FSI coupling studies}\label{appendix_FSI_coupling_studies}

\par The algorithm utilized for the estimation of displacements, as well as the added mass effect and the coupling method are thoroughly analyzed. Since the utilization of the estimation of displacements together with a loosely coupled algorithm requires the smallest computational time, the studies primarily aim at the analysis of the accuracy of this method through a comparison with other algorithms. Thereby, the best compromise between accuracy and numerical effort can be determined and further utilized for the FSI simulations.

\subsection{Estimation of displacements}
\label{subsec:investigation_estimation_displacement}

\par The linear extrapolation of the displacements at the beginning of each new time step is implemented in order to improve the convergence properties and therefore to reduce the required computational time. Two strongly coupled simulations of the fluid-structure interaction between the NACA0012 airfoil and the flow are executed for the $m{-}L_3^{min}{-}y_{max}^+$ mesh utilizing the mapping strategy 4. The FSI is computed for an undamped system composed of two degrees of freedom (pitch and up and down) with linear and  torsional stiffnesses of $k_{l,\,2}=144\,\text{N}{\cdot}\text{m}^{-1}$ and $k_{t,\,3}=0.3832\,\text{N}{\cdot}\text{m}{\cdot}\text{rad}^{-1}$, respectively. Moreover, the chord Reynolds number is $Re=30{,}000$, the FSI convergence criterion is $\epsilon_{FSI,\,disp}<1{\cdot}10^{-6}$ and the computation runs until a dimensionless time of $t^*=4.47$ is achieved. Further properties of these simulations are summarized in Table \ref{table:comp_est_disp}:
\begin{table}[H]
	\centering
	\begin{tabular}{p{3cm} p{2.7cm} p{2.5cm} p{3cm} p{2cm}}
		\hline
		\multicolumn{1}{c}{\multirow{3}{*}{\centering{\bf{Mesh}}}} & \centering{\bf{Estimation of displacement}} & \multicolumn{1}{c}{\multirow{3}{2.5cm}{\centering{\bf{Computed time steps}}}} & \centering{\bf{Maximal number of sub-iterations}} & \centering{\bf{Time step size (s)}}\tabularnewline \hline
		\centering{$m{-}L_3^{min}{-}y^{+}_{max}$} & \centering{Yes} & \centering{$10{,}000$} & \centering{20} & \centering{$1\cdot10^{-5}$} \tabularnewline
		\centering{$m{-}L_3^{min}{-}y^+_{max}$} & \centering{No} & \centering{$14{,}286$} &\centering{20} & \centering{$7\cdot10^{-6}$} \tabularnewline			
		\hline	
	\end{tabular}
	\caption{\label{table:comp_est_disp}Simulation properties of the test cases aimed at the study of the advantages achieved by the  utilization of the prediction of the displacements.}
\end{table}
\par The simulation without prediction of the displacements requires a smaller time step, otherwise the computation of the FSI case diverges since the utilization of 20 sub-iterations is not enough to obtain a converged result for each time step.  Therefore, in order to predict the same dimensionless time interval, this simulation requires also the calculation of a greater number of time steps, increasing the necessary numerical effort.
\par
\begin{figure}[H]
	\centering
	\subfigure[Simulation with estimation of displacement.]{\includegraphics[width=0.495\textwidth]{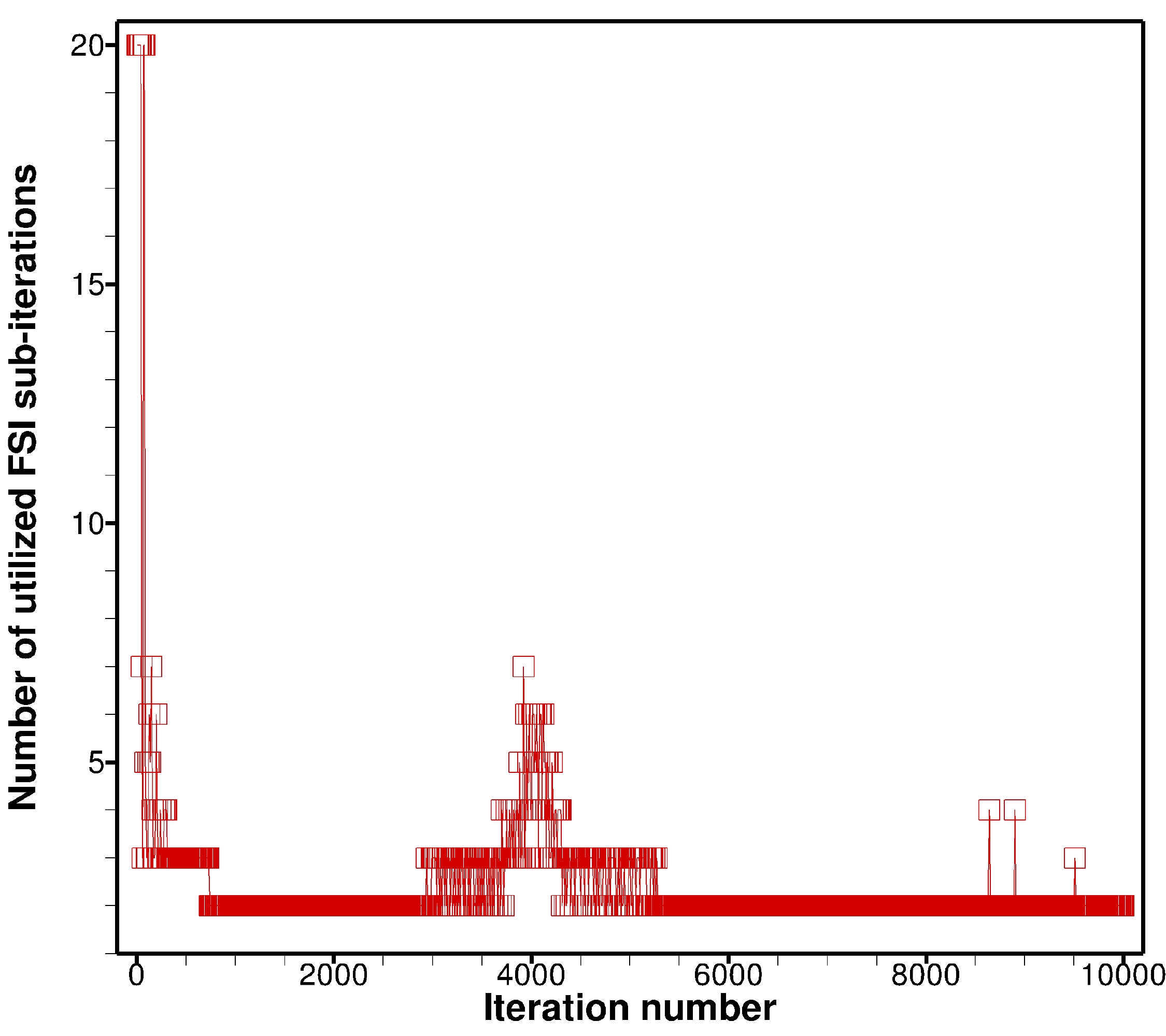} \label{fig:used_iterout_est_disp}}\hfill
	\subfigure[Simulation without estimation of displacement.]{\includegraphics[width=0.495\textwidth]{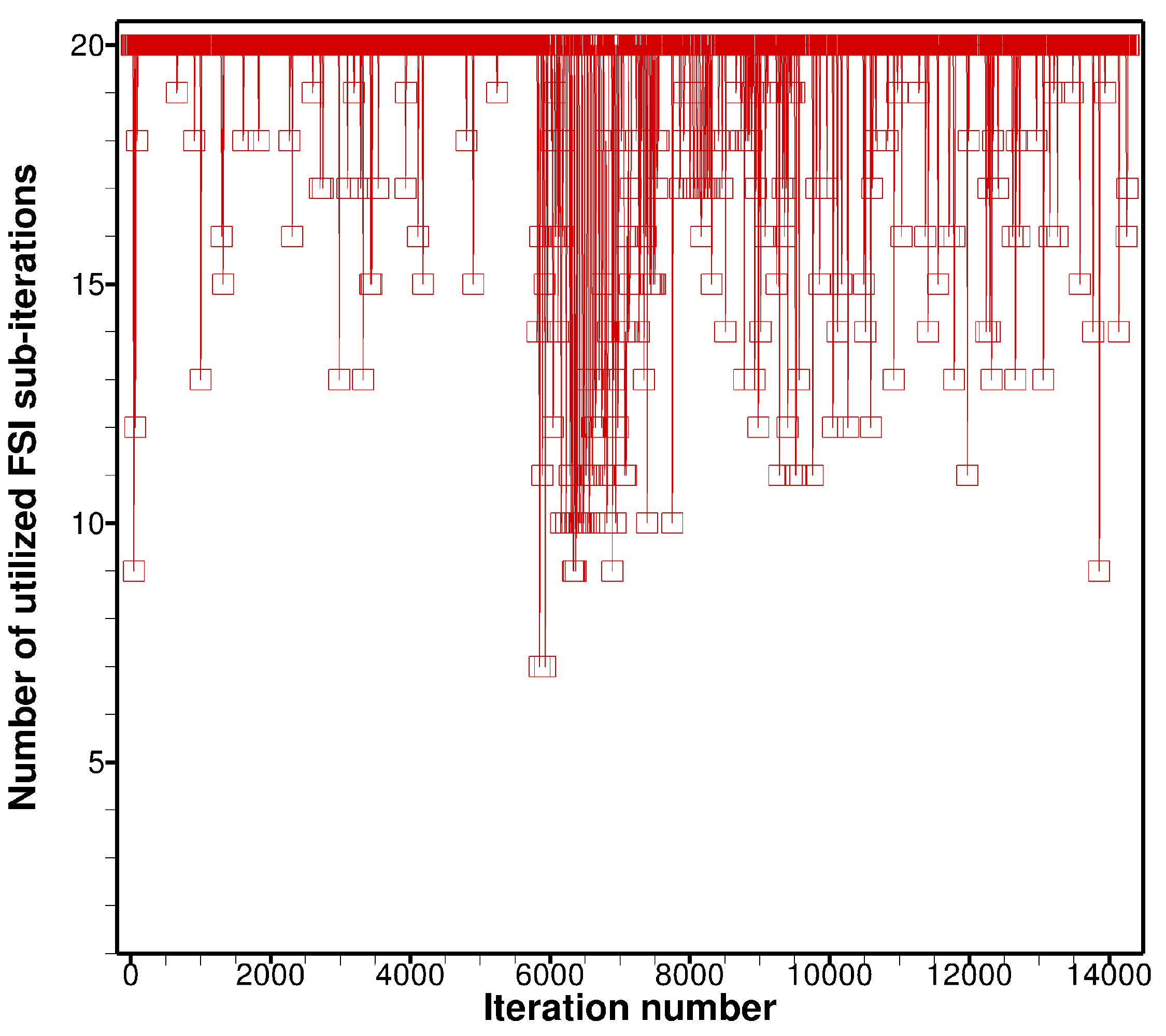}\label{fig:used_iterout_no_est_disp}}\hfill
	\caption{FSI sub-iterations for the test cases with and without the prediction of the displacements.}
	\label{fig:used_iterout_comparison}
\end{figure}

\begin{figure}[H]
	\centering
	\subfigure[Simulation with estimation of displacement.]{\includegraphics[width=0.495\textwidth]{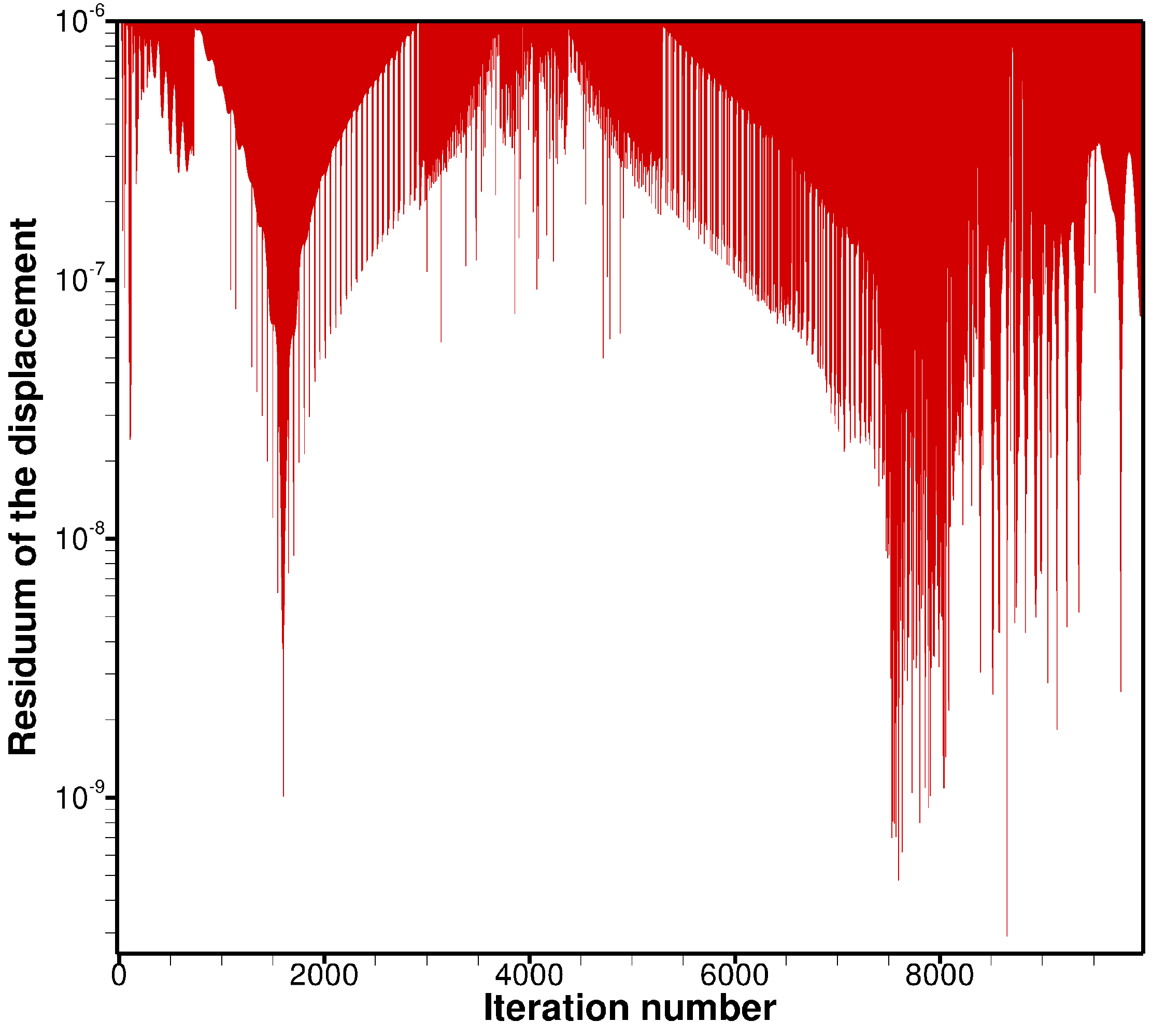} \label{fig:residuum_convergence_est_disp}}\hfill
	\subfigure[Simulation without estimation of displacement.]{\includegraphics[width=0.495\textwidth]{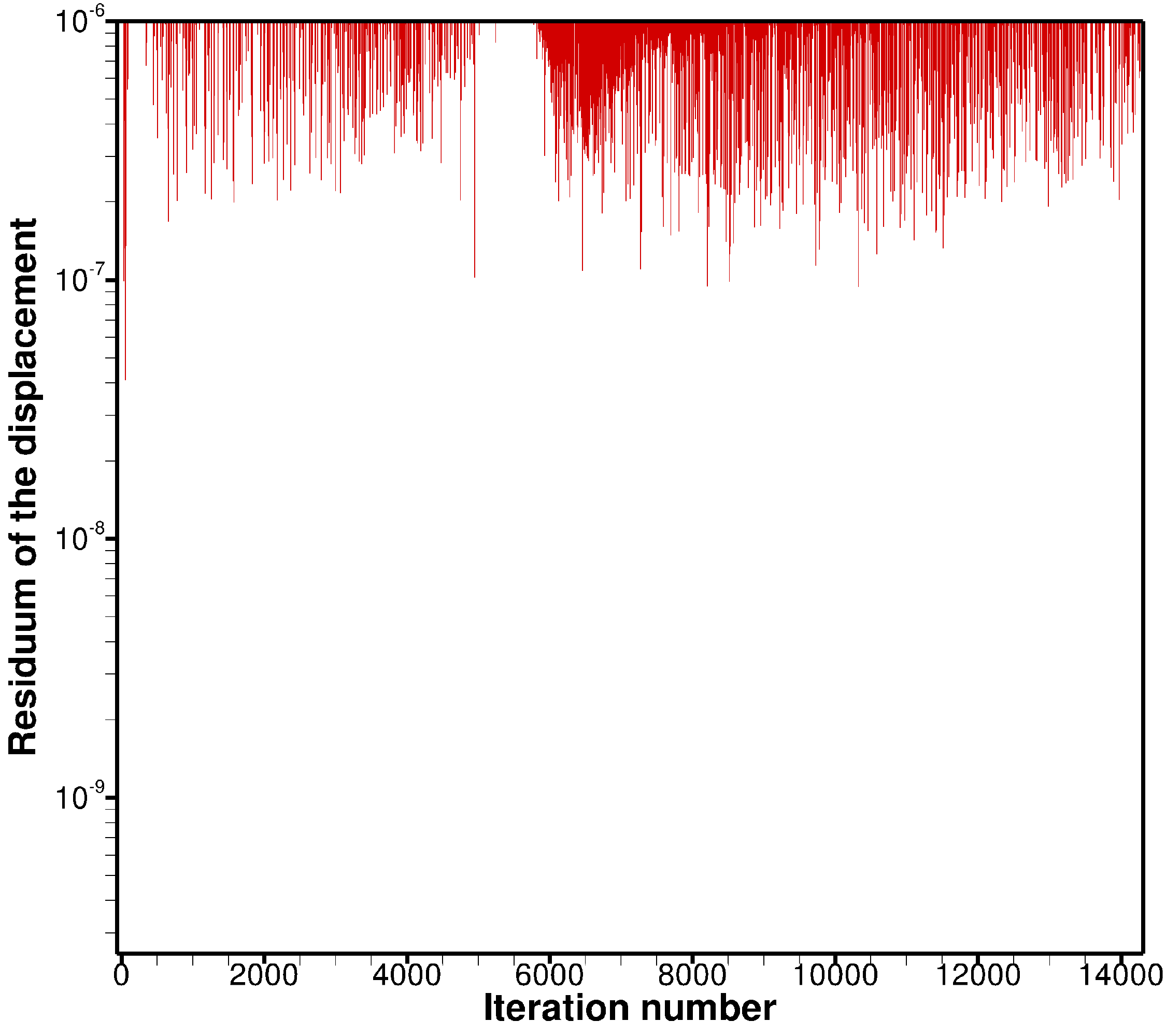}\label{fig:residuum_convergence_no_est_disp}}\hfill
	\caption{History of the FSI convergence ($\epsilon_{FSI,\,disp}<1{\cdot}10^{-6}$) for the test cases with and without estimation of the displacements.}
	\label{fig:residuum_convergence_comparison}
\end{figure}

 After the initialization phase, the simulation that utilizes the estimation of displacements generally requires the calculation of only two sub-iterations in order to converge, while the simulation that does not predict the displacements frequently requires all twenty sub-iterations without always achieving the convergence criterion, as illustrated in\break \mbox{Figs.\ \ref{fig:used_iterout_comparison} and \ref{fig:residuum_convergence_comparison}.} Although the convergence criterion is $\epsilon_{FSI,\,disp}<1{\cdot}10^{-6}$, many time steps of the simulations with the prediction of the displacements achieve residua in the order of magnitude of $\epsilon_{FSI,\,disp}\leq\mathcal{O}(10^{-7})$. This occurs, since the previous sub-iteration $n_{FSI}$ has not yet achieved the convergence criterion, while the current sub-iteration $n_{FSI}+1$ achieves it and also reduces the residuum of displacement up to three orders of magnitude. 
\par
\begin{figure}[H]
	\centering
	\subfigure[Simulation with estimation of displacement (\mbox{$9{,}996\leq n\leq10{,}000$}).]{\includegraphics[width=0.48\textwidth]{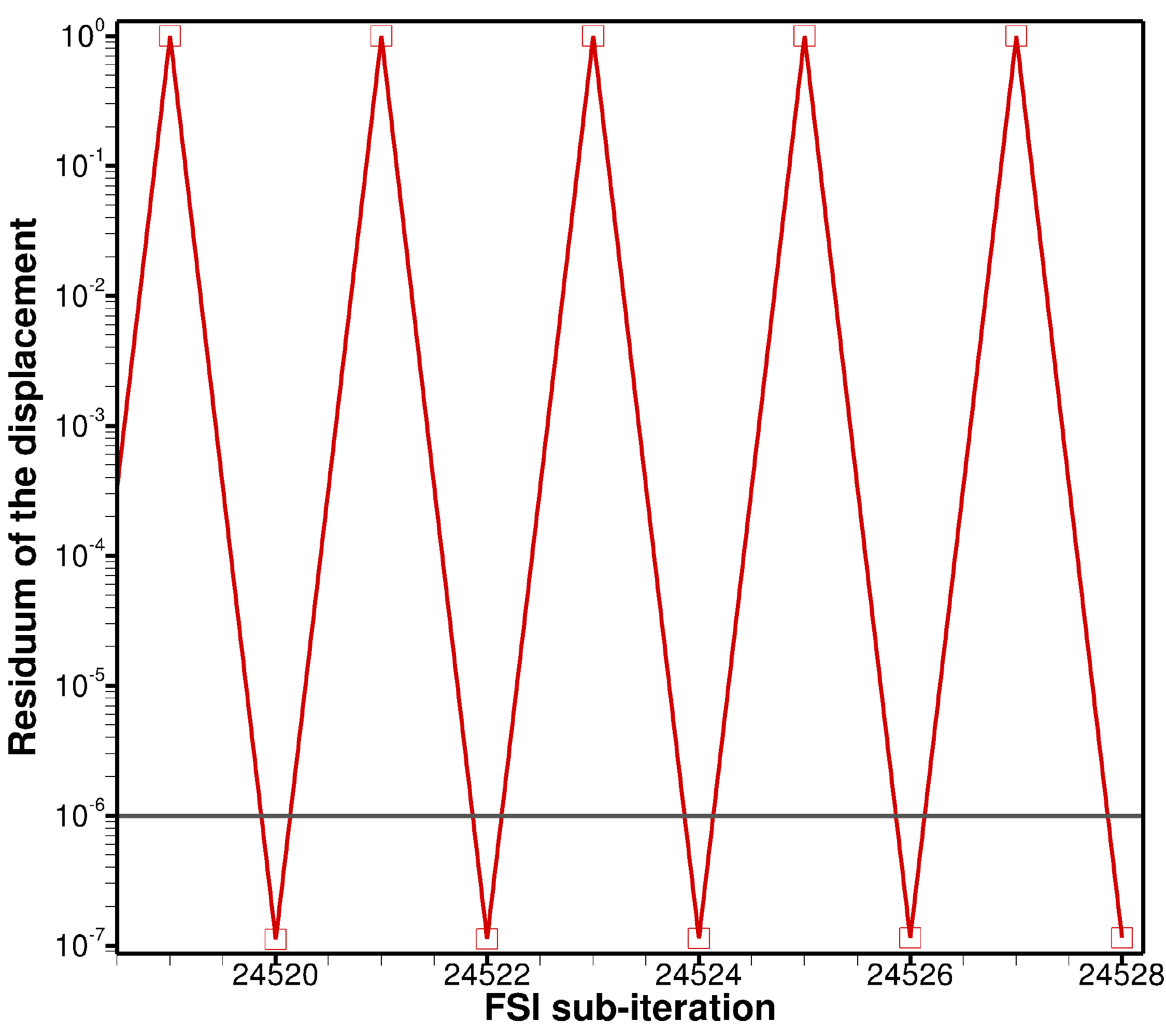} \label{fig:residuum_est_disp}}\hfill
	\subfigure[Simulation without estimation of displacement (\mbox{$14{,}282\leq n\leq14{,}286$}).]{\includegraphics[width=0.48\textwidth]{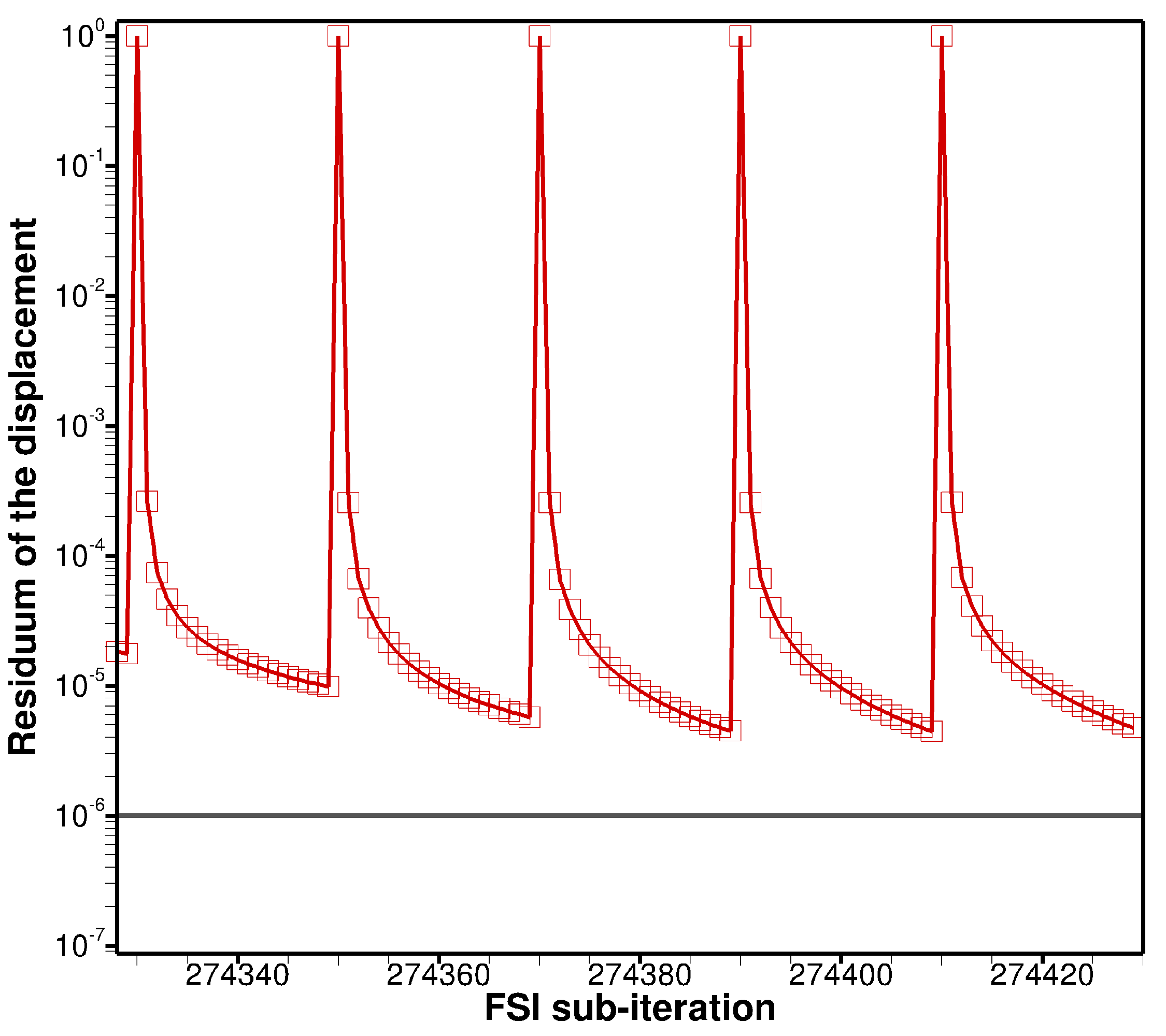}\label{fig:residuum_no_est_disp}}\hfill
	\caption{Residuum of displacement at the last five time steps of the simulations with and without estimation of displacement.}
	\label{fig:residuum_comparison}
\end{figure}

 Furthermore, since Fig.\ \ref{fig:residuum_convergence_comparison} does not show in detail the residuum of the displacements at each sub-iteration, this is meticulously illustrated in Fig.\ \ref{fig:residuum_comparison} for the last five time steps of each simulation. These once again indicate that although all twenty sub-iterations are computed, the FSI convergence criterion is not achieved for the simulations without the estimation of displacements, resulting in a loss of accuracy. 
\par
\begin{figure}[H]
	\centering
	\subfigure[Translational displacement $X_2$.]{\includegraphics[width=0.48\textwidth]{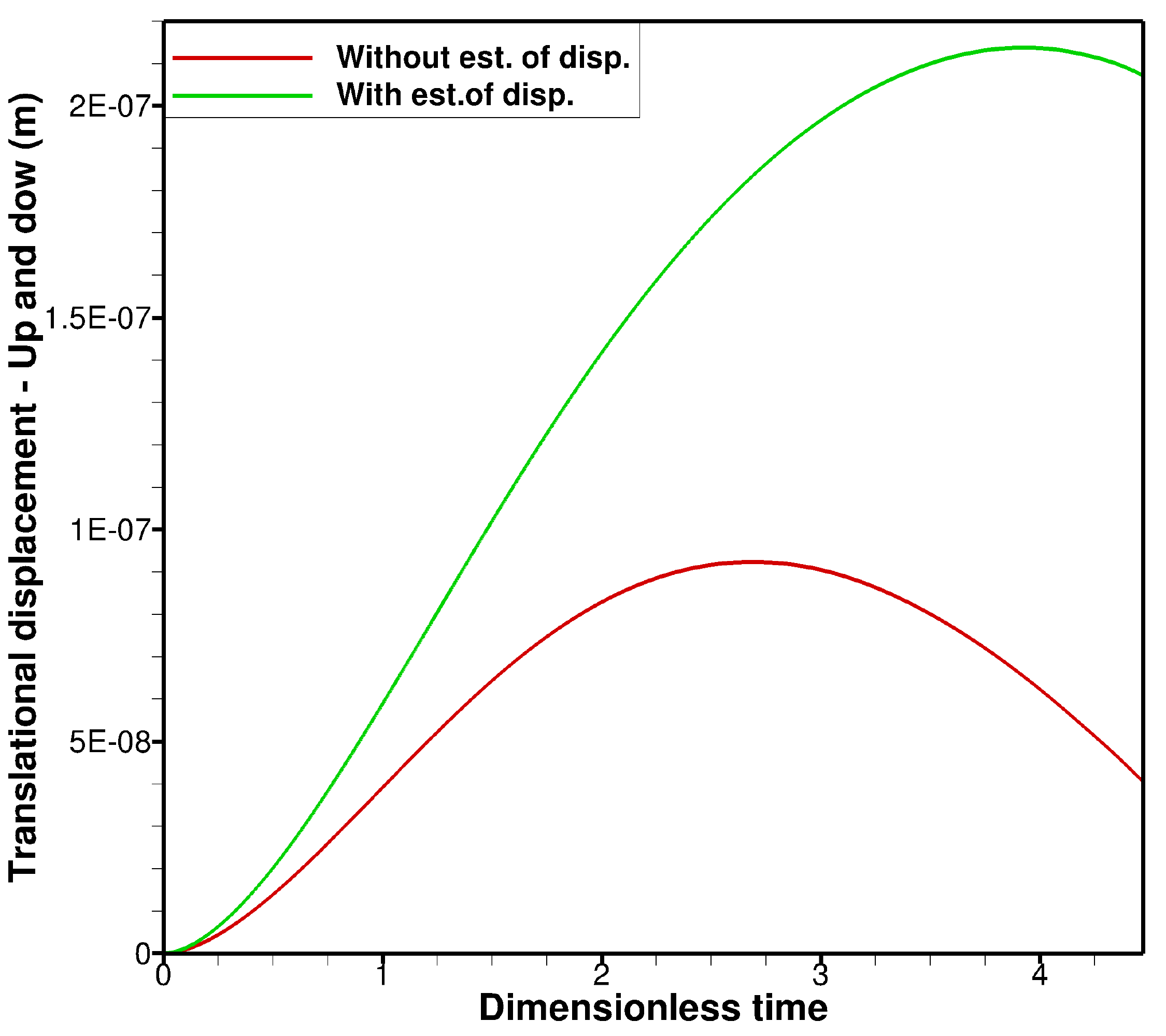} \label{fig:ldift_est_disp}}\hfill
	\subfigure[Rotational displacement $\varphi_3$.]{\includegraphics[width=0.48\textwidth]{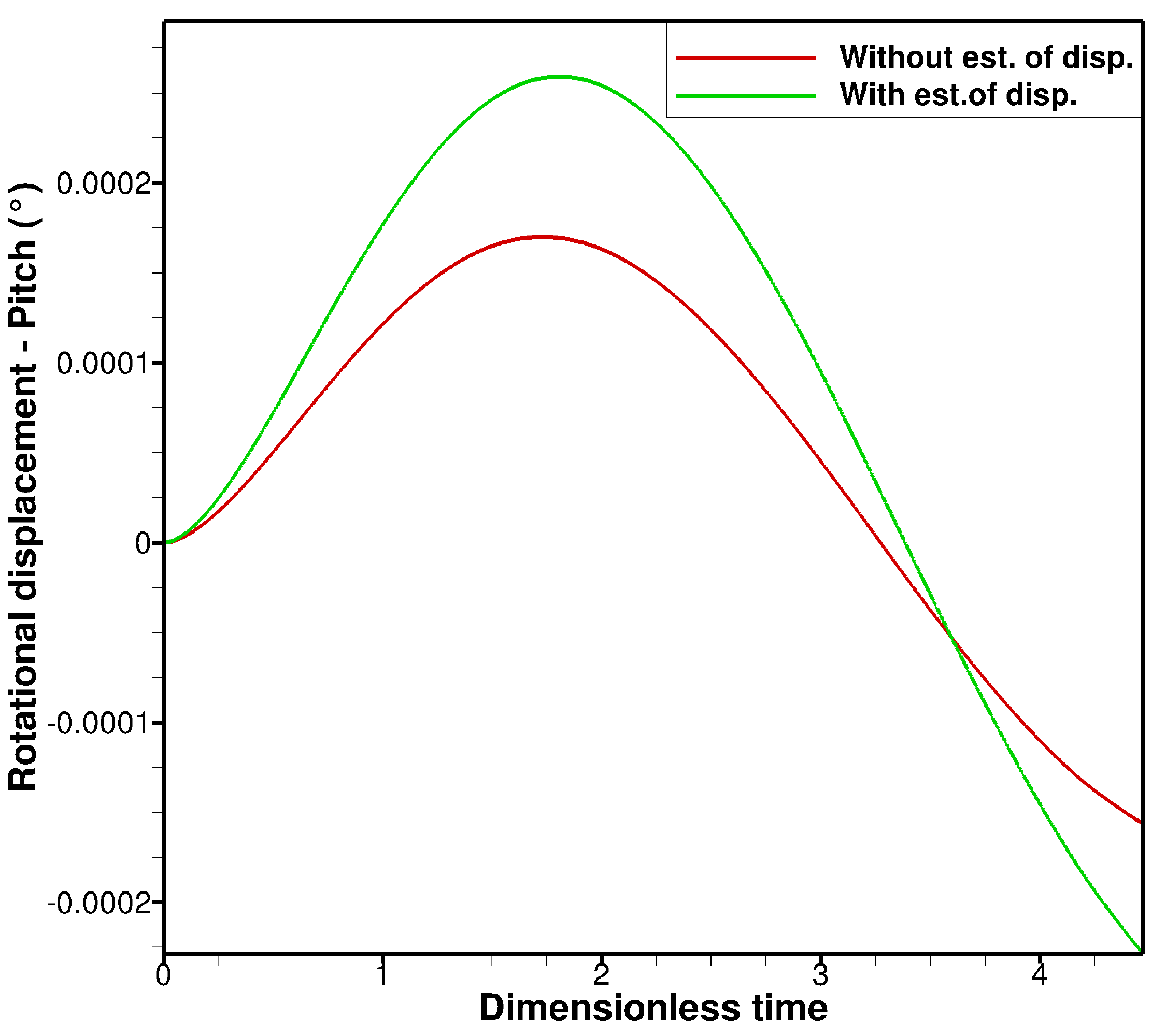}\label{fig:lidft_no_est_disp}}\hfill
	\caption{Time history of the displacements: Cases with and without the prediction of displacements.}
	\label{fig:disp_comparison}
\end{figure}

 The displacements achieved by both simulations are illustrated in Fig.\ \ref{fig:disp_comparison}. Since these test cases are computed only until a dimensionless time of $t^*=4.47$ and a fully developed stated is not achieved, only a qualitative comparison of the results is possible.\break 
\par
\begin{figure}[H]
	\centering
	\subfigure[Lift force (\mbox{$14{,}282\leq n\leq14{,}286$}).]{\includegraphics[width=0.495\textwidth]{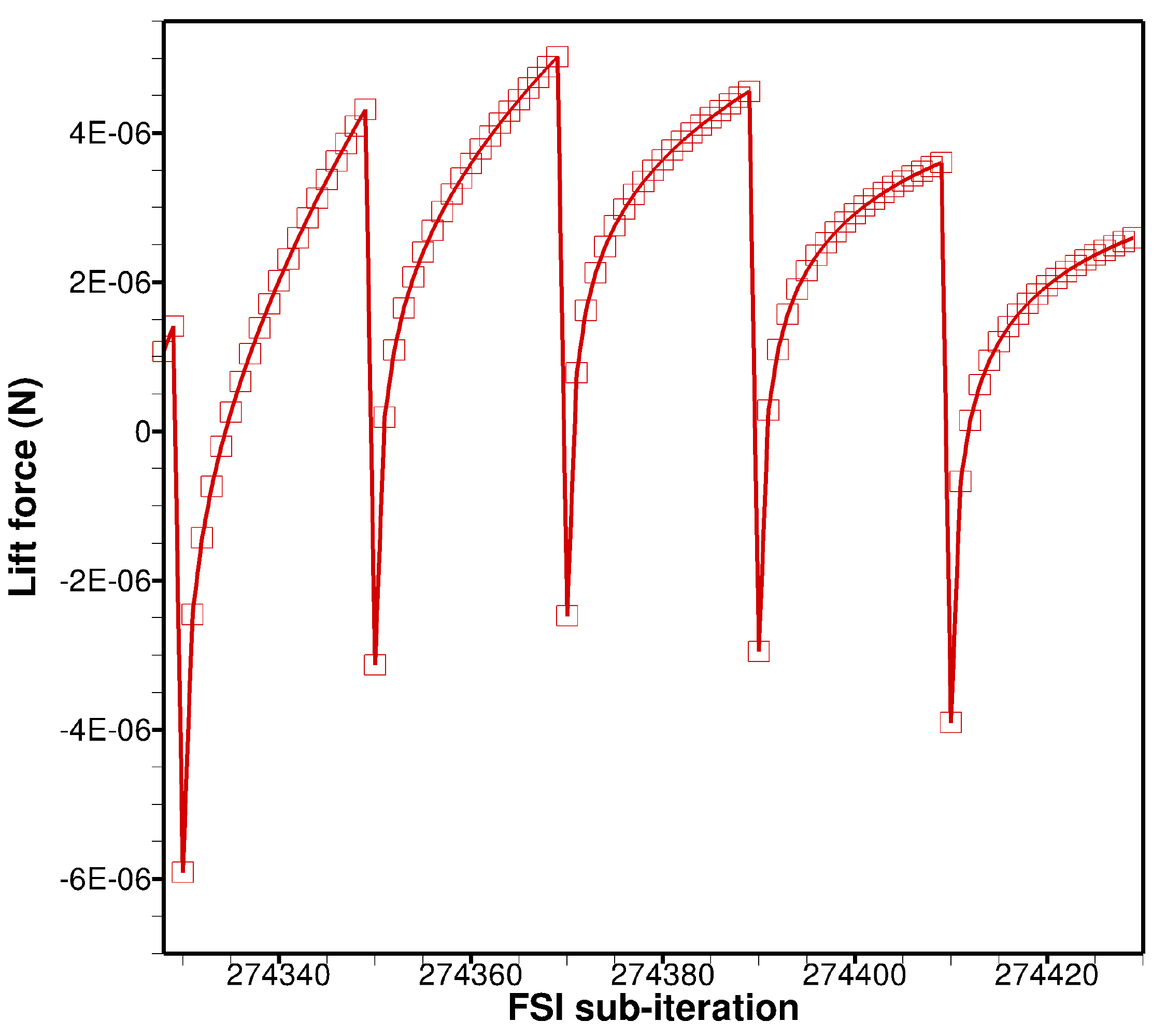} \label{fig:ldift_esst_disp}}\hfill
	\subfigure[Pitching moment (\mbox{$14{,}282\leq n\leq14{,}286$}).]{\includegraphics[width=0.495\textwidth]{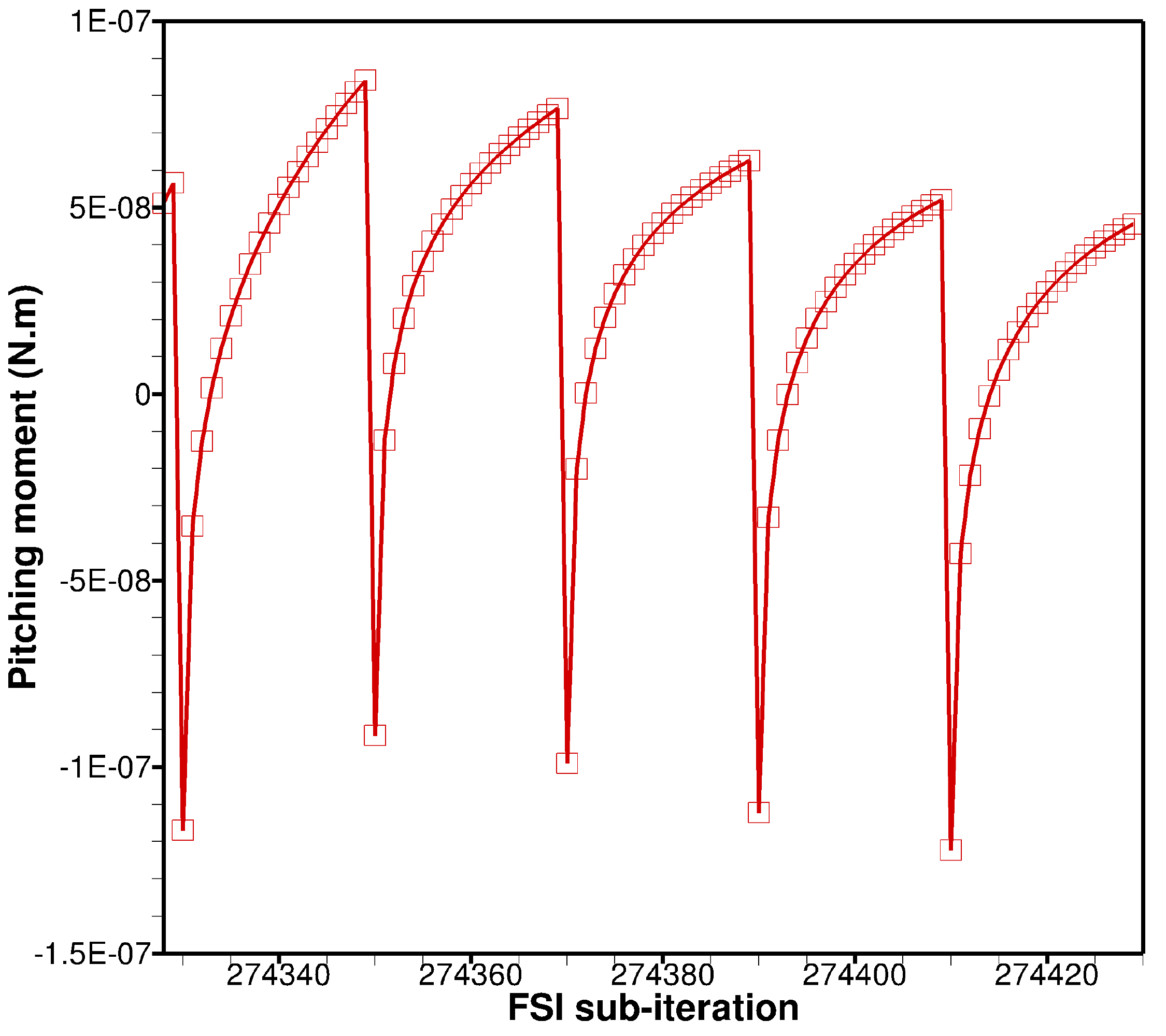}\label{fig:slidft_no_est_disp}}\hfill
	\caption{Lift forces and pitching moment at the last five time steps of the simulations without the prediction of the displacements.}
	\label{fig:lift_moment_no_est}
\end{figure}

 The amplitudes of oscillation of both translation and rotational motions are lower for the case that does not predict the displacements. Since the lift force and the pitching moment increases with the FSI sub-iteration (see Fig.\ \ref{fig:lift_moment_no_est}) and these are directly proportional to the translational and rotational displacements, this under-estimation is caused by the not achievement of the convergence criterion at every time step \mbox{(see Fig.\ \ref{fig:residuum_no_est_disp}).} Therefore, the results achieved without the prediction of the displacements and the computation of a maximal of 20 sub-iterations are not reliable. 

\par Finally, the simulation of the fluid-structure interaction with the prediction of the displacements is approximately $6.7$ times faster, since it requires the computation of less sub-iterations in order to converge. Thus, the estimation of displacements is applied for all simulations in order to enable a faster convergence of the FSI solution and consequently save computational time.

\subsection{Added mass effect}
\label{subsec:added_mass_effect}
\par Added mass is the additional inertia added to a system composed of a fluid and a body, which is submitted to a relative velocity. The deflection of the fluid due to the movement of the body generates additional fluid forces, which act on the surfaces in contact with the flow. This added mass acts as an extra mass on the structural degrees of freedom at the coupling interface and has destabilizing effects on loosely coupled algorithms due to their explicit nature. Moreover, the numerical effect of this added mass also varies with the time discretization methods utilized by the structural and fluid subproblems (see \mbox{Förster et al.\ \cite{Foerster_2007})}.
\par The ratio between the mass of the body and the mass of the displaced fluid\break \mbox{(see Eq.\ (\ref{eq:mass_ratio}))} indicates if the added mass effect must be considered or not.  The higher this ratio, the smaller the effect of the added mass and therefore sequentially staggered schemes (loose coupling) are numerically stable and can be utilized (see Causin et al.\ \cite{Causin_2005}).
\begin{equation}
\label{eq:mass_ratio}
m^*=\frac{m^{body}}{m^{fluid}_{disp}}
\end{equation}
\par Song et al.\ \cite{Song_2013} investigated the influence of the added mass effect on the flow around stiff plate structures of different materials utilizing a partitioned approach based on a loose coupling algorithm. The case with $m^*=10$ exhibited no convergence problems, while under-relaxation factors were used in the case with $m^*=1$ in order to guarantee the convergence of the solution. Therefore, the influence of the added mass effect for $m^*=10$ was negligible, while this effect was responsible for numerical difficulties for the $m^*=1$ case.  
\par In the present work, the mass ratio between the airfoil and the displaced fluid is calculated according to Eqs.\ (\ref{eq:displaced_mass}) and (\ref{eq:mass_ratio_naca}), regarding that the supports fixed on the airfoil are not considered. $V_N$, $m_N$ and $\rho_N$ are the airfoil volume, mass and density. $\rho_f$ is the fluid density. 
\begin{eqnarray}
\label{eq:displaced_mass}
m^{fluid}_{disp}&=&V_N\,\rho_f=\frac{m_N}{\rho_N}\rho_f\\
\label{eq:mass_ratio_naca}
m^*=\frac{\rho_N}{\rho_f}&=&\frac{700}{1.225}=571.43
\end{eqnarray}
\par Due to the high mass ratio, i$.$e$.$, $m^*\approx571$, a loose coupling method ($n_{FSI}=1$) can be utilized due to its numerical stability and lower required computational time (compared to strong coupling approaches).

\subsection{FSI coupling algorithms}
\label{subsec:investigation_coupling_algorithm}

\par Although the added mass effect is negligible for the NACA0012 FSI test cases\break \mbox{(see Section \ref{subsec:added_mass_effect})}, the accuracy of the loose coupling algorithm is tested by a comparison of the airfoil responses computed with both explicit and implicit coupling approaches. Three simulations with two degrees of freedom (pitch and up and down) are performed with the $m{-}L_3^{min}{-}y_{max}^+$ mesh, the mapping strategy 4 (see Section \ref{subsubsec:study_mapping_strategy}), the TFI re-meshing method and the prediction of displacements until a dimensionless time of $t^*=1564.5$ is achieved. The simulated Reynolds number is $Re=30{,}000$ and the utilized linear and torsional stiffnesses are  $k_{l,\,2}=144\,\text{N}{\cdot}\text{m}^{-1}$ and $k_{t,\,3}=0.3832\text{N}\,\text{m}{\cdot}\text{rad}^{-1}$, respectively. No damping is applied.
\par The setups of the simulations are summarized in Table \ref{table:comp_FSI_coupling}. That includes a test case that solves the fluid domain with an implicit coupling algorithm for the first \mbox{$n=100{,}000$} time steps and afterwards with an explicit algorithm.
\begin{table}[H]
	\centering
	\begin{tabular}{p{2.8cm} p{4.1cm} p{2.0cm} p{3cm} p{1.8cm}}
		\hline
		\multicolumn{1}{c}{\multirow{3}{*}{\centering{\bf{Mesh}}}} & \multicolumn{1}{c}{\multirow{3}{*}{\centering{\bf{Coupling algorithm}}}} & \multicolumn{1}{c}{\multirow{3}{*}{\centering{\bf{Time steps}}}} & \centering{\bf{Maximal number of FSI sub-iterations}} & \centering{\bf{Time step size (s)}}\tabularnewline \hline
		\centering{$m{-}L_3^{min}-y^{+}_{max}$} & \centering{Loose} & \centering{$3{,}500{,}000$} & \centering{1} & \centering{$1\cdot10^{-5}$} \tabularnewline \hline
		\centering{$m{-}L_3^{min}{-}y^+_{max}$} & \centering{Strong} & \centering{$3{,}500{,}000$} &\centering{20} & \centering{$1\cdot10^{-5}$} \tabularnewline \hline
		\multicolumn{1}{c}{\multirow{2}{*}{\centering{$m{-}L_3^{min}{-}y^+_{max}$}}} & \centering{Strong for $n\leq100{,}000$} & \centering{$100{,}000$} & \multicolumn{1}{c}{\centering{20}} & \multicolumn{1}{c}{\multirow{2}{*}{\centering{$1\cdot10^{-5}$}}} \tabularnewline	\cline{2-4}
		& \centering{Loose for $n>100{,}000$} & \centering{$3{,}500{,}000$} & \centering{1} & \tabularnewline
		\hline	
	\end{tabular}
	\caption{\label{table:comp_FSI_coupling}Simulation properties of the test cases aimed at the comparison of the coupling algorithms.}
\end{table}
\par The maximal number of sub-iterations is set to $n_{FSI}=20$ for the cases with strong coupling, since this is sufficient and necessary to achieve the convergence criterion of all computed time steps, especially during the initialization phase. After this phase, however, only two sub-iterations are in general utilized by the FSI solver (see Section \ref{subsec:investigation_estimation_displacement}), due to the fact that the convergence criterion is achieved before all sub-iterations are performed.
\par
\begin{figure}[H]
	\subfigure[Strong coupling algorithm: Translational displacement $X_2$.]{\includegraphics[width=0.48\textwidth]{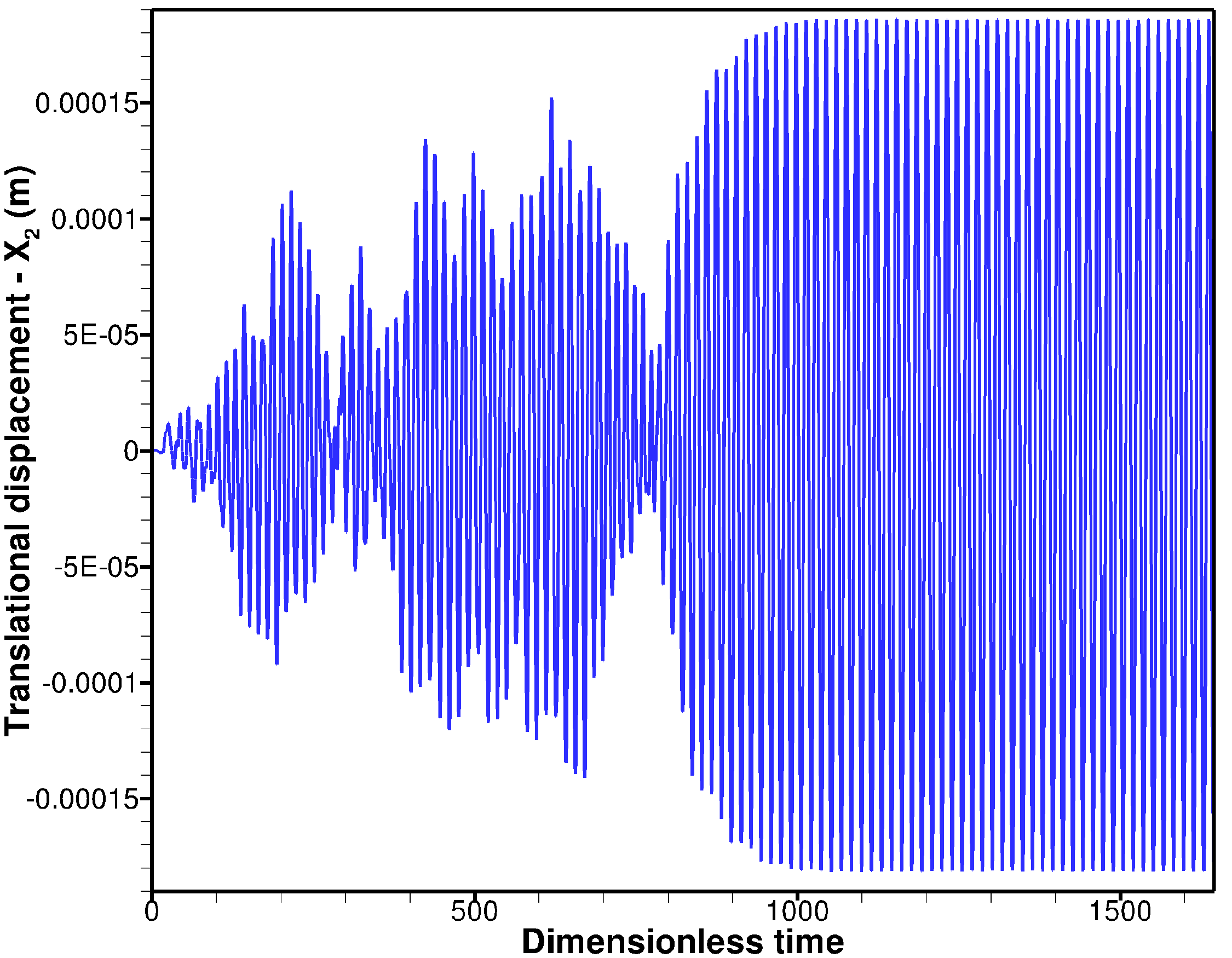} \label{fig:frequencies_translatsiosn}}\hfill
	\subfigure[Strong coupling algorithm: Rotational displacement $\varphi_3$.]{\includegraphics[width=0.48\textwidth]{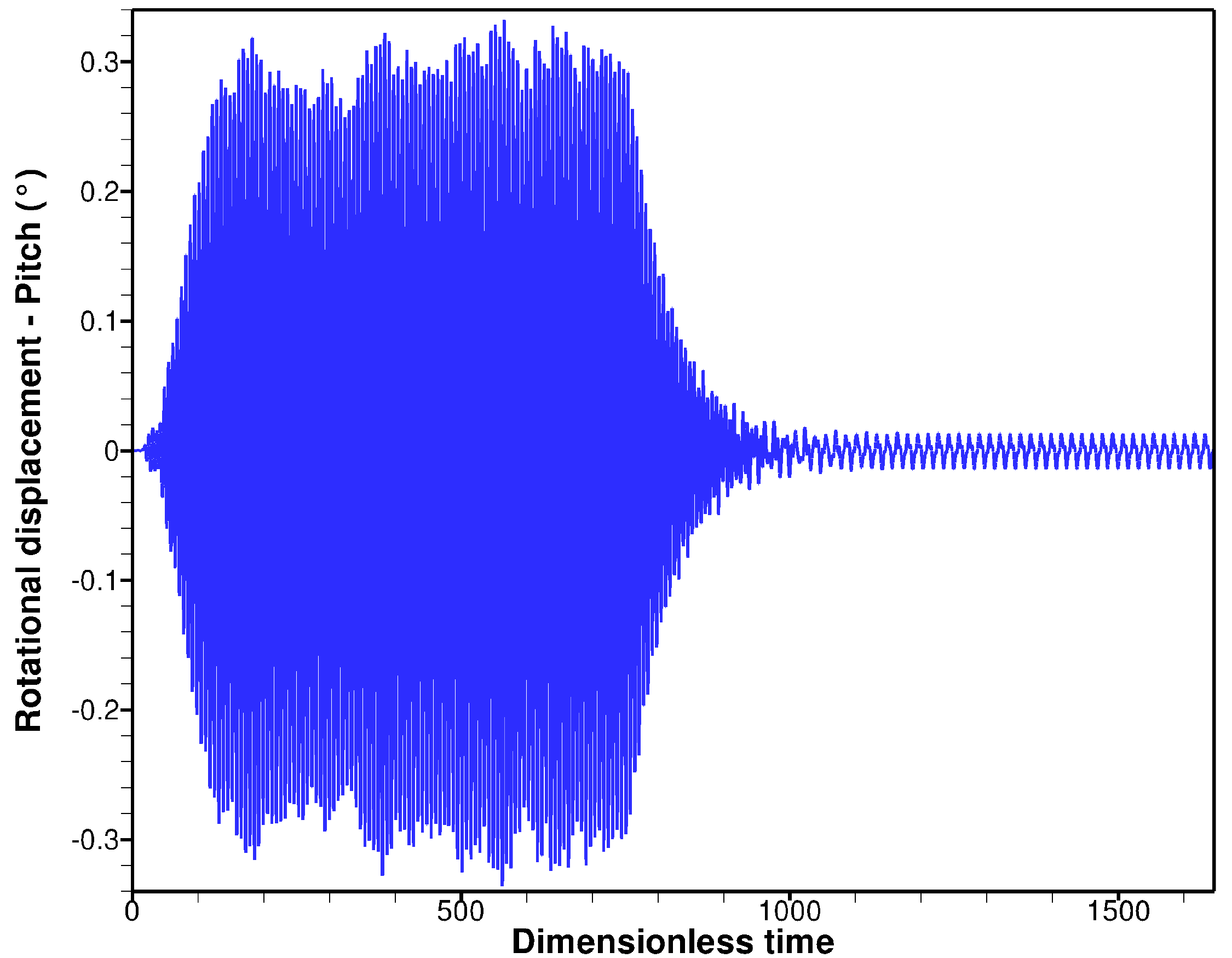} \label{fig:frequencies_stranslation}}\hfill
	\centering
	\subfigure[Strong coupling algorithm for $t\leq1\,s$ and loose coupling algorithm for $t>1\,s$: Translational displacement $X_2$.]{\includegraphics[width=0.48\textwidth]{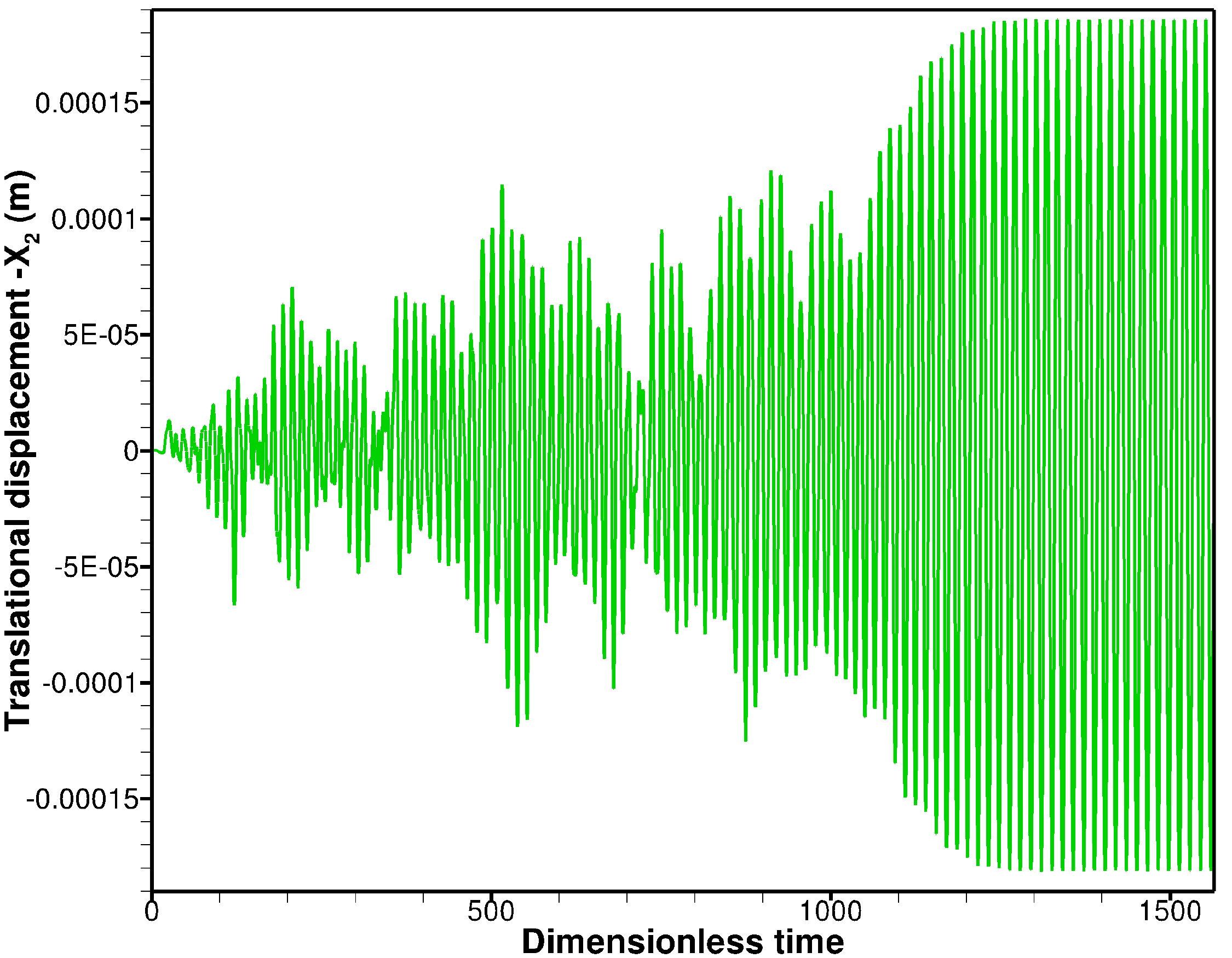} \label{fig:frequencies_translatison}}\hfill
	\subfigure[Strong coupling algorithm for $t\leq1\,s$ and loose coupling algorithm for $t>1\,s$: Rotational displacement $\varphi_3$.]{\includegraphics[width=0.48\textwidth]{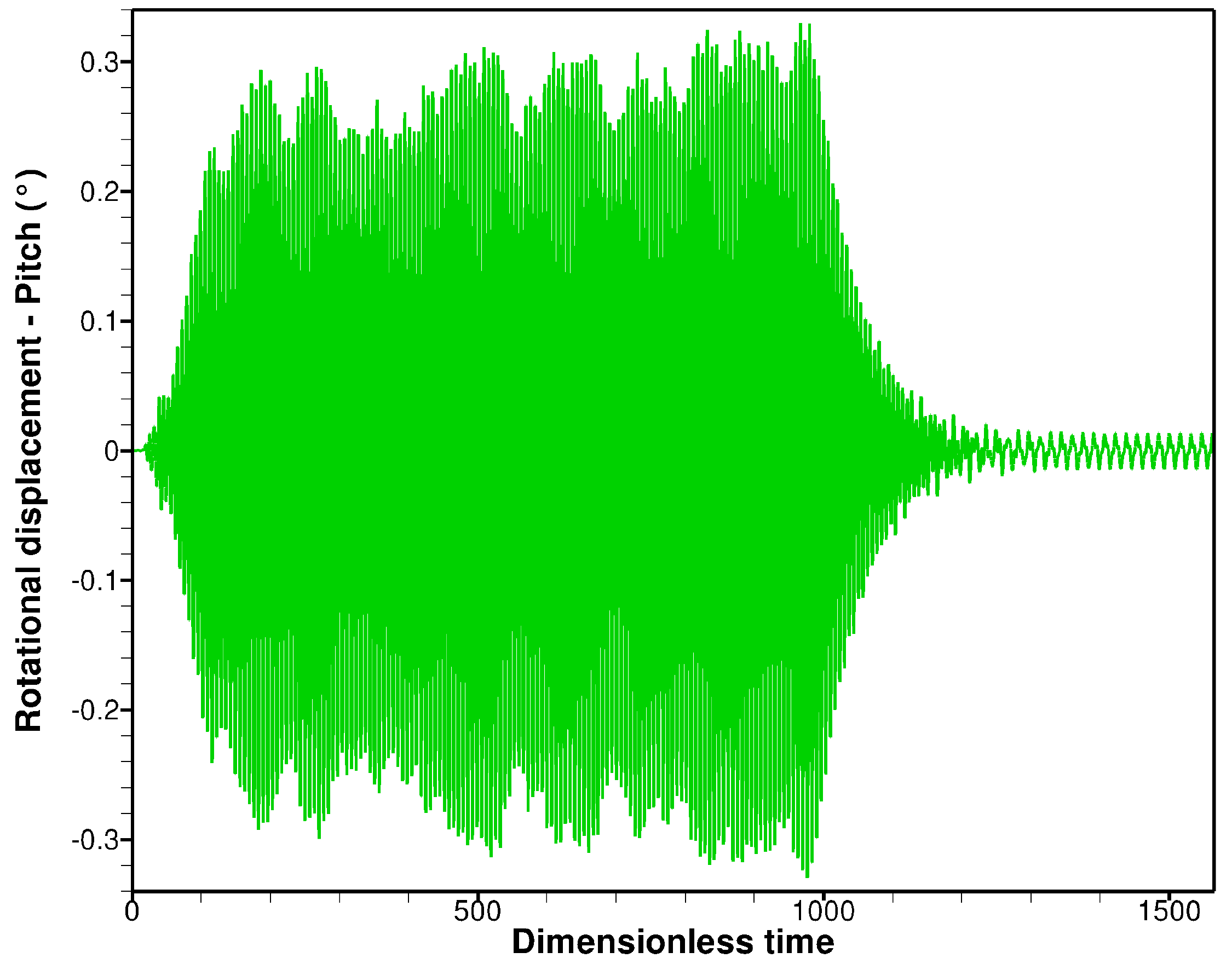} \label{fig:frequencies_translatiosn}}\hfill
	\caption{Comparison of the coupling algorithms: Translational ($X_2$) and rotational ($\varphi_3$) displacements.}
	\label{fig:displacements_coupling_investigation}
\end{figure}

 The structural responses of the airfoil in the form of translational displacements in the \mbox{$x_2$-direction} and rotational displacements around the $x_3$-axis are illustrated in Fig.\ \ref{fig:displacements_coupling_investigation} as a function of the time. Although the initialization phase and the fully developed state are illustrated, the results are analyzed based only on the fully developed state, which is characterized by limit-cycle oscillations (LCO), i$.$e$.$, oscillations with a limited amplitude.
\par Although the various coupling algorithms require different initialization times in order to achieve a steady-state oscillation, the amplitudes of these oscillations are the same for all coupling approaches, that is, \mbox{$X_{2}=1.8{\cdot}10^{-4}\,\text{m}$} and \mbox{$\varphi_{3}=0.0135^\circ$} for the translational and rotational motions, respectively. These displacements indicate that the airfoil hardly moves, even though this has two degrees of freedom. 
\begin{figure}[H]
	\centering
	\subfigure[Loose coupling algorithm: Translational displacement $X_2$.]{\includegraphics[width=0.48\textwidth]{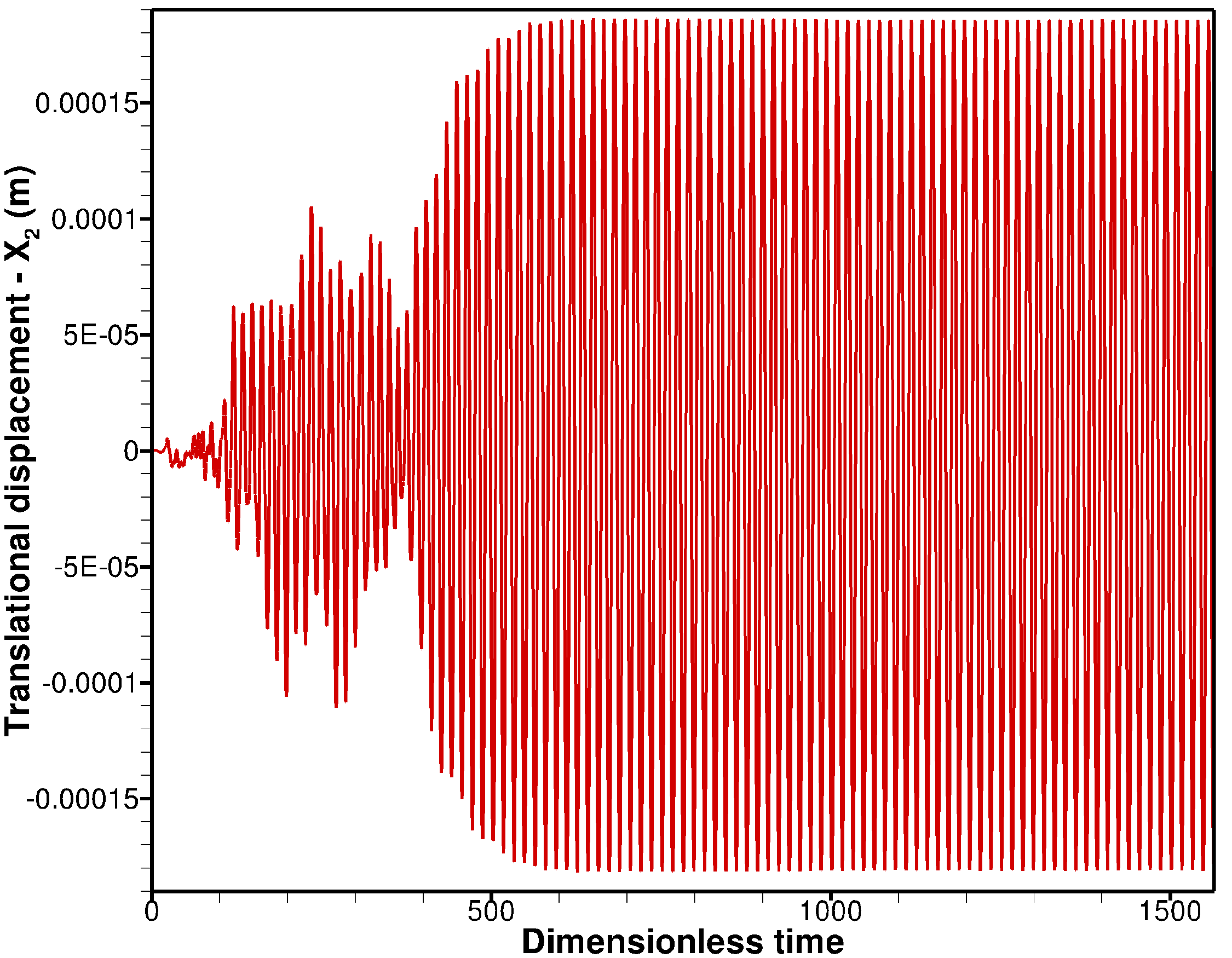} \label{fig:frequenciess_translation}}\hfill
	\subfigure[Loose coupling algorithm: Rotational displacement $\varphi_3$.]{\includegraphics[width=0.48\textwidth]{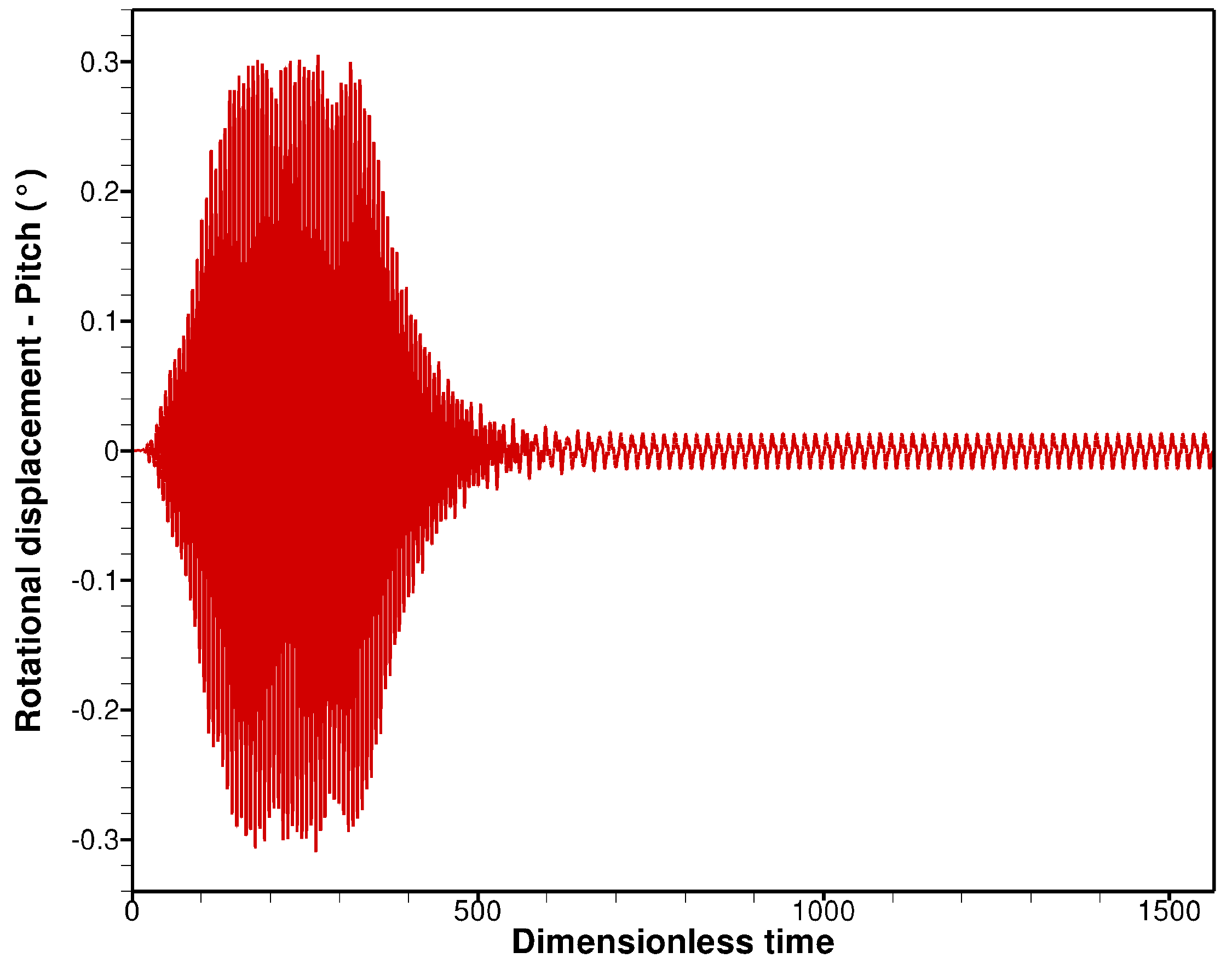}\label{fig:frequencies_rsotation}}\hfill
\end{figure}
\par Since the final amplitudes achieved with by loose coupling algorithm are in a good agreement with the ones achieved by both other approaches, the coupling scheme does not influence the computation of the coupled problem and therefore is applicable to further simulate the fluid-structure interaction between the NACA0012 airfoil and the flow. This result is expected, since this FSI case is characterized by a negligible added mass effect (see \mbox{Section \ref{subsec:added_mass_effect}).} Nevertheless, the frequencies generated by the fluid-structure interaction are also investigated.
\par
\begin{figure}[H]
	\centering
	\subfigure[Loose coupling.]{\includegraphics[width=0.48\textwidth]{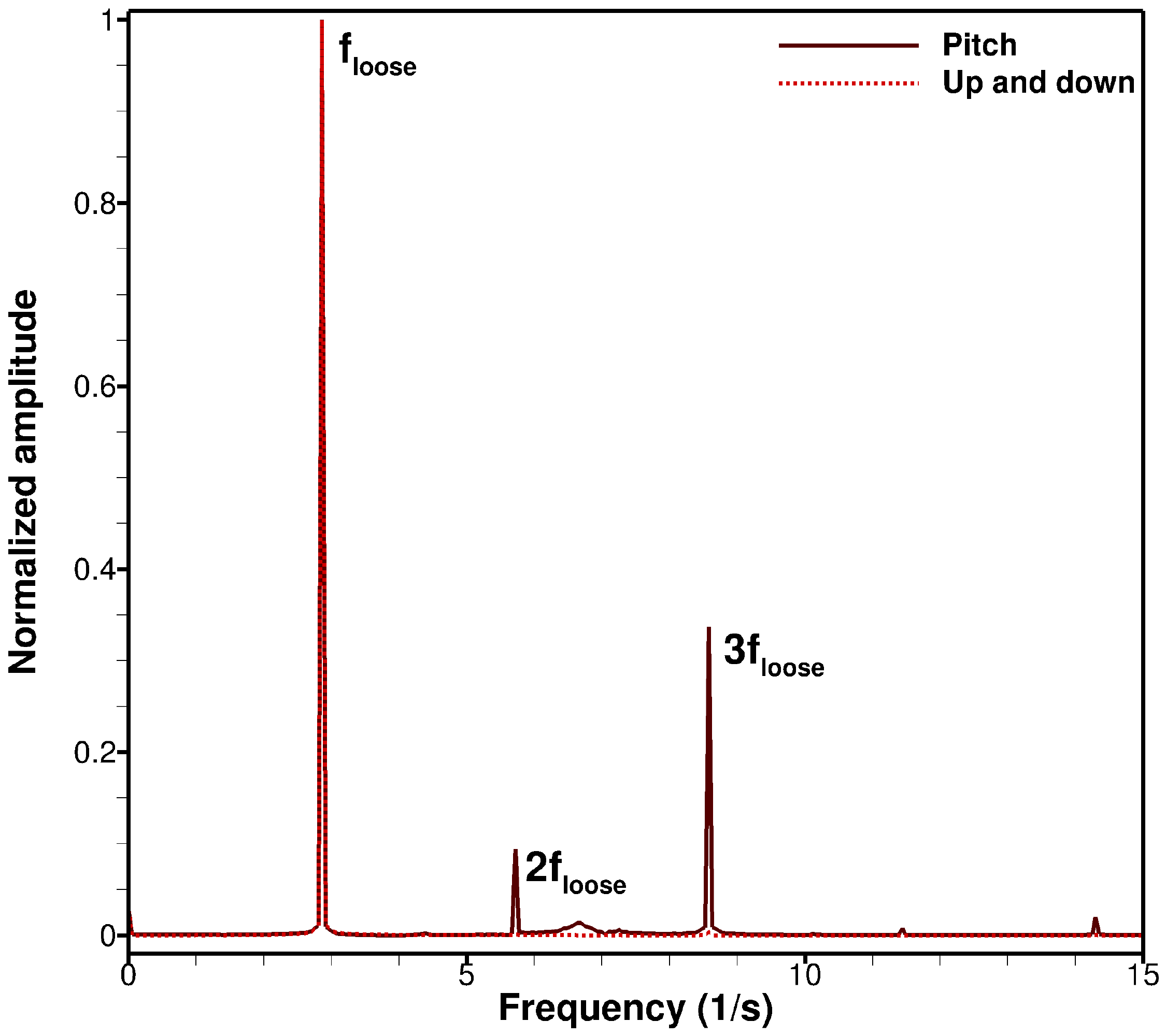} \label{fig:frequencies_loose}}\hfill
	\subfigure[Strong coupling.]{\includegraphics[width=0.48\textwidth]{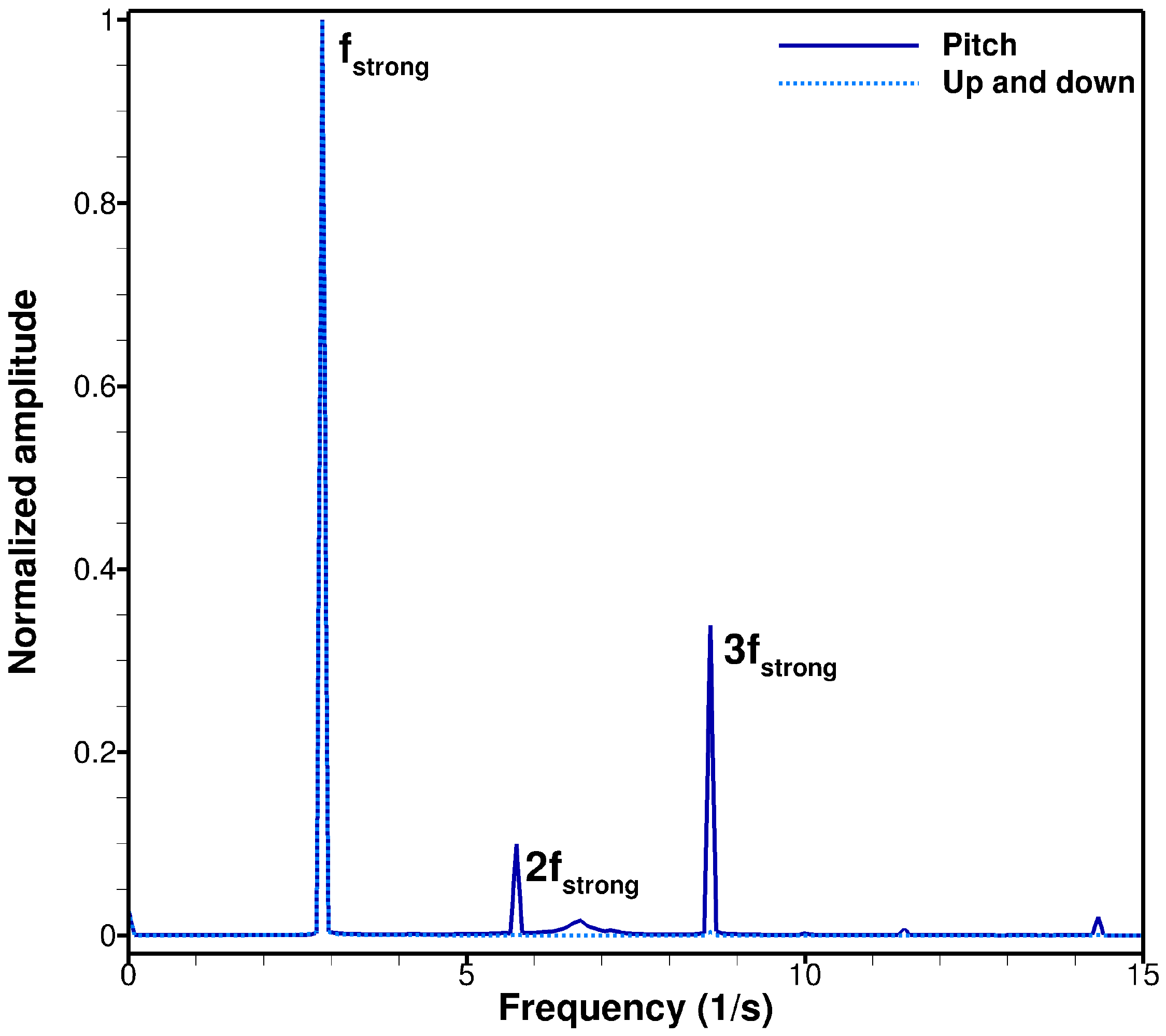}\label{fig:frequencies_strong}}\hfill
	\caption{Frequency domain of the up and down ($X_2$) and pitch ($\varphi_3$) displacements.}
	\label{fig:frequencies_displacements_coupling}
\end{figure}

 In order to analyze the frequency domain, a Fourier transform with a Hamming window is carried out, considering that the independent variable is the time $t$ and that the dependent variables are the up and down and pitch displacements, i$.$e$.$, respectively $X_2$ and $\varphi_3$. The latter are utilized as parameters for the Fourier transformation since these generate more distinct peaks in the frequency domain compared to the lift and drag forces, facilitating the comparison of the coupling algorithms. The frequency domain of the translational and rotational displacements are illustrated in Fig.\ \ref{fig:frequencies_displacements_coupling} for the time interval of the steady-state oscillation. The amplitudes are normalized by the maximal value, which is related to the initialization time: The higher the time needed to reach the fully developed state, the lower the amplitudes and the broader the peaks. This occurs due to the fact that less data  of the fully developed state is available. Therefore, the frequency domain of the simulation characterized by the utilization of both loose and strong coupling algorithms is not displayed.
\par The frequencies of the three peaks characterized by the highest amplitudes are described in \mbox{Table \ref{table:frequencies_coupling}} for both strong and loose coupling algorithms. These indicate the vortex shedding frequency as well as the second and third super-harmonics. Therefore, the Strouhal number in relation to the airfoil chord length is also calculated. Since only minor deviations are present concerning the frequencies generated by both coupling algorithms, it is confirmed that the coupling method does not influence the calculation of the coupled problem. 
\begin{table}[H]
	\centering
	\begin{tabular}{p{3cm} p{4cm} p{4cm}}
		\hline
		\centering{\bf{FSI coupling}} & \centering{\bf{Frequency ($\text{Hz}$)}} & \centering{\bf{Strouhal number}} \tabularnewline \hline
		\centering{Strong} & \centering{2.886} & \centering{0.0646} \tabularnewline
		\centering{Loose} & \centering{2.859} & \centering{0.0639} \tabularnewline		
		\hline	
	\end{tabular}
	\caption{\label{table:frequencies_coupling}Characteristic frequencies: Simulations with different coupling schemes.}
\end{table}
\par Finally, the computational times of all simulations are compared, in \mbox{Table \ref{table:compt_time_coupling}}. The loose coupling algorithm requires about $85\%$ and $56\%$ of the computational time of the mixed and the strong couplings, respectively. Since it is also proven that the explicit coupling is accurate, this is utilized for the further investigations of the fluid-structure interaction.
\begin{table}[H]
	\centering
	\begin{tabular}{p{4.2cm} p{6cm}}
		\hline
		\centering{\bf{FSI coupling}} & \centering{\bf{Computational time (days)}} \tabularnewline \hline
		\centering{Loose} &  \centering{23.07} \tabularnewline	
		\centering{Strong for $n\leq100{,}000$} & \multicolumn{1}{c}{\multirow{2}{*}{{$27.05$}}} \tabularnewline	
		\centering{Loose for $n>100{,}000$} & \tabularnewline
		\centering{Strong} & \centering{$41.23$} \tabularnewline
		\hline	
	\end{tabular}
	\caption{\label{table:compt_time_coupling}Computational time: Simulations with different coupling schemes.}
\end{table}

\subsection{Conclusions}
\label{conclusions_FSI_coupling}

\par The thorough analysis of the implemented estimation of the displacements at the beginning of each time step points out that this prediction improves the convergence properties of coupled problems. For the NACA0012 FSI test cases, the application of this estimation is responsible for a reduction of $85\%$ of the required computational time compared with a computation that does not utilize this algorithm.
\par The investigation of the mass ratio proves that the added mass effect is negligible and therefore an explicit coupling method can be applied. Nevertheless, a loose, a strong and a mixed coupling scheme were tested.  Due to the fact that the loose coupling scheme requires only $56\%$ of the time required by the strong coupling algorithm, this explicit coupling together with the prediction of the displacements is further applied in the FSI test cases.
\par Although none of the investigated system configurations is characterized by large displacements, the results achieved are also applicable for these cases. While the estimation of displacements is not influenced by the airfoil response, the negligible added mass effect ensures that a loose coupling algorithm can be applied.
\chapter{Results and discussion}
\label{chap:results}

\par When a relative velocity is present in a system composed of fluid and body, aerodynamic forces arise. These are proportional to the pressure and skin friction forces and are affected by the geometry of the body, the fluid properties (density, viscosity, speed and compressibility), as well as the flow characteristics such as the presence of boundary layer detachments and separation bubbles (see Jones et al.\ \cite{Jones_2008}).
\par The forces generated by the flow on the body can lead to a structural displacement. The lift and drag forces, as well as the pitching moment have a major influence on the displacements and resulting movements. The pitching moment is around an axis perpendicular to both the drag and the lift forces.
\par Specific system configurations are responsible for the generation of aeroelastic instabilities, which can cause excessive body deformations and lead to a structural failure. This occurs mainly by an increase in the relative flow speed, which leads to a raise in the aerodynamic forces, while the material and geometry-dependent structural stiffness remains constant. 
\par Limit-cycle oscillation (LCO), torsional divergence and the flutter phenomenon represent major problems generated by aeroelastic instabilities. The former is characterized by the presence of self-sustained oscillations with a limited amplitude, which may cause a structural failure due to fatigue after a determined number of cycles. 
Torsional divergence is a steady-state instability characterized by an infinite frequency of oscillation that leads to a rapid twist and failure of the airfoil. The latter, i$.$e$.$, flutter, is a dynamic instability that presents self-sustained oscillations, which grow exponentially and also lead to a rapid structural failure. 

\par To be more specific, LCO is induced by the presence of complex structural and/or aerodynamic nonlinearities, which can limit the amplitude of oscillation even when the free-stream velocity is higher than the critical speed calculated by linear aeroleastic \mbox{theories \cite{Fung_2002}}, such as the theory presented in the work of Theodorsen et al.\ \cite{Theodorsen_1935}. The structural nonlinearity arises due to material and geometrical nonlinearities, which can be caused by plastic deformations and distortions, respectively. The aerodynamic nonlinearity results from the fluid compressibility and viscous effects and are greatly influenced by the Reynolds number \cite{Ramesh_2015}. Torsional divergence, on the other hand, occurs when the system stiffness is not high enough in order to counter-act the presence of an aerodynamic moment generated either by an accidental deformation or by the fluid-structure interaction. Flutter, is induced when the vortex shedding frequency $f_v$ achieves the critical flutter frequency, which lies between the two natural frequencies at zero airspeed for systems with pitch and up and down degrees of freedom (see Fung \cite{Fung_2002}).
\par The current work aims primarily at the analysis of the limit-cycle oscillations. Since the simulated NACA0012 is rigid and submitted to an incompressible flow, only the nonlinear aerodynamic effects are relevant for the LCO, such as the separation of the laminar boundary layer, either in the form of  trailing-edge separation or in the form of a laminar separation bubble (see Poirel et al.\ \cite{Poirel_2014} and Lapointe and Dumas \cite{Lapointe_2012}).
\par Moreover, the torsional divergence phenomenon is also investigated. However, only results with a TFI mesh adaptation algorithm are currently available and these are not reliable due to the quality of the adapted mesh.
\par Furthermore, a study of the structural parameters of the NACA0012 airfoil is carried out in order to possibly achieve a case characterized by the flutter phenomenon. However, since this simulation requires large computational times, not enough data are currently available in order to thoroughly analyze this test case. 
\par The achieved results are studied in a dimensionless manner, which is characterized by an $^*$, i$.$e$.$, $t^*$, $u_i^*$, etc$.$. The reference quantities are the airfoil chord $c$ and free-stream velocity $u_{in,\,1}$. A comparison with experimental data in order to validate the simulations is not possible, since the complete experimental setup is currently not available.

\section{Configuration 1: Limit-cycle oscillation}\markboth{CHAPTER 4.$\quad$RES. AND DISC.}{4.1$\quad$CONFIGURATION 1: LCO}
\label{sec:LCO}

\par Eleven system configurations characterized by the same torsional stiffness of\break  \mbox{$k_{t\,3,\,eq}=0.3832\,\text{N}{\cdot}\text{m}{\cdot}\text{rad}^{-1}$}, and by eleven different linear stiffnesses varying between\break \mbox{$k_{l\,2,\,eq}=40\,\text{N}{\cdot}\text{m}^{-1}$} and \mbox{$k_{l\,2,\,eq}=144\,\text{N}{\cdot}\text{m}^{-1}$} (see Table \ref{table:investigated_stiffnesses}), which are based on the springs available in the experiments (see Appendix \ref{appendix_experimental_setup}), are simulated. The torsional stiffness is not varied due to the fact that preliminary experimental investigations observed that the other available torsional springs are not stiff enough in order to acquire a limit-cycle oscillation. These coupled simulations are carried out for the $m{-}L_3^{min}{-}y_{max}^+$ mesh applying an explicit coupling algorithm at a chord Reynolds number of $Re=30{,}000$. The applied time step size is $\Delta t=1{\cdot}10^{-5}\,\text{s}$ and the simulations run until a dimensionless time step of $t^*=1564$ is achieved. The transfinite interpolation mesh adaption method is applied due to the low required computational time. Later on, TFI proves to be the best choice for these simulations, since only small displacements are present for all eleven configurations \mbox{(see Section \ref{sec:airfoil_response}).}
\begin{table}[H]
	\centering
	\begin{tabular}{p{2.4cm} p{2.4cm} p{2.4cm} p{4cm}}
		\hline
		\multicolumn{3}{c}{\multirow{2}{*}{\centering{\bf{Linear stiffness $k_{l\,2,\,eq}$ ($\text{N}{\cdot}\text{m}^{-1}$)}}}} &  \centering{\bf{Torsional stiffness $k_{t\,3,\,eq}$ ($\text{N}{\cdot}\text{m}{\cdot}\text{rad}^{-1}$)}} \tabularnewline \hline
		\centering{$40$} & \centering{$50$} & \centering{$60$} & \centering{$0.3832$} \tabularnewline
		\centering{$70$} & \centering{$80$} & \centering{$92$} & \tabularnewline
		\centering{$104$} & \centering{$114$} &  		\centering{$124$} & \tabularnewline
		\centering{$134$} & \centering{$144$} & & \tabularnewline
		\hline			
	\end{tabular}
	\caption{\label{table:investigated_stiffnesses}Limit-cycle oscillation: Investigated system stiffnesses.}
\end{table}
\par The results achieved by all computations present the same pattern: After diverse initialization times, a limit-cycle oscillation (LCO) characterized by extremely small amplitudes, which are approximately equal for all simulations, is achieved. Since the test case with $k_{l\,2,\,eq}=92\,\text{N}{\cdot}\text{m}^{-1}$ requires the smallest computational time in order to achieve a fully developed state, this is thoroughly analyzed in the present work.
\par The results are studied in a dimensionless manner, when the flow is fully developed. Firstly, the aerodynamic properties of the body are analyzed according to the time history of the drag and lift coefficients. Then, the airfoil displacements, as well as the aeroelastic properties are investigated. The latter is studied according to the frequencies related to the flow and structure. Finally, the instantaneous and time-averaged flow fields are analyzed.

\subsection{Aerodynamic properties}\markboth{CHAPTER 4.$\quad$RES. AND DISC.}{4.1$\quad$TEST CASE 1: LCO}
\label{sec:flow_forces}
\par The forces generated by the coupled problem are investigated according to the lift $C_L$ and drag $C_D$ coefficients. These non-dimensional numbers, which are proportional to the pressure and shear forces, are calculated according to Eqs.\ (\ref{eq:lift_coefficient}) and (\ref{eq:drag_coefficient}). The pitching moment coefficient is not calculated in the present work, since this is zero for the NACA0012 airfoil up to an angle of attack of $\alpha=14^\circ$ \mbox{(see Fung \cite{Fung_2002})} and the current test case is characterized by extremely small rotation displacements \mbox{(see Section \ref{sec:airfoil_response})}, which maintain the airfoil at an incidence of approximately $\alpha=0^\circ$. This occurs due to the airfoil symmetry, regarding that the pitching moment is the moment in relation to the $x_3$-axis.
\begin{eqnarray}
\label{eq:lift_coefficient}
C_L&=&\frac{2\,F_L}{\rho_f\,S_{ref}\,u_{in,\,1}^2},\\
\label{eq:drag_coefficient}
C_D&=&\frac{2\,F_D}{\rho_f\,S_{ref}\,u_{in,\,1}^2}.
\end{eqnarray}
\par $F_D$, $F_L$, $u_{in,\,1}$ and $\rho_f$ are respectively the drag and lift forces, the free-stream velocity and the fluid density. The reference area $S_{ref}$ is calculated according to Eq.\ (\ref{eq:reference_area}). $L_{3}$ and $c$ are the span-wise length of the computational domain and the airfoil chord length, i$.$e$.$, $L_{3}=0.025\,\text{m}$ and $c=0.1\,\text{m}$.
\begin{equation}
S_{ref}=c\;L_{3} \label{eq:reference_area}
\end{equation}
\par The time history of the lift and drag coefficients are depicted in Fig.\ \ref{fig:drag_lift_history} for a dimensionless time interval of $1470\leq t^*\leq1500$. The initialization phase, which is present until about $t^*=600$, is not shown for the sake of brevity. In general, the frequency of the lift is lower than the frequency of the drag. However, the lift signal is also characterized by the presence of high frequencies, which might be caused by the Newmark method. 
\begin{figure}[H]
	\centering
	\subfigure[Lift coefficient.]{\includegraphics[width=0.49\textwidth]{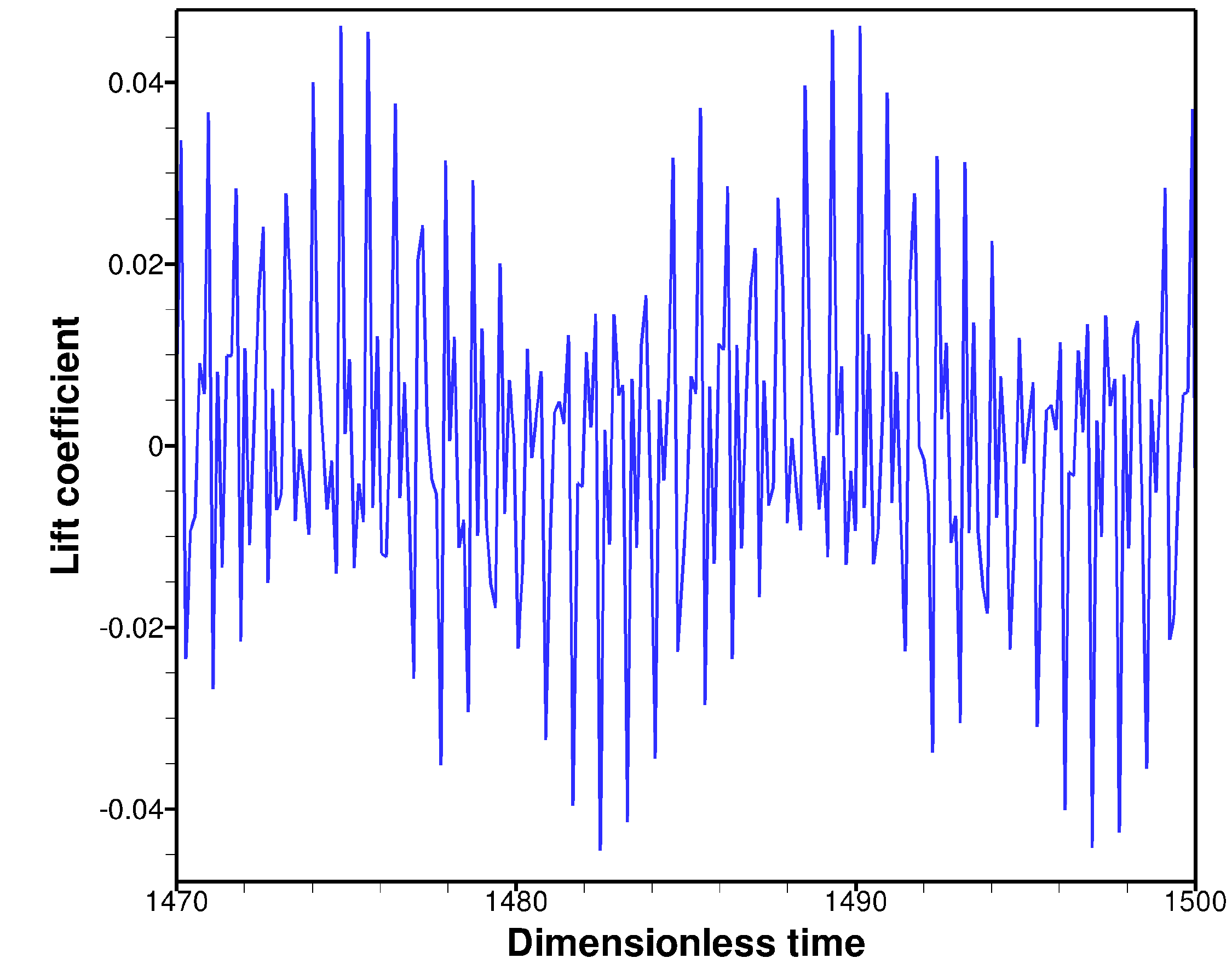}\label{fig:lift_history}}\hfill
	\subfigure[Drag coefficient.]{\includegraphics[width=0.49\textwidth]{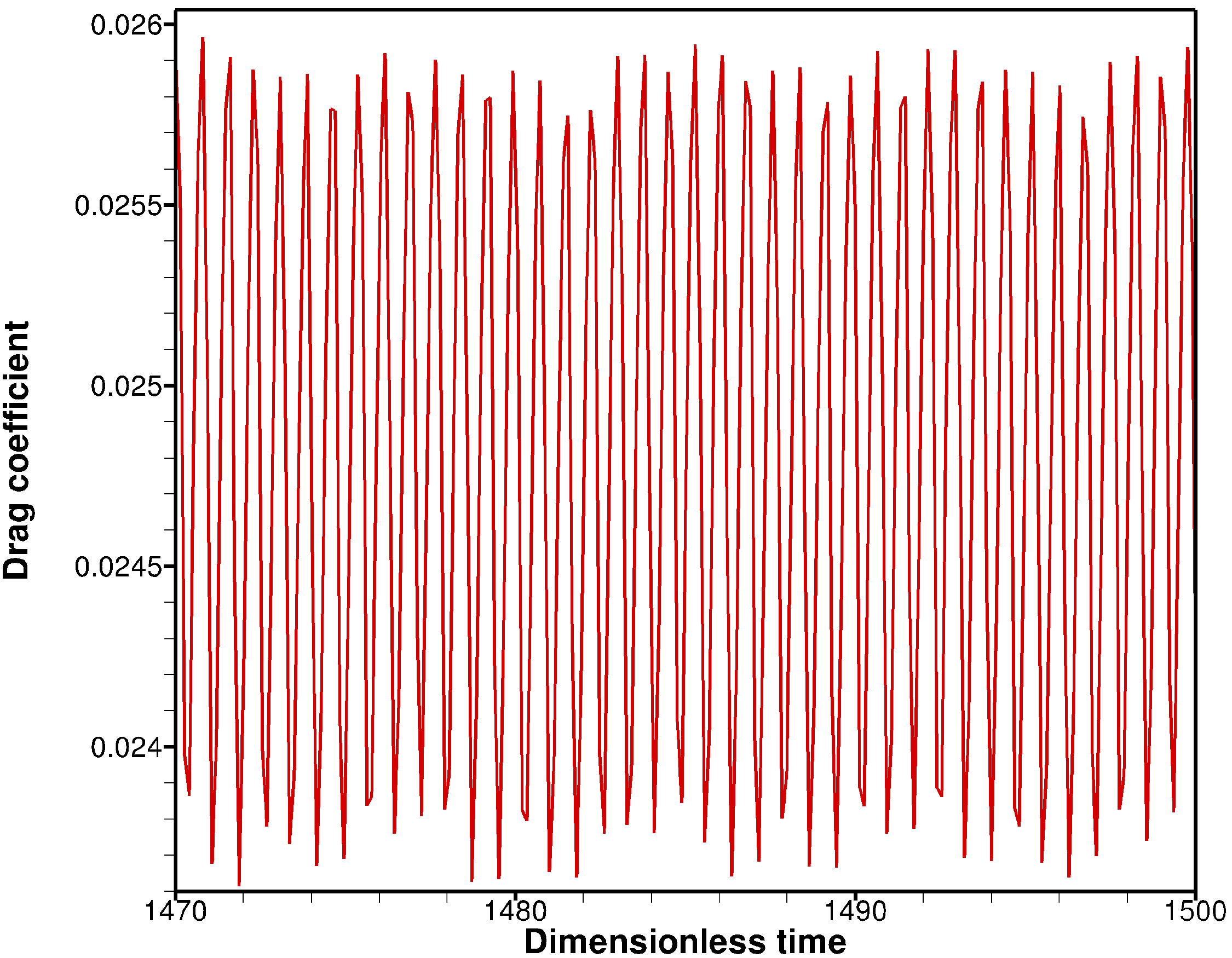} \label{fig:drag_history}}\hfill
	\caption{Time history of the lift and drag coefficients: Test case characterized by $Re=30{,}000$, \mbox{$k_{t\,3,\,eq}=0.3832\,N{\cdot}m{\cdot}rad^{-1}$} and \mbox{$k_{l\,2,\,eq}=92\,N{\cdot}m^{-1}$}.}
	\label{fig:drag_lift_history}
\end{figure}
\par The mean values and standard deviations of the lift and drag coefficients are summarized in Table \ref{table:mean_standard_aerodynamic_coefficients} considering a fully developed state. Although the standard deviation of the lift coefficient is about two orders of magnitude higher than the standard deviation of the drag coefficient, both aerodynamic coefficients weakly oscillate around a mean value. For the case of the lift coefficient, the mean value is approximately zero, which indicates that the symmetric NACA0012 airfoil remains at an incidence of $\alpha=0^\circ$. Moreover, the lift and drag coefficients for the coupled problem are approximately equal to the ones calculated for the fixed airfoil (see Section \ref{sec:study_span_wise_length}). This indicates that although the NACA0012 has two degrees of freedom, it barely moves.
\begin{table}[H]
	\centering
	\begin{tabular}{p{2.4cm} p{4cm} p{0.2cm} p{2.4cm} p{4cm}}
		\hline
		\multicolumn{2}{c}{\textbf{Lift coefficient}} & & \multicolumn{2}{c}{\textbf{Drag coefficient}} \tabularnewline \cline{1-2} \cline{4-5}
		\centering{\textbf{Mean value}} & \centering{\textbf{Standard deviation}} & & \centering{\textbf{Mean value}} & \centering{\textbf{Standard deviation}} \tabularnewline
		\hline
		\centering{$4.95{\cdot}10^{-4}$} & \centering{$1.79{\cdot}10^{-2}$} & & 		\centering{$2.48{\cdot}10^{-2}$} & \centering{$8.02{\cdot}10^{-4}$} \tabularnewline
		\hline			
	\end{tabular}
	\caption{\label{table:mean_standard_aerodynamic_coefficients}Mean values and standard deviations of the aerodynamic coefficients. Test case: $Re=30{,}000$, \mbox{$k_{t\,3,\,eq}=0.3832\,\text{N}{\cdot}\text{m}{\cdot}\text{rad}^{-1}$} and \mbox{$k_{l\,2,\,eq}=92\,\text{N}{\cdot}\text{m}^{-1}$}.}
\end{table}
\subsection{Airfoil displacements}\markboth{CHAPTER 4.$\quad$RESULTS AND DISCUSSION.}{4.1$\quad$TEST CASE 1: LCO}
\label{sec:airfoil_response}
\par
\begin{figure}[H]
	\centering
	\subfigure[Translational displacement $X_2$.]{\includegraphics[width=0.49\textwidth]{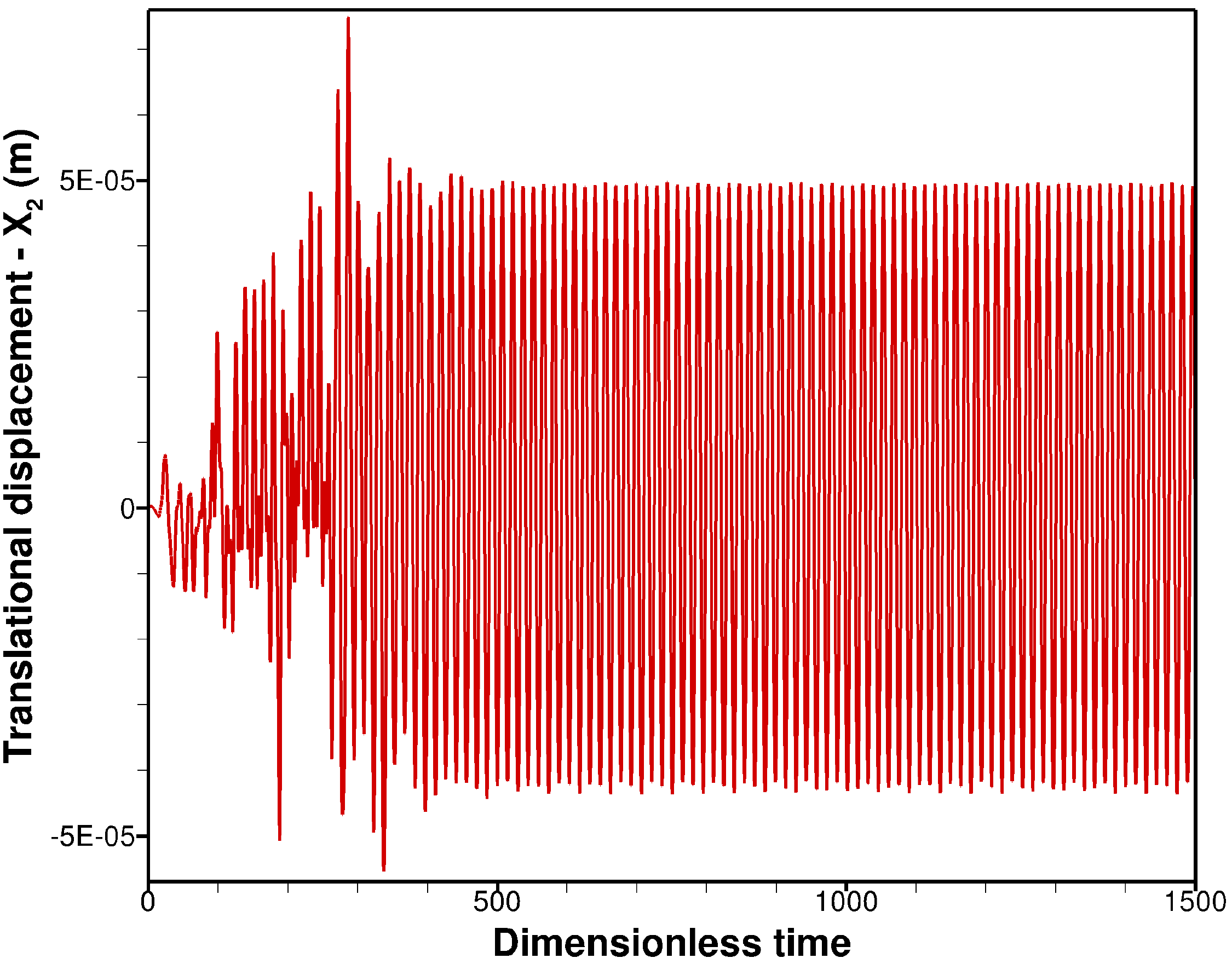} \label{fig:plunge_history}}\hfill
	\subfigure[Translational displacement $X_2$: Close up.]{\includegraphics[width=0.49\textwidth]{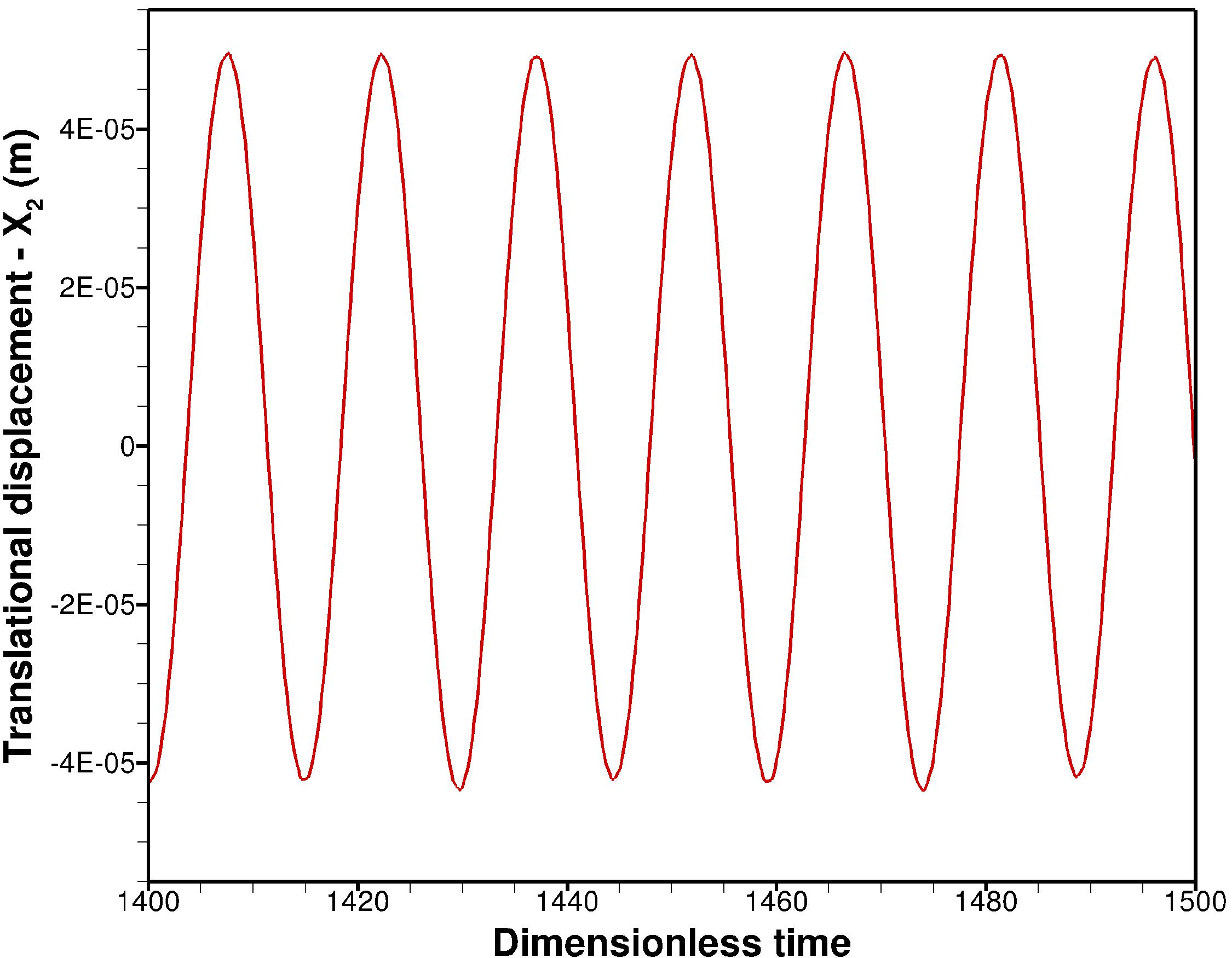} \label{fig:plunge_zoom_history}}\hfill
	\subfigure[Rotational displacement $\varphi_3$]{\includegraphics[width=0.49\textwidth]{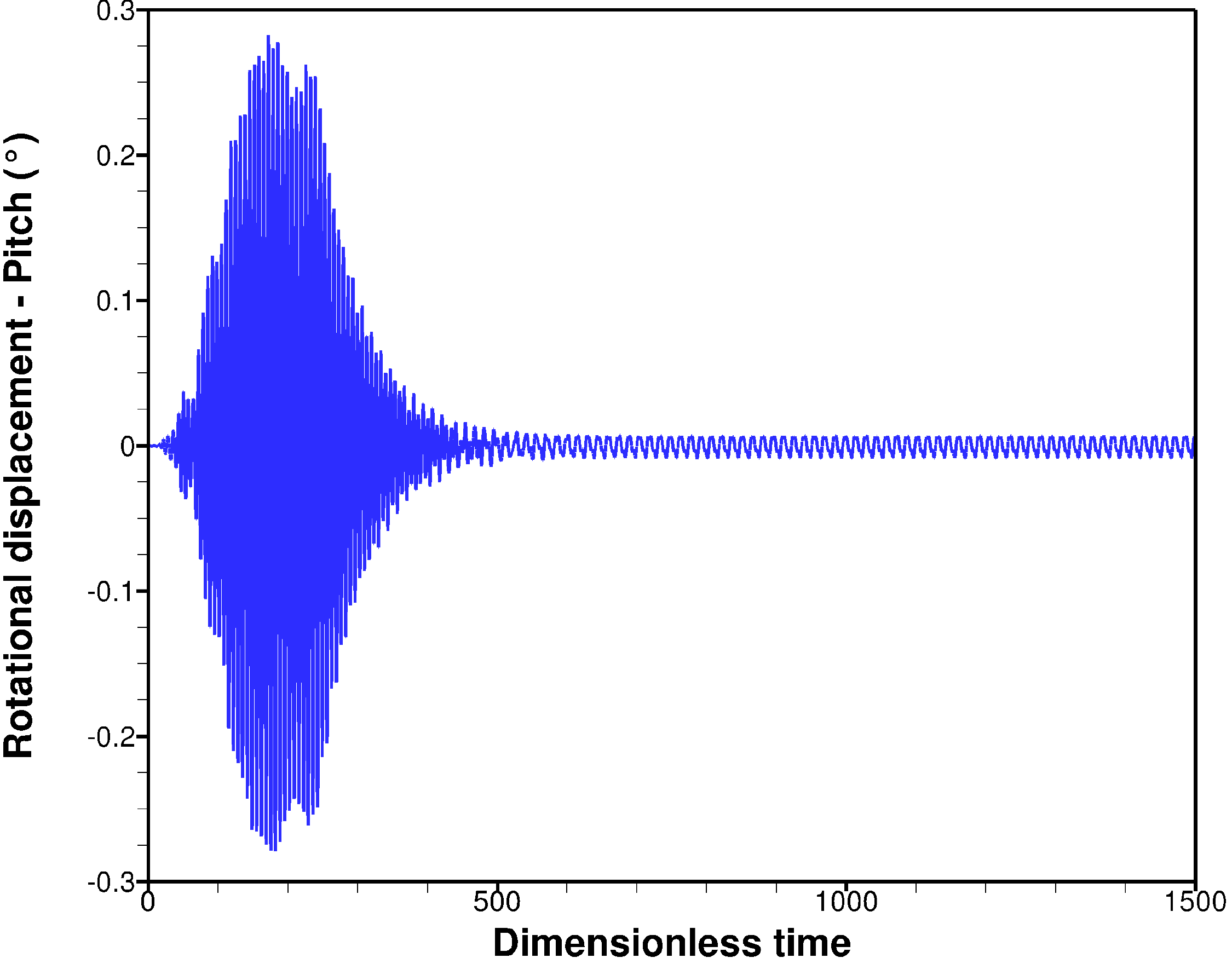}\label{fig:pitch_history}}\hfill
	\subfigure[Rotational displacement $\varphi_3$: Close up.]{\includegraphics[width=0.49\textwidth]{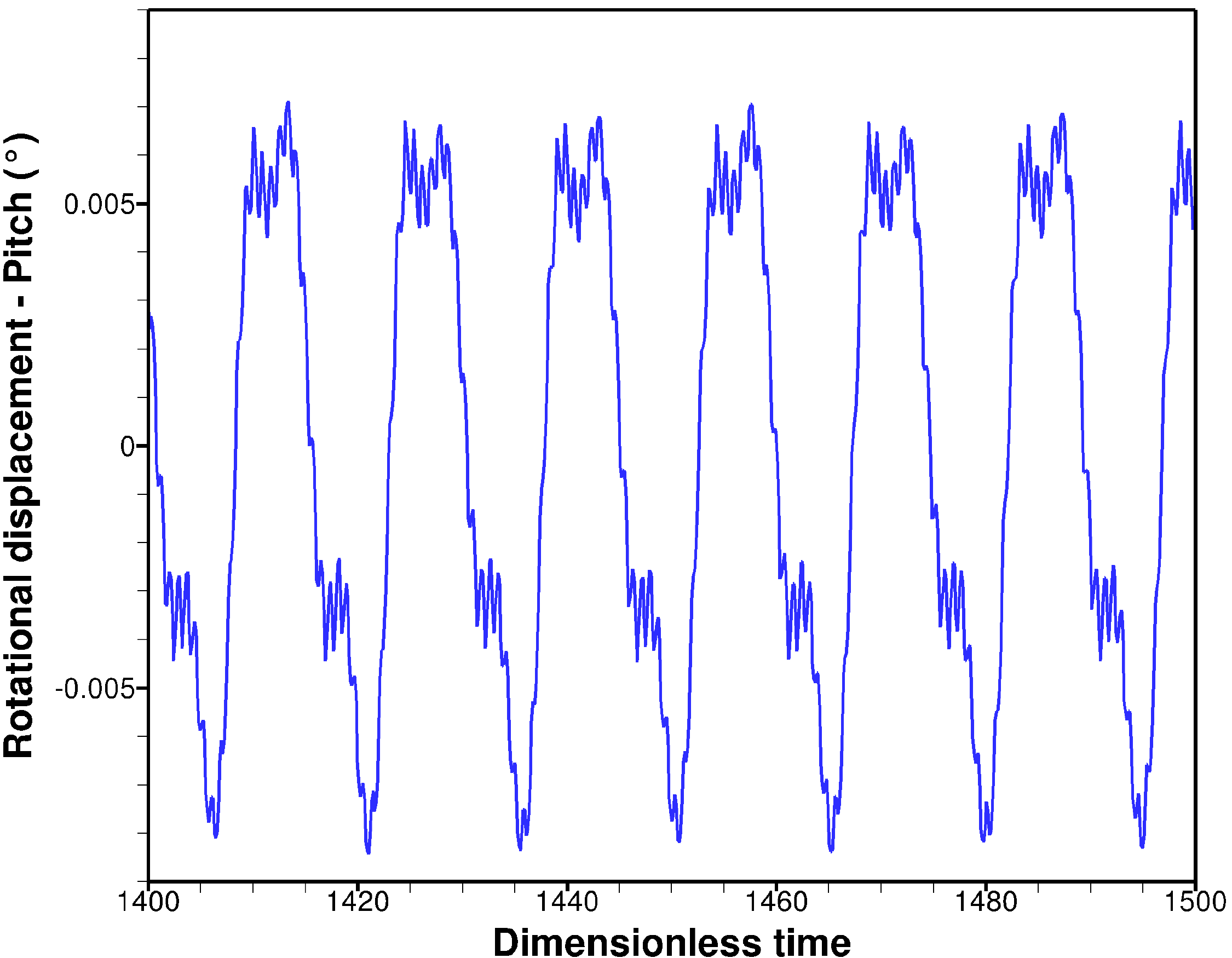}\label{fig:pitch_zoom_history}}\hfill
	\caption{Time history of the translational and rotational displacements: \mbox{$Re=30{,}000$}, \mbox{$k_{t\,3,\,eq}=0.3832\,\text{N}{\cdot}\text{m}{\cdot}\text{rad}^{-1}$} and \mbox{$k_{l\,2,\,eq}=92\,\text{N}{\cdot}\text{m}^{-1}$}.}
	\label{fig:pitch_plunge_history}
\end{figure}

 The fluid forces and moments act on the airfoil and generate displacements in relation to the up and down and pitch degrees of freedom. These displacements are illustrated in \mbox{Fig.\ \ref{fig:pitch_plunge_history}} in relation to the dimensionless time $t^*$. High frequencies are present in the pitch motion. After the transitional phase, which is present until $t^*\approx600$, a limit-cycle oscillation is generated, which is characterized by weak oscillations around a mean value. This value, as well as the standard deviation of both displacements are described in \mbox{Table \ref{table:mean_standard_displacements}}.
\begin{table}[H]
	\centering
	\begin{tabular}{p{2.4cm} p{4cm} p{0.2cm} p{2.4cm} p{4cm}}
		\hline
		\multicolumn{2}{c}{\textbf{Translational displacement $X_2$ (m)}} & & \multicolumn{2}{c}{\textbf{Rotational displacement $\varphi_3$ ($^\circ$)}} \tabularnewline \cline{1-2} \cline{4-5}
		\centering{\textbf{Mean value}} & \centering{\textbf{Standard deviation}} & & \centering{\textbf{Mean value}} & \centering{\textbf{Standard deviation}} \tabularnewline
		\hline
		\centering{$3.44{\cdot}10^{-6}$} & \centering{$3.27{\cdot}10^{-5}$} & & \centering{$2.70{\cdot}10^{-6}$} & \centering{$8.62{\cdot}10^{-5}$} \tabularnewline
		\hline			
	\end{tabular}
	\caption{\label{table:mean_standard_displacements}Mean values and standard deviations of the airfoil displacements. Case: \mbox{$Re=30{,}000$}, \mbox{$k_{t\,3,\,eq}=0.3832\,\text{N}{\cdot}\text{m}{\cdot}\text{rad}^{-1}$} and \mbox{$k_{l\,2,\,eq}=92\,\text{N}{\cdot}\text{m}^{-1}$}.}
\end{table}
\par The limit-cycle amplitudes of the pitch and up and down motions are extremely small, which is expected for the current test case. This phenomenon arises for fluid-structure interaction of the NACA0012 airfoil at low to moderate chord Reynolds numbers \mbox{($15{,}000 \leq Re\leq 500{,}000$)} either when no angular and up and down velocities are initially available and the airfoil is at an incidence of \mbox{$\alpha=0^\circ$} \mbox{(see Lapointe and Dumas \cite{Lapointe_2012})}, or when the linear stiffness is small, i$.$e$.$, roughly $k_{l\,2,\,eq}\leq300\,N{\cdot}m^{-1}$ (see Poirel and\break Mendes \cite{Poirel_2012} and Lapointe and Dumas \cite{Lapointe_2012}). In the former configuration, i$.$e$.$, airfoil at $\alpha=0^\circ$ and no initial angular and up and down velocities, the small LCO is triggered by a break of the natural symmetry due to the separation of the laminar boundary layer and the vortex shedding at the trailing-edge (see Section \ref{sec:instationary_flow_field} and \ref{sec:time_avg_flow_field}). In the latter, i$.$e$.$, $k_{l\,2,\,eq}\leq300\,N{\cdot}m^{-1}$, small LCO arises due to the structural pitch-up/down coupling itself, since the energy provided by this coupling feeds the LCO and is not high enough in order to generate large amplitudes of oscillation \cite{Poirel_2012}.  
\begin{figure}[H]
	\centering
	\subfigure[Translational displacement $X_2$.]{\includegraphics[width=0.49\textwidth]{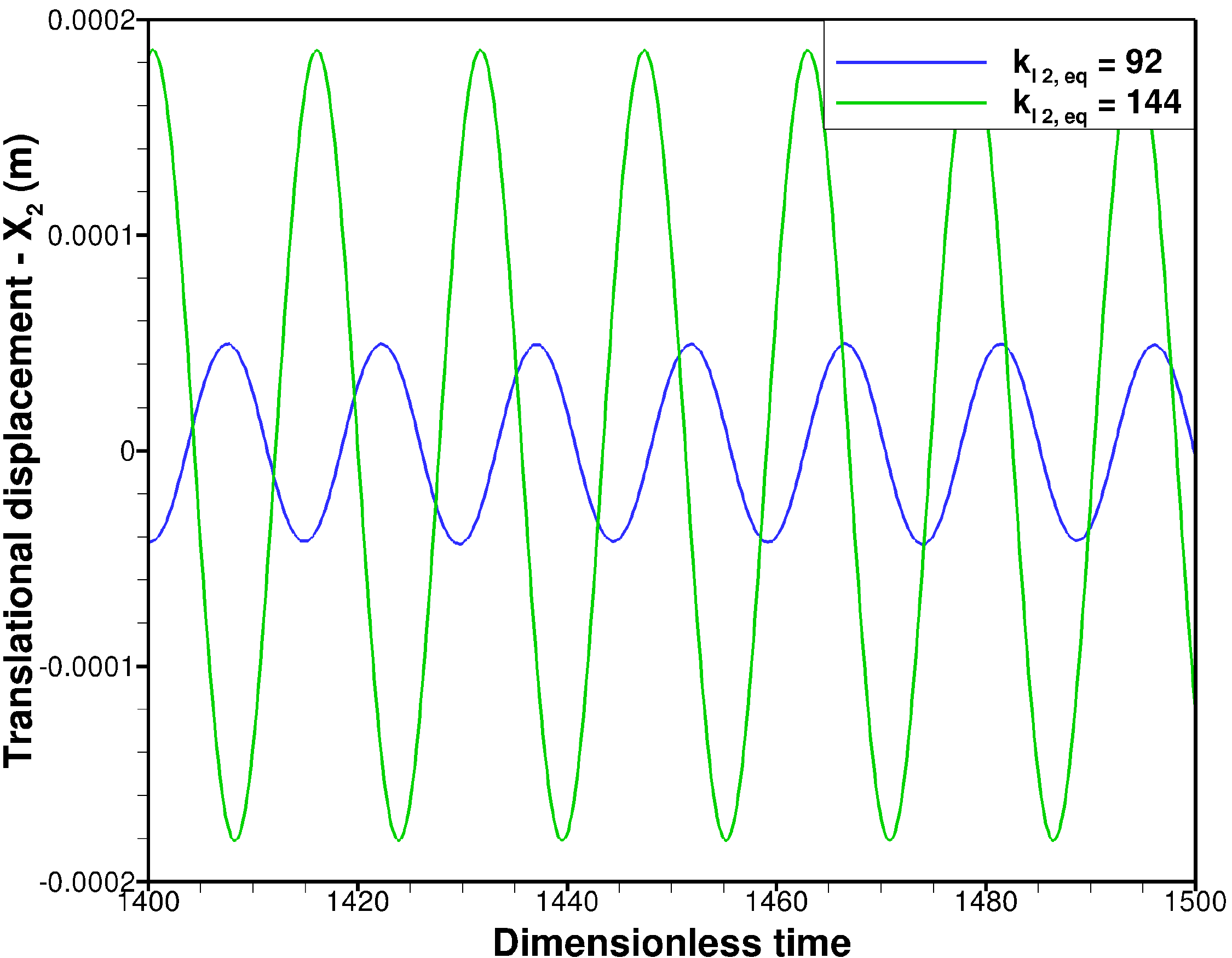} \label{fig:plunge_comp_kl_history}}\hfill
	\subfigure[Rotational displacement $\varphi_3$.]{\includegraphics[width=0.49\textwidth]{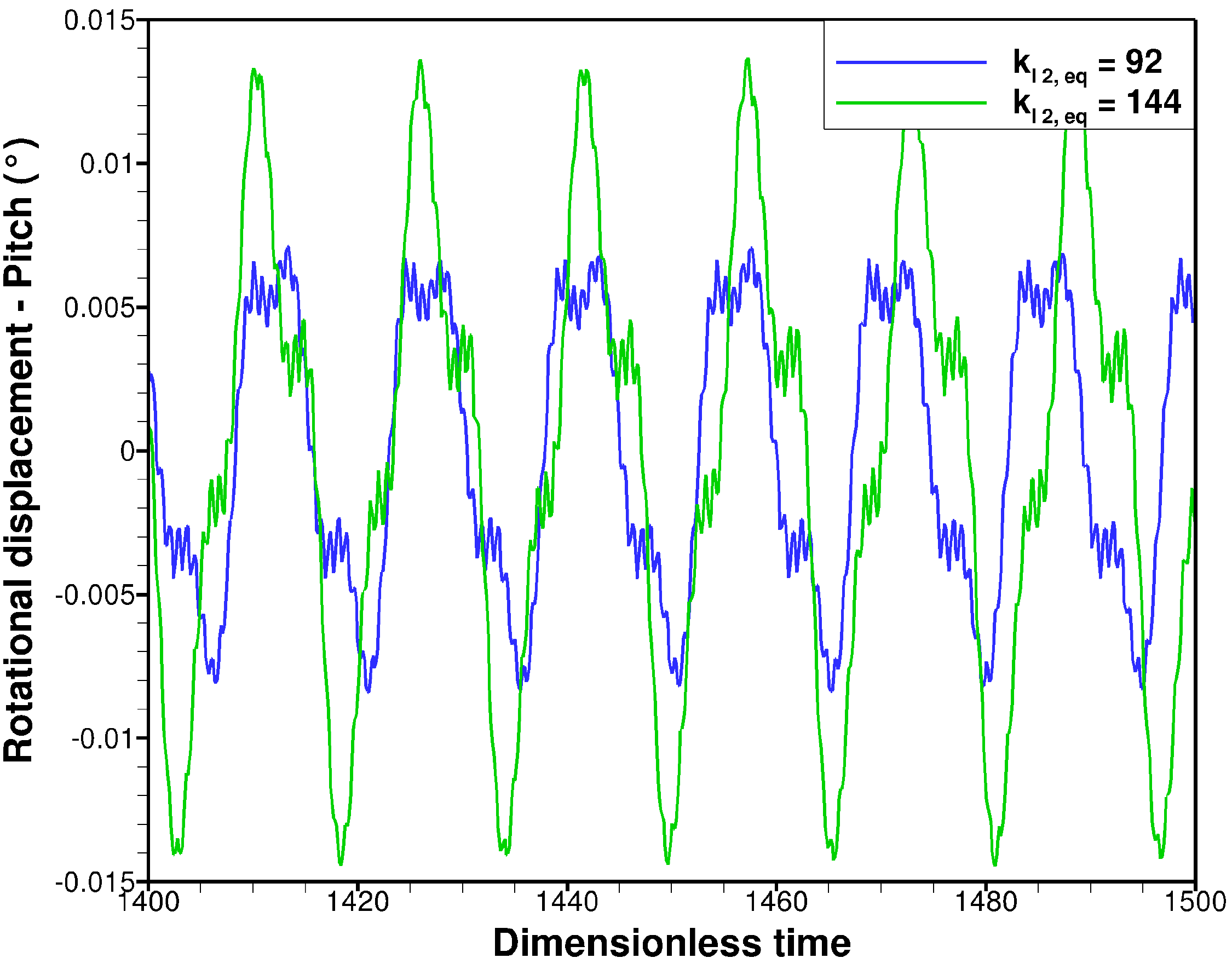} \label{fig:pitch_comp_kl_history}}\hfill
	\caption{Time history of the translational and rotational displacements: Comparison of the test cases characterized by \mbox{$k_{l\,2,\,eq}=92\,\text{N}{\cdot}\text{m}^{-1}$} and \mbox{$k_{l\,2,\,eq}=144\,\text{N}{\cdot}\text{m}^{-1}$}.}
	\label{fig:plunge_pitch_comp_kl}
\end{figure}
\par An increase of the linear stiffness up to $k_{l\,2,\,eq}=144\,N{\cdot}m^{-1}$, which is the maximal linear stiffness available for the experiments (see Appendix \ref{appendix_experimental_setup}), leads to an increase in the amplitudes of the pitch and up and down motions. However, the limit-cycle oscillation is still characterized by extremely small amplitudes, as illustrated in Fig.\ \ref{fig:plunge_pitch_comp_kl}. 
\begin{figure}[H]
	\centering
	\centering
	\includegraphics[width=0.49\textwidth,draft=\drafttype]{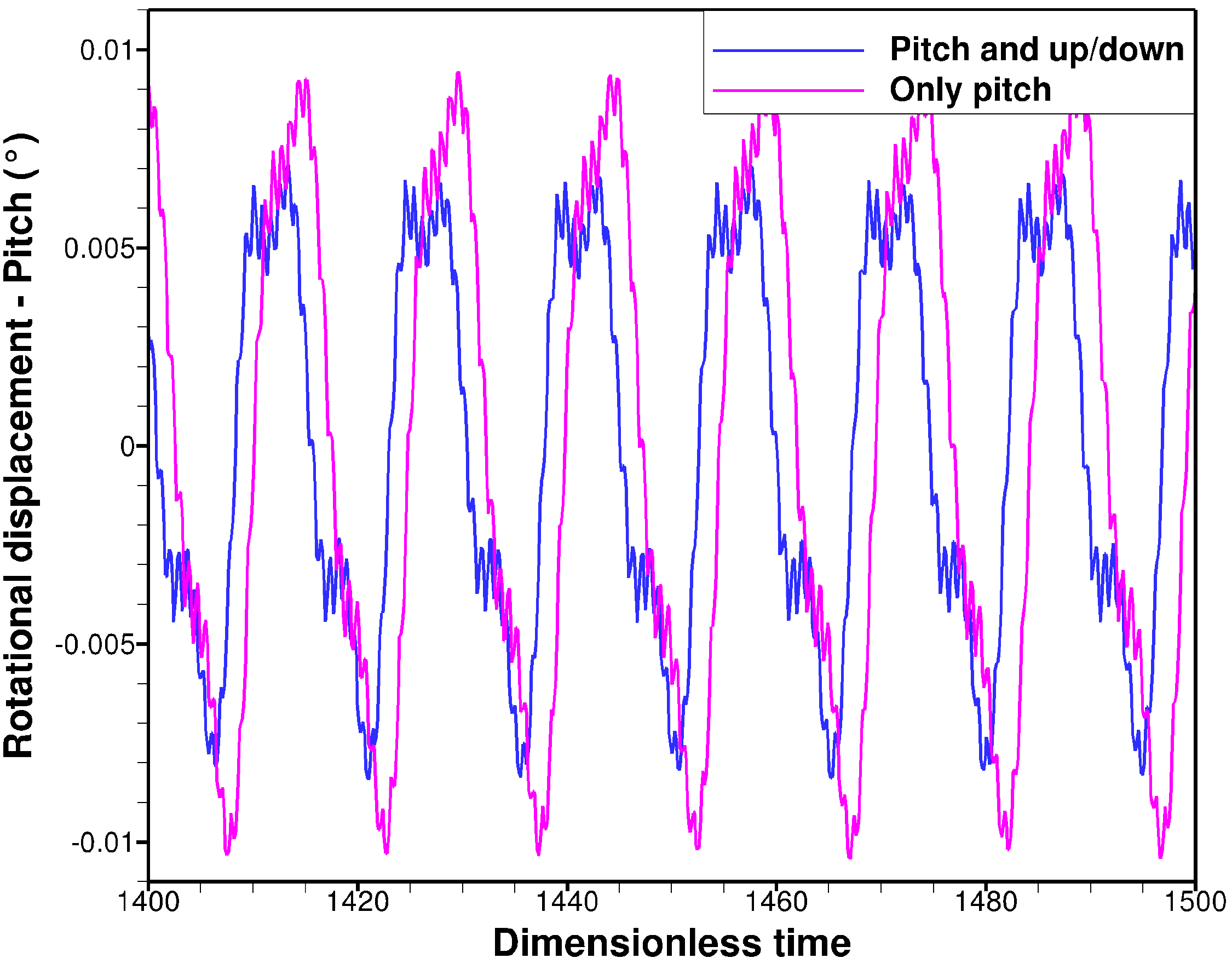}
	\caption{\label{fig:pitch_comp_history}Time history of the rotational displacements regarding a torsional stiffness of \mbox{$k_{t\,3,\,eq}=0.3832\,\text{N}{\cdot}\text{m}{\cdot}\text{rad}^{-1}$}: Cases with and without up and down degree of freedom.}
\end{figure}
\par Moreover, Lapointe and Dumas \cite{Lapointe_2012} stated that if a fluid-structure interaction is characterized by LCO with small amplitudes, then these amplitudes are similar to the ones achieved with cases characterized by only the pitch degree of freedom. Therefore, Fig.\ \ref{fig:pitch_comp_history} illustrates a comparison of the rotational displacements achieved by the current test case (\mbox{$k_{t\,3,\,eq}=0.3832\,\text{N}{\cdot}\text{m}{\cdot}\text{rad}^{-1}$} and \mbox{$k_{l\,2,\,eq}=92\,\text{N}{\cdot}\text{m}^{-1}$}) and by the computation with only the pitch degree of freedom (\mbox{$k_{t\,3,\,eq}=0.3832\,\text{N}{\cdot}\text{m}{\cdot}\text{rad}^{-1}$}). As expected, the achieved pitching amplitudes are very similar in both cases.

\subsection{Frequencies}
\label{sec:frequencies}

\par The aerodynamic forces, as well as the displacements are characterized by an\break oscillatory behavior. Therefore, Fourier transforms are applied in order to analyze the frequency domain of these variables. In order to acquire accurate results the available data is firstly post-processed in order to obtain oscillations composed of only complete periods. The Fourier transform is then carried out. Rectangular, triangular, Hann and Hamming window functions are investigated and deliver exactly the same result.
\par The frequency domain is analyzed aiming primarily at the identification of the\break vortex shedding frequency $f_v$, which is required in order to calculate the Strouhal number $St$. This is a dimensionless number that characterizes unsteady flow mechanisms,\break i$.$e$.$, it describes the oscillatory phenomenon available in a flow. In the current work,\break this dimensionless number is a function of the vortex shedding frequency $f_v$, the\break airfoil chord $c=0.1\,\text{m}$ and the free-stream velocity $u_{1,\,in}=4.47\,\text{m}{\cdot}\text{s}^{-1}$, as described in Eq.\ (\ref{eq:strouhal_number}):
\begin{equation}
\label{eq:strouhal_number}
St=\frac{f_{v}\,c}{u_{1,\,in}}.
\end{equation}
\par
\begin{figure}[H]
	\subfigure[Overall frequency domain: Lift.]{\includegraphics[width=0.49\textwidth]{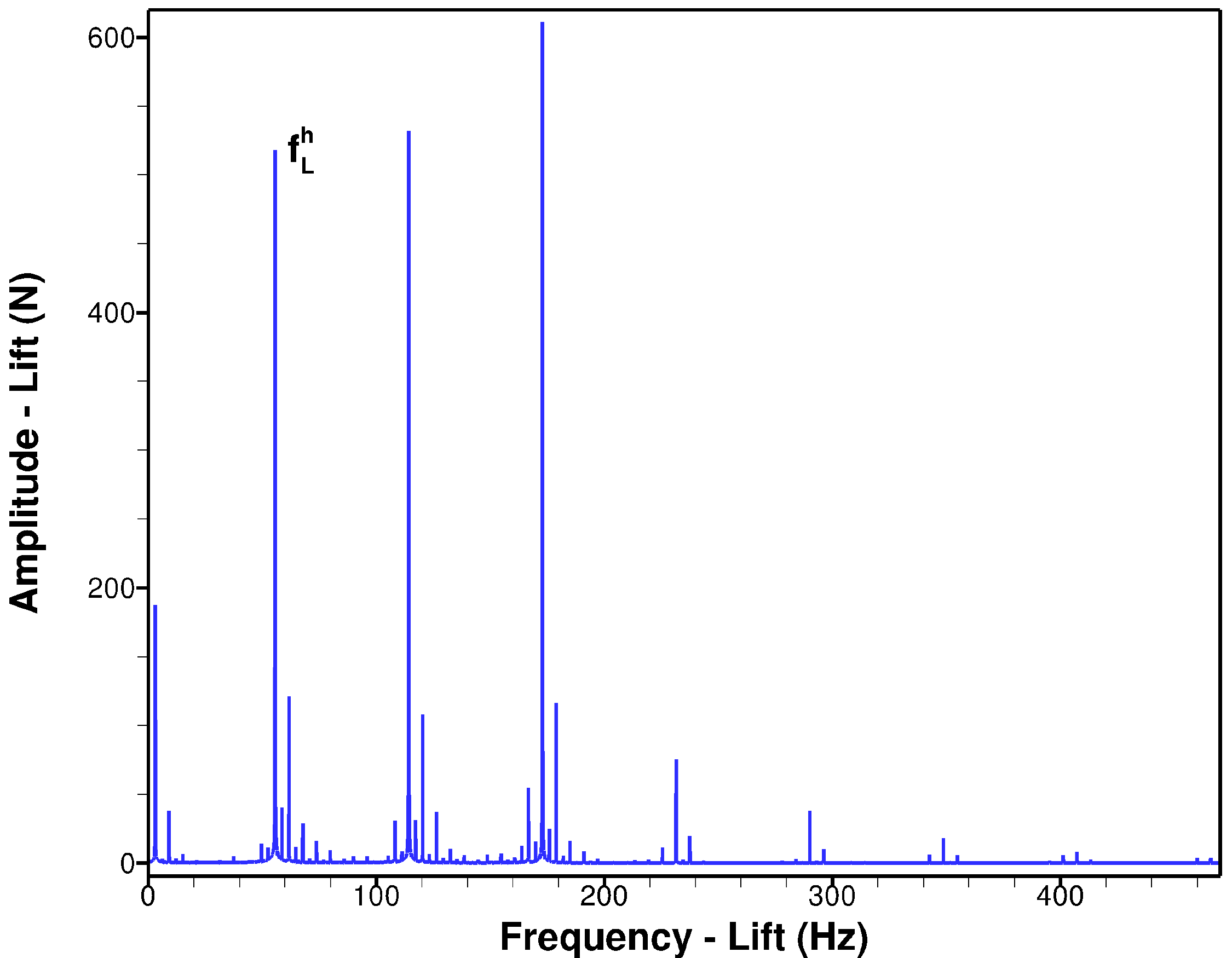} \label{fig:frequency_lift}}\hfill
	\subfigure[Focus on the low frequencies: Lift.]{\includegraphics[width=0.49\textwidth]{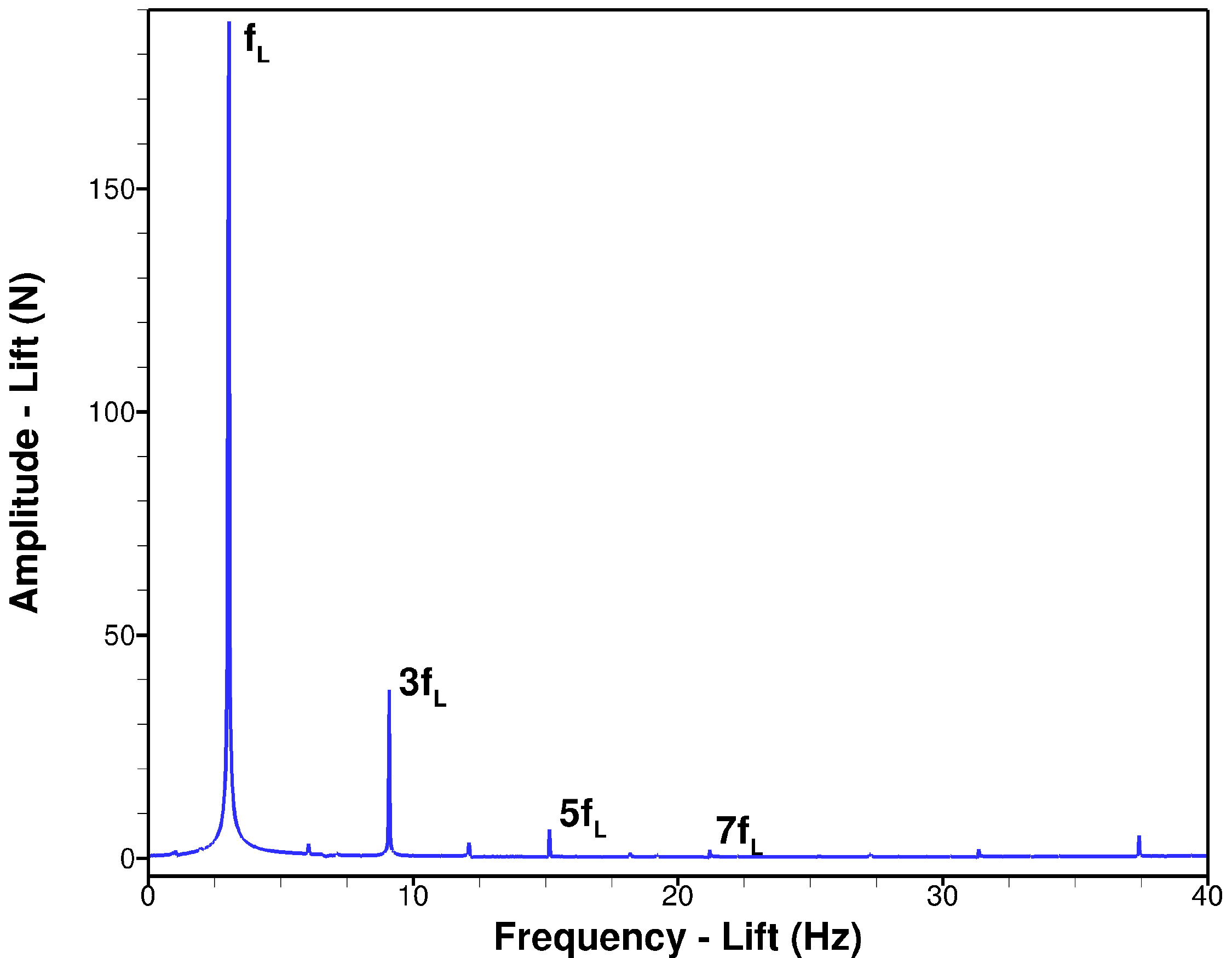} \label{fig:frequency_zoom_lift}}\hfill
	\centering
	\subfigure[Overall frequency domain: $X_2$.]{\includegraphics[width=0.49\textwidth]{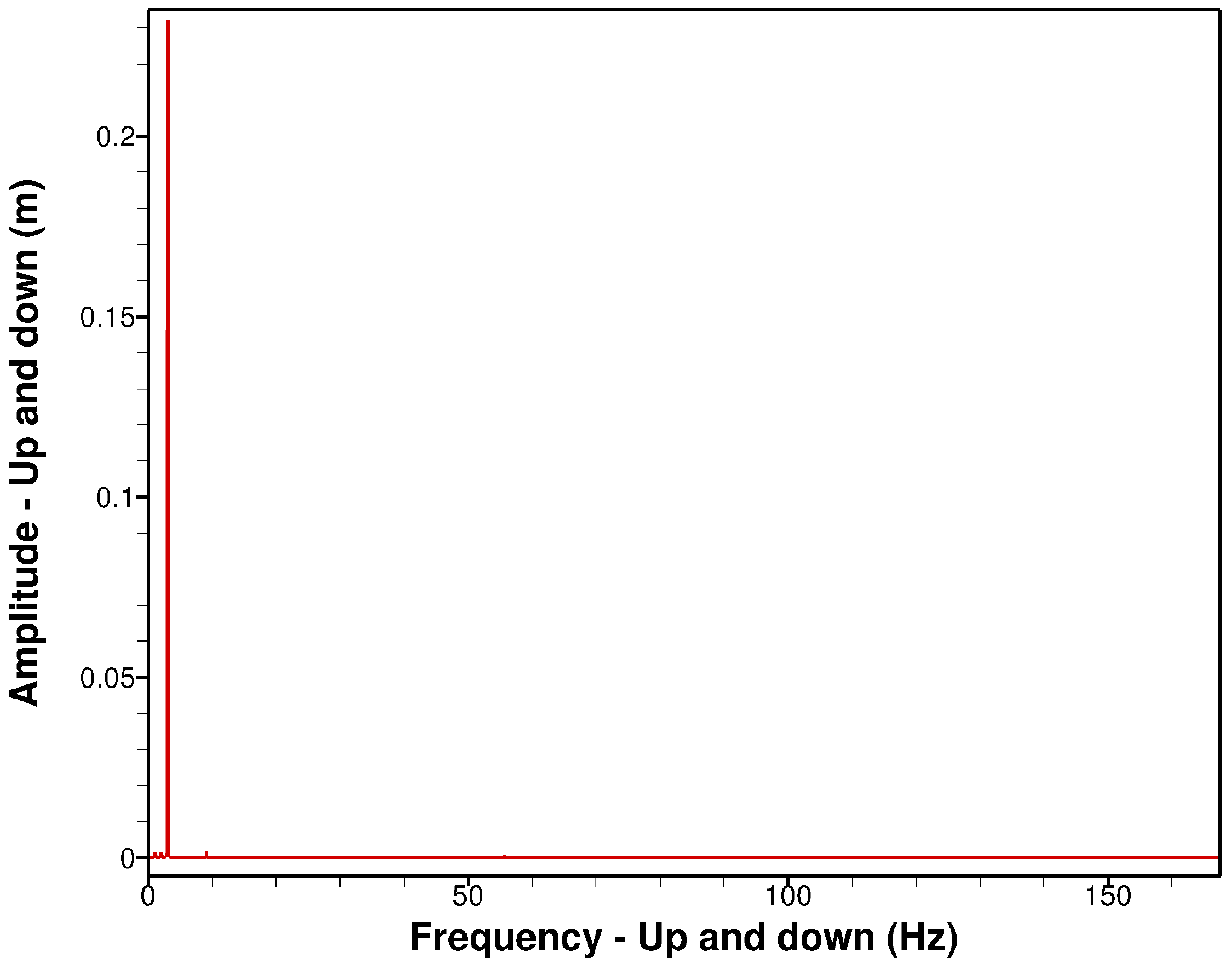} \label{fig:plunge_comps_kl_history}}\hfill
	\subfigure[Focus on the low frequencies: $X_2$.]{\includegraphics[width=0.49\textwidth]{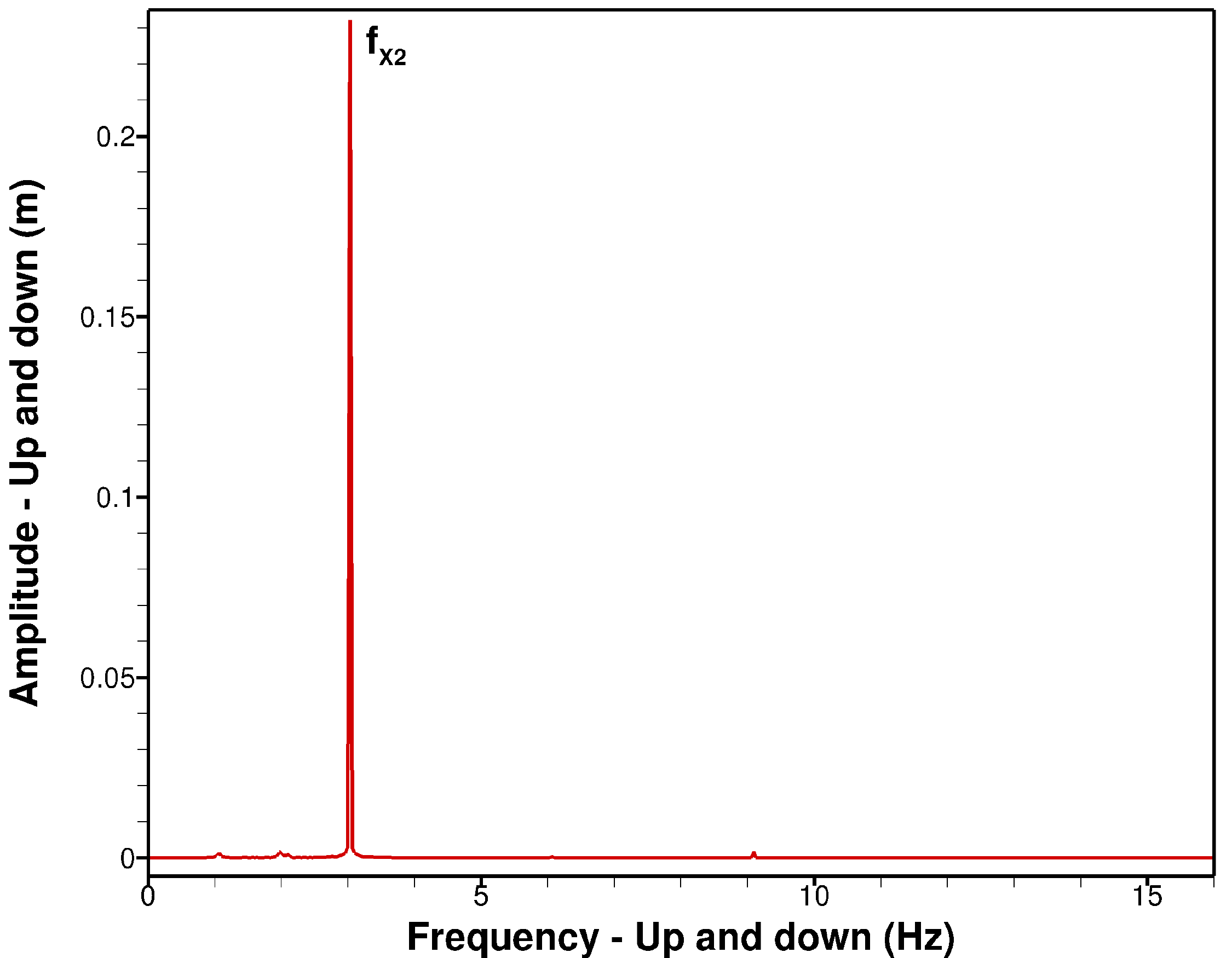} \label{fig:pitch_comp_kls_history}}\hfill
	\subfigure[Overall frequency domain: $\varphi_3$.]{\includegraphics[width=0.49\textwidth]{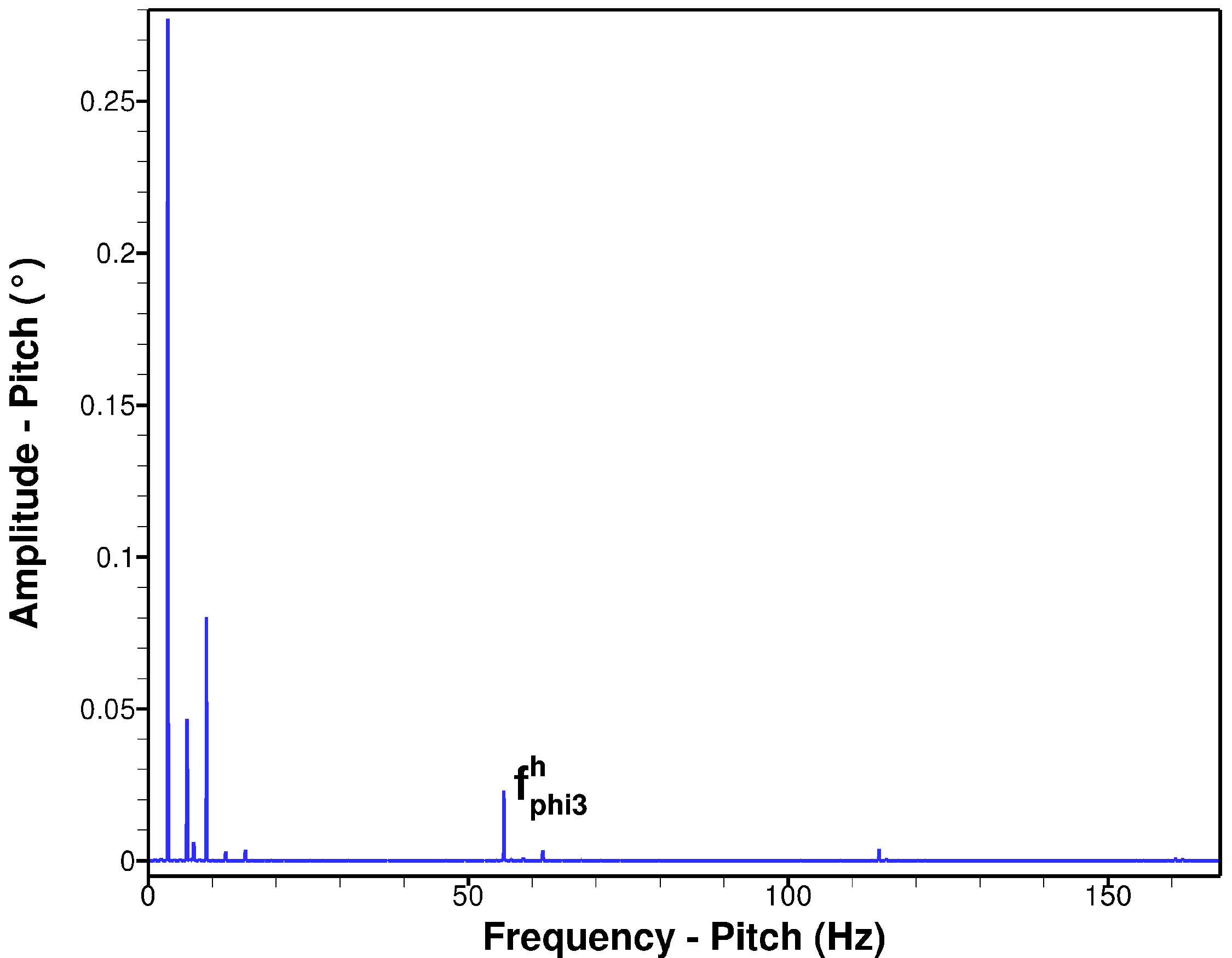} \label{fig:plunge_comp_ksl_history}}\hfill
	\subfigure[Focus on the low frequencies: $\varphi_3$.]{\includegraphics[width=0.49\textwidth]{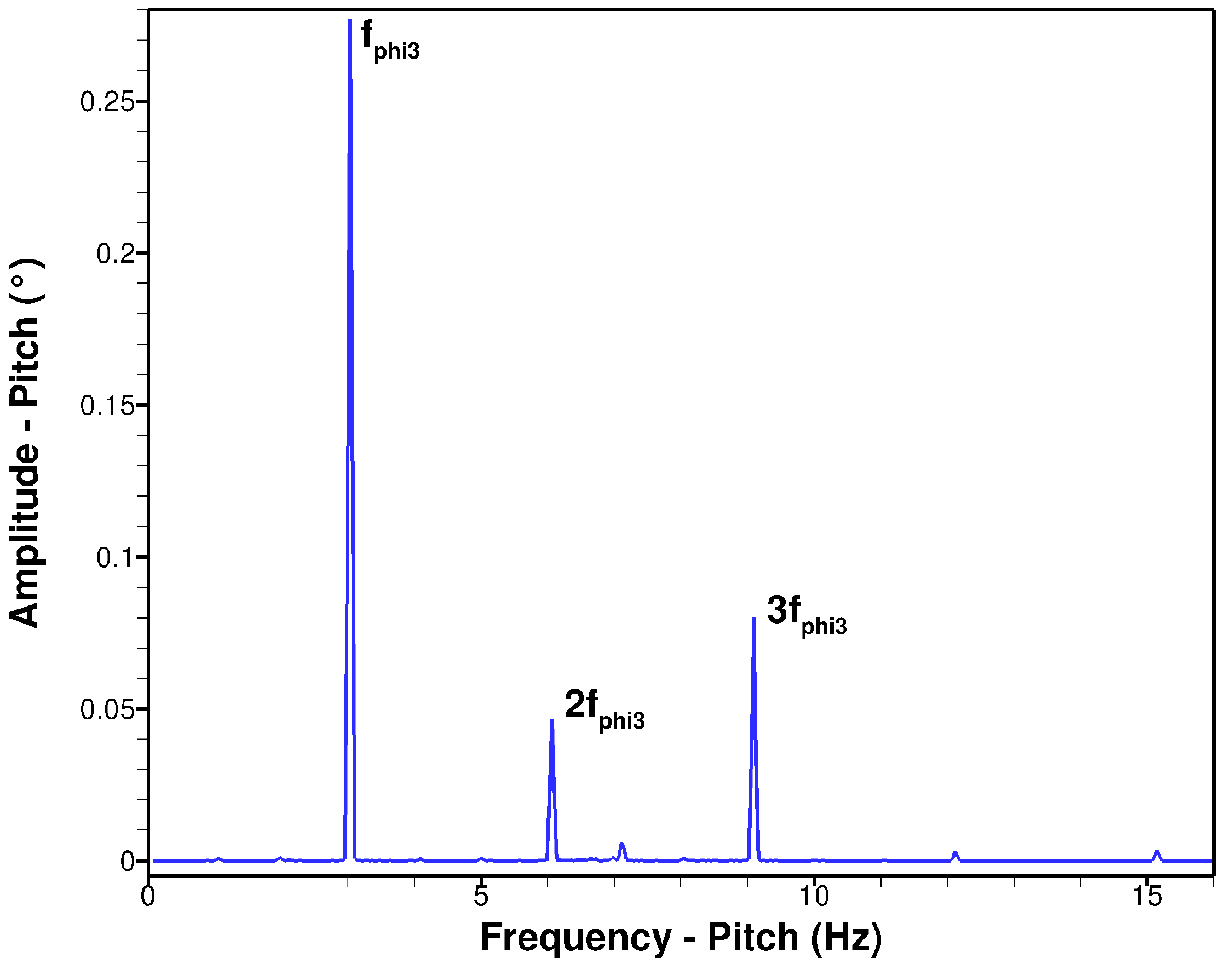} \label{fig:pitch_comp_kl_hsistory}}\hfill
	\caption{Frequency domain of the drag force, lift force and translational $X_2$ and rotational $\varphi_3$ displacements: Test case characterized by $Re=30{,}000$, \mbox{$k_{t\,3,\,eq}=0.3832\,\text{N}{\cdot}\text{m}{\cdot}\text{rad}^{-1}$} and \mbox{$k_{l\,2,\,eq}=92\,\text{N}{\cdot}\text{m}^{-1}$}.}
	\label{fig:frequency_domain}
\end{figure}

 The frequency domains of the drag and lift forces as well as of the translational and rotational displacements are depicted in Figs.\ \ref{fig:frequency_domain}. An overall view of these domains as well as a view focused on the low frequencies are available.
\begin{figure}[H]
	\centering
	\subfigure[Overall frequency domain: Drag.]{\includegraphics[width=0.49\textwidth]{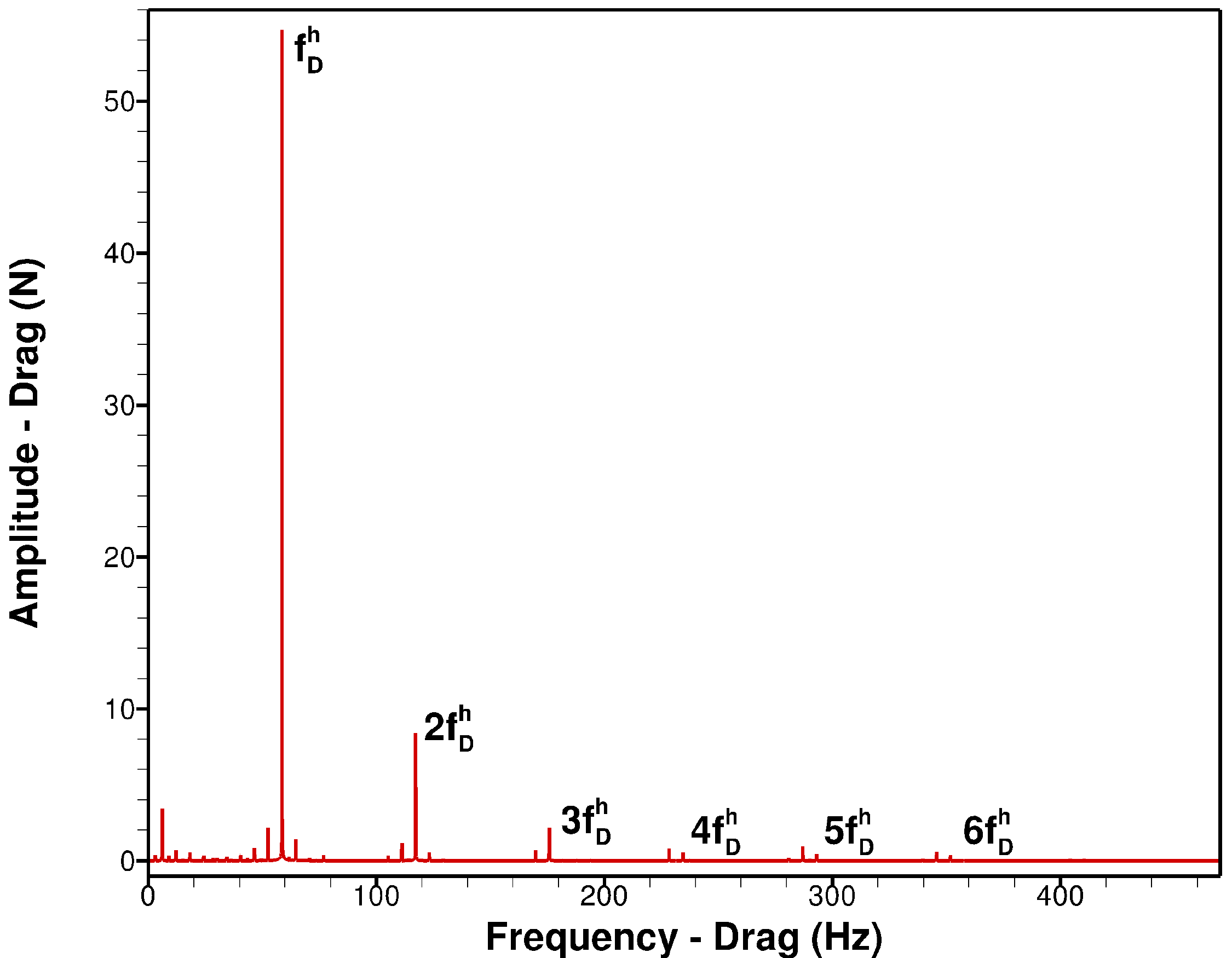} \label{fig:frequency_drag}}\hfill
	\subfigure[Focus on the low frequencies: Drag.]{\includegraphics[width=0.49\textwidth]{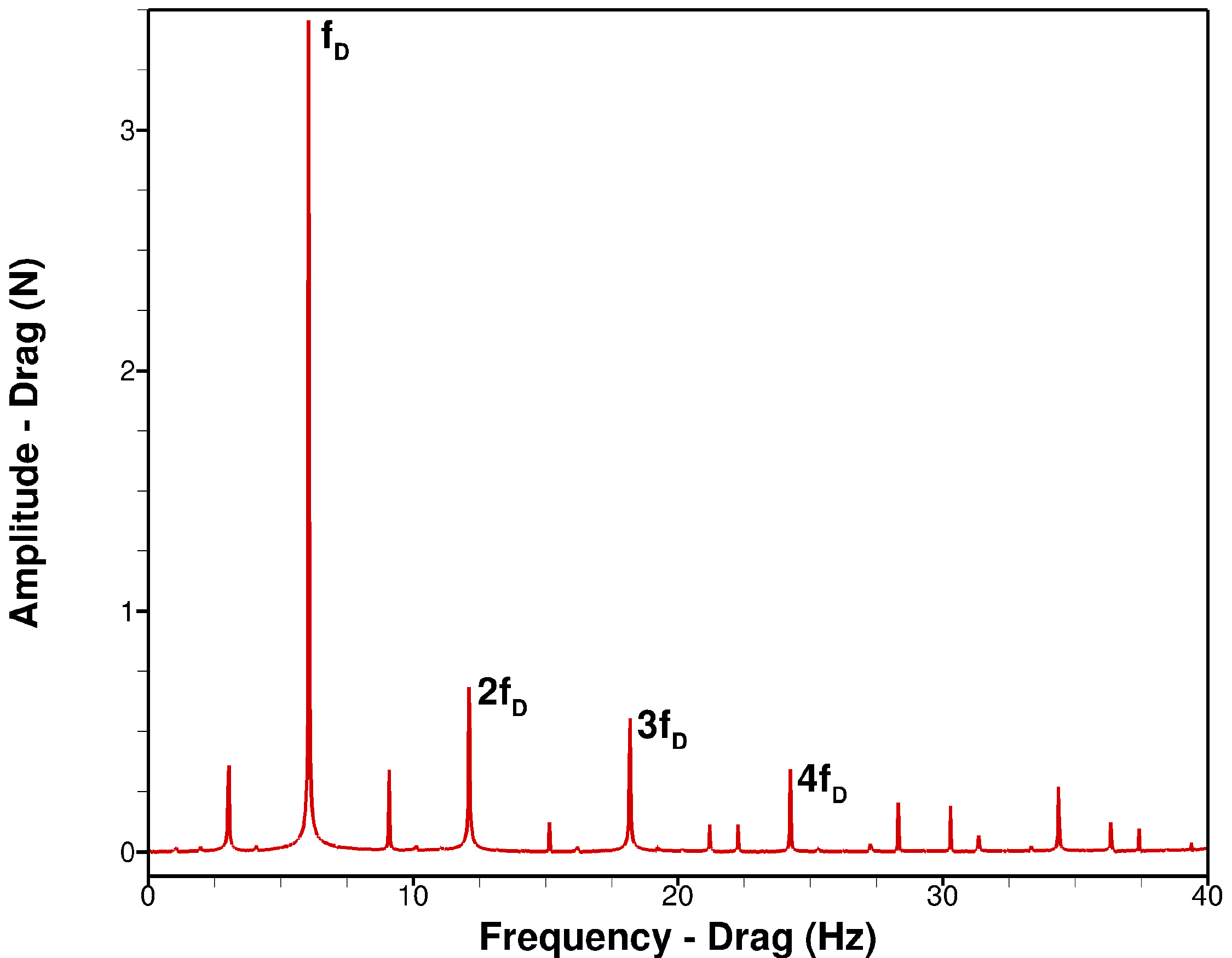} \label{fig:frequency_zoom_drag}}\hfill
\end{figure}
\par In the case of high frequencies, the drag frequency domain is characterized by a frequency of $f_D^h=58.61\,\text{Hz}$, which super-harmonics at $2f_D^h$, $3f_D^h$, $4f_D^h$, $5f_D^h$ and $6f_D^h$ are present. The high frequencies generated by the lift are characterized mainly by the frequency $f_L^h=55.6\,\text{Hz}$. Although no exact super-harmonic of this frequency is present, many peaks at frequencies near the super-harmonic ones are present, i$.$e$.$, at $f=114.21\,\text{Hz}$, $f=172.81\,\text{Hz}$, $f=231.42\,\text{Hz}$, $f=290.03\,\text{Hz}$, $f=348.64\,\text{Hz}$ and $f=401.19\,\text{Hz}$. For the displacements, only the pitch motion presents high frequencies. This domain is characterized mainly by a high frequency of $f_{\varphi_3}^h=55.54\,\text{Hz}$, which is similar to the high frequency generated by the lift force. 
\par The high frequencies present in the investigated domains are not related to the vortex shedding. These are possibly related to some instability generated by the geometry of the structure or by the structure itself, since its natural frequencies are high, i$.$e$.$, respectively $w_{n,\,x_{2}}=18.25\,\text{Hz}$ and $w_{n,\,\varphi_{3}}=44.08\,\text{Hz}$ for the up and down and pitch degrees of freedom. The exact nature of these frequencies is, however, not thoroughly investigated in the current work. 
\par The low frequency domain is characterized by frequencies related uniquely to the flow, which are illustrated in Table \ref{table:flow_frequencies}. The frequencies generated by the lift, drag and rotational displacement present the same pattern: a strong peak at $f_L$, $f_D$  and $f_{\varphi_{3}}$ is available and these are followed by the super-harmonics, which points out the strength of the flow nonlinearities. The main frequencies of the lift, translation and rotational displacements are equal. Moreover, the drag frequency is approximately twice this value, indicating the presence of a von K\'arm\'an vortex street \mbox{(see Manhart \cite{Manhart_1998})}. This instability is characterized by a periodic detachment of pairs of alternating vortices and is thoroughly explored in \mbox{Section \ref{sec:instationary_flow_field}}. Due to the presence of this instability, the vortex shedding frequency is equal to the lift frequency. 
\begin{table}[H]
	\centering
	\begin{tabular}{p{3.5cm} p{3.4cm} p{3.6cm} p{3.5cm}}
		\hline
		\centering{\textbf{Drag frequency $f_D$ (Hz)}} & \centering{\textbf{Lift frequency $f_L$ (Hz)}} & \centering{\textbf{Plunge frequency $f_{X_{2}}$ (Hz)}} & \centering{\textbf{Pitch frequency $f_{\varphi_{3}}$ (Hz)}} \tabularnewline \hline
		\centering{6.05} & \centering{3.03} & \centering{3.03} & \centering{3.03} \tabularnewline
		\hline
	\end{tabular}
	\caption{\label{table:flow_frequencies}Flow frequencies. Test case with $Re=30{,}000$, \mbox{$k_{t\,3,\,eq}=0.3832\,\text{N}{\cdot}\text{m}{\cdot}\text{rad}^{-1}$} and \mbox{$k_{l\,2,\,eq}=92\,\text{N}{\cdot}\text{m}^{-1}$}.}
\end{table}
\par The vortex shedding frequency and the calculated Strouhal number in relation to the chord length are summarized in Table \ref{table:vortex_shedding_strouhal}. The achieved Strouhal number is in agreement with the experimental studies of Poirel et al.\ \cite{Poirel_2008}.
\begin{table}[H]
	\centering
	\begin{tabular}{p{7.7cm} p{4.6cm}}
		\hline
		\centering{\textbf{Vortex shedding frequency $f_v$ (Hz)}} & \centering{\textbf{Strouhal number $St$}}  \tabularnewline \hline
		\centering{3.03} & \centering{0.068} \tabularnewline
		\hline
	\end{tabular}
	\caption{\label{table:vortex_shedding_strouhal}Vortex shedding frequency and dimensionless Strouhal number. Case: $Re=30{,}000$, \mbox{$k_{t\,3,\,eq}=0.3832\,\text{N}{\cdot}\text{m}{\cdot}\text{rad}^{-1}$} and \mbox{$k_{l\,2,\,eq}=92\,\text{N}{\cdot}\text{m}^{-1}$}.}
\end{table}

\subsection{Instantaneous flow field}
\label{sec:instationary_flow_field}
\par The instantaneous flow field is studied according to the velocity, pressure, pressure fluctuation, as well as the normalized span-wise vorticity and vorticity magnitude. Since the achieved LCO is characterized by extremely small displacements and the frequency of the structural displacement is equal to the vortex shedding frequency (see Section \ref{sec:frequencies}), this investigation aims specially at the study of the vortex shedding.
\par The temporal progress of the fully developed instantaneous flow field for a complete cycle of the structural displacement is thoroughly analyzed by figures at $t$, \mbox{$t+0.25\,T_v$}, $t+0.5\,T_v$ and $t+0.75\,T_v$. The former illustrates the beginning of the oscillation, considering a sinusoidal wave. The second depicts the crest of the wave and is followed by an image at the oscillation mean value. Finally, the wave trough is illustrated. $t$ represents an arbitrary time and $T_v$ is the vortex shedding period, i$.$e$.$, $T_v\approx0.33\,\text{s}$.
\par
\begin{figure}[H]
	\centering
	\subfigure[$t$]{\includegraphics[width=0.490\textwidth]{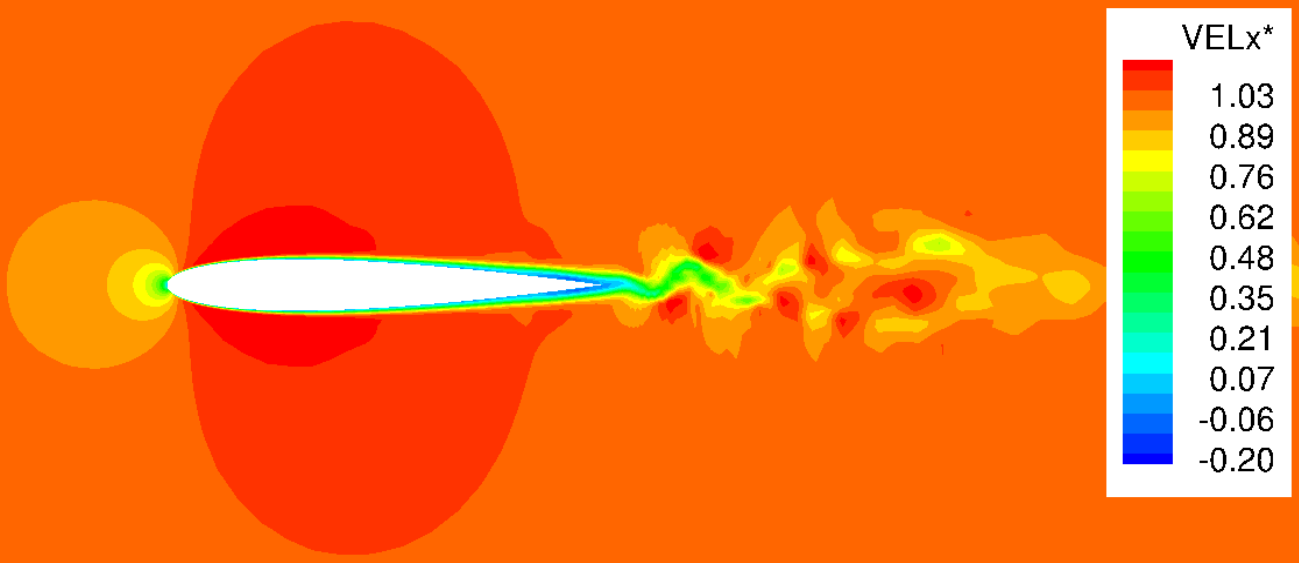} \label{fig:velocity_1}}\hfill
	\subfigure[$t+0.25\,T_v$]{\includegraphics[width=0.490\textwidth]{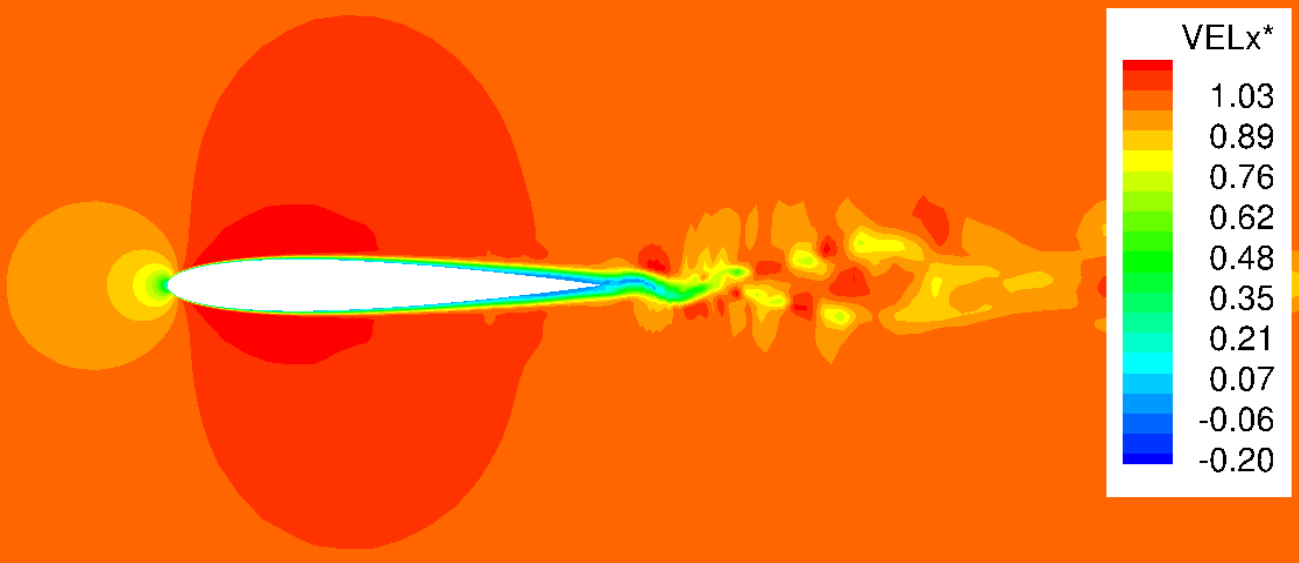}\label{fig:velocity_2}}\hfill
	\subfigure[$t+0.5\,T_v$]{\includegraphics[width=0.490\textwidth]{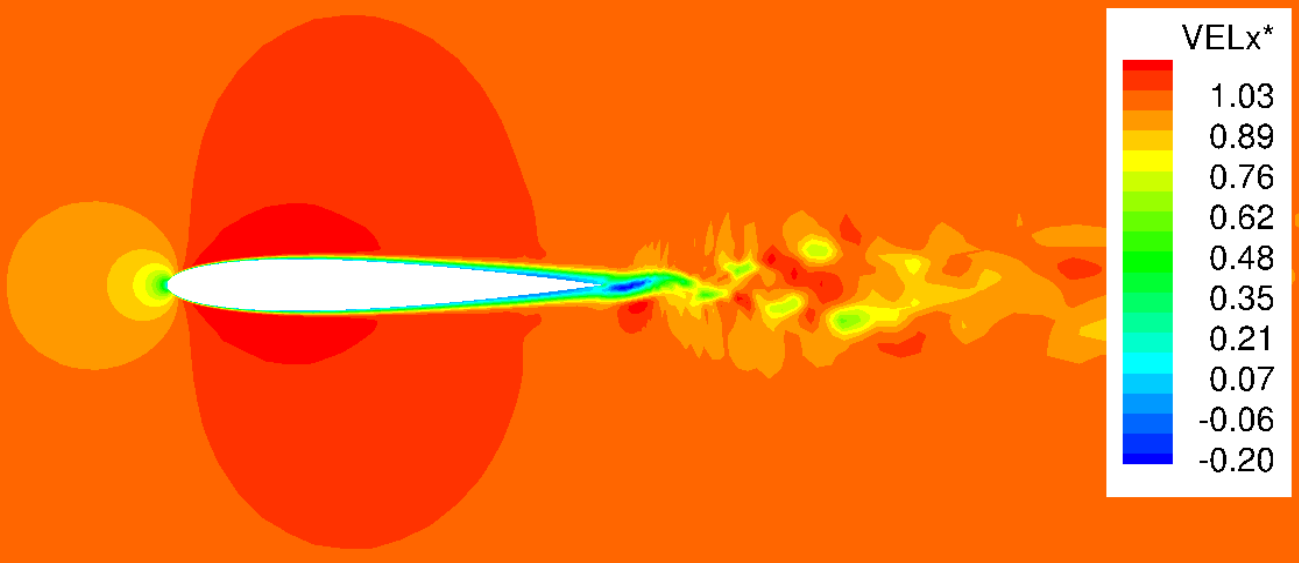}\label{fig:velocity_3}}\hfill
	\subfigure[$t+0.75\,T_v$]{\includegraphics[width=0.490\textwidth]{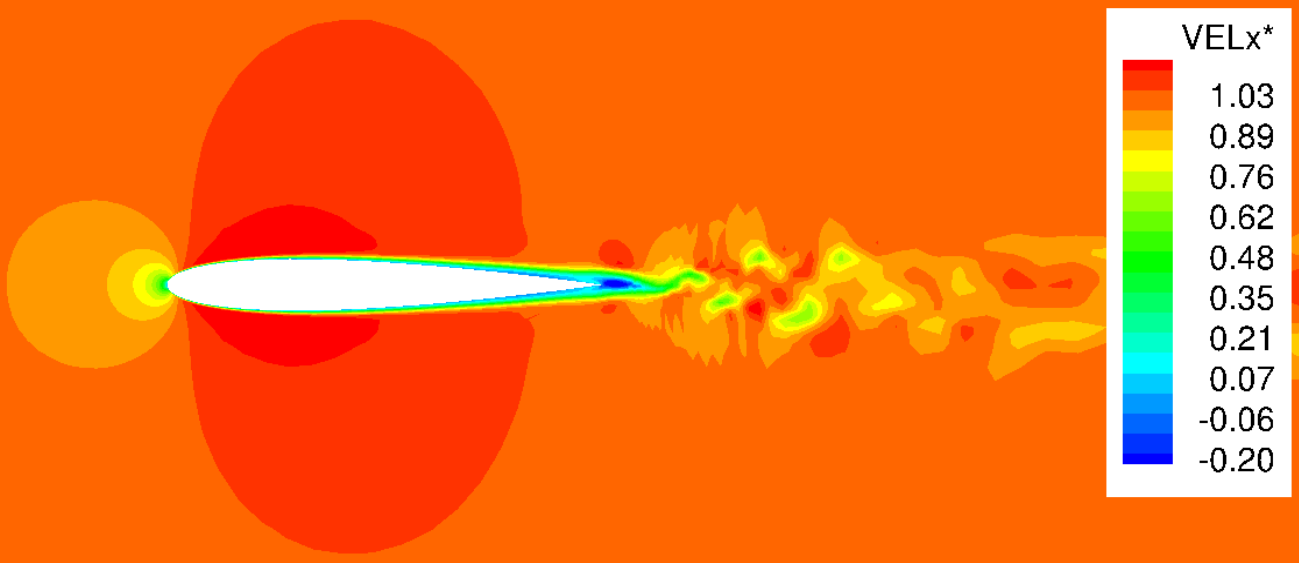}\label{fig:velocity_4}}\hfill
	\caption{Temporal development of the dimensionless stream-wise velocity field. Test case: \mbox{$Re=30{,}000$,} \mbox{$k_{t\,3,\,eq}=0.3832\,\text{N}{\cdot}\text{m}{\cdot}\text{rad}^{-1}$} and \mbox{$k_{l\,2,\,eq}=92\,\text{N}{\cdot}\text{m}^{-1}$}.}
	\label{fig:velocity_development}
\end{figure}

 The dimensionless instantaneous velocity in the stream-wise direction is depicted in \mbox{Fig.\ \ref{fig:velocity_development}}. A boundary layer detachment is present at $x^*=0.8$ (see Section \ref{sec:time_avg_flow_field}). This is responsible for the shedding of vortices that travel downstream and interact with the structure leading to extremely small structural oscillations, which cannot be distinguished in the figures.   
\par
\begin{figure}[H]
	\centering
	\subfigure[$t$]{\includegraphics[width=0.490\textwidth]{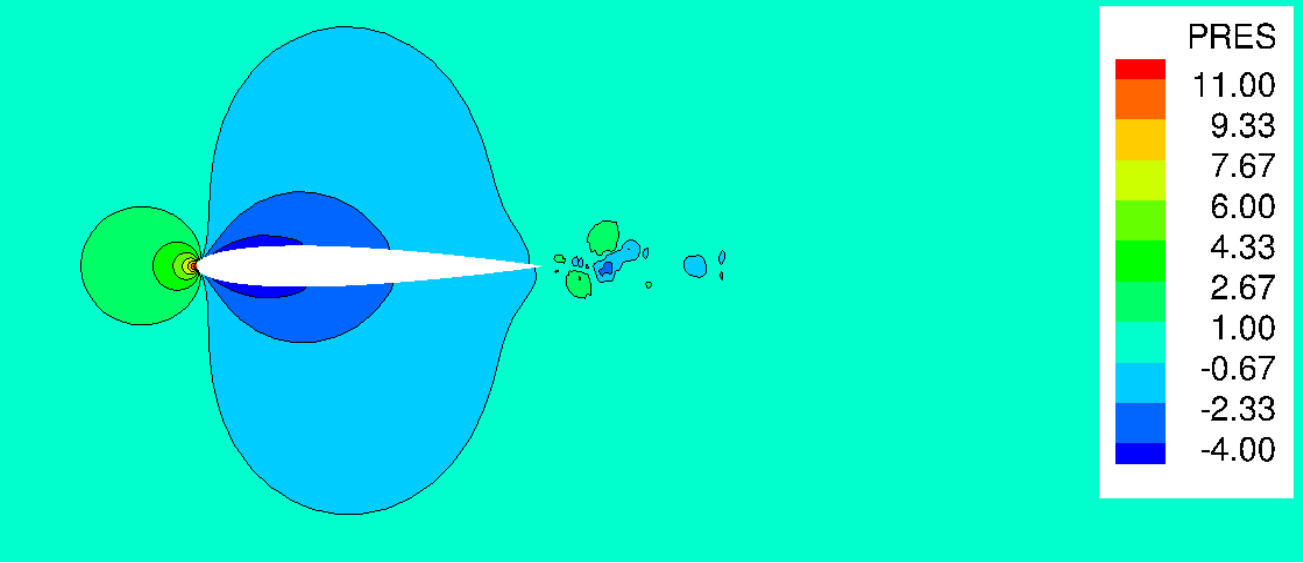} \label{fig:pressure_1}}\hfill
	\subfigure[$t+0.25\,T_v$]{\includegraphics[width=0.490\textwidth]{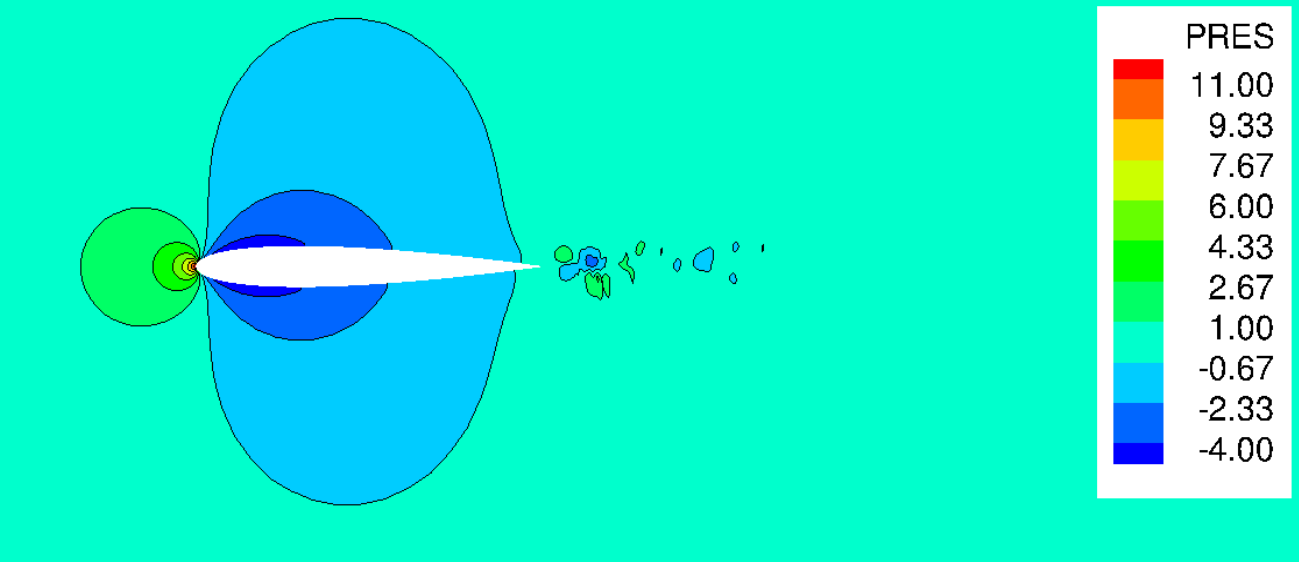}\label{fig:pressure_2}}\hfill
	\subfigure[$t+0.5\,T_v$]{\includegraphics[width=0.490\textwidth]{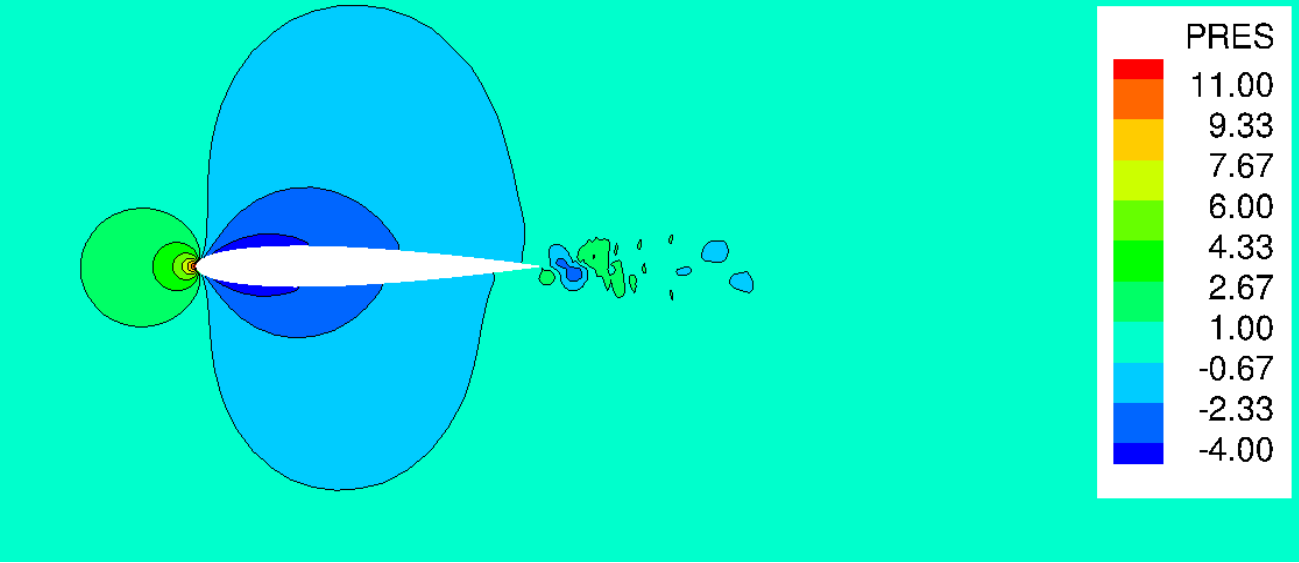}\label{fig:pressure_3}}\hfill
	\subfigure[$t+0.75\,T_v$]{\includegraphics[width=0.490\textwidth]{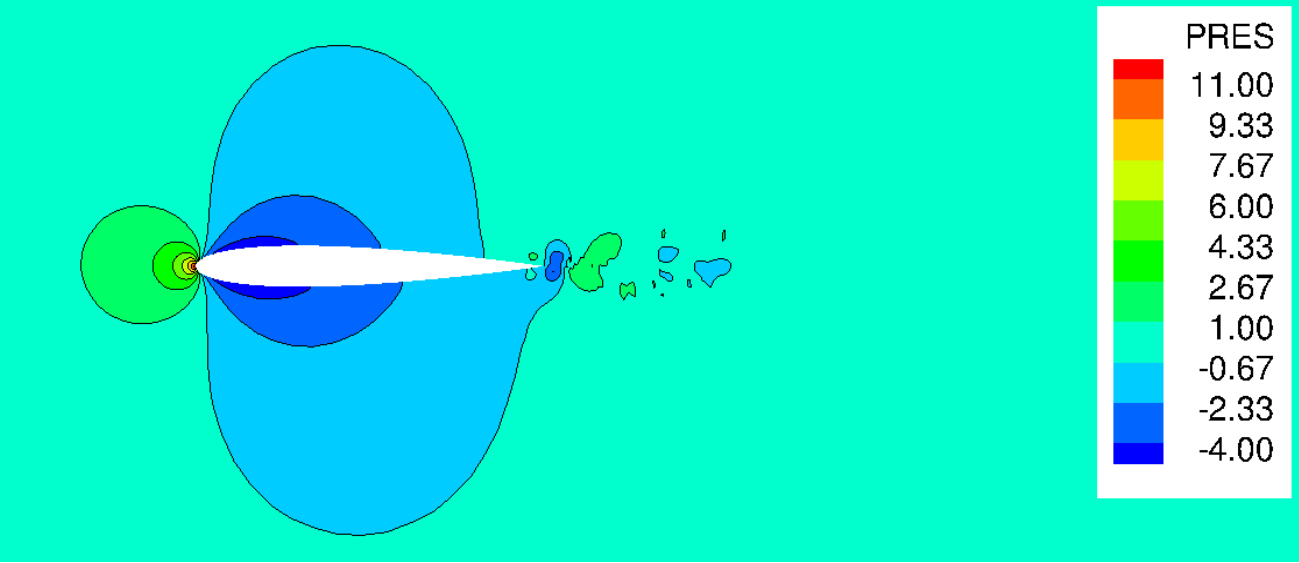}\label{fig:pressure_4}}\hfill
	\caption{Temporal development of the dynamic pressure field: Test case characterized by $Re=30{,}000$, \mbox{$k_{t\,3,\,eq}=0.3832\,\text{N}{\cdot}\text{m}{\cdot}\text{rad}^{-1}$} and \mbox{$k_{l\,2,\,eq}=92\,\text{N}{\cdot}\text{m}^{-1}$}.}
	\label{fig:pressure_development}
\end{figure}

 The unsteady pressure field is depicted in Fig.\ \ref{fig:pressure_development}. This is approximately symmetric in the first half of the profile due to the symmetry of the airfoil and the applied Reynolds number, which is characterized by a laminar flow until the boundary layer is detached (see Section \ref{sec:time_avg_flow_field}). After the detachment point, the flow is turbulent and characterized by the formation of vortices. Moreover, the unbalanced distribution of the pressure near the airfoil trailing-edge leads to the minimal oscillation of the airfoil.
\par Since the swirling direction of the vortices, as well as the vortex cores cannot be identified solely by the unsteady velocity and pressure fields, the pressure fluctuations $p'$, the span-wise vorticity $\omega_{f,\,3}$ and the vorticity magnitude $|\omega_f|$ are also displayed, considering that these values are calculated according to Eqs.\ (\ref{eq:pressure_fluctuation}) through (\ref{eq:vorticity_mag}). ${<}p{>}$ represents the time and spatial (in the $x_3$-direction) averaged pressure.
\begin{eqnarray}
\label{eq:pressure_fluctuation}
&&\quad\quad\quad\quad\quad\quad p'\quad=\quad p\,\,-\,{<}p{>},\\
\label{eq:vorticity_z}
&&\quad\quad\quad\quad\quad \omega_{f,\,3}\quad=\quad\frac{\partial u_2}{\partial x_1}- \frac{\partial u_1}{\partial x_2},\\ \label{eq:vorticity_mag}
|\omega_{f}|&=&\sqrt{\left(\frac{\partial u_3}{\partial x_2}- \frac{\partial u_2}{\partial x_3}\right)^2+\left(\frac{\partial u_1}{\partial x_3}- \frac{\partial u_3}{\partial x_1}\right)^2+\left(\frac{\partial u_2}{\partial x_1}- \frac{\partial u_1}{\partial x_2}\right)^2}.
\end{eqnarray} 
\par
\begin{figure}[H]
	\centering
	\subfigure[$t$]{\includegraphics[width=0.490\textwidth]{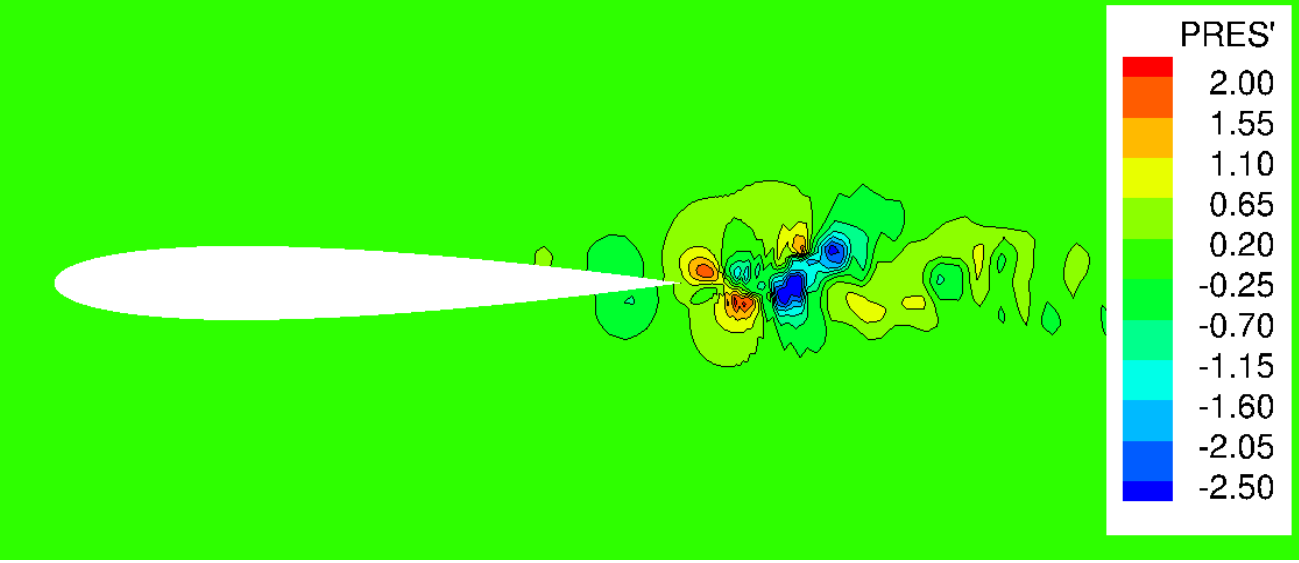} \label{fig:pressure_fluc_1}}\hfill
	\subfigure[$t+0.25\,T_v$]{\includegraphics[width=0.490\textwidth]{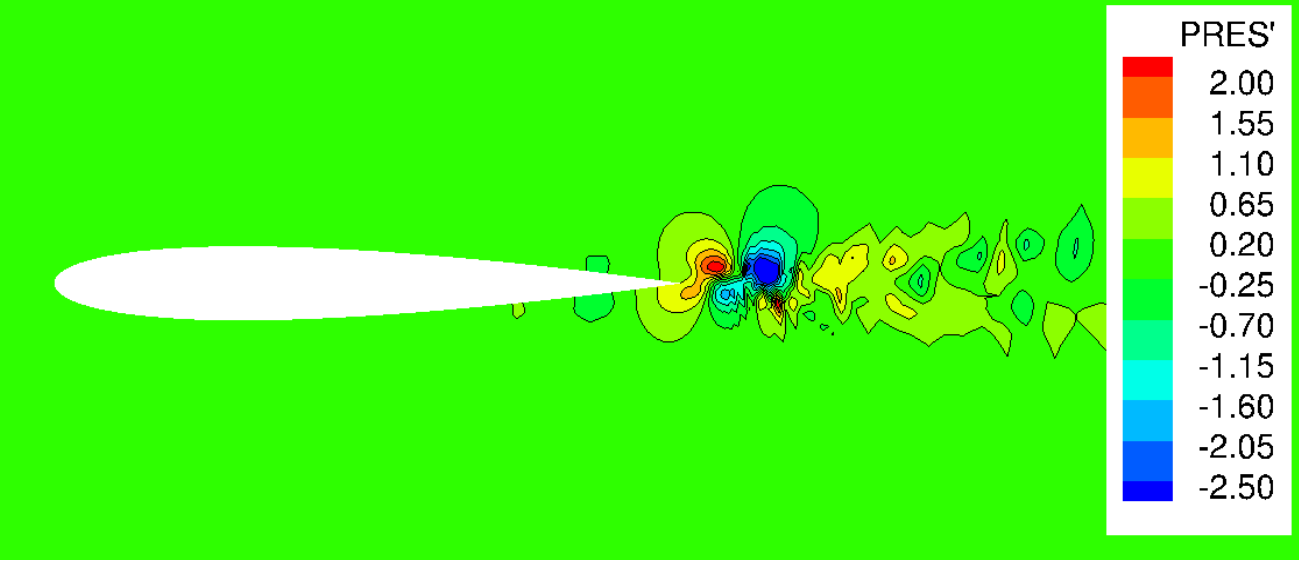}\label{fig:pressure_fluc_2}}\hfill
	\subfigure[$t+0.5\,T_v$]{\includegraphics[width=0.490\textwidth]{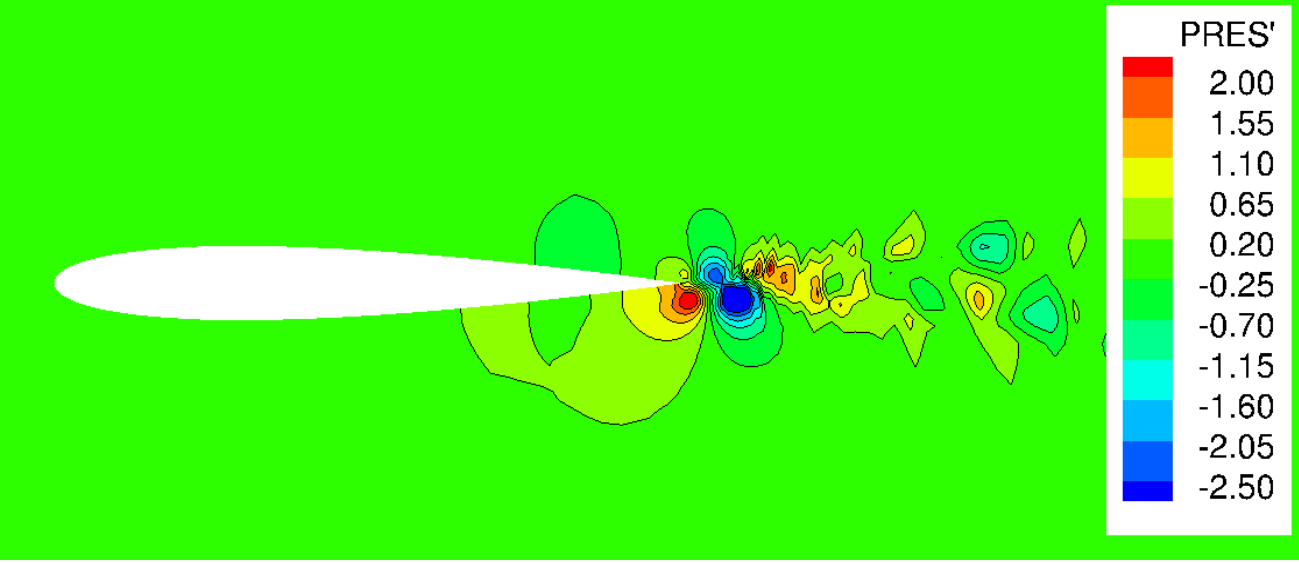}\label{fig:pressure_fluc_3}}\hfill
	\subfigure[$t+0.75\,T_v$]{\includegraphics[width=0.490\textwidth]{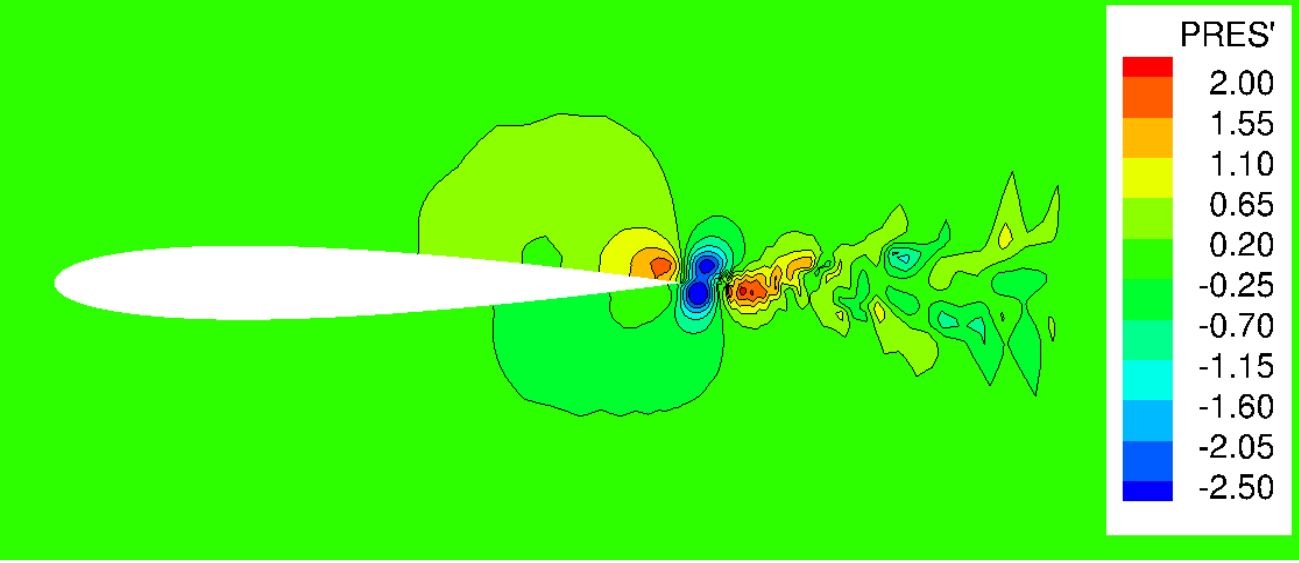}\label{fig:pressure_fluc_4}}\hfill
	\caption{Temporal development of the pressure fluctuation field: Test case characterized by $Re=30{,}000$, \mbox{$k_{t\,3,\,eq}=0.3832\,\text{N}{\cdot}\text{m}{\cdot}\text{rad}^{-1}$} and \mbox{$k_{l\,2,\,eq}=92\,\text{N}{\cdot}\text{m}^{-1}$}.}
	\label{fig:pressure_fluctuation_delevopment}
\end{figure}

 The position of vortex cores, as well as stagnation points can be investigated according to the pressure fluctuations $p'$, considering that regions with negative pressure fluctuations correspond to vortex cores, whereas regions of positive pressure fluctuations indicate the position of stagnation points in the turbulent flow field \cite{Manhart_1998}. Figure \ref{fig:pressure_fluctuation_delevopment} shows the pressure fluctuation field, which is characterized by the presence of vortex cores in the wake.
\par
\begin{figure}[H]
	\centering
	\subfigure[$t$]{\includegraphics[width=0.490\textwidth]{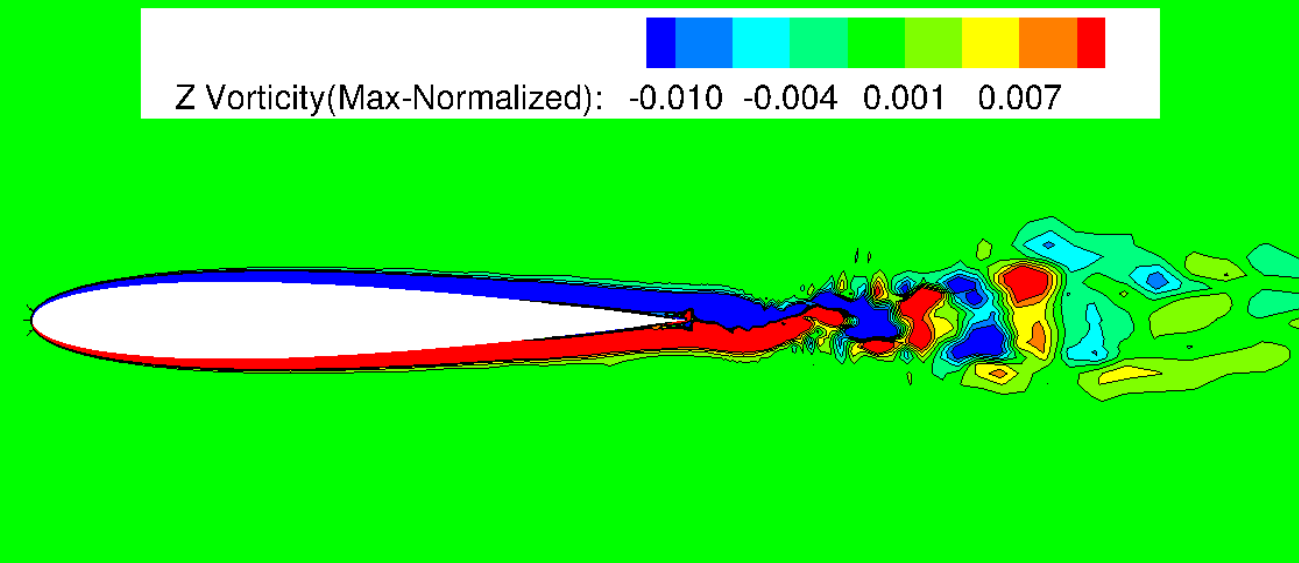} \label{fig:vorticity_z_1}}\hfill
	\subfigure[$t+0.25\,T_v$]{\includegraphics[width=0.490\textwidth]{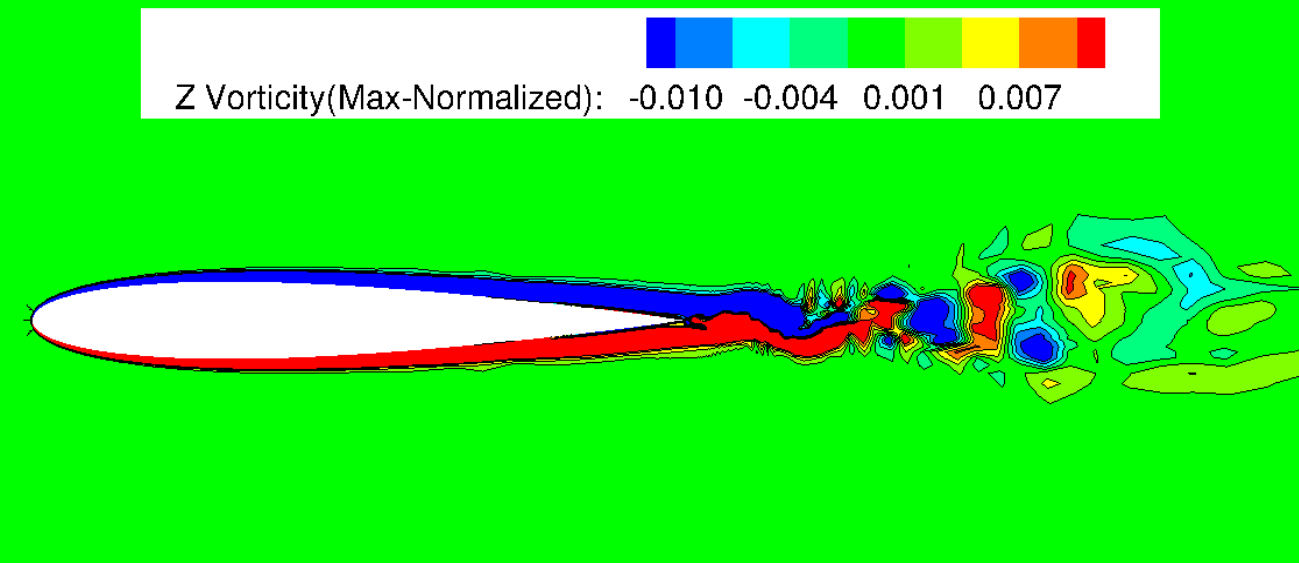}\label{fig:vorticity_z_2}}\hfill
	\subfigure[$t+0.5\,T_v$]{\includegraphics[width=0.490\textwidth]{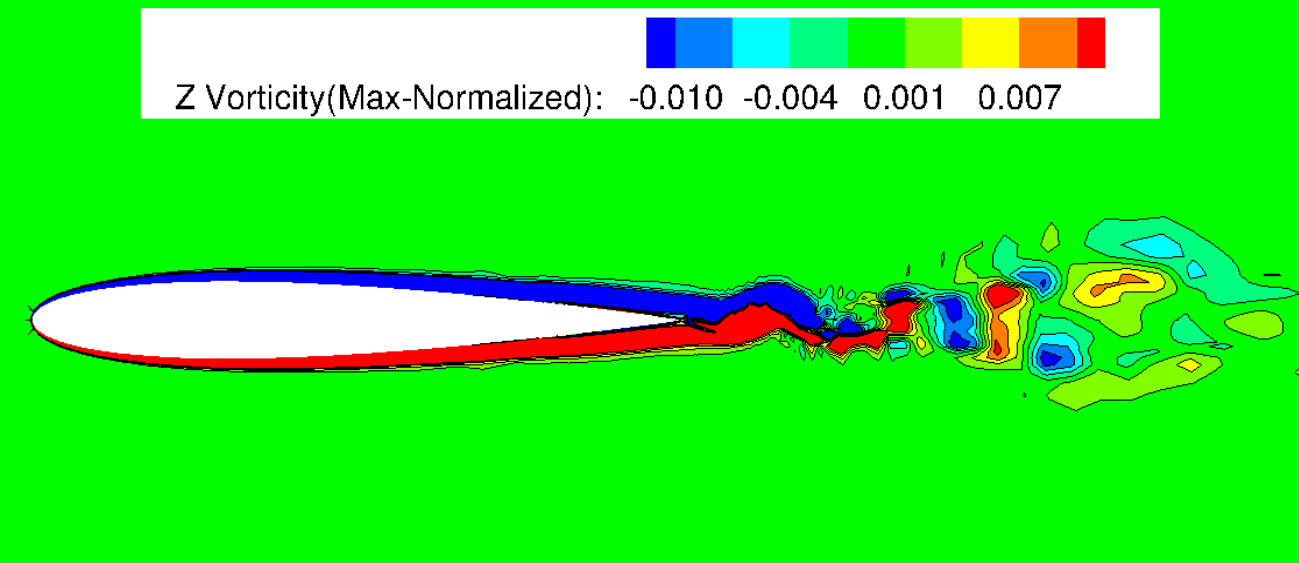}\label{fig:vorticity_z_3}}\hfill
	\subfigure[$t+0.75\,T_v$]{\includegraphics[width=0.490\textwidth]{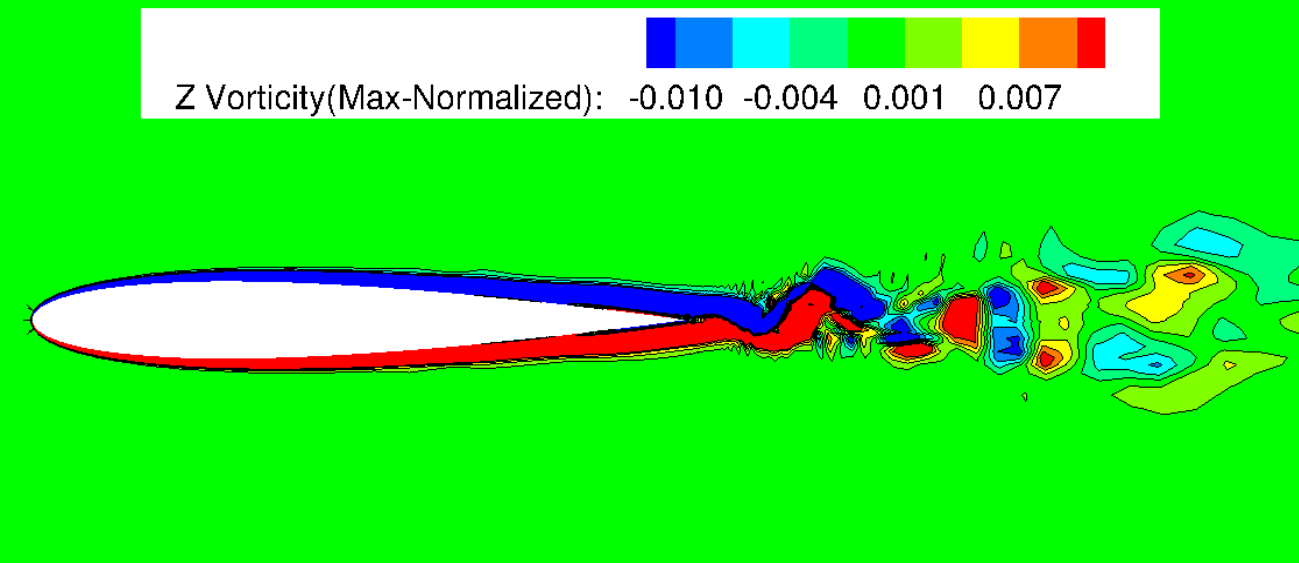}\label{fig:vorticity_z_4}}\hfill
	\caption{Temporal development of the normalized span-wise vorticity field. Test case: \mbox{$Re=30{,}000$,} \mbox{$k_{t\,3,\,eq}=0.3832\,\text{N}{\cdot}\text{m}{\cdot}\text{rad}^{-1}$} and \mbox{$k_{l\,2,\,eq}=92\,\text{N}{\cdot}\text{m}^{-1}$}.}
	\label{fig:vorticity_z_development}
\end{figure}

 In order to precisely analyze the art of vortex shedding present near the airfoil trailing-edge, the vorticity in the span-wise direction is studied, as illustrated in Fig.\ \ref{fig:vorticity_z_development}. A negative vorticity indicates that the vortex rolls in the clockwise direction, while a positive vorticity evinces a counter-clockwise rotation \cite{Streher_2017}. Therefore, a von K\'arm\'an vortex street, i$.$e$.$, a repeating pattern of alternating swirling vortices, can be identified. The vortices formed on the suction side rotate in the clockwise direction, while the vortices shed on the pressure side rotate in the counter-clockwise direction.
\par
\begin{figure}[H]
	\centering
	\subfigure[$t$]{\includegraphics[width=0.49\textwidth]{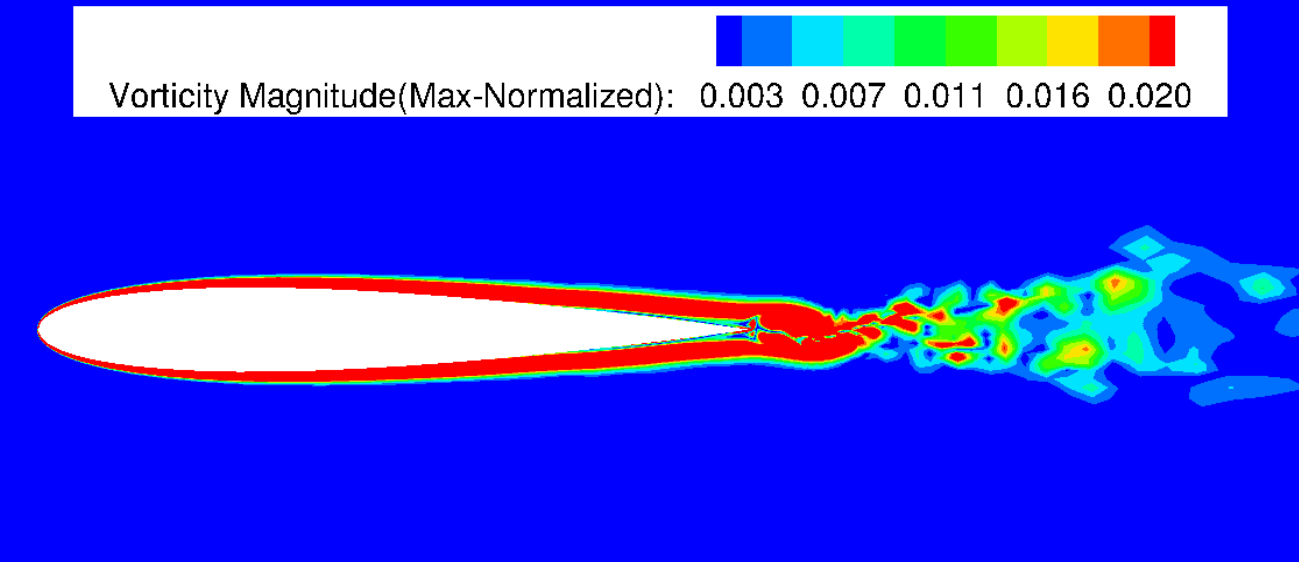} \label{fig:vorticity_mag_1}}\hfill
	\subfigure[$t+0.25\,T_v$]{\includegraphics[width=0.49\textwidth]{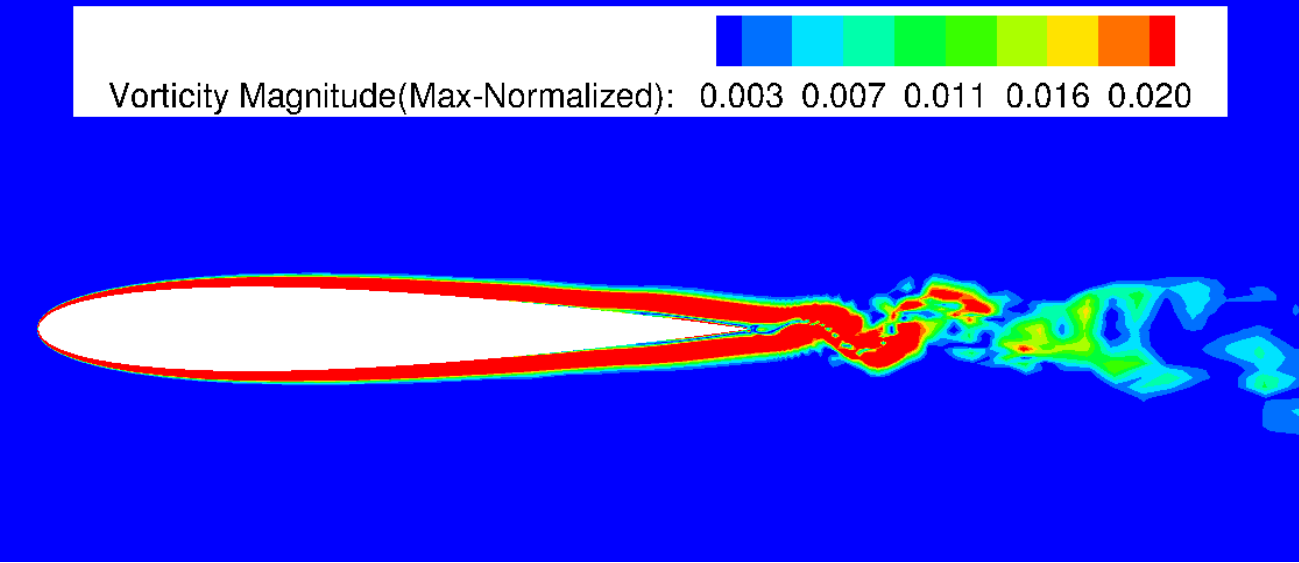}\label{fig:vorticity_mag_2}}\hfill
	\subfigure[$t+0.5\,T_v$]{\includegraphics[width=0.49\textwidth]{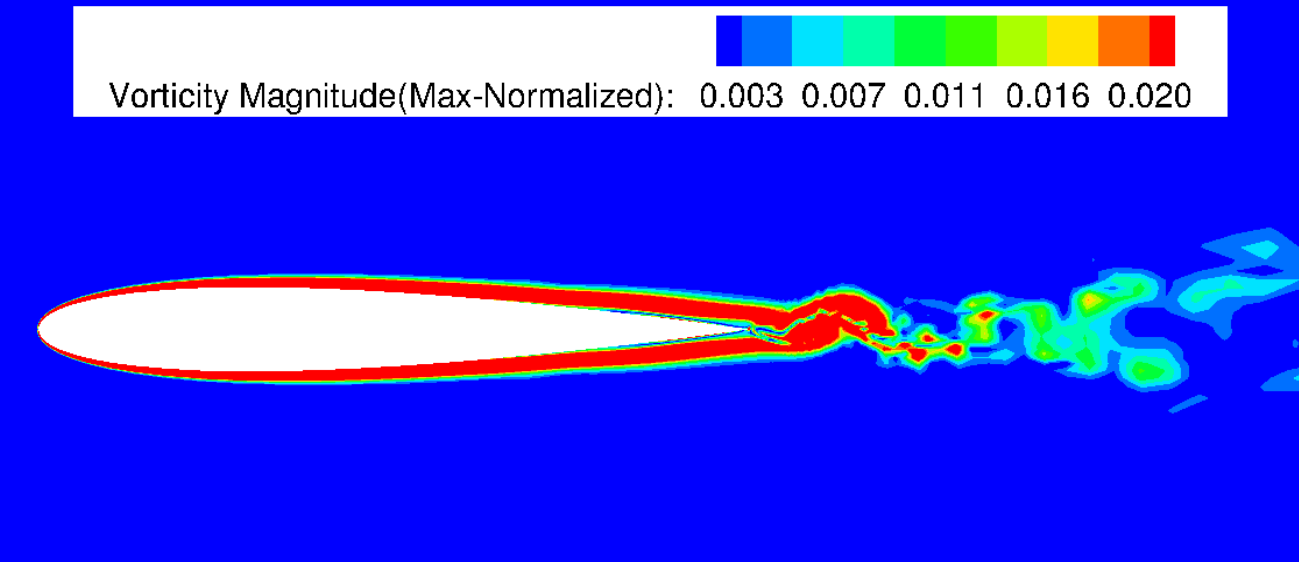}\label{fig:vorticity_mag_3}}\hfill
	\subfigure[$t+0.75\,T_v$]{\includegraphics[width=0.49\textwidth]{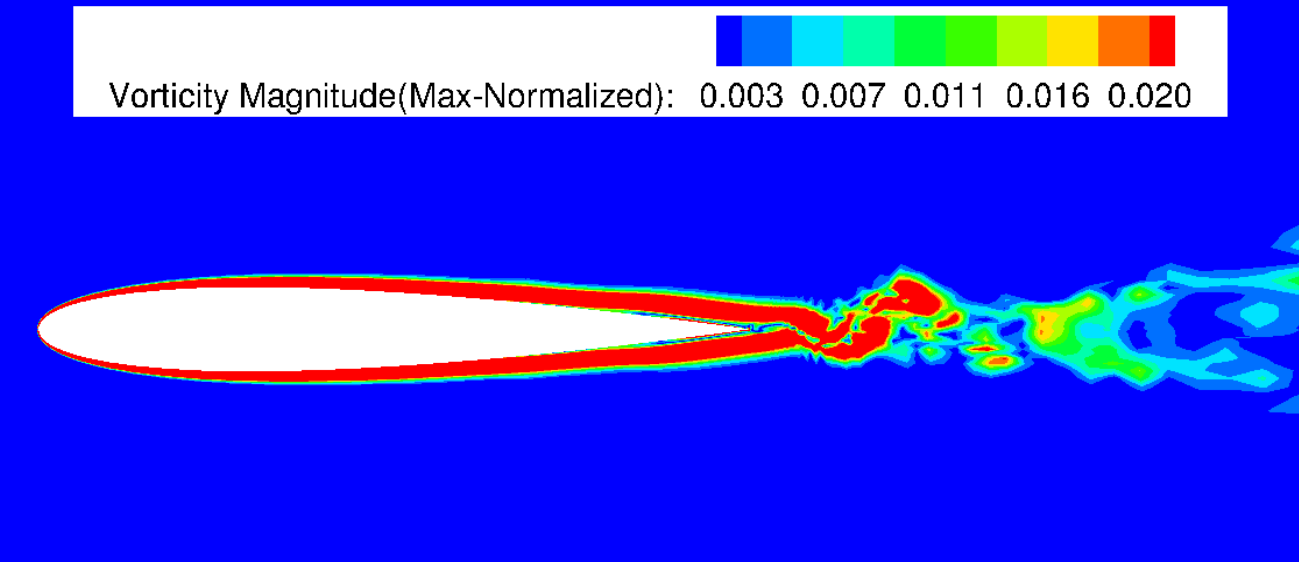}\label{fig:vorticity_mag_4}}\hfill
	\caption{Temporal development of the normalized vorticity magnitude field. Test case: \mbox{$Re=30{,}000$,} \mbox{$k_{t\,3,\,eq}=0.3832\,\text{N}{\cdot}\text{m}{\cdot}\text{rad}^{-1}$} and \mbox{$k_{l\,2,\,eq}=92\,\text{N}{\cdot}\text{m}^{-1}$}.}
	\label{fig:vorticity_mag_development}
\end{figure}

 Finally, the vorticity magnitude normalized in relation to the maximal value is depicted in Fig.\  \ref{fig:vorticity_mag_development}. This clearly illustrates the formation of the vortices. Moreover, the boundary layer detachment near the airfoil can also be identified, since this is characterized by small magnitudes of the vorticity.

\subsection{Time-averaged flow field}
\label{sec:time_avg_flow_field}

\par Although the current work describes the fluid-structure interaction of the NACA0012 profile, the amplitudes of the limit-cycle oscillations are so small that the time and spatial (in the $x_3$-direction) averaged flow field can be investigated. The unsteady flow field is averaged for a time period equivalent of about 30 vortex shedding cycles. This study aims at the investigation of complex viscous phenomena. 
\begin{figure}[H]
	\centering
	\subfigure[Overall view.]{\includegraphics[width=0.49\textwidth]{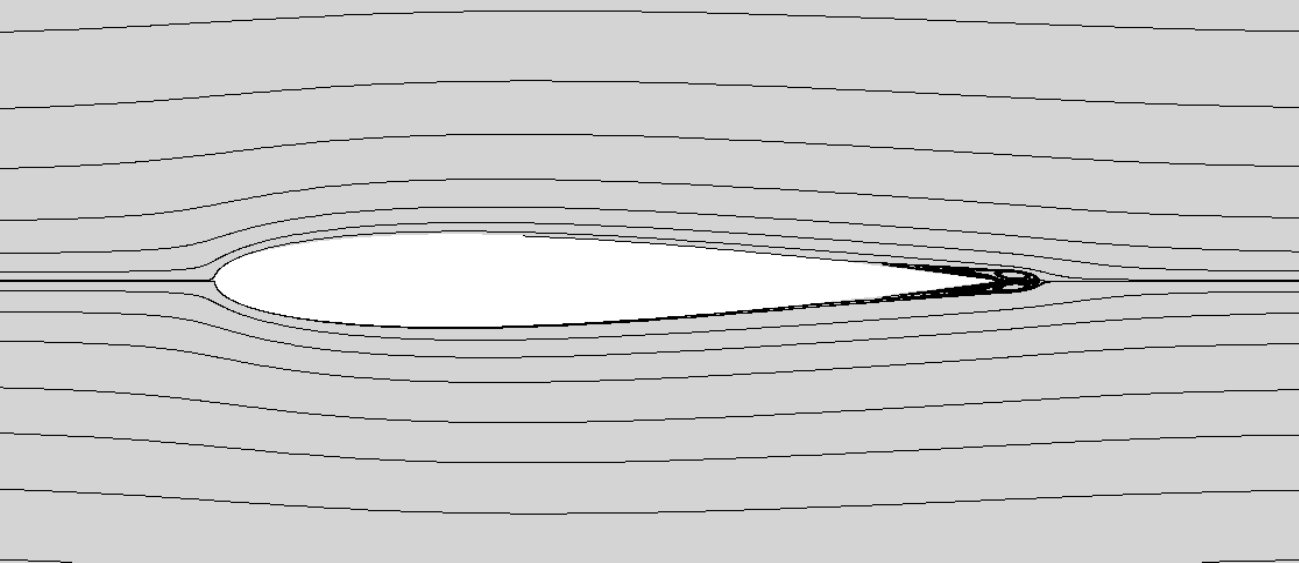} \label{fig:streamlines_overall}}\hfill
	\subfigure[Focus on the trailing-edge.]{\includegraphics[width=0.49\textwidth]{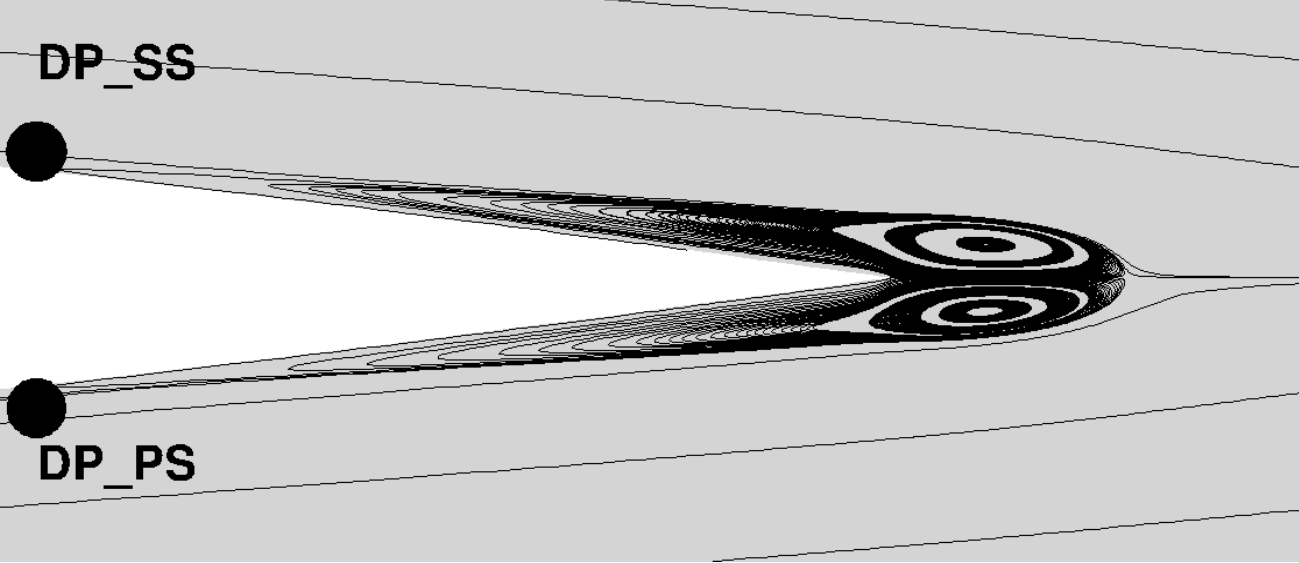}\label{fig:streamlines_zoom}}\hfill
	\caption{Time-averaged streamlines (results are averaged spatial-averaged in the span-wise direction). Case: $Re=30{,}000$, \mbox{$k_{t\,3,\,eq}=0.3832\,\text{N}{\cdot}\text{m}{\cdot}\text{rad}^{-1}$} and \mbox{$k_{l\,2,\,eq}=92\,\text{N}{\cdot}\text{m}^{-1}$}.}
	\label{fig:streamlines_avg}
\end{figure}
\par According to Fig.\ \ref{fig:streamlines_avg}, a detachment of the boundary layer is present near the trailing-edge on the suction and pressure sides of the airfoil, i$.$e$.$, at $x_{DP}^*=0.8$. This is not reattached further downstream and is responsible for the formation of a von K\'arm\'an vortex street (see Section \ref{sec:instationary_flow_field}).
\par
\begin{figure}[H]
	\centering
	\subfigure[$\tau_{11}^{turb^{*}}$]{\includegraphics[width=0.49\textwidth]{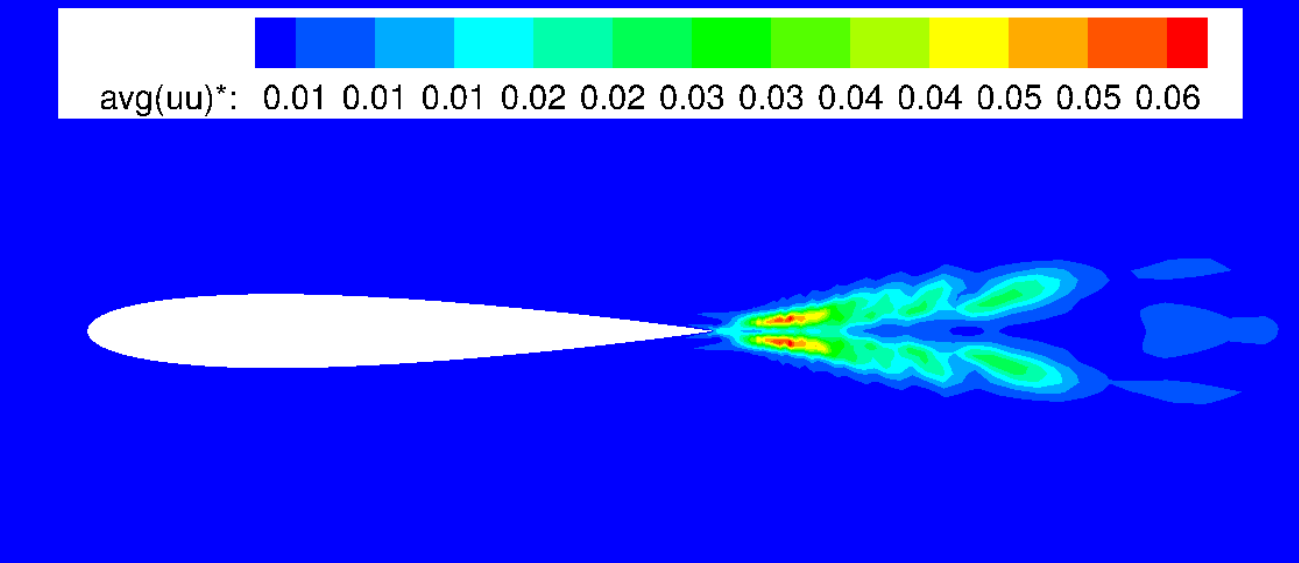} \label{fig:uu}}\hfill
	\subfigure[$\tau_{12}^{turb^{*}}$]{\includegraphics[width=0.49\textwidth]{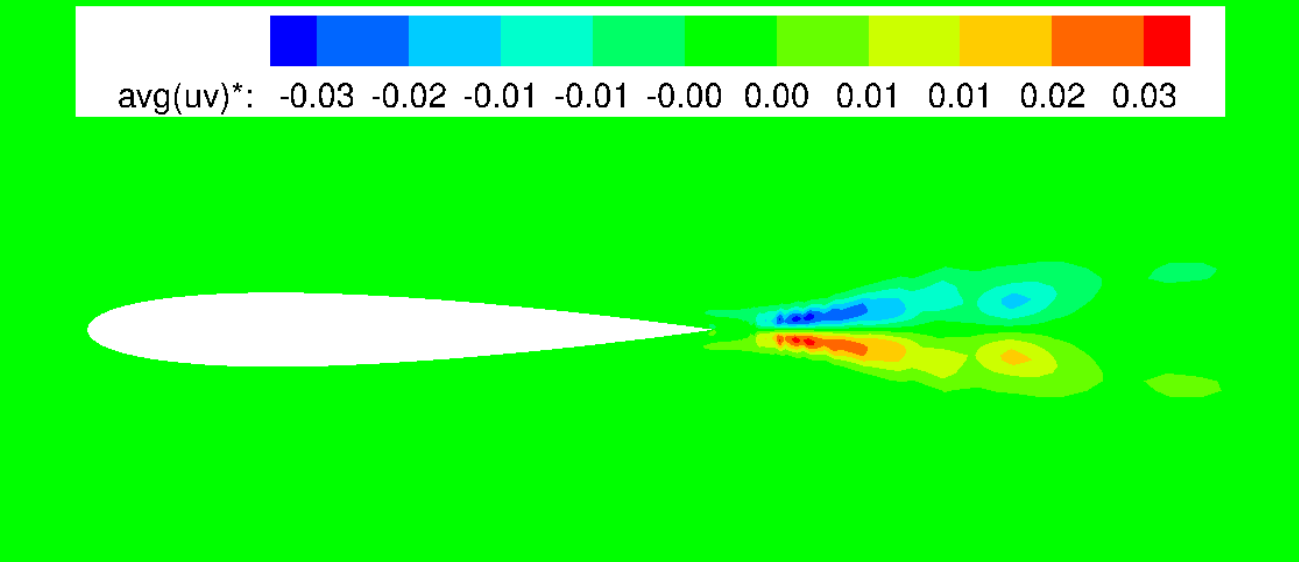}\label{fig:uv}}\hfill
	\subfigure[$\tau_{22}^{turb^{*}}$]{\includegraphics[width=0.49\textwidth]{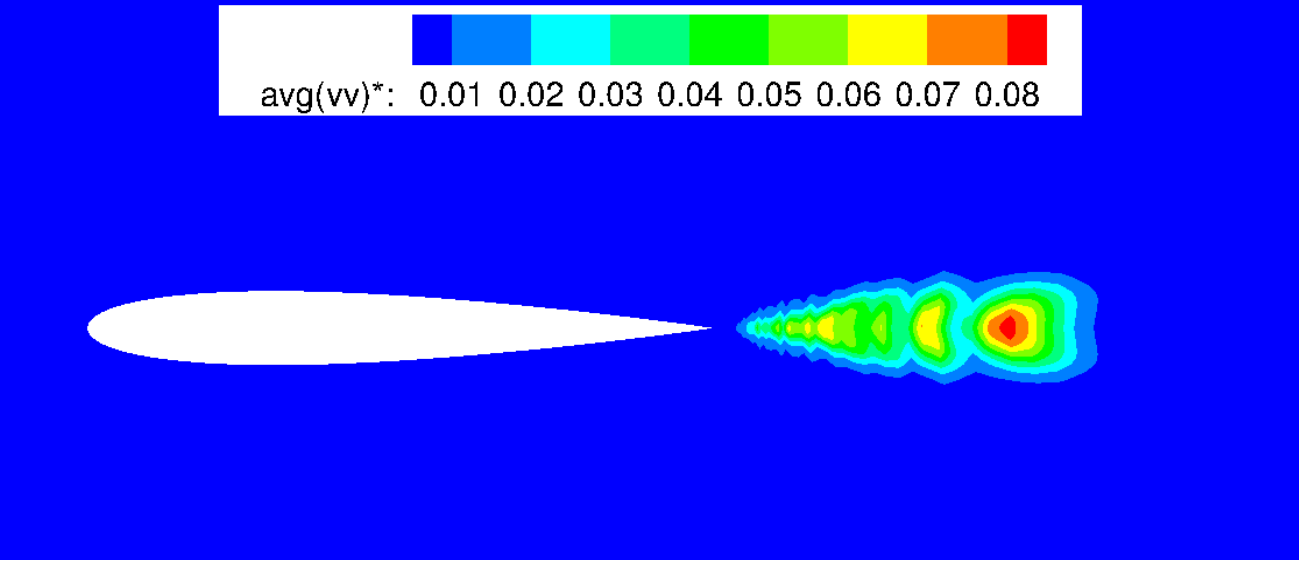}\label{fig:vv}}\hfill
	\caption{Time-averaged Reynolds stresses (results are averaged spatial-averaged in the span-wise direction). Test case: $Re=30{,}000$,  \mbox{$k_{t\,3,\,eq}=0.3832\,\text{N}{\cdot}\text{m}{\cdot}\text{rad}^{-1}$} and \mbox{$k_{l\,2,\,eq}=92\,\text{N}{\cdot}\text{m}^{-1}$}.}
	\label{fig:reynolds_stresses_avg}
\end{figure}

 The distribution of the Reynolds stresses around the NACA0012 airfoil confirms\break that the flow at $Re=30{,}000$ is in a sub-critical regime, i$.$e$.$, it is laminar. The transition to a turbulent regime occurs after the detachment point, which leads to a turbulent wake. This is illustrated in Fig.\ \ref{fig:reynolds_stresses_avg} for the $\tau_{11}^{turb^{*}}$ $\tau_{12}^{turb^{*}}$ and $\tau_{22}^{turb^{*}}$ Reynolds stresses components. 
\par Poirel et al.\ \cite{Poirel_2011} observed that a turbulent regime inhibit the existence of limit-cycle oscillations. Therefore, the existing laminar flow field influence the generation of the small LCO. Moreover, Poirel and Yuan \cite{Poirel_2010} studied self-sustained pitch oscillations of a NACA0012 airfoil at transitional Reynolds numbers ($50{,}000\leq Re\leq130{,}000$) and suggested that laminar separation leading to the formation of a laminar separation bubble plays a key role in the oscillations. Although the current work computes the fluid domain at a Reynolds number of $Re=30{,}000$ and the airfoil has up and down and pitch degrees of freedom, the presence of only a laminar separation at the airfoil trailing-edge, i$.$e$.$, the absence of a laminar separation bubble, might be related to the small LCO achieved by the fluid-structure interaction of the NACA0012 airfoil.

\section{Configuration 2: Torsional divergence}\markboth{CHAPTER 4.$\quad$RES. AND DISC.}{4.2$\quad$CONFIG. 2: TORSIONAL DIVERGENCE}
\label{sec:torsional_divergence}

\par Torsional divergence is a phenomenon caused by a deformation of the lift distribution over a lifting surface (see Fung \cite{Fung_2002}). This is evoked by the displacement or deformation of an airfoil, which generally induces an aerodynamic moment since the lift forces act on the aerodynamic center while the airfoil center of rotation is located at the center of mass. This moment tends to twist the wing if the system stiffness is not high enough to counter-balance this movement. The vortex shedding that occurs on the NACA0012 trailing-edge is responsible for a minimal displacement of the airfoil (see Section \ref{sec:LCO}). This can trigger the twist of this airfoil if the system is characterized by a low stiffness. 
\par A configuration characterized by a linear stiffness of \mbox{$k_{l\,2,\,eq}={3}\,\text{N}{\cdot}\text{m}^{-1}$} and a torsional stiffness of  \mbox{$k_{t\,3,\,eq}=0.005\,\text{N}{\cdot}\text{m}{\cdot}\text{rad}^{-1}$}, is then established aiming at the analysis of this steady-state aeroelastic instability. The simulation is performed for a Reynolds number of $Re=30{,}000$ according to a loose coupling scheme. The applied time step size is $\Delta t=1{\cdot}10^{-5}\,\text{s}$ and the displacements at the next time step, i$.$e$.$, $n+1$, are predicted. Firstly, the TFI mesh algorithm is applied in order to provide a reasonable computational time. However, this simulation diverges at $t^*=71$. Nevertheless, the cause of this divergence is thoroughly studied as well as the generated structural displacements. A simulation applying the hybrid IDW-TFI mesh adaption algorithm is also started. However, this is not investigated in the present work since not enough data are currently available due to high CPU-time requirements. 

\subsubsection{Investigation of the mesh quality}
\label{subsubsec:flutter_mesh_quality}

\par
\begin{figure}[H]
	\centering
	\subfigure[Adapted mesh - Overall view.]{\includegraphics[width=0.48\textwidth]{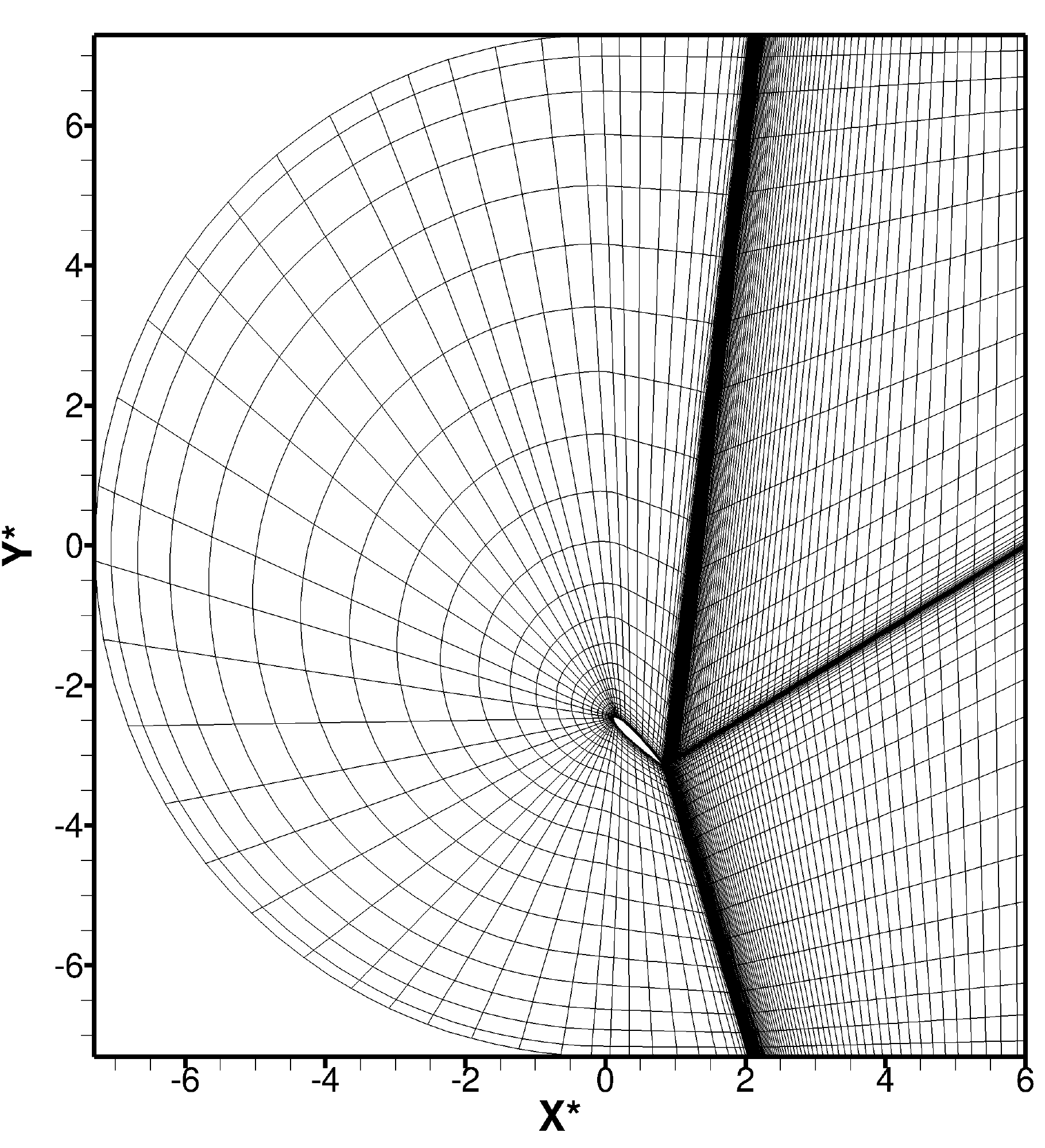} \label{fig:mesh_overall_flutter_tfi}}\hfill
	\subfigure[Adapted mesh: Focus on the trailing-edge.]{\includegraphics[width=0.48\textwidth]{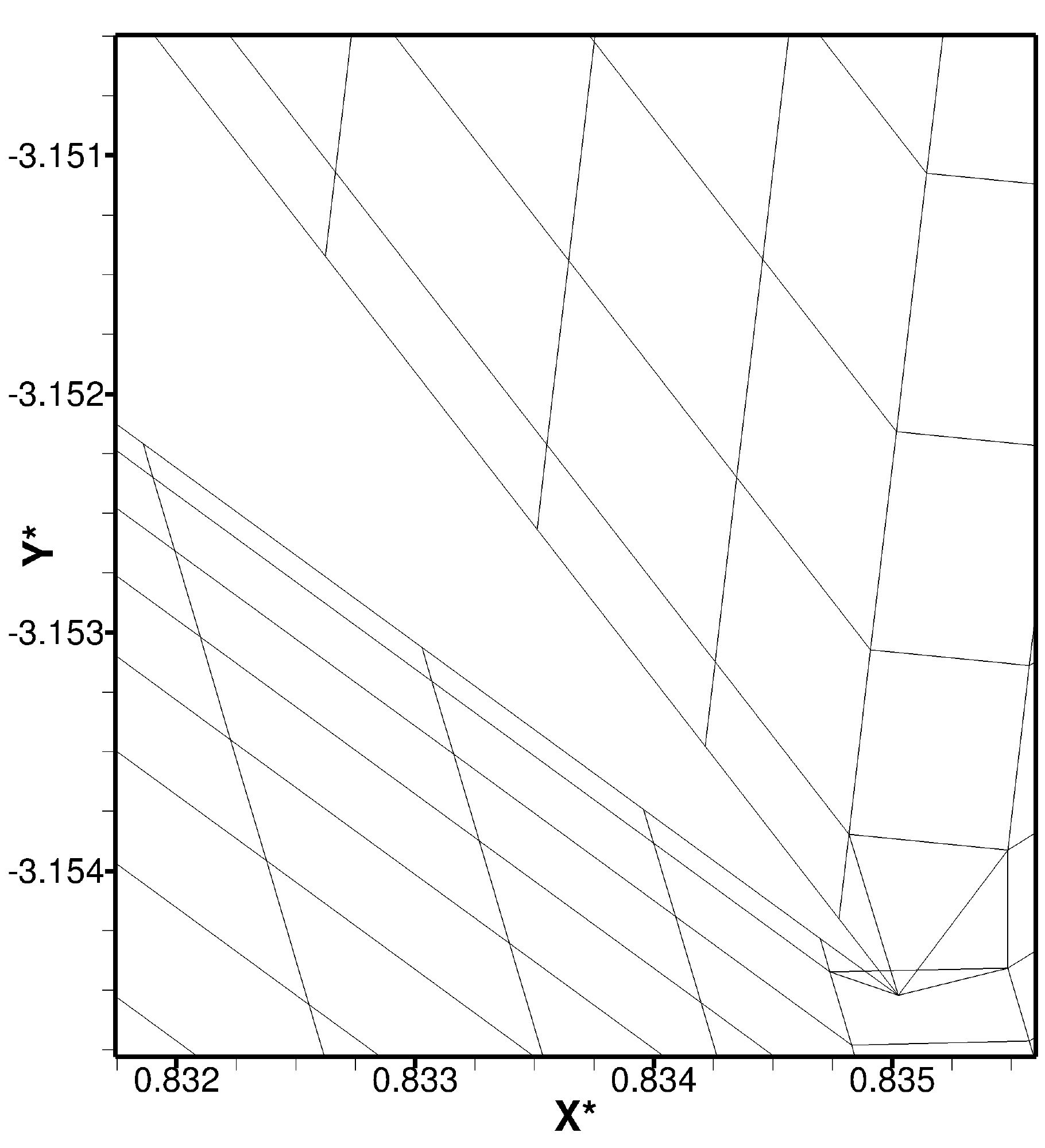}\label{fig:mesh_zoom_flutter_tfi}}\hfill
	\centering
	\subfigure[Mesh skew quality metric: Focus on the airfoil.]{\includegraphics[width=0.48\textwidth]{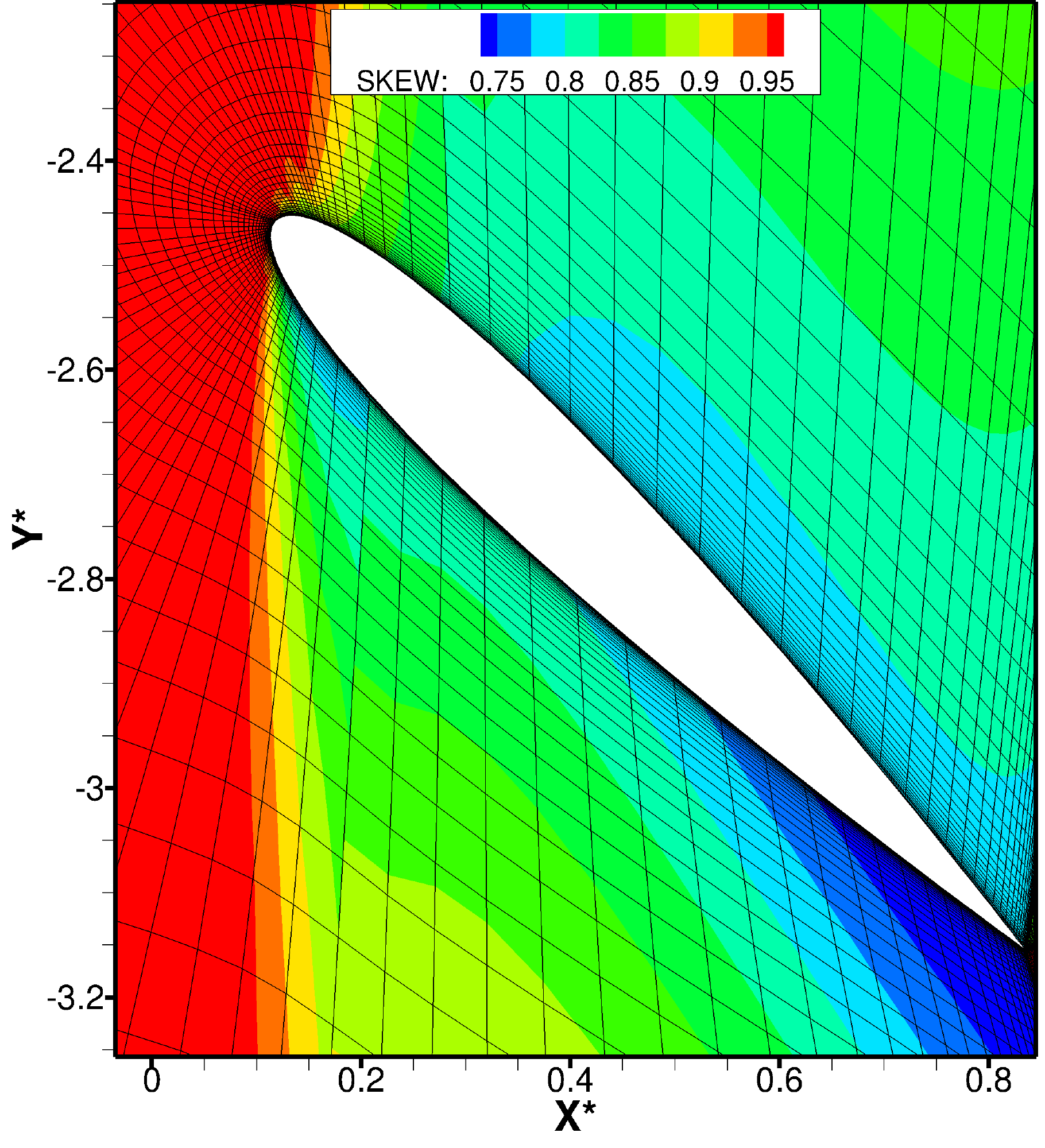}\label{fig:mesh_skew_flutter_tfi}}\hfill
	\caption{Transfinite interpolation (TFI) adapted mesh (connection of cell centers) and grid quality at \mbox{$n=160{,}000$}. Case: \mbox{$Re=30{,}000$}, \mbox{$k_{l\,2,\,eq}={3}\,\text{N}{\cdot}\text{m}^{-1}$} and \mbox{$k_{t\,3,\,eq}=0.005\,\text{N}{\cdot}\text{m}{\cdot}\text{rad}^{-1}$}.}
	\label{fig:mesh_flutter_tfi}
\end{figure}

 The quality of the mesh at the last computed time step before the solution divergence ($n=160{,}000$) is investigated in order to establish the cause of this divergence. \mbox{Figure \ref{fig:mesh_flutter_tfi}} illustrates the adapted mesh. Figure \ref{fig:mesh_overall_flutter_tfi} displays one out of two mesh lines in the $\xi$ and $\eta$ axes for the front domain, and one out of two mesh lines in the $\xi$-axis for the wake domain.
\par The TFI mesh adaption yields the formation of cells that are not orthogonal to the boundaries, specially near the airfoil trailing-edge, which leads to a loss in accuracy. The further away from the airfoil, the less the cell orthogonality is affected. Although the skew quality metric (see \mbox{Sen et al.\ \cite{Sen_2017})} indicates a reduction of the mesh quality, this is not the cause of the divergence. For instance, the skewness varies between \mbox{$0.993\leq f_{skew}\leq 1$} for the undeformed mesh, i$.$e$.$, \mbox{$m{-}L_3^{min}{-}y^+_{max}$}, and between \mbox{$0.717\leq f_{skew}\leq 1$} for the deformed mesh at \mbox{$n=160{,}000$}, regarding that $f_{skew}=1$ indicates a perfect orthogonality of \mbox{the cells.} 
\par Degenerated cells with cross-overs are formed at the trailing-edge of the airfoil as well as first cells with different heights (see Fig.\ \ref{fig:mesh_zoom_flutter_tfi}). The former leads to divergence due to the generation of unphysical negative volumes, while the latter leads to numerical difficulties due to the applied time step size. This is fixed at $\Delta t=1{\cdot}10^{-5}\,\text{s}$, while the first cell height is not maintained at $\Delta y^{first\,cell}=5{\cdot}10^{-5}\,\text{m}$. Indeed, this height decreases up to about $\Delta y^{first\,cell}=1.8{\cdot}10^{-5}\,\text{m}$, which would require much smaller time step sizes in order to obtain a converged solution.
\par Due to the presence of degenerated cells, the alteration of the first cell height and the reduction of the skew quality metric, a simulation applying the hybrid IDW-TFI mesh adaption algorithm is started. However, since this requires high computational times \mbox{(see Section \ref{appendix_mesh_adaption})} not enough results are currently available and therefore this simulation cannot be analyzed.

\subsubsection{Investigation of the displacements}
\label{subsubsec:flutter_displacements}

\par
\begin{figure}[H]
	\centering
	\subfigure[Translational displacements in the $x_2$-direction.]{\includegraphics[width=0.48\textwidth]{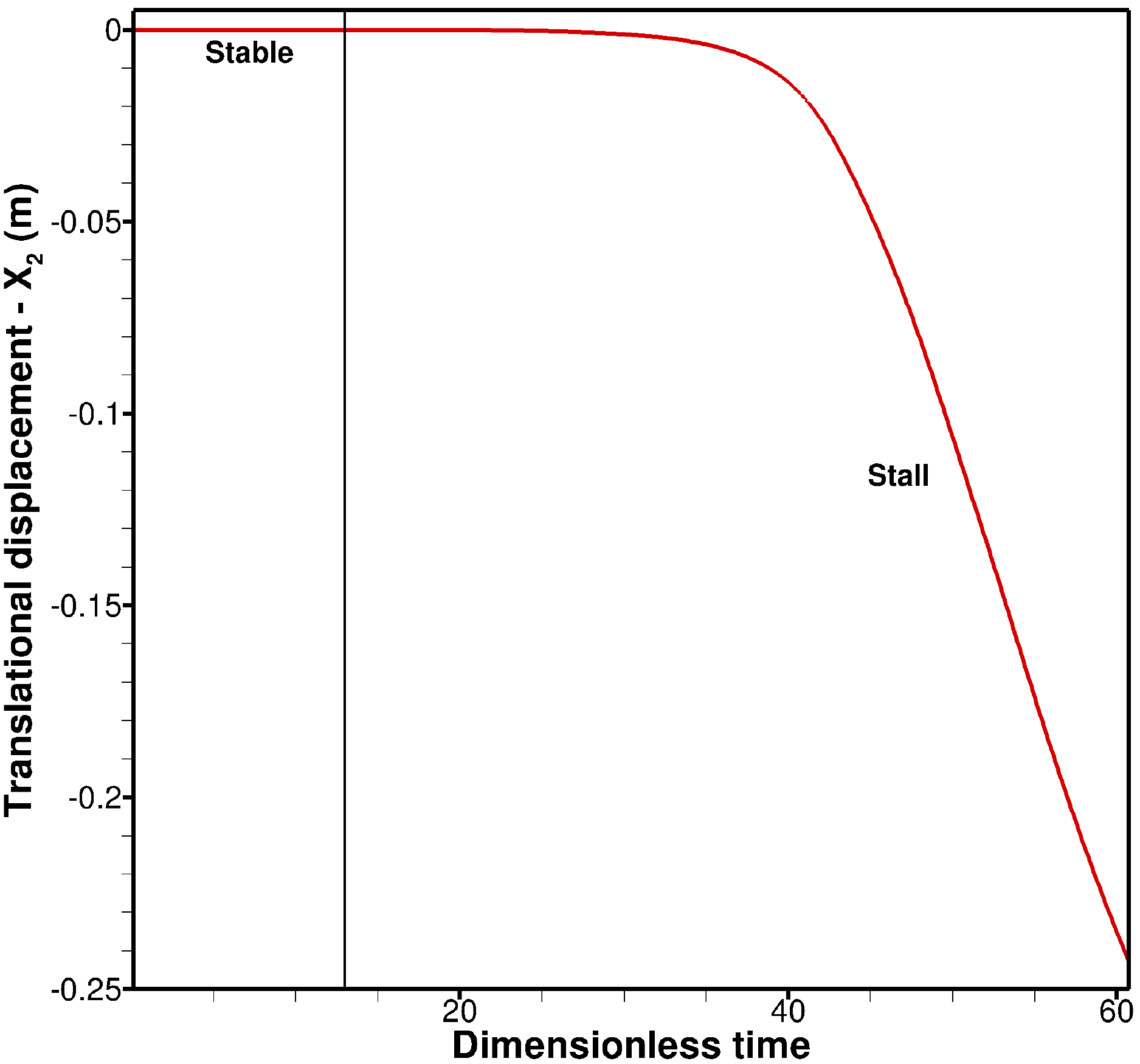} \label{fig:trans_disp_flutter_tfi}}\hfill
	\subfigure[Rotational displacements about the $x_3$-axis.]{\includegraphics[width=0.48\textwidth]{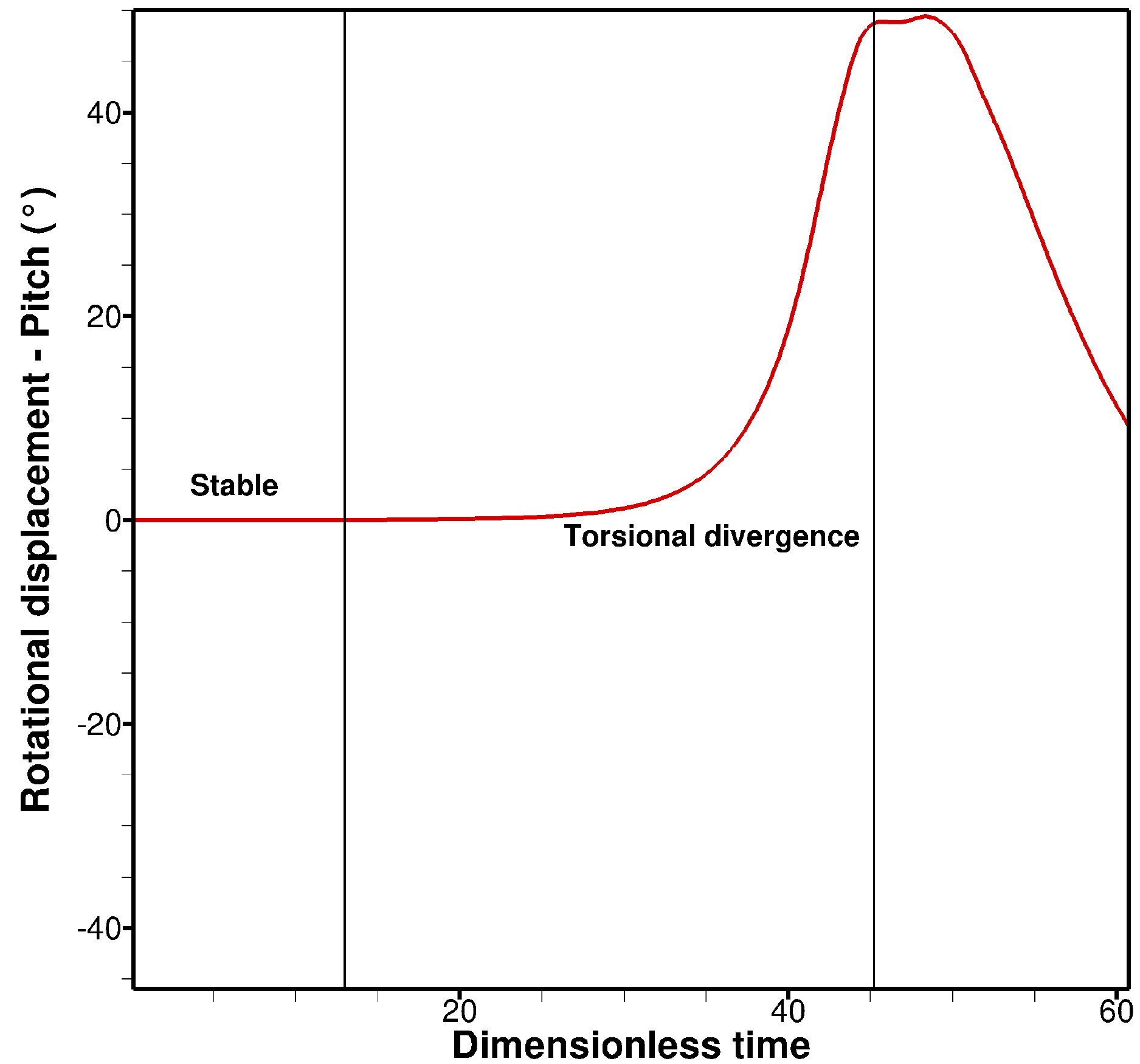}\label{fig:rot_disp_flutter_tfi}}\hfill
	\caption{Time history of the displacements. Test case: \mbox{$Re=30{,}000$}, \mbox{$k_{l\,2,\,eq}={3}\,\text{N}{\cdot}\text{m}^{-1}$} and \mbox{$k_{t\,3,\,eq}=0.005\,\text{N}{\cdot}\text{m}{\cdot}\text{rad}^{-1}$}.}
	\label{fig:disp_flutter_tfi}
\end{figure}

 Although the computation with the TFI algorithm is not accurate, a rough analysis of the displacements is carried out. Figure \ref{fig:disp_flutter_tfi} illustrates the time history of the displacements only up to $t^*=60.74$ due to the fact that the mesh quality is so degenerated that the calculations might not represent the reality. Moreover, in a real case, as soon as the airfoil suffers a major twist it would fail. The maximal translational and rotational displacements achieved are $|X_2|=0.2875\,\text{m}$ and \mbox{$\varphi_3=49.42^\circ$}, respectively. 
\par
\begin{figure}[H]
	\centering
	\includegraphics[width=0.58\textwidth]{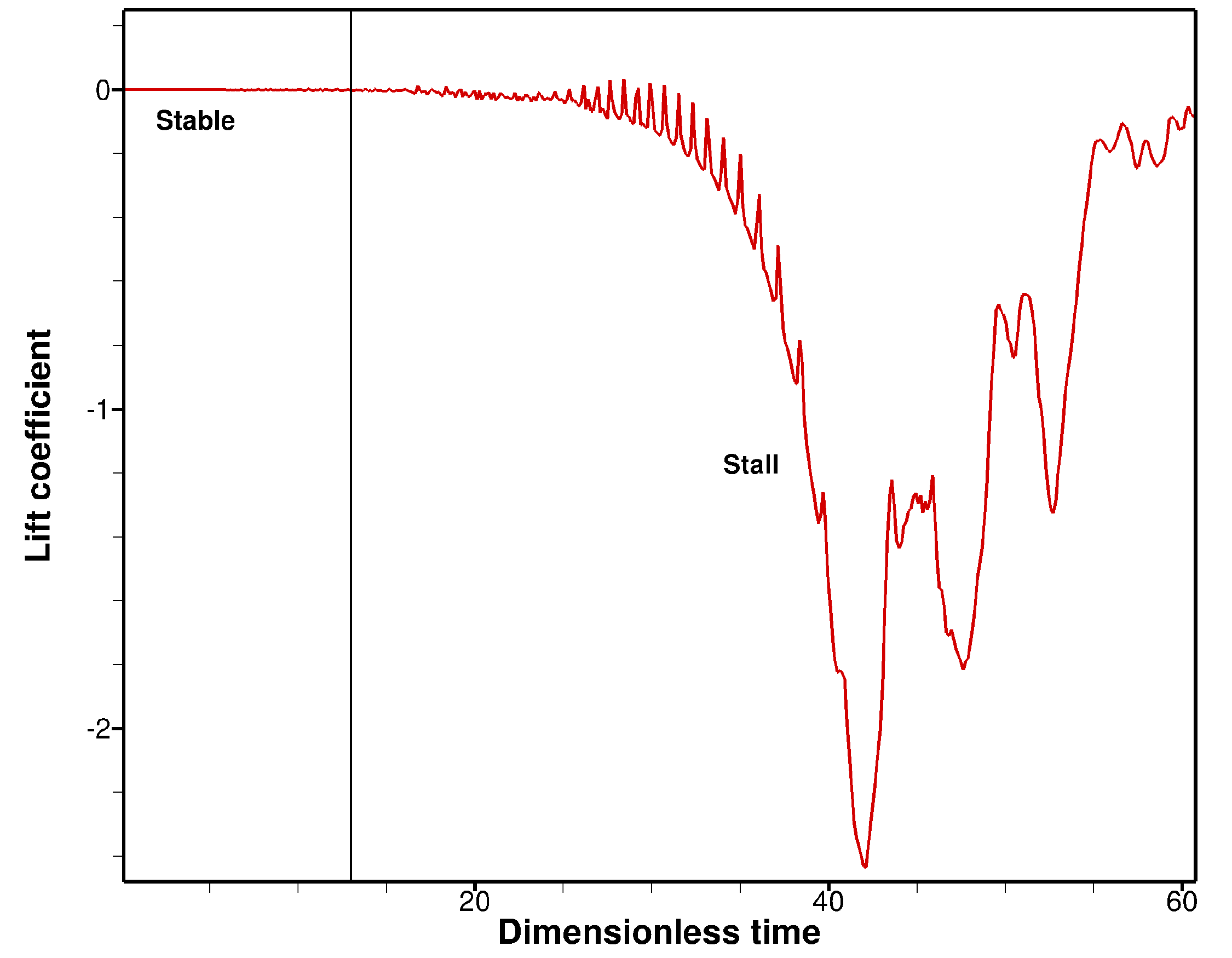} 
	\caption{Time history of the lift coefficient: Torsional divergence.}
	\label{fig:cl_torsional_divergence}
\end{figure}

 The airfoil is stable until $t^*\approx13$. Afterwards, the vortex shedding on the trailing edge interacts with the airfoil, generating fluid forces that act on the NACA0012. When the aerodynamic moment generated by these forces cannot be counter-acted by the structural stiffness anymore, the NACA0012 suffers a major and rapid twist, which leads to a stall due to the disturbance of the flow around the NACA0012, as illustrated in Fig.\ \ref{fig:cl_torsional_divergence}).
\par The airfoil twist generates a complex vortex system, which might be responsible (together with the system structural stiffness) for the following increase of the lift forces. However, this computation is not reliable due to the poor mesh quality and therefore this lift increase might also not be physical.

\section{Towards flutter}\markboth{CHAPTER 4.$\quad$RES. AND DISC.}{4.3$\quad$TOWARDS FLUTTER}
\label{sec:towards_flutter}

\par In order to analyze the flutter phenomenon, a system configuration that causes these structural responses must be established. Since this occurs when the vortex shedding frequency stays between the natural frequencies of the degrees of freedom (see \mbox{Fung \cite{Fung_2002})}, the frequencies generated by the flow of various coupled systems with up and down and pitch degrees of freedom and different system stiffnesses are primarily investigated. Since the analyzed Reynolds number is fixed at $Re=30{,}000$, minimal variations of the vortex shedding frequencies are expected. Afterwards, the natural frequencies of the up and down and pitch degrees of freedom are adjusted through the alteration of the linear and torsional stiffnesses, aiming at the generation of a natural frequency interval which contains the previously investigated flow frequencies. 

\subsection{Investigation of the frequency domain}
\label{sec:flow_frequencies_re_meshing}  

\par An analysis of the vortex shedding frequencies is carried for the eleven system configurations characterized by the generation of a limit-cycle oscillation, according to \mbox{Section 
\ref{sec:LCO}.} While the linear stiffness is varied (see Table \ref{table:investigated_stiffnesses}), the torsional stiffness is maintained constant at \mbox{$k_{t\,3,\,eq}=0.3832\,\text{N}{\cdot}\text{m}{\cdot}\text{rad}^{-1}$}, since preliminary experimental investigations observed that the other torsional springs available for the experiments are too soft and therefore could lead to torsional divergence. 
\begin{figure}[H]
	\centering
	\subfigure[$k_{l\,2,\,eq}={\left[104,124,144\right]}\,\text{N}{\cdot}\text{m}^{-1}$.]{\includegraphics[width=0.48\textwidth]{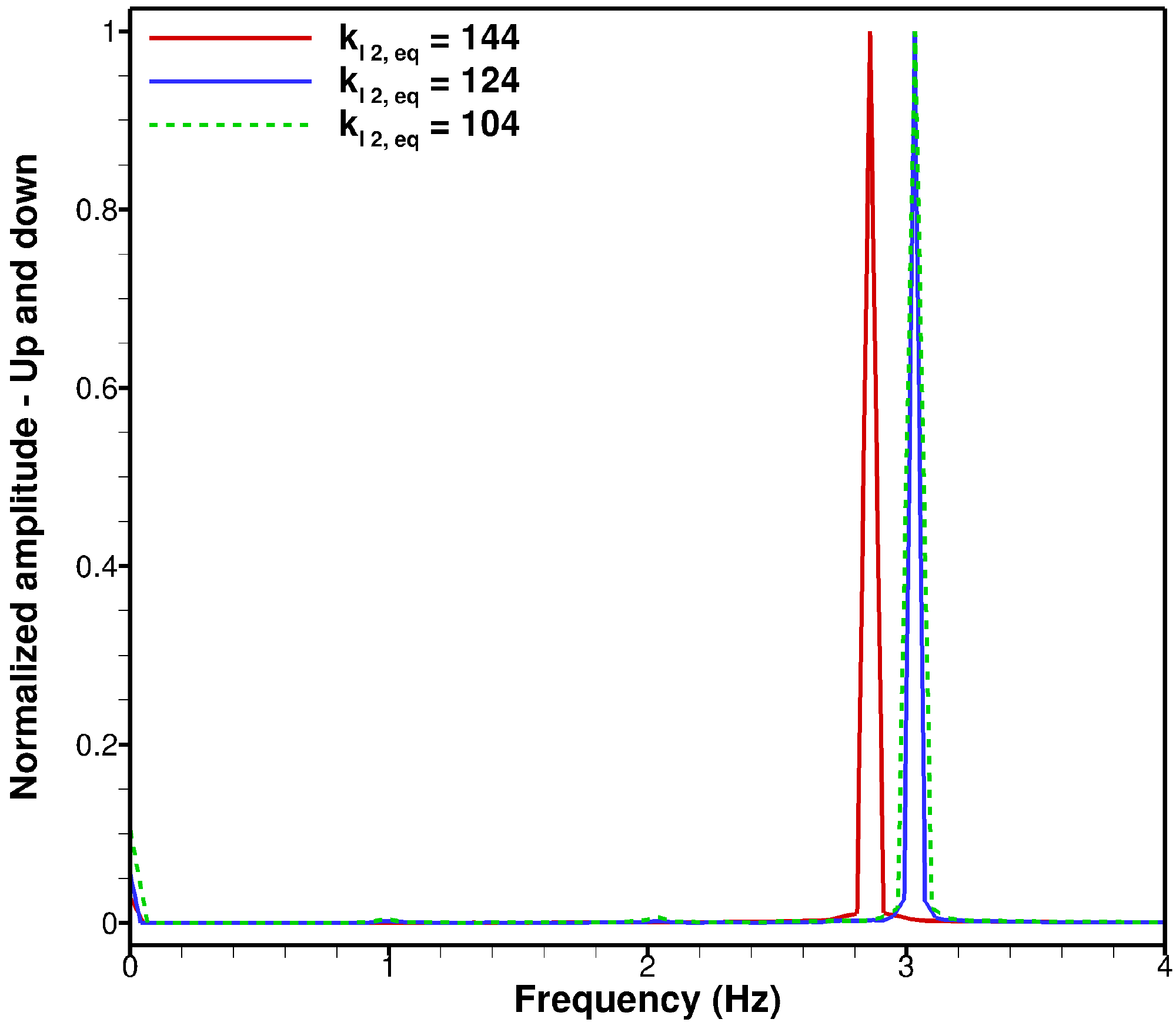} \label{fig:frequencies_high_high_kl}}\hfill
	\subfigure[$k_{l\,2,\,eq}={\left[50,60,92\right]}\,\text{N}{\cdot}\text{m}^{-1}$.]{\includegraphics[width=0.48\textwidth]{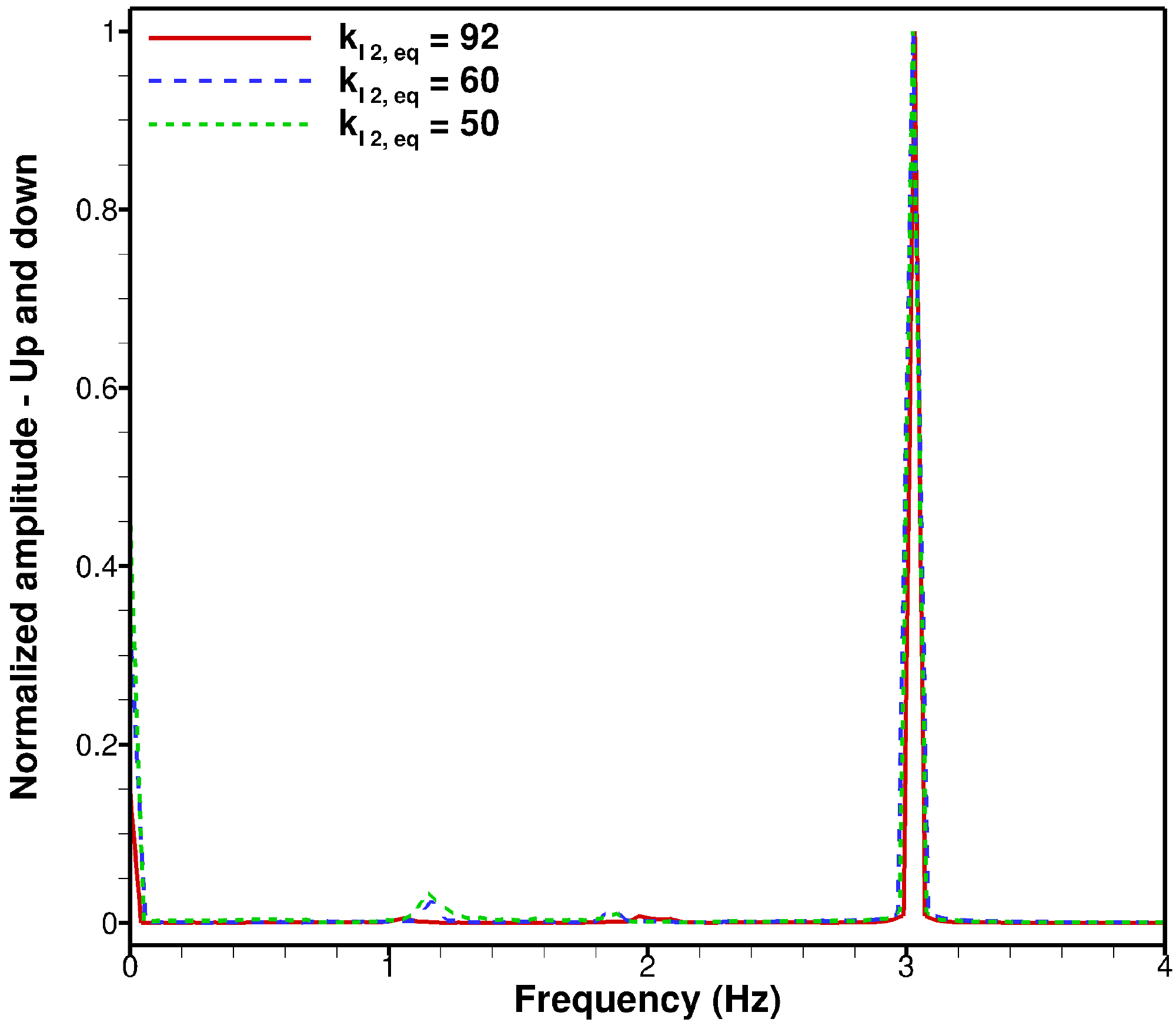}\label{fig:oifrsequencies_high_kl}}\hfill
	\caption{Frequency domain of the translational displacement for various system configurations: constant torsional stiffness of \mbox{$k_{t\,3,\,eq}=0.3832\,\text{N}{\cdot}\text{m}{\cdot}\text{rad}^{-1}$} and varying linear stiffnesses.}
	\label{fig:frequencies_all_kl}
\end{figure}
\par The frequencies generated by the translational displacement of the airfoil are acquired through the application of a Fourier transform with a Hamming window. This utilizes the time $t$ as the independent variable and the translational displacement $X_2$ as the dependent variable. The achieved frequencies are then illustrated in Fig.\ \ref{fig:frequencies_all_kl} for the test cases with \mbox{$k_{l\,2,\,eq}=50\,\text{N}{\cdot}\text{m}^{-1}$}, \mbox{$k_{l\,2,\,eq}=60\,\text{N}{\cdot}\text{m}^{-1}$}, \mbox{$k_{l\,2,\,eq}=92\,\text{N}{\cdot}\text{m}^{-1}$}, \mbox{$k_{l\,2,\,eq}=104\,\text{N}{\cdot}\text{m}^{-1}$}, \mbox{$k_{l\,2,\,eq}=124\,\text{N}{\cdot}\text{m}^{-1}$} and \mbox{$k_{l\,2,\,eq}=144\,\text{N}{\cdot}\text{m}^{-1}$}. The amplitudes are normalized in relation to the maximal amplitude achieved by each case, regarding that these values vary due to the fact that each simulation requires a different initialization time in order to achieve the fully developed state. The frequency domain of the simulations characterized by the linear stiffnesses \mbox{$k_{l\,2,\,eq}=40\,\text{N}{\cdot}\text{m}^{-1}$}, \mbox{$k_{l\,2,\,eq}=70\,\text{N}{\cdot}\text{m}^{-1}$}, \mbox{$k_{l\,2,\,eq}=80\,\text{N}{\cdot}\text{m}^{-1}$}, \mbox{$k_{l\,2,\,eq}=114\,\text{N}{\cdot}\text{m}^{-1}$} and \mbox{$k_{l\,2,\,eq}=134\,\text{N}{\cdot}\text{m}^{-1}$} are not illustrated in the present work, since these require longer initialization times and therefore only unclear frequency peaks are available.

\par The frequency spectrum of all test cases are characterized by the formation of a distinct peak, which is similar for all simulations due to the fact that the Reynolds number is fixed at $Re=30{,}000$. These peaks indicate the vortex shedding frequency, which varies within the interval of \mbox{$2.86\,\text{Hz}\leq f_{X_{2}}\leq 3.03\,\text{Hz}$}. The Strouhal number in relation to the airfoil chord  $c=0.1\,\text{m}$ vary within the interval of $0.064\leq St\leq0.068$. 

\subsection{Definition of the system configuration}
\label{sec:case_great_airfoil_disp}

\par The linear and torsional stiffnesses of the system are adapted in order to ensure that the vortex shedding frequency range calculated in Section \ref{sec:flow_frequencies_re_meshing} is contained within the interval given by the natural frequencies of the pitch and up and down degrees of freedom.
\par Two new system configurations, i$.$e$.$, $3a$ and $3b$, are established, as stated in \mbox{Table \ref{table:new_configurations}.} For both configurations, the vortex shedding frequencies evaluated in Section \ref{sec:flow_frequencies_re_meshing} are contained inside the interval created by the natural frequencies, as summarized in \mbox{Table \ref{table:frequencies_new_configurations}.}
\begin{table}[H]
	\centering
	\begin{tabular}{p{3.4cm} p{4.5cm} p{5cm}}
		\hline
		\multicolumn{1}{c}{\multirow{2}{*}{\centering{\bf{Configuration}}}} & \centering{\bf{Linear stiffness 
		$k_{l\,2,\,eq}$ ($\textbf{N}{\cdot}\text{m}^{-1}$)}} & \centering{\bf{Torsional stiffness 
		$k_{t\,3,\,eq}$ ($\textbf{N}{\cdot}\text{m}{\cdot}\text{rad}^{-1}$)}} \tabularnewline
		\hline
		\centering{3a} & \centering{135} & \centering{0.075} \tabularnewline
		\centering{3b} & \centering{160} & \centering{0.060} \tabularnewline
		\hline
	\end{tabular}
	\caption{\label{table:new_configurations}Towards flutter: New system configurations.}
\end{table}
\begin{table}[H]
	\centering
	\begin{tabular}{p{3.4cm} p{1.2cm} p{4.2cm} p{4.2cm}}
		\hline
		\multirow{2}{3.4cm}{\centering{\bf{Vortex shedding frequencies (Hz)}}} &  \multicolumn{1}{c}{\multirow{2}{*}{\centering{\bf{Config.}}}} &\multicolumn{2}{c}{\bf{Natural frequencies (Hz)}} \tabularnewline \cline{3-4}
		& & \centering{\bf{$x_2$ DOF}} & \centering{\bf{$\varphi_3$ DOF}} \tabularnewline\hline
		\multicolumn{1}{c}{\multirow{4}{*}{\centering{$2.86\leq f_v\leq 3.03$}}} & \multicolumn{1}{c}{\multirow{2}{*}{\centering{3a}}} & \multicolumn{1}{c}{\multirow{2}{*}{\centering{$f_{n,\,x_{2}}=\frac{1}{2\pi}\sqrt{\frac{k_{l\,2,\,eq}}{m_{tot}}}=2.81$}}} & \multicolumn{1}{c}{\multirow{2}{*}{\centering{$f_{n,\,\varphi_{3}}=\frac{1}{2\pi}\sqrt{\frac{k_{l\,2,\,eq}}{m_{tot}}}=3.10$}}} \tabularnewline
		& &  \tabularnewline
		& \multicolumn{1}{c}{\multirow{2}{*}{\centering{3b}}} & \multicolumn{1}{c}{\multirow{2}{*}{\centering{$f_{n,\,x_{2}}=\frac{1}{2\pi}\sqrt{\frac{k_{l\,2,\,eq}}{m_{tot}}}=3.06$}}} & \multicolumn{1}{c}{\multirow{2}{*}{\centering{$f_{n,\,\varphi_{3}}=\frac{1}{2\pi}\sqrt{\frac{k_{l\,2,\,eq}}{m_{tot}}}=2.78$}}} \tabularnewline
		& &  \tabularnewline
		\hline	
	\end{tabular}
	\caption{\label{table:frequencies_new_configurations}Towards flutter: Vortex shedding and natural frequencies.}
\end{table}
\par These new FSI configurations are started for a Reynolds number of $Re=30{,}000$ with a time step size of $\Delta t=1{\cdot}10^{-5}\,\text{s}$, regarding that a loose coupling algorithm together with the prediction of the displacements is applied. The utilization of the hybrid IDW-TFI mesh adapation method is mandatory for these test cases since large airfoil displacements are expected. Due to the high CPU-time requirements, not enough results are currently available. Therefore, an analysis of these computations is not possible in the present work.
\chapter*{Conclusions and outlook\markboth{CONCLUSIONS AND OUTLOOK}{}}
\label{general_conclusion}

\addcontentsline{toc}{chapter}{Conclusions and outlook}

\par The fluid-structure interaction of a moving rigid NACA0012 at a Reynolds number of $Re=30{,}000$ is computed in order to enable a thorough analysis of the aeroelastic properties of this airfoil when submitted to different system configurations. These computations are carried out based on the experimental setup of the rigid NACA0012 model available at the Laboratory of Fluid Mechanics located at the Helmut-Schmidt-University.

\par A partitioned approach based on two separate solvers and a FSI coupling scheme is applied in the current work. The in-house CFD solver FASTEST-3D (see Breuer et al.\ \cite{Breuer_2012}) computes the solution of the fluid domain according to the wall-resolved LES technique combined with the Smagorinsky \mbox{model \cite{Smagorinsky_1963}}. The solution of the structural sub-problem is achieved by the rigid movement solver implemented by Viets \cite{Viets_2013}, which relies on the equations of motion for rigid bodies. The FSI coupling is responsible for the exchange of information between both solvers.

\par Firstly, the experimental setup present at the Laboratory of Fluid Mechanics is thoroughly analyzed. The computational setup is then established in order to provide a direct comparison of the experimental and computational results. The CFD test section geometry and mesh are based on the work of Almutari \cite{Almutari_2010} and Schmidt \cite{Schmidt_2016}, considering that this geometry has already been successfully tested in the work of Streher \cite{Streher_2017}. The center of rotation of the airfoil is set to the center of mass in order to achieve an uncoupled system in relation to the governing equations of the CSD solver. The structural properties of the NACA0012 are then calculated, i$.$e$.$, total mass, location of the mass center and the mass moment of inertia in relation to this center. No damping is utilized in the current work while various linear and torsional stiffnesses, which are mainly based on the springs available in the experiments, are applied.  

\par Afterwards, the fluid setup, the CSD solver and the FSI coupling are studied and validated. This aims at the establishment of the computational configuration that represents the best compromise between accuracy and CPU-time requirements. 

\par The three generated meshes, i$.$e$.$, $m{-}L_3^{max}{-}y_{min}^+$, $m{-}L_3^{min}{-}y_{min}^+$ and $m{-}L_3^{min}{-}y_{max}^+$ are investigated according to the span-wise length, the mapping strategy and the first cell height. The $m{-}L_3^{min}{-}y_{max}^+$ mesh, which has a span-wise length of $L_3=0.25\,c$ and a first cell height of $\Delta y^{first\,cell}=5{\cdot}10^{-5}\,\text{m}$, together with the mapping strategy 4 is then selected and further applied in the test cases.

\par The TFI, IDW and IDW-TFI mesh adaption algorithms are investigated for large translational displacements. Although the TFI algorithm is about $482$ and $152$ times faster than the IDW and IDW-TFI schemes, degenerated cells are formed in the airfoil trailing-edge as well as the height of the cells located on the airfoil surface are not maintained. These problems are overcome by the utilization of the IDW or the hybrid IDW-TFI methods. Since the latter is less computational intensive, this  is selected to simulate cases characterized by large airfoil displacements, while the TFI algorithm is selected to simulate cases characterized by small displacements.

\par The in-house CSD solver is validated for cases characterized by time-dependent external forces and coupled translational and rotational motions. This validation is required since \mbox{Viets \cite{Viets_2013}} tested pure CSD problems only for the case of time-independent external forces, while the current work is characterized by time-dependent forces due the coupled nature of the problem. 

\par The FSI coupling is studied according to the estimation of displacements, added mass effect and coupling schemes. The implemented algorithm for the prediction of displacements is tested and leads to a reduction of $85\%$ of the computational time required by the coupled simulation. Although the added mass effect is proved to be negligible, the strong coupling algorithm is tested in order to assure that the loose coupling algorithm can be used in the test cases.

\par Finally, many test cases are computed based mainly on the physically available springs. The results of these simulations are divided in three parts: The former is characterized by the presence of small limit-cycle oscillations. This is followed by the test case that reproduces the torsional divergence aeroelastic instability. The latter aims at the prediction of the flutter phenomenon.

\par The first test case is characterized by diverse configurations of the physically available linear and torsional stiffnesses. Since the eleven different simulations present similar results, i$.$e$.$, limit-cycle oscillations characterized by an extremely small amplitude, only the case characterized by the linear stiffness of \mbox{$k_{l\,2,\,eq}=92\,\text{N}{\cdot}\text{m}^{-1}$} and torsional stiffness of \mbox{$k_{t\,3,\,eq}=0.3832\,\text{N}{\cdot}\text{m}{\cdot}\text{rad}^{-1}$} is thoroughly analyzed. This requires the lowest CPU-time in order to achieve the fully developed state. The time history of the aerodynamic coefficients shows a limit-cycle oscillation characterized by weak oscillations around a mean value. The achieved displacements are extremely small. This behavior is expected since the work of Lapointe and Dumas \cite{Lapointe_2012} and Poirel and Mendes \cite{Poirel_2012} pointed out that fluid-structure interaction of the NACA0012 airfoil at the studied Reynolds number $Re=30{,}000$ is characterized by small LCO either when no angular and plunging velocities are initially available or when the linear stiffness is small (roughly \mbox{$k_{l\,2,\,eq}\leq300\,\text{N}{\cdot}\text{m}^{-1}$}). The frequency domain of the lift and drag forces as well as of the rotational displacement show the presence of a high frequency, which might be related to the structure itself. Moreover, the formation of a von K\'arm\'an vortex street is evidenced, since the frequency of the drag is twice the lift frequency. Thus, the vortex shedding frequency is equal to the lift frequency, i$.$e$.$, $f_v=3.03\,\text{Hz}$ and the flow is characterized by a Strouhal number of $St=0.068$ in relation to the airfoil chord, i$.$e$.$, $c=0.1\,\text{m}$. This achieved Strouhal number is in agreement with the work of \mbox{Poirel et al.\ \cite{Poirel_2008}.} The instantaneous flow field clearly points out the formation of a von K\'arm\'an vortex street due to the boundary layer detachment. The time-averaged flow field is analyzed due to the extremely small amplitudes of oscillation and indicates a laminar flow field until the detachment point, i$.$e$.$, \mbox{$x^*\approx0.8$}.

\par The second test case aims at the analysis of the torsional divergence aeroelastic instability. Therefore, a new system configuration characterized by low stiffnesses, i$.$e$.$, \mbox{$k_{l\,2,\,eq}=3\,\text{N}{\cdot}\text{m}^{-1}$} and \mbox{$k_{t\,3,\,eq}=0.005\,\text{N}{\cdot}\text{m}{\cdot}\text{rad}^{-1}$}, is tested with the TFI mesh adaption algorithm in order to get results in a reasonable computational time. However, due to fact that the first cell height is not maintained, this simulation diverges at a dimensionless time of $t^*=71$. Nevertheless, the achieved results are studied according to the translational and rotational displacements. Since the results of this simulation indicates the presence of torsional divergence, a new simulation is started applying the hybrid IDW-TFI algorithm. Due to high CPU-time requirements, not enough results are currently available and therefore these are not studied in the current work.  

\par The third test case aims at the prediction of the flutter phenomenon. Firstly, eleven simulations that utilize the stiffnesses of the springs available in the experiments are analyzed in order to establish an interval of the frequencies, at which the vortex shedding occurs. Then, based on the work of Fung \cite{Fung_2002}, linear and torsional stiffnesses are carefully chosen in order to provide an interval of the natural frequencies, in which the vortex shedding frequencies are contained. Two new system configurations are proposed and simulated according to the hybrid IDW-TFI mesh adaption algorithm. However, since not enough computational results are available, this test case is not investigated in the present work.

\par The current work does not compare the achieved computational results with experimental ones, since the complete experimental setup is not currently available. Nevertheless, the preliminary studies as well as the analysis of three coupled FSI test cases characterized by different aeroelastic instabilities serve as a base for future works. Moreover, it also provides three high-quality grids: The grid for a Reynolds number of $Re=30{,}000$, which is applied in this work, and two other grids aimed at flows at $Re=50{,}000$ and $Re=100{,}000$. For the future, FSI simulations involving large displacements can rely on the hybrid IDW-TFI algorithm developed by Sen et al.\ \cite{Sen_2017}. However, some data reduction algorithm, such as the "greedy" method (see Sen et al.\ \cite{Sen_2017}), must be implemented for IDW and IDW-TFI in order to reduce the required CPU-time. Moreover, the test cases can be expanded through the application of a damping, which should reduce the difficulties encountered by the mesh adaption algorithms. 
\appendix


\chapter{Experimental Setup}\label{appendix_experimental_setup}
\par
\begin{figure}[H]
	\centering
	\includegraphics[width=1\textwidth]{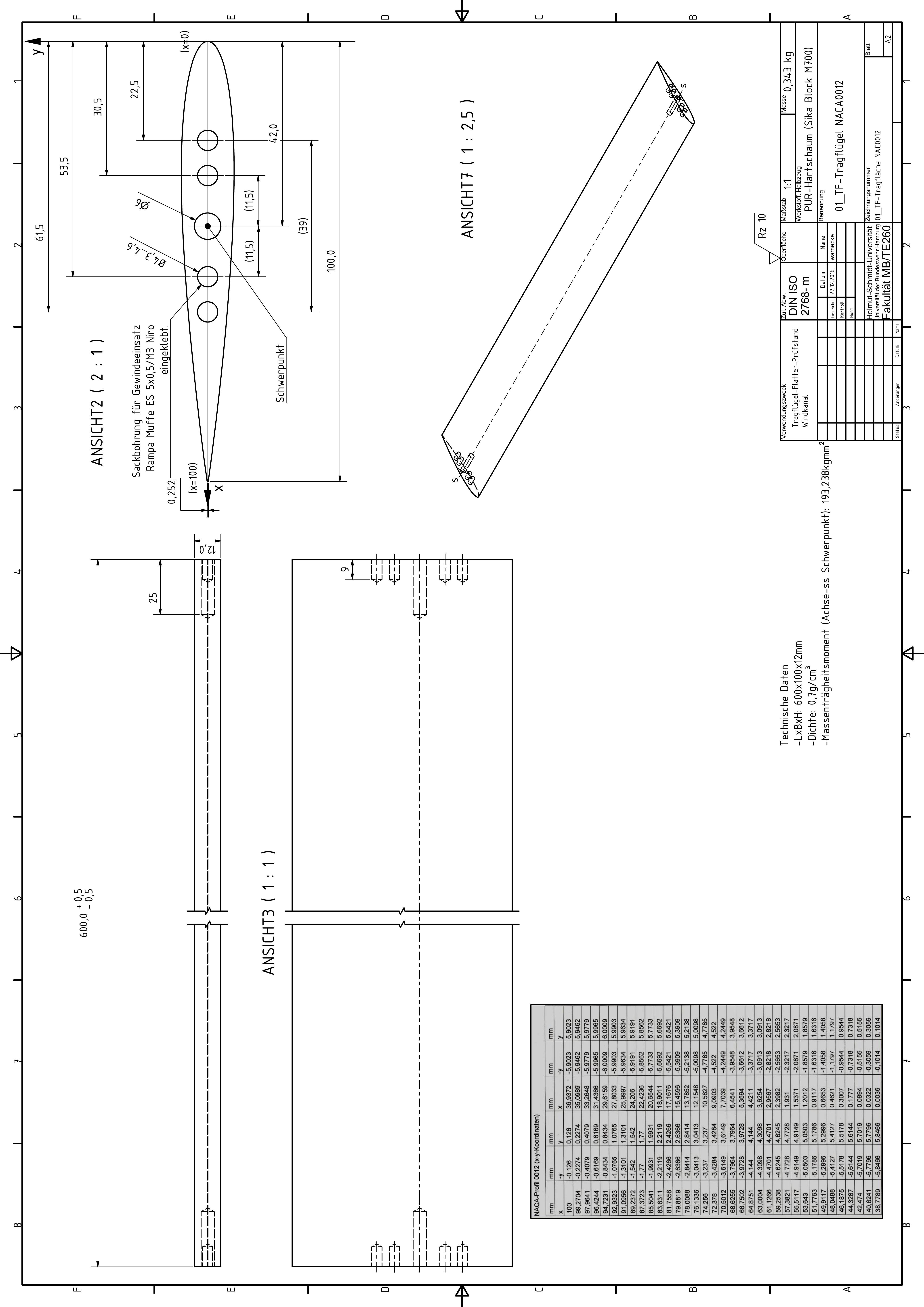}
	\caption{NACA0012 airfoil model: Technical specifications.}
	\label{fig:airfoil_model_specifications}
\end{figure}

\begin{figure}[H]
	\centering
	\includegraphics[width=1\textwidth]{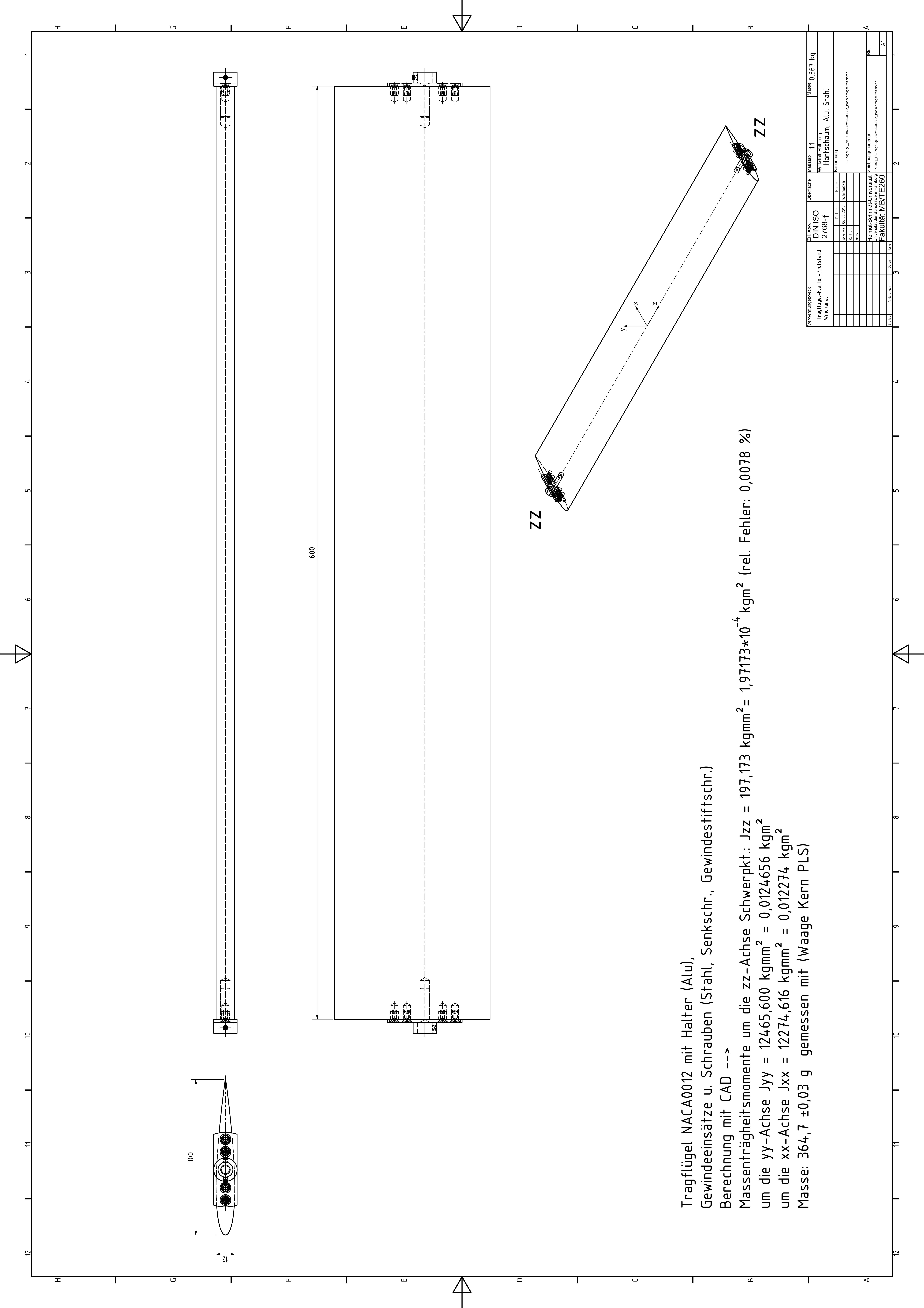}
	\caption{NACA0012 airfoil model with fixed supports: Structural properties.}
	\label{fig:airfoil_model_properties}
\end{figure}

 The airfoil in the experimental setup has a chord length of $c=0.1\,\text{m}$, and a span-wise length of \mbox{$L_{3,\,N}=0.6\,\text{m}$}. It is composed of the material Sika Block M700, which is a very fine pored polyurethane-based rigid foam special for model making and characterized by the density \mbox{$\rho_{N}=700\,\text{kg}{\cdot}\text{m}^{-3}$}. A thorough description of the airfoil and its structural properties is provided in Figs.\ \ref{fig:airfoil_model_specifications} and \ref{fig:airfoil_model_properties}. 
\par The movement of the rigid NACA0012 profile is restricted to two degrees of freedom, i$.$e$.$, pitch and up and down. Supports and guiding rods of aluminum are implemented in the front and back of the airfoil in order to guide its movement in the $x_2$-direction \mbox{(see Fig.\ \ref{fig:guiding_rod})}. The mass-spring system is composed of four linear springs ($k_{l\,2,\,1}$, $k_{l\,2,\,2}$, $k_{l\,2,\,3}$ and $k_{l\,2,\,4}$) and two torsional springs ($k_{t\,3,\,1}$ and $k_{t\,3,\,2}$) for the $x_2$ and the $\varphi_3$ degrees of freedom, respectively (see Fig.\ \ref{fig:springs_NACA0012}). Torsional and linear springs with stiffnesses of respectively $k_{t,\,3}=0.1916\,\text{N}{\cdot}\text{m}{\cdot}\text{rad}^{-1}$, as well as $k_{l,\,2}=10\,\text{N}{\cdot}\text{m}^{-1}$, $k_{l,\,2}=15\,\text{N}{\cdot}\text{m}^{-1}$, \mbox{$k_{l,\,2}=20\,\text{N}{\cdot}\text{m}^{-1}$},\break $k_{l,\,2}=26\,\text{N}{\cdot}\text{m}^{-1}$, $k_{l,\,2}=31\,\text{N}{\cdot}\text{m}^{-1}$ and $k_{l,\,2}=36\,\text{N}{\cdot}\text{m}^{-1}$ are available. The four linear springs that characterize the experimental setup can either have the same stiffness or be a combination of different springs. 
\par The center of rotation of the airfoil is located at the center of mass, assuring a torque-free rotation \mbox{(see Eq.\ (\ref{eq:fsirigidmvt_rotation}))} characterized by an uncoupled system (the $J_{tot,\,12}$, $J_{tot,\,13}$ and $J_{tot,\,23}$ variables vanish - see Section \ref{subsec:moment_inertia}). Hence, the springs are mounted at \mbox{$x_{CM,\,1}=0.0417\,\text{m}$} and $x_{CM,\,2}=0\,\text{m}$, regarding that two groups of springs are present: One at the front and other at the back of the airfoil, i$.$e$.$, at respectively $x_3=0\,\text{m}$ and $x_3=0.6\,\text{m}$ (see Fig.\ \ref{fig:springs_NACA0012}).
\par The span-wise length of the airfoil is larger than the wind tunnel width, i$.$e$.$, respectively $L_{3,\,N}=0.6\,\text{m}$ and $L_{3,\,WT}=0.5\,\text{m}$. Therefore, wingtip turbulence does not influence the measurements and the experimental results correspond to an infinite airfoil.
\par
\begin{figure}[H]
	\centering
	\includegraphics[width=1\textwidth]{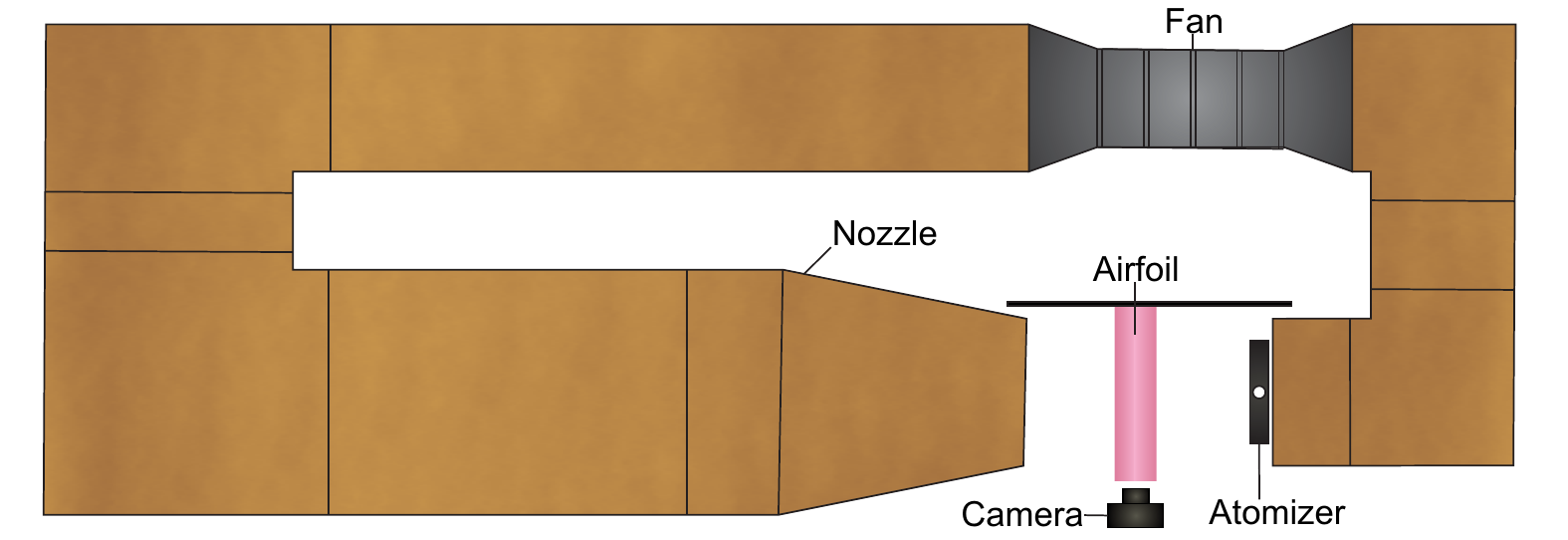}
	\caption{Simplified top view of the wind tunnel utilized for the PIV measurements of the fluid-structure interaction between flow and airfoil.}
	\label{fig:experimental_setup}
\end{figure}

 Particle-image velocimetry (PIV) is utilized in order to measure the velocity field around the moving airfoil. This method is based on the reflection of the light and is applicable for the measurement of instationary velocity fields due to its high spatial and time resolution (see \mbox{Brossard et al.\ \cite{Brossard_2009})}. A pulsed solid-state neodymium-doped yttrium aluminum garnet (Nd:YAG) laser is utilized to generate pulsed high-energy monochromatic light, which is reflected by the particles present on the flow. Since the experimented fluid (air) does not have reflective properties, artificially generated particles are introduced in the flow (at the end of the test section). These are generated by an atomizer and are composed of di-ethylhexyl sebacate (DEHS) with a diameter range of \mbox{$0.2\,\mu \text{m}\leq d_p\leq 0.3\,\mu \text{m}$}, which assures the particle capability of following the flow (see Fig.\ \ref{fig:experimental_setup}). Two laser pulses illuminate the measuring plane within a determined time interval, usually in the order of magnitude $\Delta t=\mathcal{O}(10^{-6})\,\text{s}$. The light scattered by the particles at each pulse is then recorded by a 29 Mega pixel ($6600\times4400$ pixels) digital camera. The generated pictures are afterwards divided in many elements, which contain information about a determined group of particles. The recordings are then searched by a raster in order to find matches between the particle groups present on the first and second images. When the images are matched, the displacement is calculated and consequently the two-dimensional velocity vector at the investigated plane is computed in relation to the known time interval between the two laser pulses. The PIV results are in the form of vector maps, which are usually post-processed by means of interpolation and/or smoothing techniques. Finally, the experimental results are compared with the numerical ones, in order to possibly validate the simulations.
\par The previously described PIV experiments are performed in the wind tunnel located at the Institute of Fluid Mechanics of the Helmut-Schmidt-University. This is characterized by its Göttinger design and its opened test section with a length of $780\,\text{mm}$ and a cross-section of $500\,\text{mm}\times375\,\text{mm}$. The simplified top view of the experimental setup is illustrated in \mbox{Fig.\ \ref{fig:experimental_setup}}. The laser, airfoil supports, guiding rods and springs are not illustrated for the sake of simplicity.
\par 
\par
\par 
\par
\par 
\par
\par 
\par
\par 
\par
\par 
\par
\par 
\par
\par 
\par
\par 
\par
\par 
\par
\par 
\par
\par 
\par
\par 
\par
\par 
\par
\par 
\par$\quad$
\pagebreak





\chapter{Simulations summary}\label{appendix_simulations_summary}
\par A total of 42 simulations are carried out in the current work. These are summarized in Tables \ref{table:summary_span_wise_length} to \ref{table:summary_FSI}. The variable symbols and names can be found in the nomenclature. The cases characterized by pure CFD and FSI apply a Reynolds number of $Re=30{,}000$. The $m{-}L_3^{min}{-}y_{min}^+$ and $m{-}L_3^{min}{-}y_{max}^+$ meshes require time step sizes of $\Delta t=6{\cdot}10^{-6}\,\text{s}$ and $\Delta t=1{\cdot}10^{-5}\,\text{s}$, respectively. If the units of the variables are not specified in the Tables, these are given according to the International System of \mbox{Units SI.}

\begin{table}[H]
	\centering
	\begin{tabular}{p{3.4cm} p{3cm} p{3.5cm} p{3.5cm}}
		\hline
		\multicolumn{1}{c}{\multirow{2}{*}{\centering{\bf{Mesh}}}} & \centering{\bf{Span-wise length}} & \centering{\bf{Mapping strategy}} & \centering{\bf{Time step size (s)}} \tabularnewline
		\hline
		\centering{$m{-}L_3^{min}{-}y_{min}^+$} & \centering{$0.25\,c$} & \centering{3} & \centering{$6{\cdot}10^{-6}$} \tabularnewline
		\centering{$m{-}L_3^{max}{-}y_{min}^+$} & \centering{$0.5\,c$}& \centering{3} & \centering{$6{\cdot}10^{-6}$} \tabularnewline
		\hline
	\end{tabular}
	\caption{\label{table:summary_span_wise_length}Simulations summary. Pure CFD: Span-wise length investigations.}
\end{table}

\begin{table}[H]
	\centering
	\begin{tabular}{p{3.4cm} p{3cm} p{3.5cm} p{3.5cm}}
		\hline
		\multicolumn{1}{c}{\multirow{2}{*}{\centering{\bf{Mesh}}}} & \centering{\bf{First cell height (m)}} & \centering{\bf{Mapping strategy}} & \centering{\bf{Time step size (s)}} \tabularnewline
		\hline
		\centering{$m{-}L_3^{min}{-}y_{min}^+$} & \centering{$1.8{\cdot}10^{-5}$} & \centering{3} & \centering{$6{\cdot}10^{-6}$} \tabularnewline
		\centering{$m{-}L_3^{min}{-}y_{max}^+$} & \centering{$5.0{\cdot}10^{-5}$}& \centering{3} & \centering{$1{\cdot}10^{-5}$} \tabularnewline
		\hline
	\end{tabular}
	\caption{\label{table:summary_first_cell_height}Simulations summary. Pure CFD: First cell height investigations.}
\end{table}

\begin{table}[H]
	\centering
	\begin{tabular}{p{2.7cm} p{2.5cm} p{2cm} p{0.5cm} p{0.5cm} p{0.5cm} p{0.5cm} p{0.5cm} p{0.5cm} p{0.5cm}}
		\hline
		\multicolumn{1}{c}{\multirow{2}{*}{\centering{\bf{Solver}}}} &  \centering{\bf{Time disc. method}}  & \multicolumn{1}{c}{\multirow{2}{*}{\centering{\bf{$\Delta t$}}}} &
		\multicolumn{1}{c}{\multirow{2}{*}{\centering{\bf{$m$}}}} & \multicolumn{1}{c}{\multirow{2}{*}{\centering{\bf{$D$}}}} & \multicolumn{1}{c}{\multirow{2}{*}{\centering{\bf{$k_l$}}}} & \multicolumn{1}{c}{\multirow{2}{*}{\centering{\bf{$m_{ex}$}}}} & \multicolumn{1}{c}{\multirow{2}{*}{\centering{\bf{$e_{m_{ex}}$}}}} & \multicolumn{1}{c}{\multirow{2}{*}{\centering{\bf{$\omega$}}}} &  \multicolumn{1}{c}{\multirow{2}{*}{\centering{\bf{$\omega_n$}}}} \tabularnewline
		\hline
		\centering{Rigid movement \cite{Viets_2013}} & \centering{Standard Newmark} & \multicolumn{1}{c}{\multirow{2}{*}{\centering{$10^{-3}$}}} & \multicolumn{1}{c}{\multirow{2}{*}{\centering{$10$}}} & \multicolumn{1}{c}{\multirow{2}{*}{\centering{$0$}}} & \multicolumn{1}{c}{\multirow{2}{*}{\centering{250}}} & \multicolumn{1}{c}{\multirow{2}{*}{\centering{1}}}  & \multicolumn{1}{c}{\multirow{2}{*}{\centering{0.45}}} & \multicolumn{1}{c}{\multirow{2}{*}{\centering{4.5}}} & \multicolumn{1}{c}{\multirow{2}{*}{\centering{5}}} \tabularnewline
		\centering{ode45 - Matlab Simulink} & \multicolumn{1}{c}{\multirow{3}{2.5cm}{\centering{Dormand-Prince}}} & \multicolumn{1}{c}{\multirow{3}{*}{\centering{$\left[10^{-4},10^{-3}\right]$}}} &
		\multicolumn{1}{c}{\multirow{3}{*}{\centering{$10$}}} & \multicolumn{1}{c}{\multirow{3}{*}{\centering{25}}} & \multicolumn{1}{c}{\multirow{3}{*}{\centering{250}}} & \multicolumn{1}{c}{\multirow{3}{*}{\centering{1}}}  & \multicolumn{1}{c}{\multirow{3}{*}{\centering{0.45}}} & \multicolumn{1}{c}{\multirow{3}{*}{\centering{4.5}}} & \multicolumn{1}{c}{\multirow{3}{*}{\centering{5}}} \tabularnewline
		\hline
	\end{tabular}
	\caption{\label{table:summary_mapping strategy}Simulations summary. Pure CSD: Rotating unbalance.}
\end{table}

\begin{table}[H]
	\centering
	\begin{tabular}{p{5cm} p{4cm} p{2cm} p{0.7cm} p{0.7cm} p{0.7cm}}
		\hline
		\multicolumn{1}{c}{\multirow{1}{*}{\centering{\bf{Solver}}}} &  \centering{\bf{Time disc. method}}  &\centering{\bf{$\Delta t$}} & \centering{\bf{$k_l$}} & \centering{\bf{$m$}} & \centering{\bf{$r$}} \tabularnewline
		\hline
		\centering{Rigid movement \cite{Viets_2013}} & \centering{Standard Newmark} & \centering{$10^{-3}$} & \centering{6}  & \centering{1} & \centering{0.5} \tabularnewline
		\centering{ode45 - Matlab Simulink} & \centering{Dormand-Prince} & \centering{$\left[10^{-4},10^{-3}\right]$} & \centering{6}  & \centering{1} & \centering{0.5} \tabularnewline
		\hline
	\end{tabular}
	\caption{\label{table:summary_CSD_coupled}Simulations summary. Pure CSD: Coupled translation and rotation.}
\end{table}

\begin{table}[H]
	\centering
	\begin{tabular}{p{3.4cm} p{3cm} p{3.5cm} p{3.5cm}}
		\hline
		\multicolumn{1}{c}{\multirow{3}{*}{\centering{\bf{Mesh}}}} & \centering{\bf{Mesh adaption algorithm}} & \multicolumn{1}{c}{\multirow{3}{*}{\centering{\bf{Parameters}}}} & \multicolumn{1}{c}{\multirow{3}{3.5cm}{\centering{\bf{Computational time (h)}}}} \tabularnewline
		\hline
		\centering{$m{-}L_3^{min}{-}y_{max}^+$} & \centering{TFI} & \centering{-} & \centering{0.020} \tabularnewline
		\multicolumn{1}{c}{\multirow{2}{*}{\centering{$m{-}L_3^{min}{-}y_{max}^+$}}} & \multicolumn{1}{c}{\multirow{2}{*}{\centering{IDW}}} & \centering{$\alpha_{fxd}=0.1$} & \multicolumn{1}{c}{\multirow{2}{*}{\centering{9.639}}} \tabularnewline
		& & \centering{$\alpha_{mv}=0.3$} & \tabularnewline
		\multicolumn{1}{c}{\multirow{2}{*}{\centering{$m{-}L_3^{min}{-}y_{max}^+$}}} & \multicolumn{1}{c}{\multirow{2}{*}{\centering{IDW-TFI}}} & \centering{$\alpha_{fxd}=0.05$} & \multicolumn{1}{c}{\multirow{2}{*}{\centering{3.035}}} \tabularnewline
		& & \centering{$\alpha_{mv}=0.3$} & \tabularnewline
		\multicolumn{1}{c}{\multirow{2}{*}{\centering{$m{-}L_3^{min}{-}y_{max}^+$}}} & \multicolumn{1}{c}{\multirow{2}{*}{\centering{IDW-TFI}}} & \centering{$\alpha_{fxd}=0.1$} & \multicolumn{1}{c}{\multirow{2}{*}{\centering{3.030}}} \tabularnewline
		& & \centering{$\alpha_{mv}=0.3$} & \tabularnewline
		\multicolumn{1}{c}{\multirow{2}{*}{\centering{$m{-}L_3^{min}{-}y_{max}^+$}}} & \multicolumn{1}{c}{\multirow{2}{*}{\centering{IDW-TFI}}} & \centering{$\alpha_{fxd}=0.13$} & \multicolumn{1}{c}{\multirow{2}{*}{\centering{3.031}}} \tabularnewline
		& & \centering{$\alpha_{mv}=0.3$} & \tabularnewline
		\multicolumn{1}{c}{\multirow{2}{*}{\centering{$m{-}L_3^{min}{-}y_{max}^+$}}} & \multicolumn{1}{c}{\multirow{2}{*}{\centering{IDW-TFI}}} & \centering{$\alpha_{fxd}=0.15$} & \multicolumn{1}{c}{\multirow{2}{*}{\centering{3.053}}} \tabularnewline
		& & \centering{$\alpha_{mv}=0.3$} & \tabularnewline
		\multicolumn{1}{c}{\multirow{2}{*}{\centering{$m{-}L_3^{min}{-}y_{max}^+$}}} & \multicolumn{1}{c}{\multirow{2}{*}{\centering{IDW-TFI}}} & \centering{$\alpha_{fxd}=0.2$} & \multicolumn{1}{c}{\multirow{2}{*}{\centering{3.034}}} \tabularnewline
		& & \centering{$\alpha_{mv}=0.3$} & \tabularnewline
		\multicolumn{1}{c}{\multirow{2}{*}{\centering{$m{-}L_3^{min}{-}y_{max}^+$}}} & \multicolumn{1}{c}{\multirow{2}{*}{\centering{IDW-TFI}}} & \centering{$\alpha_{fxd}=0.5$} & \multicolumn{1}{c}{\multirow{2}{*}{\centering{3.032}}} \tabularnewline
		& & \centering{$\alpha_{mv}=0.3$} & \tabularnewline
		\multicolumn{1}{c}{\multirow{2}{*}{\centering{$m{-}L_3^{min}{-}y_{max}^+$}}} & \multicolumn{1}{c}{\multirow{2}{*}{\centering{IDW-TFI}}} & \centering{$\alpha_{fxd}=0.1$} & \multicolumn{1}{c}{\multirow{2}{*}{\centering{3.033}}} \tabularnewline
		& & \centering{$\alpha_{mv}=0.1$} & \tabularnewline
		\multicolumn{1}{c}{\multirow{2}{*}{\centering{$m{-}L_3^{min}{-}y_{max}^+$}}} & \multicolumn{1}{c}{\multirow{2}{*}{\centering{IDW-TFI}}} & \centering{$\alpha_{fxd}=0.13$} & \multicolumn{1}{c}{\multirow{2}{*}{\centering{3.069}}} \tabularnewline
		& & \centering{$\alpha_{mv}=0.1$} & \tabularnewline
		\multicolumn{1}{c}{\multirow{2}{*}{\centering{$m{-}L_3^{min}{-}y_{max}^+$}}} & \multicolumn{1}{c}{\multirow{2}{*}{\centering{IDW-TFI}}} & \centering{$\alpha_{fxd}=0.13$} & \multicolumn{1}{c}{\multirow{2}{*}{\centering{3.031}}} \tabularnewline
		& & \centering{$\alpha_{mv}=0.5$} & \tabularnewline
		\hline	
	\end{tabular}
	\caption{\label{table:summary_mesh_adaption}Simulations summary. Pure CFD: Mesh adaption algorithm investigations.}
\end{table}

\begin{table}[H]
	\centering
	\begin{tabular}{p{2.5cm} p{1cm} p{2cm} p{2.5cm} p{2cm} p{1cm} p{1cm}}
		\hline
		\multicolumn{1}{c}{\multirow{2}{*}{\centering{\bf{Mesh}}}} & \centering{\bf{Mapp. strat.}} & \centering{\bf{Coup. method}} & \centering{\bf{Mesh adap. alg.}} & \centering{\bf{Est. of disp.}} & \centering{\bf{$k_{l\,2,\,eq}$}} &  \centering{\bf{$k_{t\,3,\,eq}$}} \tabularnewline
		\hline
		\centering{$m{-}L_3^{min}{-}y_{min}^+$} & \centering{$1$} & \centering{Loose} & \centering{TFI}  & \centering{Yes} & \centering{144} & \centering{0.3832} \tabularnewline
		\centering{$m{-}L_3^{min}{-}y_{min}^+$} & \centering{$2$} & \centering{Loose} & \centering{TFI}  & \centering{Yes} & \centering{144} & \centering{0.3832} \tabularnewline
		\centering{$m{-}L_3^{min}{-}y_{min}^+$} & \centering{$3$} & \centering{Loose} & \centering{TFI}  & \centering{Yes} & \centering{144} & \centering{0.3832} \tabularnewline
		\centering{$m{-}L_3^{min}{-}y_{min}^+$} & \centering{$4$} & \centering{Loose} & \centering{TFI}  & \centering{Yes} & \centering{144} & \centering{0.3832} \tabularnewline
		\end{tabular}
	\end{table}
	\begin{table}[H]
		\centering
		\begin{tabular}{p{2.5cm} p{1cm} p{2cm} p{2.5cm} p{2cm} p{1cm} p{1cm}}
		\centering{$m{-}L_3^{min}{-}y_{max}^+$} & \centering{$4$} & \centering{Strong} & \centering{TFI}  & \centering{No} & \centering{144} & \centering{0.3832} \tabularnewline
		\centering{$m{-}L_3^{min}{-}y_{max}^+$} & \centering{$4$} & \centering{Strong} & \centering{TFI}  & \centering{Yes} & \centering{144} & \centering{0.3832} \tabularnewline
		\centering{$m{-}L_3^{min}{-}y_{max}^+$} & \centering{$4$} & \centering{Mixed} & \centering{TFI}  & \centering{Yes} & \centering{144} & \centering{0.3832} \tabularnewline
		\centering{$m{-}L_3^{min}{-}y_{max}^+$} & \centering{$4$} & \centering{Loose} & \centering{TFI}  & \centering{Yes} & \centering{144} & \centering{0.3832} \tabularnewline
		\centering{$m{-}L_3^{min}{-}y_{max}^+$} & \centering{$4$} & \centering{Loose} & \centering{TFI}  & \centering{Yes} & \centering{134} & \centering{0.3832} \tabularnewline
		\centering{$m{-}L_3^{min}{-}y_{max}^+$} & \centering{$4$} & \centering{Loose} & \centering{TFI}  & \centering{Yes} & \centering{124} & \centering{0.3832} \tabularnewline
		\centering{$m{-}L_3^{min}{-}y_{max}^+$} & \centering{$4$} & \centering{Loose} & \centering{TFI}  & \centering{Yes} & \centering{114} & \centering{0.3832} \tabularnewline
		\centering{$m{-}L_3^{min}{-}y_{max}^+$} & \centering{$4$} & \centering{Loose} & \centering{TFI}  & \centering{Yes} & \centering{104} & \centering{0.3832} \tabularnewline
		\centering{$m{-}L_3^{min}{-}y_{max}^+$} & \centering{$4$} & \centering{Loose} & \centering{TFI}  & \centering{Yes} & \centering{92} & \centering{0.3832} \tabularnewline
		\centering{$m{-}L_3^{min}{-}y_{max}^+$} & \centering{$4$} & \centering{Loose} & \centering{TFI}  & \centering{Yes} & \centering{80} & \centering{0.3832} \tabularnewline
		\centering{$m{-}L_3^{min}{-}y_{max}^+$} & \centering{$4$} & \centering{Loose} & \centering{TFI}  & \centering{Yes} & \centering{70} & \centering{0.3832} \tabularnewline
		\centering{$m{-}L_3^{min}{-}y_{max}^+$} & \centering{$4$} & \centering{Loose} & \centering{TFI}  & \centering{Yes} & \centering{60} & \centering{0.3832} \tabularnewline
		\centering{$m{-}L_3^{min}{-}y_{max}^+$} & \centering{$4$} & \centering{Loose} & \centering{TFI}  & \centering{Yes} & \centering{50} & \centering{0.3832} \tabularnewline
		\centering{$m{-}L_3^{min}{-}y_{max}^+$} & \centering{$4$} & \centering{Loose} & \centering{TFI}  & \centering{Yes} & \centering{40} & \centering{0.3832} \tabularnewline
		\centering{$m{-}L_3^{min}{-}y_{max}^+$} & \centering{$4$} & \centering{Loose} & \centering{TFI}  & \centering{Yes} & \centering{$\infty$} & \centering{0.3832} \tabularnewline
		\centering{$m{-}L_3^{min}{-}y_{max}^+$} & \centering{$4$} & \centering{Loose} & \centering{TFI}  & \centering{Yes} & \centering{3} & \centering{0.0050} \tabularnewline
		\centering{$m{-}L_3^{min}{-}y_{max}^+$} & \centering{$4$} & \centering{Loose} & \centering{IDW-TFI}  & \centering{Yes} & \centering{3} & \centering{0.0050} \tabularnewline
		\centering{$m{-}L_3^{min}{-}y_{max}^+$} & \centering{$4$} & \centering{Loose} & \centering{IDW-TFI}  & \centering{Yes} & \centering{135} & \centering{0.0600} \tabularnewline
		\centering{$m{-}L_3^{min}{-}y_{max}^+$} & \centering{$4$} & \centering{Loose} & \centering{IDW-TFI}  & \centering{Yes} & \centering{160} & \centering{0.0750} \tabularnewline
		\hline
	\end{tabular}
	\caption{\label{table:summary_FSI}Simulations summary: FSI.}
\end{table}

\pagestyle{fancy}

\bibliographystyle{acm}

\bibliography{report}




\label{end} \end{document}